\documentclass[11pt]{article}
\usepackage{verbatim,amsmath,amssymb}
\usepackage{epsfig,float,color}
\usepackage[font=small,labelfont=bf]{caption}
\usepackage{geometry}
\usepackage{setspace}
\usepackage{natbib}
\usepackage{amsthm,mathrsfs,amsfonts,dsfont} 
\usepackage{bm}
\usepackage{cprotect}
\usepackage{booktabs}
\usepackage{hyperref}
\usepackage[labelfont=bf]{caption}

\definecolor{darkblue}{rgb}{0,0,1}
\definecolor{col1}{rgb}{1,0,1}		
\definecolor{col2}{rgb}{0,0.5,0}
\definecolor{col2a}{rgb}{0,0.8,0}        
\definecolor{col3}{rgb}{0.5,0,1}
\definecolor{col4}{rgb}{0.1,.75,0}

\hypersetup{pdftex=true, colorlinks=true, breaklinks=true, linkcolor=darkblue, menucolor=darkblue, pagecolor=darkblue, citecolor=darkblue, urlcolor=darkblue}

\newtheoremstyle{rem}
{6pt}
{6pt}
{\small}
{}
{\bf}
{:}
{.5em}
{}

\theoremstyle{rem}
\newtheorem{remark}{Remark}[section]

\newcommand{\bitm}{\begin{itemize}}
\newcommand{\eitm}{\end{itemize}}
\newcommand{\bnumr}{\begin{enumerate}}
\newcommand{\enumr}{\end{enumerate}}

\newcommand {\eqb}[1]{\begin{equation}\begin{array}{#1}}
\newcommand {\eqe}{\end{array}\end{equation}}

\newcommand {\esb}[1]{\begin{equation*}\begin{array}{#1}}
\newcommand {\ese}{\end{array}\end{equation*}}
\newcommand {\ds}{\displaystyle}

\newcommand {\pa}[2]{\frac{\partial{#1}}{\partial{#2}}}

\newcommand {\paqq}[3]{\frac{\partial^2{#1}}{\partial{#2}\,\partial{#3}}}
\newcommand {\back}{\! \! \! \!}
\newcommand {\is}{\back &=& \back}
\newcommand {\dis}{\back &:=& \back}

\newcommand {\ais}{\back &\approx& \back}
\newcommand {\eis}{\back &\equiv& \back}

\newcommand {\plus}{\back &+& \back}

\newcommand {\dif}{\mathrm{d}}

\newcommand {\II}{{I\kern-.3em I}}
\newcommand {\III}{{I\kern-.3em I\kern-.3em I}}

\newcommand {\mra}{\mathrm{a}}
\newcommand {\mrb}{\mathrm{b}}
\newcommand {\mrc}{\mathrm{c}}

\newcommand {\mre}{\mathrm{e}}
\newcommand {\mrf}{\mathrm{f}}

\newcommand {\mrh}{\mathrm{h}}
\newcommand {\mri}{\mathrm{i}}

\newcommand {\mrm}{\mathrm{m}}
\newcommand {\mrn}{\mathrm{n}}

\newcommand {\mrp}{\mathrm{p}}

\newcommand {\mrr}{\mathrm{r}}
\newcommand {\mrs}{\mathrm{s}}
\newcommand {\mrt}{\mathrm{t}}

\newcommand {\mrx}{\mathrm{x}}

\newcommand{\mrT}{\mathrm{T}}

\newcommand {\mcc}{\mathbf{c}}

\newcommand {\mf}{\mathbf{f}}

\newcommand {\mk}{\mathbf{k}}

\newcommand {\mm}{\mathbf{m}}

\newcommand {\mss}{\mathbf{s}}

\newcommand {\muu}{\mathbf{u}}
\newcommand {\mv}{\mathbf{v}}

\newcommand {\mx}{\mathbf{x}}

\newcommand {\ba}{\boldsymbol{a}}

\newcommand {\bg}{\boldsymbol{g}}

\newcommand {\bi}{\boldsymbol{i}}

\newcommand {\bn}{\boldsymbol{n}}

\newcommand {\bt}{\boldsymbol{t}}
\newcommand {\bu}{\boldsymbol{u}}
\newcommand {\bv}{\boldsymbol{v}}

\newcommand {\bx}{\boldsymbol{x}}

\newcommand {\bphi}{\mbox{\boldmath$\phi$}}

\newcommand {\mC}{\mathbf{C}}

\newcommand {\mK}{\mathbf{K}}

\newcommand {\mM}{\mathbf{M}}
\newcommand {\mN}{\mathbf{N}}

\newcommand {\mT}{\mathbf{T}}

\newcommand {\bF}{\boldsymbol{F}}

\newcommand {\bI}{\boldsymbol{I}}

\newcommand {\bX}{\boldsymbol{X}}

\newcommand {\bone}{\mathbf{1}}

\newcommand {\IR}{{\rm\kern.24em
   \vrule width.02em height1.53ex depth-.05ex
   \kern-.3em R}}
\newcommand {\ic}{{\rm\kern.20em
   \vrule width.02em height1.0ex depth-.05ex
   \kern-.22em c}}
\newcommand {\ia}{{\rm\kern.20em
   \vrule width.02em height1.05ex depth-.0ex
   \kern-.25em a}}
\newcommand {\IC}{{\rm\kern.24em
   \vrule width.02em height1.4ex depth-.05ex
   \kern-.26em C}}
\newcommand {\ID}{{\rm\kern.34em
   \vrule width.02em height1.5ex depth-.05ex
   \kern-.36em D}}
\newcommand {\IS}{{\rm\kern.24em
   \vrule width.02em height1.6ex depth.05ex
   \kern-.26em S}}
\newcommand {\IT}{{\rm\kern.50em
   \vrule width.02em height1.55ex depth-.05ex
   \kern-.52em T}}

\newcommand {\IE}{{\rm\kern.24em
   \vrule width.02em height1.55ex depth-.05ex
   \kern-.33em E}}
\newcommand {\IEa}{{\rm\kern.24em
   \vrule width.02em height1.55ex depth-.05ex
   \kern-.33em E}^{1}_{ijkl}}
\newcommand {\IEb}{{\rm\kern.24em
   \vrule width.02em height1.55ex depth-.05ex
   \kern-.33em E}^{2}_{ijkl}}

\newcommand {\sB}{\mathcal{B}}

\newcommand {\sS}{\mathcal{S}}

\newcommand {\Ass}[2]{\kern 0.9ex \vrule width0.45em height0.2ex depth0ex \kern -2.1ex \bigwedge_{#1}^{#2}}
\newcommand {\ASS}[2]{\kern 1.45ex \vrule width0.5em height0.2ex depth0ex \kern -2.65ex \bigwedge_{#1}^{#2}}

\newcommand{\unde}[1]{\mathsf{#1}}	
\newcommand{\undet}{\bm{\unde{t}}}
\newcommand{\undetc}{\bm{\unde{t}}_\unde{c}}

\newcommand{\undeEc}{\bm{\unde{E}}_\unde{c}}

\newcommand{\Jc}{J_\mrc}
\newcommand{\nc}{n_\mrc}

\newcommand{\Lgi}{\mathring\bg_\mri}

\begin{document}

\begin{center}
\Large{\bf{A~coupled~finite~element~formulation~for~chemo-mechano-thermo-\\
dynamical contact and its application to bonding and debonding}}

\end{center}

\renewcommand{\thefootnote}{\fnsymbol{footnote}}

\begin{center}
\large{Roger A. Sauer$^{\mra,\mrb,\mrc,}$\footnote[1]{corresponding author, email: roger.sauer@rub.de}}
\vspace{3mm}

\small{\textit{
$^\mra$Institute for Structural Mechanics, Ruhr University Bochum, 44801 Bochum, Germany \\[1mm]
$^\mrb$Department of Structural Mechanics, Gda\'{n}sk University of Technology, 80-233 Gda\'{n}sk, Poland \\[1mm]
$^\mrc$Mechanical Engineering, Indian Institute of Technology Guwahati, Assam 781039, India}}

\vspace{3mm}

\small{Published\footnote[2]{This pdf is the personal version of an article whose journal version is available at \href{https://doi.org/10.1016/j.cma.2026.119454}{www.sciencedirect.com}} 
in \textit{Comput.~Methods Appl.~Mech.~Eng.}, \href{https://doi.org/10.1016/j.cma.2026.119454}{DOI: 10.1016/j.cma.2026.119454} \\
Submitted on 12 June 2026; Revised on 26 August 2026; Accepted on 22 September 2026}

\end{center}

\renewcommand{\thefootnote}{\arabic{footnote}}

\vspace{-4mm}

\rule{\linewidth}{.15mm}
{\bf Abstract:}
This work presents a finite element formulation for coupled chemo-mechano-thermo-dynamical large deformation contact.
The formulation is based on the contact theory of \citet{cobo} that contains six coupled (but separate) fields:
the deformation and temperature of the two contacting bodies, as well as an interfacial bonding field and interfacial temperature.
The latter is governed by the chemical and mechanical energy dissipation at the interface.
Here the focus is placed on the evolution of bonding and debonding, and how it is coupled to the mechanical and thermal contact state.
Several elementary models are proposed for this based on a quadratic contact potential.
The resulting contact formulation becomes very general and versatile, which is illustrated by several challenging examples.
They include pressure- and gap-dependent bonding, exothermic bonding reactions, thermal hardening and thermal expansion, as well as simultaneous bonding and debonding.
They are based on a monolithic finite element implementation using classical and isogeometric shape functions together with implicit time integration.
Its full linearization, required for the Newton-Raphson solution method, is also provided for the six-field problem, as well as for several of its simplifications.
If bonding sites are material points, the bonding variable can be condensed-out locally.


\vspace{-5mm}
\rule{\linewidth}{.15mm}

\section{Introduction}\label{s:intro}

The bonding and debonding of two bodies plays an important role in many areas of science and technology.
Examples are natural bonding mechanisms -- such as adhesion, interlocking and osseointegration -- 
man-made joining technologies -- such as glueing, sintering, soldering and welding -- and subsequent failure mechanisms -- such as delamination, fracture and peeling. 
Before bonding occurs, the bodies have to be brought into mechanical contact.
Bonding is then a chemical reaction that can release or consume heat.
The entire process is therefore a coupled chemo-mechano-thermodynamical one, in general.
In \citet{cobo} a corresponding contact theory was developed in the framework of large deformations that explores the general coupling in the balance laws and constitutive equations.
The following paper now presents a monolithically coupled finite element implementation of the theory, using both Lagrange and NURBS shape functions.
The formulation is very general -- accounting for pressure- and temperature-dependent bonding, exo- and endothermic bonding reactions, and separation- and temperature-dependent debonding, as well as the classical coupling inherent in bulk thermo-mechanics such as thermal expansion, thermal softening and thermal hardening.
Here, the focus is placed on thermoelastic bodies, but the proposed contact formulation can be readily applied to any other bulk materials.

When bonding is absent, the present formulation contains existing approaches for thermo-mechanical contact, chemo-mechanical debonding, and cohesive zone modeling as special cases, and so a brief account of those approaches is given in the following.

The first finite element formulations for thermo-mechanical contact have restricted themselves to frictionless contact \citep{zavarise92},
linear elasticity \citep{johansson93}
and staggered coupling based on an isothermal split \citep{wriggers94,saracibar98}.
The first monolithically coupled formulation appeared by \citet{oancea97}, and it was compared to the isothermal split.
A thermo-mechanical formulation can also be split adiabatically, which was presented by \citet{laursen99}. 
A comparison of different coupling algorithms was also conducted by \citet{ireman02}.
Subsequent works have advanced thermo-mechanical contact formulations to
shells \citep{bergman04},
mesh adaptivity \citep{rieger04},
mortar formulations \citep{hueber09},
Eulerian descriptions \citep{stromberg11},
isogeometric analysis \citep{dittmann14,temizer14},
XFEM \citep{khoei18},
dual mortar methods \citep{seitz18},
Nitsche's method \citep{seitz19},
variational multiscale formulations \citep{wan21}, 
and third medium contact \citep{wriggers26}.
Applications of thermo-mechanical contact computations have studied for example
wear \citep{stromberg99,stupkiewicz99,molinari01},
plasticity \citep{pantuso00,xing02},
homogenization \citep{temizer10,temizer16},
granular media \citep{zhao20},
and electric contact \citep{andras24}.

The present approach is also related to the computational debonding model of \citet{raous99} that is based on the theory of \citet{fremond88}.
The model contains an independent field variable to characterize the state of debonding.
It is monotonically increasing and hence describes the progressive damage of the interfacial bond.
The debonding model of \citet{fremond88} and \citet{raous99} has been applied to
bone-implant interfaces \citep{rojek01a,rojek01b},
electroelastic contact \citep{sofonea06},
rubber friction \citep{wriggers08},
steel-concrete interfaces \citep{raous09},
pile-soil interaction \citep{terfaya18},
and orthotropic adhesion \citep{hu22}.
The model has been analyzed mathematically \citep{sofonea}, and extended to soft/rigid thermal contact \citep{bonetti09}, friction \citep{bonetti14} and soft/soft thermal contact \citep{roubicek19}.  
A generalization was also provided by \citet{delpiero10}, see also \citet{raous11}. 
Recently, also a dual mortar implementation has appeared \citep{sabino26}.
The debonding field of the model follows from an evolution law that is like a chemical reaction that only progresses one way (towards debonding), but not the other (towards bonding).
In contrast, the computational model proposed below allows for both reactions to occur, also simultaneously.
Further, the reaction equation is thermodynamically consistently derived from the conservation of bonding sites \citep{cobo}. \\
A state variable also appears in the description of contact ageing \citep{dieterich78,rice83,ruina83} and wear \citep{stromberg99,stupkiewicz99}.
The latter works are in the framework of thermo-mechanical contact and hence also three-field contact models as the one contained here.

A similar class of debonding models are \textit{cohesive zone models} (CZMs), such as the widely used model of \citet{xu93} and its improvement by \citet{vandenbosch06}.
They also contain a damage variable, but one that is usually defined from the surface separation instead of being an independent state variable as in the Fr\'emond model.
CZMs have been extended by a temperature field to describe thermo-mechanical failure in
fibrillar composites \citep{hattiangadi04},
composite materials \citep{willam04},
elastoplastic materials \citep{fagerstrom08},
granular microstructures \citep{ozdemir10},
rubber  \citep{fleischhauer13},
transient problems \citep{sapora14},
materials with energetic interfaces \citep{esmaeili16},
and nanocomposites \citep{shu20}.
CZMs have also been extended by a chemical field to study
hydrogen embrittlement \citep{busto17},
lithium-ion material damage \citep{bai20,singh20},
and oxygen embrittlement \citep{auth22}.
Thermo-mechanical CZMs have also been extended by a third field variable to 
include humidity diffusion in concrete \citep{wu15},
combine phase-field fracture with surface debonding \citep{dittmann18},
include fatigue \citep{springer19},
and model the influence of cement hydration \citep{fang21}. 
Recently, two chemo-mechano-thermodynamical CZMs have been presented to study
crack healing \citep{salmon26}, 
and chemical bonding \citep{dandrea26}.
They use a similar setup as the one followed here \citep{cobo}, but they do not consider an interfacial heat, e.g.~due to exothermic bonding reactions, which is a central aspect of the model proposed here.

The model proposed here is also related to computational curing models that describe the change of bulk material behavior under progressive polymerization of molecular chains.
Initial efforts have considered curing under small deformations \citep{white92,adolf96,kiasat,andre05}, while later work generalized to large deformations \citep{lion07,mahnken13}.
In those models, the curing state is described by a separate field variable, which is similar to the one used here.
Chemo-thermo-mechanical curing models, and hence three-field models, 
were considered by \citet{lion07,landgraf14,sain18}. 
They are, however, bulk models and not interface models such as the one presented here.

A final class of related models are biomechanical bonding models that have been considered for natural bonding processes such as osseointegration \citep{rojek01b,moreo09,lutz12}
and cell adhesion \citep{bell78,bell84}.
However, fully coupled chemo-mechano-thermodynamical bonding models seem to be still missing for these bonding examples.

The survey above shows that many two-field contact models exist, but not general three-field formulations.
To address this gap the coupled chemo-mechano-thermodynamical contact model of \citet{cobo} was developed.
That work, however, is purely theoretical.
Therefore the present work proposes a corresponding FE formulation, focussing in particular on bonding and debonding and their coupling to thermo-mechanical contact.
Apart from the FE formulation and its illustration by six numerical examples, three new constitutive models are presented here.
The key novelties of the present work can be summarized as follows:
\\[-8mm]
\begin{itemize}
\item It presents a nonlinear finite element model for chemo-mechano-thermodynamical contact, \\[-7mm]
\item that is based on six interacting fields: a chemical, two mechanical and three thermal fields.  \\[-7mm] 
\item It is monolithically coupled to robustly capture strongly interacting fields. \\[-7mm]
\item It provides a unified framework for separate and simultaneous bonding and debonding. \\[-7mm]
\item Its full linearization is provided using static condensation for the local bonding variable. \\[-7mm]
\item Its convergence behavior is demonstrated on several challenging examples, \\[-7mm]
\item that include osseointegration, membrane debonding and pressure-dependent exothermic bonding. \\[-7mm]
\end{itemize}

The remainder of this work is structured as follows:
Sec.~\ref{s:coupl} briefly summarizes the governing equations of the chemo-mechano-thermodynamical contact model of \citet{cobo}.
A new constitutive model, suitable for describing general contact bonding and debonding, is then presented in Sec.~\ref{s:consti}.
Corresponding monolithic finite element formulations are then presented in Sec.~\ref{s:FE}.
Sec.~\ref{s:ex} illustrates the finite element model by several different 3D applications.
Finally, Sec.~\ref{s:concl} gives a conclusion and an outlook on possible extensions.
For an overview of the notation used here, we refer to the list of variables provided in \citet{cobo}.

\section{Coupled chemo-mechano-thermodynamical contact}\label{s:coupl}

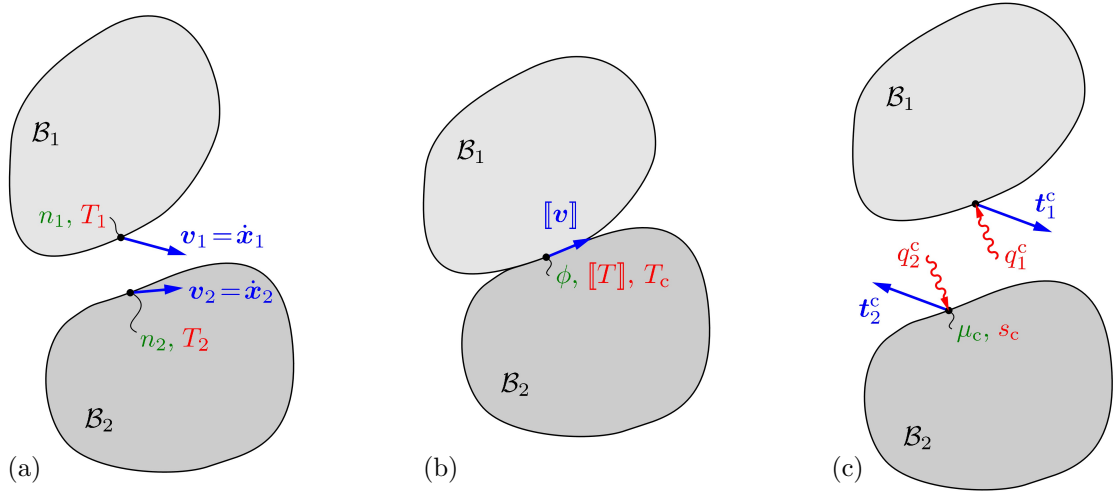
\begin{figure}[ht]
\begin{center} \unitlength1cm
\begin{picture}(0,6.3)
\put(-7.5,-.3){\includegraphics[width=40mm]{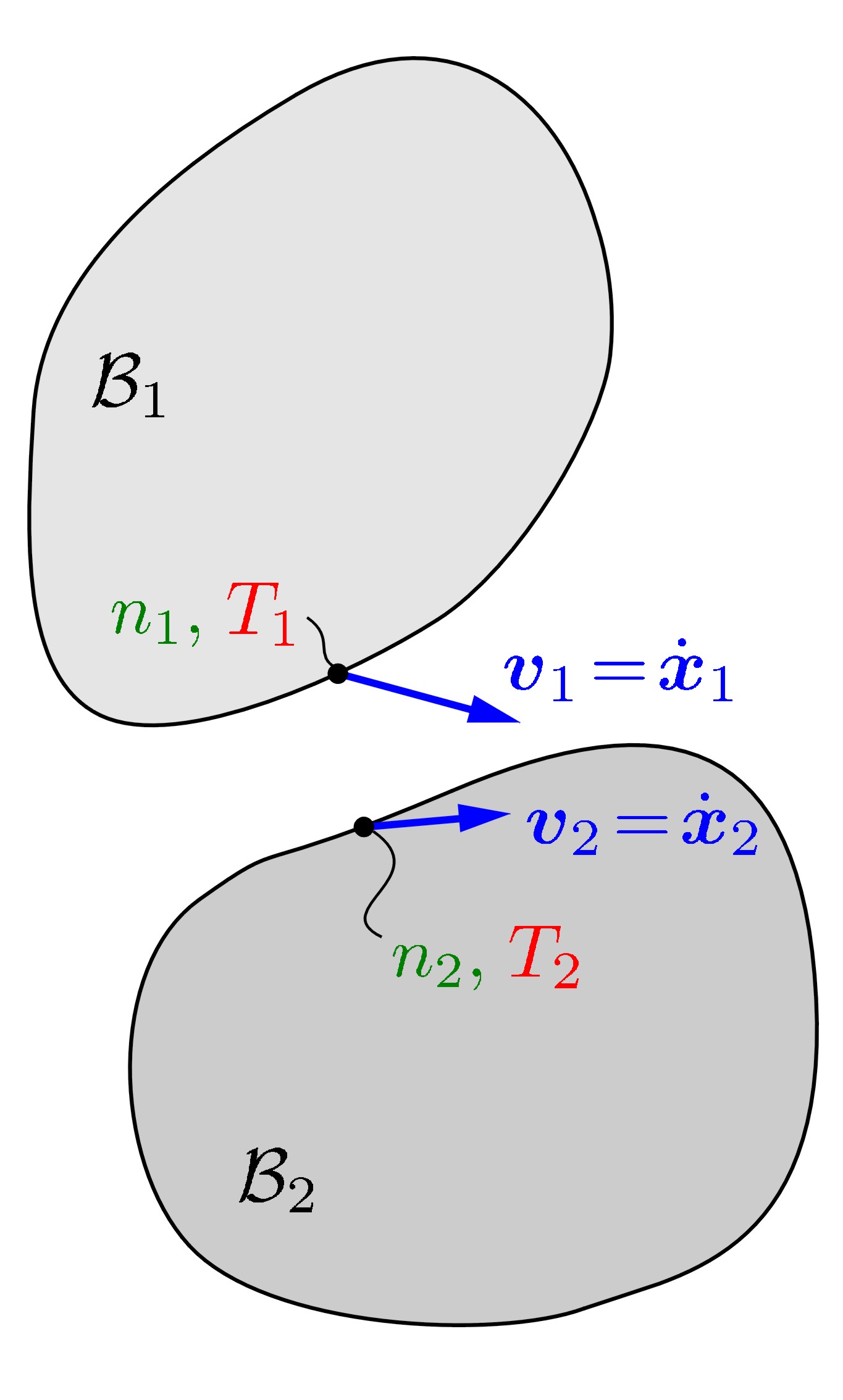}}
\put(-2,-.3){\includegraphics[width=40mm]{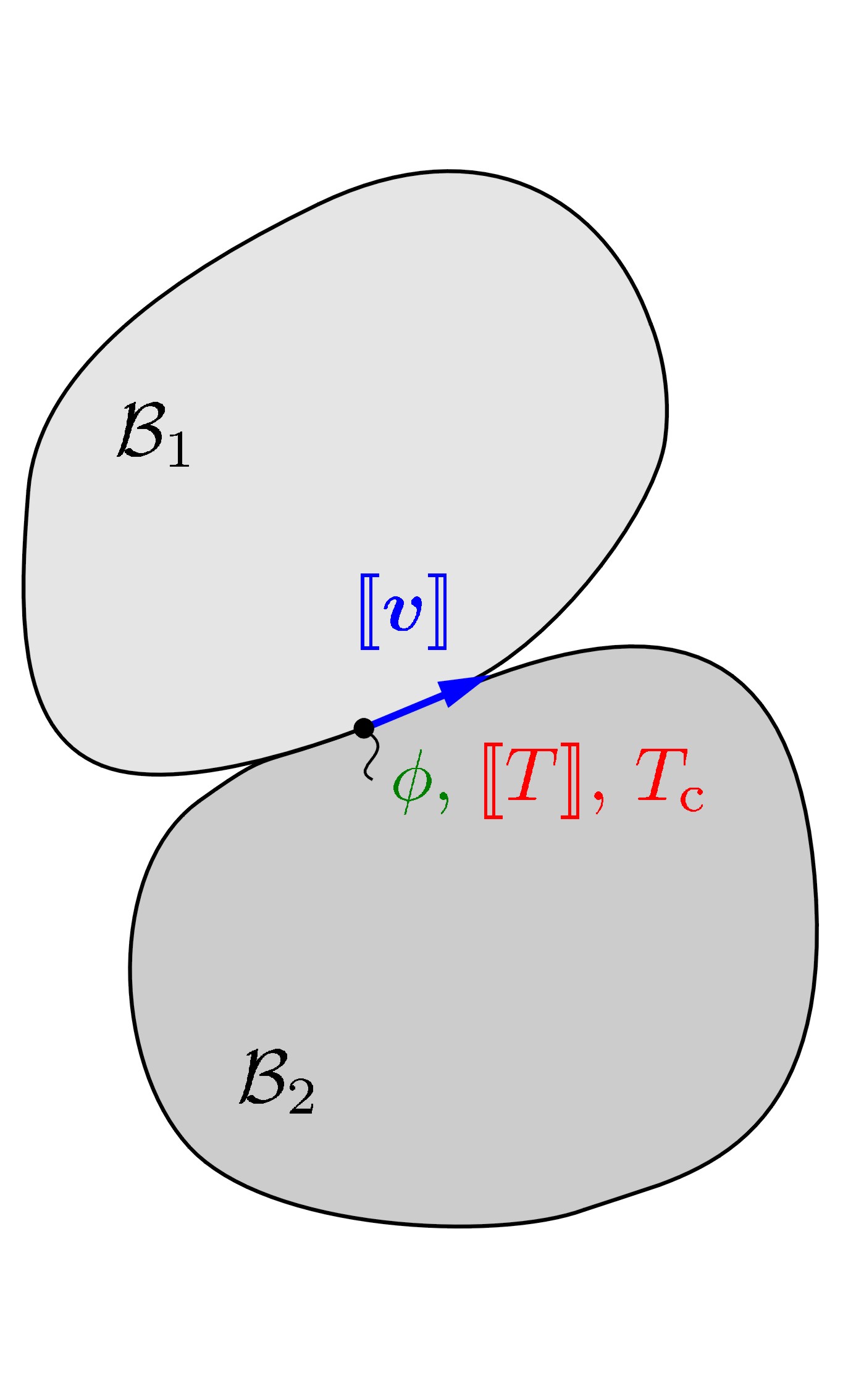}}
\put(3.5,-.3){\includegraphics[width=40mm]{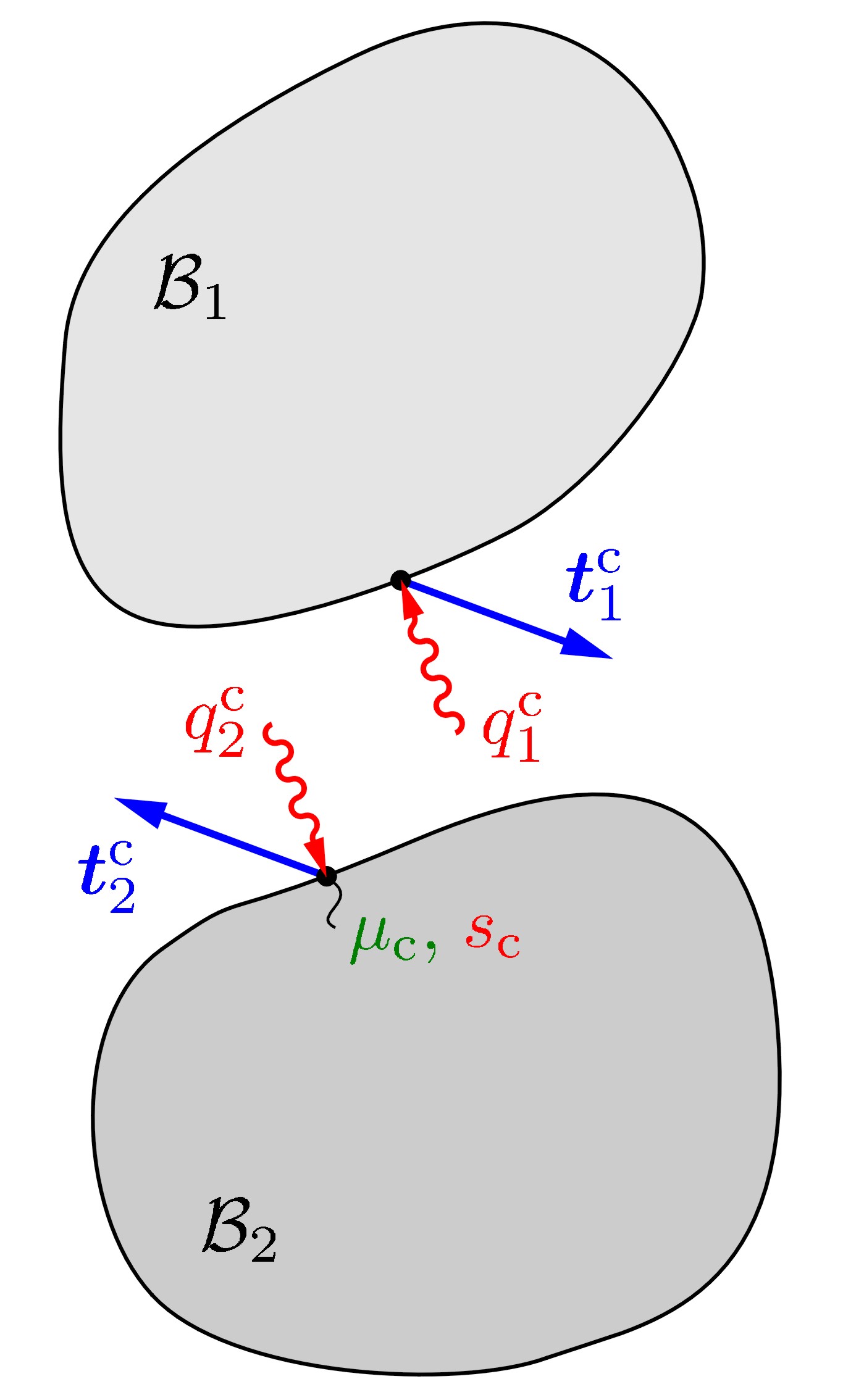}}
\put(-7.4,0){\small (a)}
\put(-1.9,0){\small (b)}
\put(3.5,0){\small (c)}
\end{picture}
\caption{Coupled chemo-mechano-thermodynamical contact setup \citep{cobo}: 
(a) bodies before contact; (b) bodies in contact; (c) free-body diagram for contact. 
Here, $n_k$, $\bx_k$ and $T_k$ denote the bonding site density, deformation and temperature, respectively, on the contact surface of body $\sB_k$ ($k=1,2$) at time $t$.
During contact, $\phi$ denotes the bonding state, $[\![\bv]\!]$ the velocity jump, $[\![T]\!]$ the temperature jump, $T_\mrc$ the contact temperature, $\bt^\mrc_k$ the contact tractions, $q^\mrc_k$ the heat influx, $\mu_\mrc$ the chemical contact potential and $s_\mrc$ the contact entropy. 
$T_\mrc$ and $s_\mrc$ are associated with an interfacial medium that may be present.} 
\label{f:cont}
\end{center}
\end{figure}

\begin{figure}[h]
\begin{center} \unitlength1cm
\begin{picture}(0,6.6)
\put(-6,-.2){\includegraphics[height=68mm]{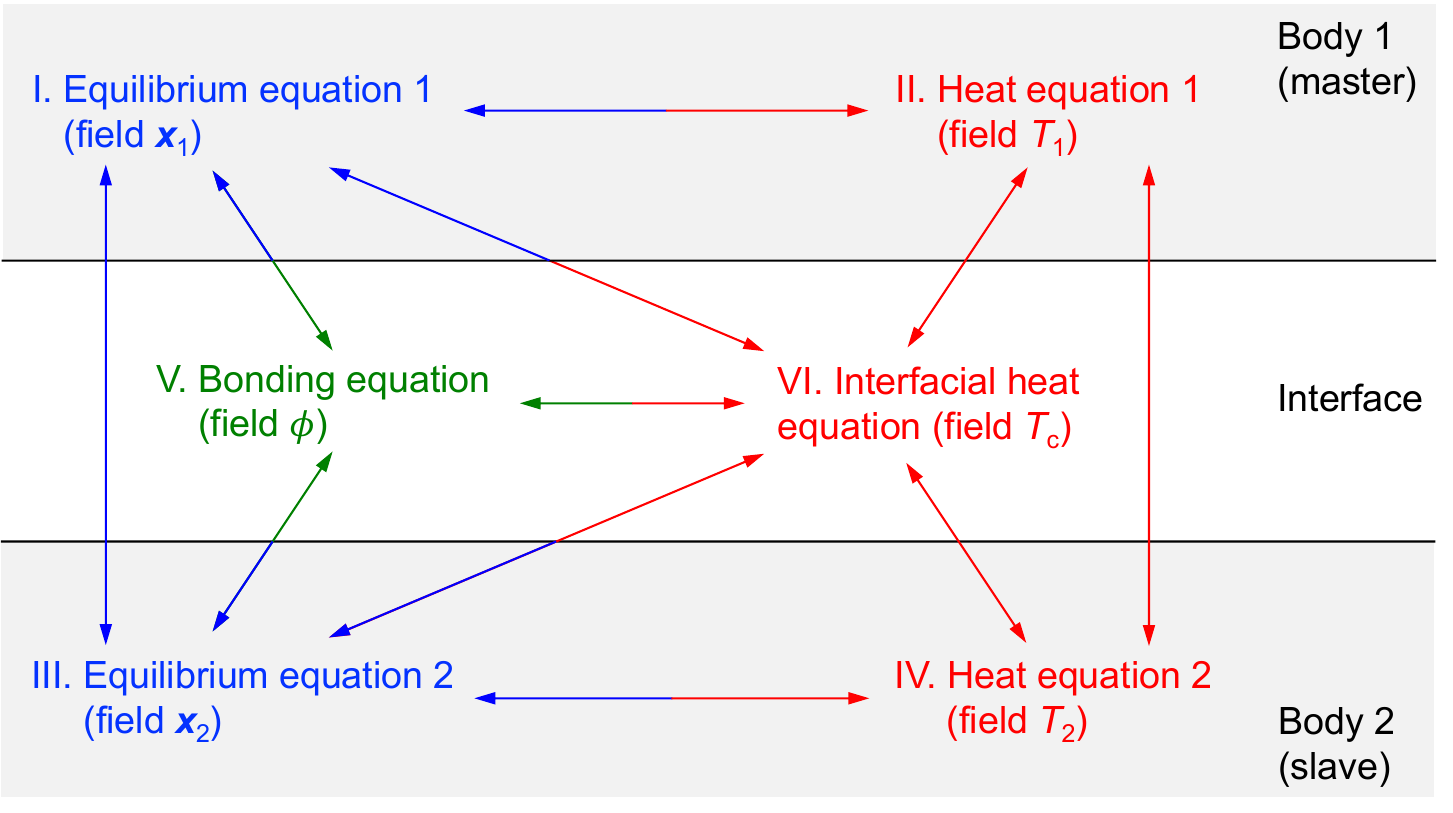}}
\end{picture}
\caption{Coupled chemo-mechano-thermodynamical contact interactions:
In the contact interface chemical bonding and thermal energy evolve together (Eqs.~V \& VI). 
This interacts with the mechanical and thermal states within the two bodies (Eqs.~I -- IV), which in turn interact with each other across the contact interface and within the bodies.
The interface is considered very thin, such that no separate interfacial equilibrium equation is needed. 
Also, no chemical state variable within each body is considered.}
\label{f:cocobo}
\end{center}
\end{figure}

This section summarizes the coupled chemo-mechano-thermodynamical continuum contact mo-del of \citet{cobo}.
Its basic setup is illustrated in Fig.~\ref{f:cont},
while its coupling is illustrated in Fig.~\ref{f:cocobo}. 
The model contains six fields: 
the deformations of body 1 and 2, $\bx_1$ and $\bx_2$, the temperature of the two bodies, $T_1$ and $T_2$, the bonding state $\phi$, and the interfacial temperature $T_\mrc$.
They are governed by six corresponding field equations (I -- VI) that interact with each other as indicated by the arrows in Fig.~\ref{f:cocobo}:
Classical thermo-mechanical coupling within each body corresponds to the coupling of Eqs.~I with II and III with IV.
Classical mechanical and thermal contact leads to the coupling of Eqs.~I with III and II with IV, respectively.
The bonding equation is influenced by contact pressures and gaps (coupling I--V and III--V) and interfacial temperatures (coupling V--VI).
The interfacial temperature in turn changes due to friction (coupling I--VI and III--VI), heat transfer (coupling II--VI and IV--VI) and the energy release of bonding reactions (coupling V--VI).

\begin{figure}[h]
\begin{center} \unitlength1cm
\begin{picture}(0,5)
\put(-3.8,-.2){\includegraphics[height=52mm]{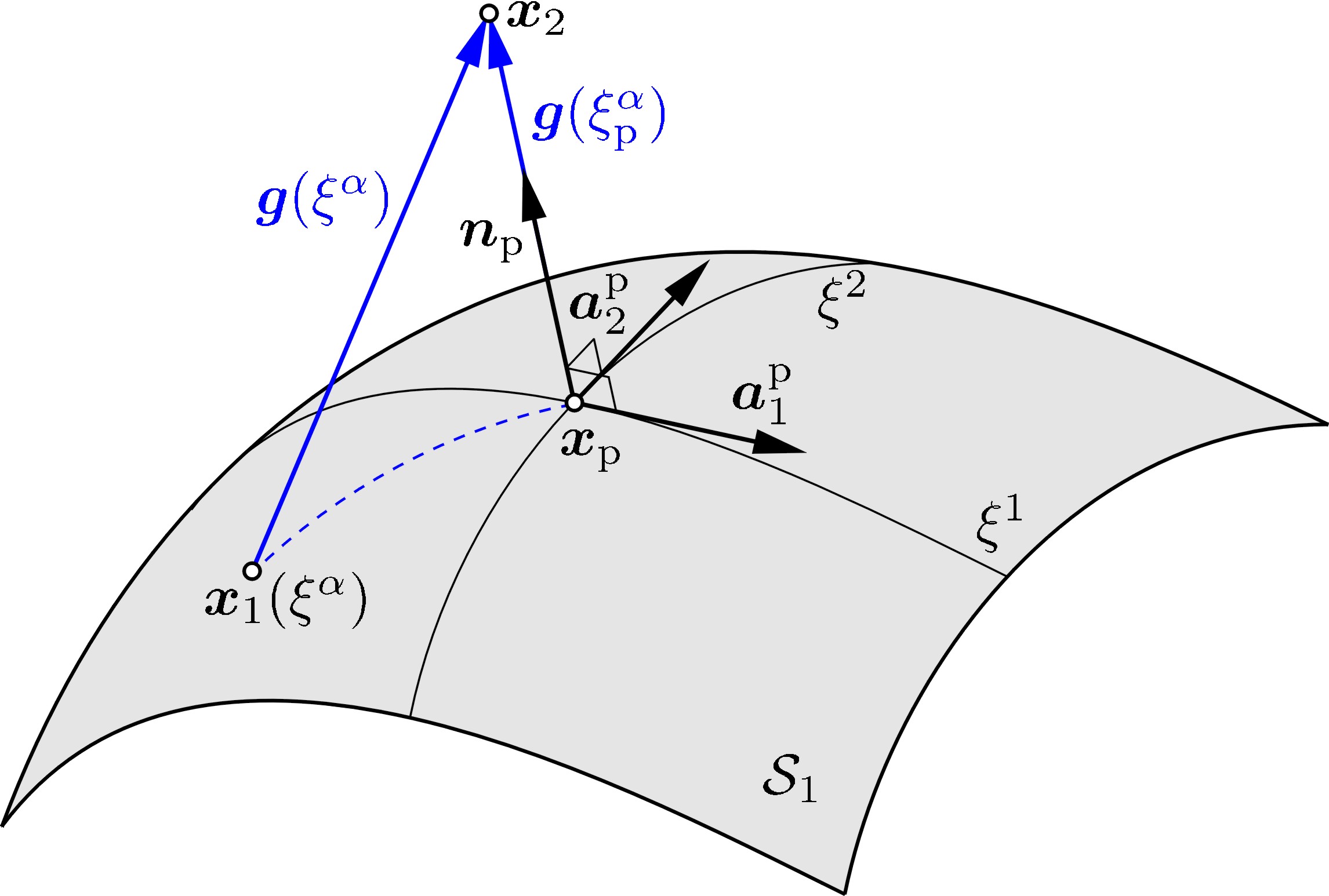}}
\end{picture}
\caption{Mechanical contact: Contact gap vector $\bg$ between \textit{slave} point $\bx_2$ and \textit{master} surface $\sS_1$ for a general master point $\bx_1$ and the closest projection point $\bx_\mrp$.
The dashed line marks a possible sliding path from $\bx_1$ to $\bx_\mrp$.
Adapted from \citet{cobo}.
}
\label{f:cpp}
\end{center}
\end{figure}

The contact state is characterized by the four contact variables and corresponding energy-conjugated flux variables listed in Table~\ref{t:conj}.
\begin{table}[h]
\centering
\begin{tabular}{|l|l|l|}
  \hline
  field & kinematic variable & conjugated flux variable \\[0mm] \hline 
  & & \\[-3.5mm]
  \textcolor{col2}{chemical} & \textcolor{col2}{bonding state $\phi\in[0, 1]$} & \textcolor{col2}{chemical contact potential $\mu_\mrc$} \\[.5mm]
  \textcolor{blue}{mechanical} & \textcolor{blue}{velocity jump $[\![\bv]\!]$} & \textcolor{blue}{contact traction $\bt_\mrc$}  \\[.5mm]
  \textcolor{red}{thermal (gap)} & \textcolor{red}{temperature jump $[\![T]\!]$} & \textcolor{red}{contact heat influx $q_\mrc$} \\[.5mm]
  \textcolor{red}{thermal (medium)} & \textcolor{red}{contact temperature $T_\mrc$} & \textcolor{red}{contact entropy $s_\mrc$} \\[.5mm]
  \hline
\end{tabular}
\vspace{-1mm}  
\caption{Energy-conjugated chemo-mechano-thermodynamical contact pairs \citep{cobo}.}
\label{t:conj}
\end{table}
The degree of bonding is characterized by the state variable $\phi$ that ranges between $\phi = 1$, for full bonding, and $\phi = 0$, for no bonding (or full debonding).
The mechanical contact state is described by the velocity jump $[\![\bv]\!] = \bv_2-\bv_1$ across the surface, that corresponds to the Lie derivative of the gap vector $\bg = \bx_2 - \bx_1$,  see Fig.~\ref{f:cpp}.
The thermal contact state is characterized by the temperature jump $[\![T]\!]:=T_2-T_1$ across the interface (from $\bx_2$ to $\bx_1$) and possibly the temperature $T_\mrc$ of an interfacial contact medium (if present).

For the case of sticking contact, where no frictional sliding occurs, as well as for the case of debonding from an initially bonded state, $\bx_2$ can be taken as the initial bond pair to $\bx_1$.
In both cases $\bg$ is a reversible (i.e.~elastic) quantity also denoted $\bg_\mre$.
The normal component of $\bg$ is obtained as $g_\mrn = \bg\cdot\bn_\mrp$, where $\bn_\mrp$ is the surface normal of the master surface at the closest projection point $\bx_\mrp$ of slave point $\bx_2$, see Fig.~\ref{f:cpp}.
In principle it is zero during contact ($g_\mrn=0$), unless we assume it to be a homogenized quantity that can become negative during contact, e.g.~due to the compression of microscale surface asperities or an interfacial contact layer such as a glue.
In this case the penetration $d_\mrn := -g_\mrn$ is assumed to be small.

Specific quantities, such as the chemical contact potential and the contact traction, can be expressed per reference area, per current area, or per bonding site.
To distinguish this, the notation presented in Table~\ref{t:not} is used: 
small italic symbols for bond-specific quantities, large italic symbols for current area-specific quantities, and large upright symbols for reference area-specific quantities.
In case the first kind is not needed -- as for tractions and heat fluxes -- the latter two kinds are also written with small symbols.
\begin{table}[h]
\centering
\begin{tabular}{|l|l|l|l|}
  \hline
  quantity & per bond & per current area & per reference area \\[0mm] \hline 
  & & & \\[-3.5mm]
  internal contact energy & $u_\mrc$ & $U_\mrc$ & $\unde{U_c}$ \\[.5mm]
  Helmholtz free contact energy & $\psi_\mrc$ & $\Psi_\mrc$ & $\unde{\Psi_c}$ \\[.5mm]
  contact entropy & $s_\mrc$ & $S_\mrc$ & $\unde{S_c}$ \\[.5mm]
  contact reaction rate & $r_\mrc$ & $R_\mrc$ & $\unde{R_c}$ \\[.5mm]
  chemical contact potential & $\mu_\mrc$ & $M_\mrc$ & $\unde{M_c}$ \\[.5mm]
  contact traction & -- & $\bt_\mrc$ & $\undetc$ \\[.5mm]
  contact heat flux & -- & $q_\mrc$ & $\unde{q_c}$ \\[.5mm]
  number of contact bonds & -- & $n_\mrc$ & $\unde{n_c}$ \\[.5mm]
   \hline
\end{tabular}
\vspace{-1mm}  
\caption{Notation for specific contact quantities \citep{cobo}. 
Given a bond-specific quantity, such as $u_\mrc$, the corresponding current-area-specific quantity follows from $U_\mrc = \nc\,u_\mrc$, while the corresponding reference-area-specific quantity is $\unde{U_c} = \unde{n_c}\,u_\mrc = \Jc\,U_\mrc$, where $J_\mrc = \unde{n_c}/n_\mrc$ denotes the local contact area change.
Note that in \citet{cobo} $\unde{n_c}$ is written $N_\mrc$.}
\label{t:not}
\end{table}

\begin{figure}[ht]
\begin{center} \unitlength1cm
\begin{picture}(0,4.5)
\put(-7.2,-.15){\includegraphics[height=45mm]{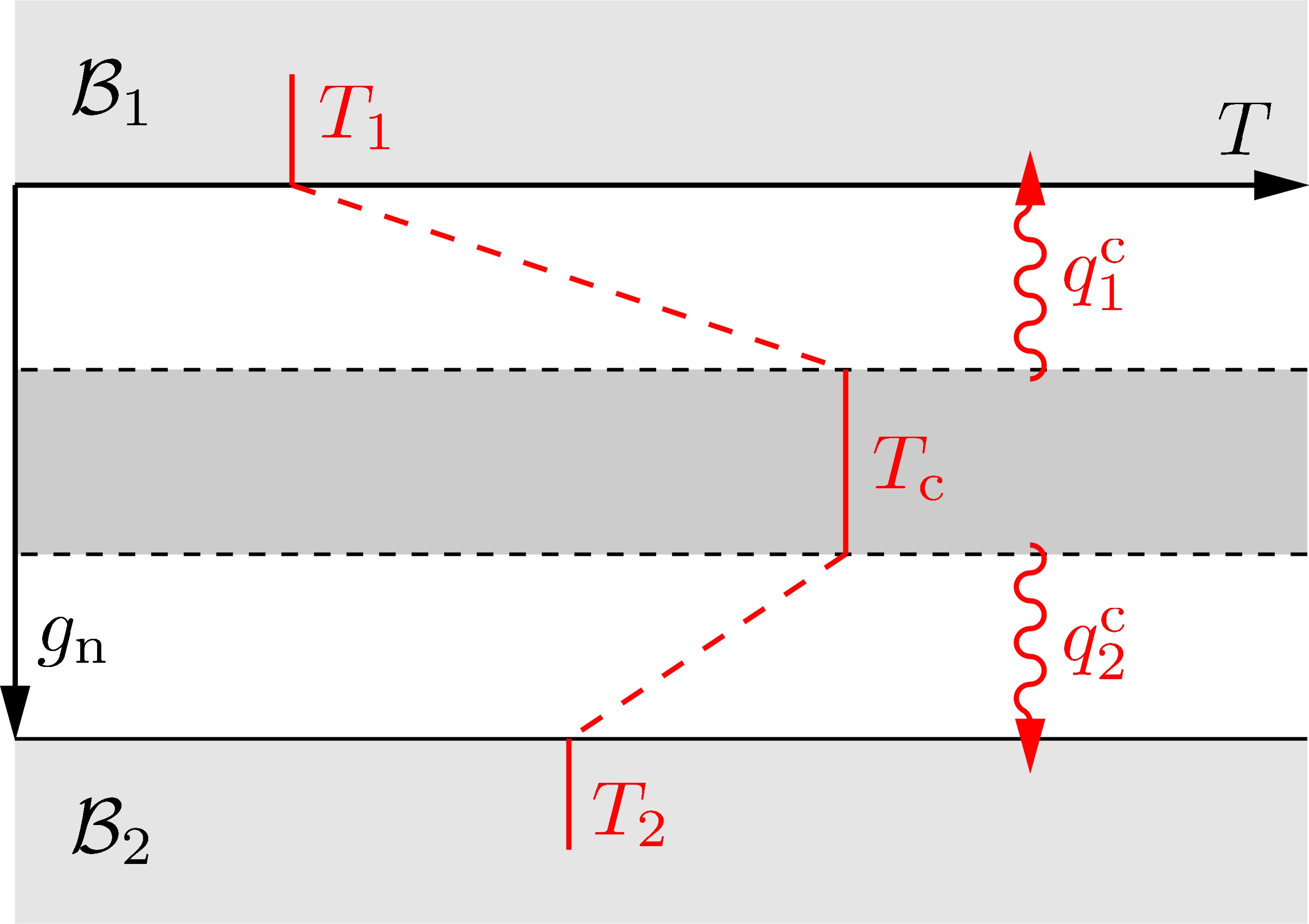}}
\put(1,-.15){\includegraphics[height=45mm]{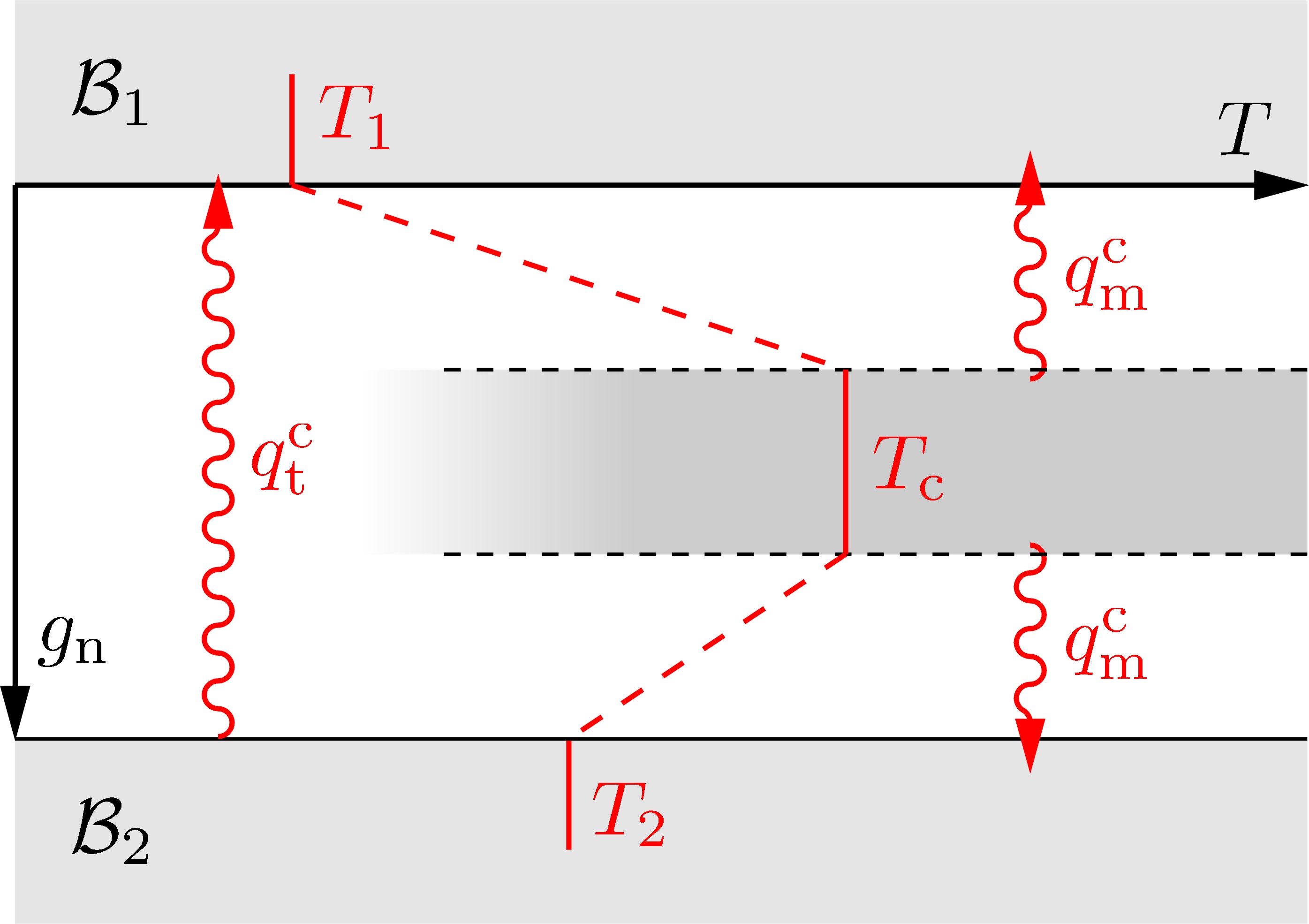}}
\put(-7.7,-.05){\footnotesize (a)}
\put(0.5,-.05){\footnotesize (b)}
\end{picture}
\caption{Heat transfer between master body $\sB_1$, slave body $\sB_2$ and interfacial medium (shown as a free body diagram):
(a) contact heat fluxes $q_1^\mrc$ and $q_2^\mrc$, (b) contact heat fluxes $q_\mrt^\mrc$ and $q_\mrm^\mrc$.}
\label{f:qc}
\end{center}
\end{figure}

Apart from the mechanical and thermal equilibrium equations inside each body
(Eqs.~I--IV),
the contact state at an interface point $\bx_\mrc$ is governed by the following equations \citep{cobo}:
Firstly, the bonding ODE
\eqb{l}
\unde{n_c}\, \dot\phi = \unde{R_c} \,,
\label{e:bondODE0}\eqe
where $\unde{n_c}$ and $\unde{R_c}$ denote the bond site density (unit [1/m$^2$]) and the bonding reaction rate (unit [1/(m$^2$s)]), respectively -- both w.r.t.~the contact area in the reference configuration;
secondly, the mechanical equilibrium between the contact tractions, i.e. 
\eqb{l}
\bt^\mrc_1=-\bt^\mrc_2=:\bt_\mrc\,;
\label{e:sf_moc}\eqe
thirdly, the thermal contact equilibrium equation
\eqb{l}
\nc\,\dot u_\mrc + q_1^\mrc + q_2^\mrc - \bt_\mrc\cdot[\![\bv]\!] = 0\,,
\label{e:sf_enc}\eqe
where $n_\mrc$ is the bond site density in the current configuration ($=\unde{n_c}/J_\mrc$, where $J_\mrc$ is the local contact area change);
and fourthly, the constitutive relations for the flux variables
\eqb{l}
\undetc = \ds\pa{\unde{\Psi_{\!c}}}{\bg_\mre}\,,\quad
\unde{M_c} = \ds\pa{\unde{\Psi_{\!c}}}{\phi}\,,\quad
\unde{S_c} = -\ds\pa{\unde{\Psi_{\!c}}}{T_\mrc}\,,
\label{e:tMS}\eqe
and
\eqb{l}
\undetc\cdot\Lgi\geq 0 \,,\quad
\unde{M_c} \unde{R_c} \leq 0 \,, \quad
q_k^\mrc \, \big(T_\mrc - T_k\big) \geq 0\,,
\label{e:tMS2}\eqe
where $\unde{\Psi_{\!c}}$ is specified in the following section.
Eq.~(\ref{e:tMS2}.1) governs sliding friction, which is not considered here,
Eq.~(\ref{e:tMS2}.2) governs the bonding reaction rate $\unde{R_c}$, and
Eq.~(\ref{e:tMS2}.3) governs the contact heat fluxes $q_k^\mrc$ for $k=1,2$.
The simplest reaction rate model satisfying condition (\ref{e:tMS2}.2) is 
\eqb{l}
\unde{R_c} = -c_\mrr\,\unde{M_c}\,,
\eqe
where $c_\mrr\geq0$ is the bonding reaction rate coefficient.
It is considered independent of $\unde{M_c}$ or $\phi$, but it can depend on the contact temperature, contact gap or contact pressure.
Inserting this into Eq.~\eqref{e:bondODE0} leads to the bonding ODE (Eq.~V)
\eqb{l}
\unde{n_c}\, \dot\phi + c_\mrr\,\unde{M_c} = 0\,.
\label{e:bondODE}\eqe
Its solution requires an initial condition for $\phi$, i.e.~the initial bonding state $\phi_0(\bx)$.
The simplest heat transfer model satisfying condition (\ref{e:tMS2}.3) is
\eqb{l}
q_k^\mrc = h_k\,(T_\mrc - T_k)\,,
\label{e:qck}\eqe
where $h_k\geq0$ is the heat transfer coefficient between body $\sB_k$ and the interfacial medium.
It is considered independent of $T_\mrc$ and $T_k$, but it can depend on the bonding state, contact gap or contact pressure.
It is instructive to decompose the heat fluxes $q_1^\mrc$ and $q_2^\mrc$ into the transfer heat flux from body $\sB_2$ to $\sB_1$,
\eqb{l}
q_\mrt^\mrc := \ds\frac{q_1^\mrc-q_2^\mrc}{2}\,,
\label{e:qct}\eqe
and the mean influx into both bodies
\eqb{l}
q_\mrm^\mrc := \ds\frac{q_1^\mrc+q_2^\mrc}{2} 
\label{e:qcm0}\eqe
according to Fig.~\ref{f:qc}.
The former describes the heat exchange between the two bodies, while the latter describes the heat flowing from the interface into the two bodies.
It follows from \eqref{e:sf_enc} as \citep{cobo}
\eqb{l}
q_\mrm^\mrc = \ds\frac{1}{2}\Big(\bt_\mrc\cdot\Lgi - \mu_\mrc\,R_\mrc - n_\mrc\,\dot s_\mrc\,T_\mrc\Big)\,,
\label{e:qcm}\eqe
i.e.~it is caused by frictional heating, chemical reactions, and entropy changes of the interface.
This work considers no mechanical dissipation ($\bt_\mrc\cdot\Lgi=0$).
The sign of $q_\mrm^\mrc$ then defines whether the contact bonding reactions are exothermic ($q_\mrm^\mrc >0$), isothermic ($q_\mrm^\mrc=0$) or endothermic ($q_\mrm^\mrc<0$).
Inserting \eqref{e:qck} and \eqref{e:qcm} into \eqref{e:qcm0} leads to the interfacial heat equation (Eq.~VI)
\eqb{l}
n_\mrc\,T_\mrc\,\dot s_\mrc = \bt_\mrc\cdot\Lgi - \mu_\mrc\,R_\mrc - (h_1+h_2)\,T_\mrc + h_1T_1 + h_2T_2\,,
\label{e:Tc_evolv}\eqe
that is exemplified for specific constitutive choices in the following section.

\begin{remark}\label{r:TcT2}
Eq.~\eqref{e:Tc_evolv} can be rewritten in temperature form using $n_\mrc\,T_\mrc\,\dot s_\mrc = C_\mrc\dot T_\mrc$ (for the case that the interfacial entropy $s_\mrc$ is not (or only negligibly) a function of $\bg_\mre$ and $\phi$).
In case the interfacial heat capacity $C_\mrc$ is zero, $T_\mrc$ can then be determined analytically, which is used in Sec.~\ref{s:FET} below.
\end{remark}

\section{Contact constitution}\label{s:consti}

This section proposes three constitutive models for the description of thermo-mechanical bonding and debonding in the framework of the new contact formulation.
The linearization of all models is also presented as this is required for implicit FE formulations, such as those in Sec.~\ref{s:FE}.
The constitutive models are based on the simple quadratic thermo-chemo-mechanical (de)bonding potential
\eqb{l}
\unde{\Psi_{\!c}}(\bg_\mre,\phi,T_\mrc) = \ds\frac{1}{2}\,\bg_\mre\cdot\undeEc\,\bg_\mre  + \frac{1}{2}\,\overrightarrow{\unde{K_c}}\,(\phi-1)^2 
	+ \frac{1}{2}\,\overleftarrow{\unde{K_c}} \,\phi^2 - \frac{\unde{C_c}}{2T_0}\,(T_\mrc-T_0)^2 \,.
\label{e:Psi0}\eqe
The four terms in \eqref{e:Psi0} characterize mechanical contact, chemical bonding, chemical debonding and thermal contact, respectively.
They are minimal at zero contact gap ($\bg_\mre = \mathbf{0}$), full bonding ($\phi = 1$), full debonding ($\phi = 0$) and isothermal contact ($T_\mrc = T_0$), respectively. 
The corresponding positive (definite) parameters $\undeEc$, $\overrightarrow{\unde{K_c}}$, $\overleftarrow{\unde{K_c}}$ and $\unde{C_c}$,
denote contact stiffness, bond energy density\footnote{that needs to be overcome during separation $\phi=1\rightarrow 0$}, unbonded surface energy density\footnote{that needs to be overcome during bonding $\phi= 0\rightarrow 1$} and contact heat capacity -- all expressed per reference surface area. 
They can be transformed into quantities per current surface area if divided by the contact area change $J_\mrc$. 
They are treated as constant w.r.t.~their primary fields, but can still be functions of the other fields, i.e.~$\undeEc = \undeEc(T_\mrc,\phi)$, $\overleftrightarrow{\unde{K_c}} = \overleftrightarrow{\unde{K_c}}(\bg_\mre,T_\mrc)$ and $\unde{C_c} = \unde{C_c}(\bg_\mre,\phi)$. 
Potential \eqref{e:Psi0} extends the potentials of \citet{johansson93}, \citet{oancea97} and \citet{cobo} to simultaneous bonding and debonding (if both $\overrightarrow{\unde{K_c}}>0$ and $\overleftarrow{\unde{K_c}}>0$).

The contact stiffness can be decomposed into normal and tangential contributions, i.e.
\eqb{l}
\undeEc = \unde{E_n}\, \bn_\mrp\otimes\bn_\mrp + \unde{E_t}\, \bi\,,
\label{e:Ec}\eqe
where
\eqb{l}
\bi := \bone - \bn_\mrp\otimes\bn_\mrp
\eqe
is the surface identity on the master surface at $\bx_\mrp$.
This makes $\unde{\Psi_{\!c}}$ an explicit function of $\bn_\mrp$ (unless $\unde{E_t} = \unde{E_n}$), which can be avoided through approach (b) discussed in Remark~\ref{r:ge}.
Two cases will be used in the subsequent examples:
(i) sticking contact with $\unde{E_t} = \unde{E_n}$, and
(ii) frictionless contact with $\unde{E_t} = 0$.  
In both cases, the first term in \eqref{e:Psi0} is treated as a penalty term that is only considered active during contact, i.e.
\eqb{l}
\left\{\begin{array}{ll}
\unde{E_n} > 0  & $for$~g_\mrn \leq 0~~($contact$)\,, \\[1mm]
\unde{E_n} = 0  & $for$~g_\mrn > 0~~($no contact$)\,,
\end{array}\right.
\label{e:ccases1}\eqe
where $g_\mrn = \bg_\mre\cdot\bn_\mrp$ is the normal gap. 

\begin{remark}\label{r:30}
A penalty formulation allows for small penetration during contact, and so $g_\mrn$ can become negative.
The penetration remains small, if $\unde{E_n}$ is taken as a large number during contact. 
The limit $\unde{E_n}\rightarrow\infty$ then leads to $g_\mrn\rightarrow0$, such that the contact pressure $p_\mrc = -E_\mrn\,g_\mrn$ (with $E_\mrn := \unde{E_n}/J_\mrc$) 
remains bounded and the contact energy $\unde{E_n}\,g_\mrn^2/2$ vanishes.
\end{remark}

\begin{remark}\label{r:ge}
Instead of (a) distinguishing normal and tangential stiffnesses in \eqref{e:Ec} for the common gap vector $\bg_\mre = \bg = \bx_2 - \bx_1$,
one can also (b) use the common stiffness $\undeEc = \unde{E_n}\bone$ and distinguish the gap vector by
\eqb{l}
\bg_\mre = \left\{\begin{array}{ll}
\bx_2 - \bx_1  & $for sticking contact (i)$\,, \\[1mm]
\bx_2 - \bx_\mrp = g_\mrn\,\bn_\mrp & $for frictionless contact (ii)$\,,
\end{array}\right.
\label{e:ccasesi}\eqe
where $\bx_\mrp = \bx_1(\bx_2)$ is the closest projection point of slave point $\bx_2$ onto the master surface, see Fig.~\ref{f:cpp},
and
\eqb{l}
g_\mrn := (\bx_2 - \bx_\mrp)\cdot\bn_\mrp
\label{e:gn}\eqe
is the normal gap defined from $\bx_\mrp$ in case (ii).
This definition is exact, whereas the one used in approach~(a) (i.e.~$g_\mrn = (\bx_2-\bx_1)\cdot\bn_\mrp$) is inaccurate when the surface is curved and $\bx_1$ is not identical to $\bx_\mrp$, see Fig.~\ref{f:cpp}. 
Therefore approach (b) is used in the following contact formulation and its linearization.
\end{remark}

The bonded and unbonded surface energy densities in \eqref{e:Psi0} are considered of the form
\eqb{lll}
\overrightarrow{\unde{K_c}}(\bg_\mre,T_\mrc) \is \overrightarrow{\unde{K_0}}~\overrightarrow{\unde{f_m}}(\bg_\mre) \, \overrightarrow{\unde{f_T}}(T_\mrc)\,, \\[1mm]
\overleftarrow{\unde{K_c}}(\bg_\mre,T_\mrc) \is \overleftarrow{\unde{K_0}}~\overleftarrow{\unde{f_m}}(\bg_\mre) \, \overleftarrow{\unde{f_T}}(T_\mrc)\,,
\label{e:Kc}\eqe
where $\overrightarrow{\unde{K_0}}$ and $\overleftarrow{\unde{K_0}}$ are positive constants, while $\overrightarrow{\unde{f_\bullet}}\!(...)$ and $\overleftarrow{\unde{f_\bullet}}\!(...)$ are positive functions that depend on the contact state and are discussed in Secs.~\ref{s:con1}-\ref{s:con4} below.
This way it can for example be ensured that bonding is only active during contact, and debonding only active if the bodies are separating.

The treatment of the last term in \eqref{e:Psi0} is more tricky.
The term is associated with an interfacial medium, like wear debris, that does not disappear during debonding.
Therefore the term should not be deactivated when contact is lost, even if it was zero before contact.
The only exception is when $\unde{C_c}$ is zero itself, which is the case of the first five examples in Sec.~\ref{s:ex}.

From \eqref{e:tMS}, \eqref{e:Psi0} and \eqref{e:Kc} follow the contact traction
\eqb{l}
\undetc 
= \undeEc\,\bg_\mre + \overrightarrow{\unde{K_0}}\,\ds\frac{1}{2}\pa{\overrightarrow{\unde{f_m}}}{\bg_\mre} \, \overrightarrow{\unde{f_T}} \,  (\phi-1)^2
+ \overleftarrow{\unde{K_0}}\,\ds\frac{1}{2}\pa{\overleftarrow{\unde{f_m}}}{\bg_\mre} \, \overleftarrow{\unde{f_T}}\,  \phi^2 \,,
\label{e:tc}\eqe
the chemical contact potential
\eqb{l}
\unde{M_c} = \overrightarrow{\unde{K_0}}\,\overrightarrow{\unde{f_m}} \, \overrightarrow{\unde{f_T}}\,(\phi-1) 
+ \overleftarrow{\unde{K_0}}\,\overleftarrow{\unde{f_m}} \, \overleftarrow{\unde{f_T}}\,\phi \,,
\label{e:Mc}\eqe
and the contact entropy
\eqb{l}
\unde{S_c} = \ds\frac{\unde{C_c}}{T_0}\big(T_\mrc - T_0\big)
- \overrightarrow{\unde{K_0}}\,\overrightarrow{\unde{f_m}} \, \ds\frac{1}{2}\pa{\overrightarrow{\unde{f_T}}}{T_\mrc} \,(\phi-1)^2 
- \overleftarrow{\unde{K_0}}\,\overleftarrow{\unde{f_m}} \, \ds\frac{1}{2}\pa{\overleftarrow{\unde{f_T}}}{T_\mrc}\,\phi^2  \,.
\label{e:Sc}\eqe
The linearization of these quantities according to Remark \ref{r:ge} leads to the increments
\eqb{llrlrlr}
\Delta\undetc 
\is \ds\pa{\undetc}{\bg_\mre}\cdot \Delta\bg_\mre \plus \ds\pa{\undetc}{\phi}\, \Delta\phi \plus \ds\pa{\undetc}{T_\mrc}\, \Delta T_\mrc\,, \\[4.5mm]
\Delta\unde{M_c}
\is \ds\pa{\unde{M_c}}{\bg_\mre}\cdot\Delta\bg_\mre \plus \ds\pa{\unde{M_c}}{\phi}\, \Delta\phi \plus \ds\pa{\unde{M_c}}{T_\mrc}\, \Delta T_\mrc \,, \\[4.5mm]
\Delta\unde{S_c} 
\is \ds\pa{\unde{S_c}}{\bg_\mre}\cdot\Delta\bg_\mre \plus \ds\pa{\unde{S_c}}{\phi}\, \Delta\phi \plus \ds\pa{\unde{S_c}}{T_\mrc}\, \Delta T_\mrc \,,
\label{e:DtMSc}\eqe
with
\eqb{rll}
\ds\pa{\undetc}{\bg_\mre} \is \undeEc 
	+ \overrightarrow{\unde{K_0}}\,\ds\frac{1}{2}\paqq{\overrightarrow{\unde{f_m}}}{\bg_\mre}{\bg_\mre} \, \overrightarrow{\unde{f_T}} \,  (\phi-1)^2
	+ \overleftarrow{\unde{K_0}}\,\ds\frac{1}{2}\paqq{\overleftarrow{\unde{f_m}}}{\bg_\mre}{\bg_\mre} \, \overleftarrow{\unde{f_T}}\,  \phi^2\,, \\[4mm]
\ds\pa{\undetc}{\phi} \is \ds\pa{\unde{M_c}}{\bg_\mre}
	= \overrightarrow{\unde{K_0}}\,\ds\pa{\overrightarrow{\unde{f_m}}}{\bg_\mre} \, \overrightarrow{\unde{f_T}} \,  (\phi-1)
	+ \overleftarrow{\unde{K_0}}\,\ds\pa{\overleftarrow{\unde{f_m}}}{\bg_\mre} \, \overleftarrow{\unde{f_T}}\,  \phi\,, \\[4mm]
\ds\pa{\undetc}{T_\mrc} \is -\ds\pa{\unde{S_c}}{\bg_\mre} 
	= \overrightarrow{\unde{K_0}}\,\ds\frac{1}{2}\pa{\overrightarrow{\unde{f_m}}}{\bg_\mre} \, \pa{\overrightarrow{\unde{f_T}}}{T_\mrc} \,  (\phi-1)^2
	+ \overleftarrow{\unde{K_0}}\,\ds\frac{1}{2}\pa{\overleftarrow{\unde{f_m}}}{\bg_\mre} \, \pa{\overleftarrow{\unde{f_T}}}{T_\mrc}\,  \phi^2\,, \\[5mm]
\ds\pa{\unde{M_c}}{\phi} \is \overrightarrow{\unde{K_0}}\,\overrightarrow{\unde{f_m}} \, \overrightarrow{\unde{f_T}} 
	+ \overleftarrow{\unde{K_0}}\,\overleftarrow{\unde{f_m}} \, \overleftarrow{\unde{f_T}}\,, \\[3mm]
\ds\pa{\unde{M_c}}{T_\mrc} \is -\ds\pa{\unde{S_c}}{\phi} 
	= \overrightarrow{\unde{K_0}}\,\overrightarrow{\unde{f_m}} \, \ds\pa{\overrightarrow{\unde{f_T}}}{T_\mrc}\,(\phi-1) 
	+ \overleftarrow{\unde{K_0}}\,\overleftarrow{\unde{f_m}} \, \ds\pa{\overleftarrow{\unde{f_T}}}{T_\mrc}\,\phi\,, \\[4mm]
\ds\pa{\unde{S_c}}{T_\mrc} \is \ds\frac{\unde{C_c}}{T_0}
	- \overrightarrow{\unde{K_0}}\,\overrightarrow{\unde{f_m}} \, \ds\frac{1}{2}\frac{\partial^2\overrightarrow{\unde{f_T}}}{\partial T_\mrc^2} \,(\phi-1)^2 
	- \overleftarrow{\unde{K_0}}\,\overleftarrow{\unde{f_m}} \, \ds\frac{1}{2}\frac{\partial^2\overleftarrow{\unde{f_T}}}{\partial T_\mrc^2}\,\phi^2  \,.
\label{e:ptMS}\eqe
The gap increment $\Delta\bg_\mre$ depends on the two cases in \eqref{e:ccasesi}, which leads to differences in the evaluation of $\Delta\undetc$ as is shown in Appendix~\ref{s:lin-ii}. 

The following subsections discuss various choices for functions $\overrightarrow{\unde{f_\bullet}}$ and $\overleftarrow{\unde{f_\bullet}}$.

\subsection{Simple mechanical debonding}\label{s:con1}

\begin{figure}[h]
\begin{center} \unitlength1cm
\begin{picture}(0,4.7)
\put(-5.67,.25){\includegraphics[height=43mm]{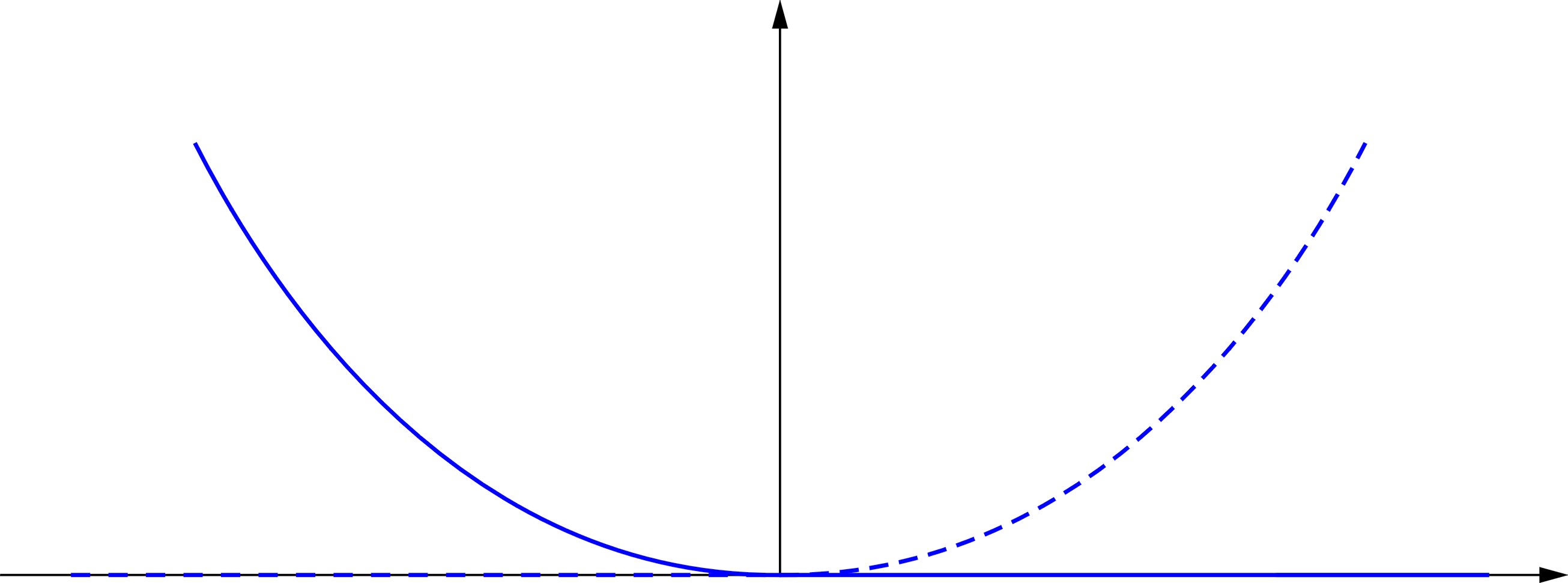}}
\put(0.25,3.95){\small $\overleftrightarrow{\unde{K_c}}$}
\put(4.45,2.9){\small $\overleftarrow{\unde{K_c}}\sim\overleftarrow{\unde{f_m}}$}
\put(-5.4,2.9){\small $\overrightarrow{\unde{K_c}}\sim\overrightarrow{\unde{f_m}}$}
\put(0,-.05){\small 0}
\put(5.5,-.05){\small $g_\mathrm{n}$}
\end{picture}
\caption{Mechanical contact constitution: Separation-dependent debonding potential $\protect\overleftarrow{\unde{K_c}}$ 
based on function \eqref{e:fm1}, and pressure-dependent bonding potential $\protect\overrightarrow{\unde{K_c}}$ based on function \eqref{e:fm3}.}
\label{f:fm1}
\end{center}
\end{figure}

A simple example for the mechanical debonding function is
\eqb{l}
\overleftarrow{\unde{f_m}}(\bg_\mre) = \ds\frac{1}{g_0^2} \left\{\begin{array}{ll}
\bg_\mre\cdot\bg_\mre & $for$~~g_\mrn \geq 0 \,, \\[1mm]
0 & $else$\,,
\end{array}\right.
\label{e:fm1}\eqe
where $g_0$ is a constant with units of length.
Fig.~\ref{f:fm1} illustrates the debonding potential resulting from this.
Model \eqref{e:fm1}  is only active during separation ($g_\mrn\geq0$) and in this case gives
\eqb{l}
\ds\frac{1}{2}\pa{\overleftarrow{\unde{f_m}}}{\bg_\mre} = \frac{\bg_\mre}{g_0^2}\,,\quad
\ds\frac{1}{2}\paqq{\overleftarrow{\unde{f_m}}}{\bg_\mre}{\bg_\mre} = \frac{\bone}{g_0^2} \,,
\eqe
such that Eqs.~\eqref{e:tc}--\eqref{e:Sc} lead to
\eqb{rll}
\undetc \is \undeEc\,\bg_\mre + \ds\frac{\overleftarrow{\unde{K_0}}}{g_0^2}\,\phi^2\,\bg_\mre\,, \\[0mm]
\unde{M_c} \is \ds\frac{\overleftarrow{\unde{K_0}}}{g_0^2}\,\bg_\mre\cdot\bg_\mre\,\phi  \,, \\[4mm]
\unde{S_c} \eis 0 \,,
\label{e:tMS1}\eqe
in case there are no other dependencies ($\overrightarrow{\unde{K_c}} = \unde{C_c}=0$ and $\overleftarrow{\unde{f_T}} = 1$).
The non-zero terms in linearization \eqref{e:DtMSc}-\eqref{e:ptMS} then follow as
\eqb{lll}
\ds\pa{\undetc}{\bg_\mre} \is \undeEc + \ds\frac{\overleftarrow{\unde{K_0}}}{g_0^2}\,\phi^2\,\bone\,, \quad
	\ds\pa{\unde{M_c}}{\phi} = \ds\frac{\overleftarrow{\unde{K_0}}}{g_0^2}\,\bg_\mre\cdot\bg_\mre\,, \\[4mm]
\ds\pa{\undetc}{\phi} \is \ds\pa{\unde{M_c}}{\bg_\mre} = 2\,\ds\frac{\overleftarrow{\unde{K_0}}}{g_0^2}\,\phi \,\bg_\mre \,.
\label{e:ptM1}\eqe
Note that in Eqs.~\eqref{e:tMS1} and \eqref{e:ptM1} the terms with $\undeEc$ are zero for $g_\mrn>0$ due to \eqref{e:ccases1}, while for $g_\mrn<0$ all other terms are zero due to Eq.~\eqref{e:fm1}.

\subsection{Thermo-mechanical bonding above a minimum temperature}\label{s:con3}

A simple example for the thermal bonding function is
\eqb{l}
\overrightarrow{\unde{f_T}}(T_\mrc)  = \ds\frac{1}{T_0^2} \left\{\begin{array}{ll}
(T_\mrc-T_\mathrm{min})^2 & $for$~~T_\mrc \geq T_\mathrm{min}\,, \\[1mm]
0 & $else$\,,
\end{array}\right.
\label{e:ft1}\eqe
where $T_0$ is a constant reference temperature and $T_\mathrm{min}$ is a constant limit temperature above which bonding starts.
This can be combined with contact pressure-dependent bonding, which can be described by the mechanical bonding function 
\eqb{l}
\overrightarrow{\unde{f_m}}(\bg_\mre) = \ds\frac{1}{g_0^2} \left\{\begin{array}{ll}
g_\mrn^2 & $for$~~g_\mrn \leq 0\,, \\[1mm]
0 & $else$\,,
\end{array}\right.
\label{e:fm3}\eqe
since the contact pressure is proportional to the normal gap as mentioned in Remark~\ref{r:30}.
The bonding potential resulting from \eqref{e:fm3} is illustrated in Fig.~\ref{f:fm1}.
The normal gap is defined in Eq.~\eqref{e:gn}, which gives $g_\mrn = \bg_\mre\cdot\bn_\mrp$ for $\bg_\mre = g_\mrn\bn_\mrp$.
Hence $g_\mrn^2 = \bg_\mre\cdot\bg_\mre$.
When active, $\overrightarrow{\unde{f_T}}(T_\mrc)$ and $\overrightarrow{\unde{f_m}}(\bg_\mre)$ lead to
\eqb{l}
\ds\frac{1}{2}\pa{\overrightarrow{\unde{f_T}}}{T_\mrc}  = \ds\frac{T_\mrc - T_\mathrm{min}}{T_0^2}\,,
\label{e:dft1}\eqe
and
\eqb{l}
\ds\frac{1}{2}\pa{\overrightarrow{\unde{f_m}}}{\bg_\mre} = \frac{\bg_\mre}{g_0^2} = \frac{g_\mrn}{g_0^2}\,\bn_\mrp\,,
\label{e:dfm3}\eqe
such that
\eqb{rll}
\undetc \is \undeEc\,\bg_\mre 
	+ \ds\frac{\overrightarrow{\unde{K_0}}}{g_0^2\,T_0^2}\,g_\mrn\,(\phi-1)^2\,(T_\mrc-T_\mathrm{min})^2\,\bn_\mrp\,, \\[4mm]
\unde{M_c} \is \qquad\qquad~~~
	\ds\frac{\overrightarrow{\unde{K_0}}}{g_0^2\,T_0^2}\,g_\mrn^2\,(\phi-1)\,(T_\mrc-T_\mathrm{min})^2\,, \\[4mm]
\unde{S_c} \is \ds\frac{\unde{C_c}}{T_0}\big(T_\mrc - T_0 \big)
	-\frac{\overrightarrow{\unde{K_0}}}{g_0^2\,T_0^2}\,g_\mrn^2\,(\phi-1)^2\,(T_\mrc-T_\mathrm{min})\,,
\label{e:fttMS}\eqe
follow from \eqref{e:tc}-\eqref{e:Sc} for $g_\mrn \leq 0$ and $T_\mrc \geq T_\mathrm{min}$ (and $\overleftarrow{\unde{K_0}}=0$).
In case $g_\mrn > 0$, all terms associated with $\undeEc$ and $\overrightarrow{\unde{K_0}}$ vanish according to Eqs.~\eqref{e:ccases1} and \eqref{e:fm3}.
In case $T_\mrc < T_\mathrm{min}$, all the rear terms in \eqref{e:fttMS} vanish, which can be also achieved in a computational implementation by setting $\overrightarrow{\unde{K_0}}$ to zero in this case.
It is noted that the first term in (\ref{e:fttMS}.1) is for both cases in \eqref{e:ccasesi}, whereas the rear term only uses $\bg_\mre = g_\mrn\bn_\mrp$ as in \eqref{e:fm3} and \eqref{e:dfm3}.

\begin{remark}\label{r:31}
If the front term in (\ref{e:fttMS}.1) also uses $\bg_\mre = g_\mrn\bn_\mrp$, the contact traction can be simply written as
\eqb{l}
\undetc = \unde{E_n^{eff}}\,\bg_\mre\,, 
\label{e:tEg}\eqe
where
\eqb{l}
\unde{E_n^{eff}} := \unde{E_n} + \overrightarrow{\unde{E_n}}\,,\quad $with$~~ 
 \overrightarrow{\unde{E_n}} := \ds\frac{\overrightarrow{\unde{K_0}}}{g_0^2\,T_0^2}\,(\phi-1)^2\,(T_\mrc-T_\mathrm{min})^2\,,
\label{e:Eneff}\eqe
is the effective contact stiffness during bonding.
Its second part depends on $\phi$ and $T_\mrc$.
Therefore, unless $\overrightarrow{\unde{K_0}}/g_0^2 \ll \unde{E_n}$, significant thermo-chemo-mechanical coupling can be expected in Eq.~(\ref{e:fttMS}.1).
Likewise, significant thermo-chemo-mechanical coupling can be expected in Eq.~(\ref{e:fttMS}.3), unless $\overrightarrow{\unde{K_0}}/T_0 \ll \unde{C_c}$.
The coupling in (\ref{e:fttMS}.1) vanishes in the penalty limit $\unde{E_n}\rightarrow\infty$.
Given \eqref{e:tEg} 
and $\undetc = J_\mrc\,\bt_\mrc$, the contact pressure $p_\mrc = -\bt_\mrc\cdot\bn_\mrp$ follows as 
\eqb{l}
p_\mrc = -\unde{E_n^{eff}}g_\mrn/J_\mrc\,.
\eqe
\end{remark}

The second derivatives of \eqref{e:ft1} and \eqref{e:fm3} for the active case are simply
\eqb{l}
\ds\frac{1}{2}\frac{\partial^2\overrightarrow{\unde{f_T}}}{\partial T^2_\mrc}  = \ds\frac{1}{T_0^2}
\label{e:ddft1}\eqe
and
\eqb{l}
\ds\frac{1}{2}\paqq{\overrightarrow{\unde{f_m}}}{\bg_\mre}{\bg_\mre} = \frac{\bone}{g_0^2}\,.
\label{e:ddfm3}\eqe
Given \eqref{e:ft1}-\eqref{e:dfm3} and \eqref{e:ddft1}-\eqref{e:ddfm3} all terms in \eqref{e:ptMS} can then be evaluated.
This then allows to evaluate the increments in \eqref{e:DtMSc}, which is elaborated further in Appendix~\ref{s:lin-ii}. 

\subsection{Mechano-sensitive osseointegration }\label{s:con4}

\begin{figure}[h!]
\begin{center} \unitlength1cm
\begin{picture}(0,4.7)
\put(-5.67,.25){\includegraphics[height=43mm]{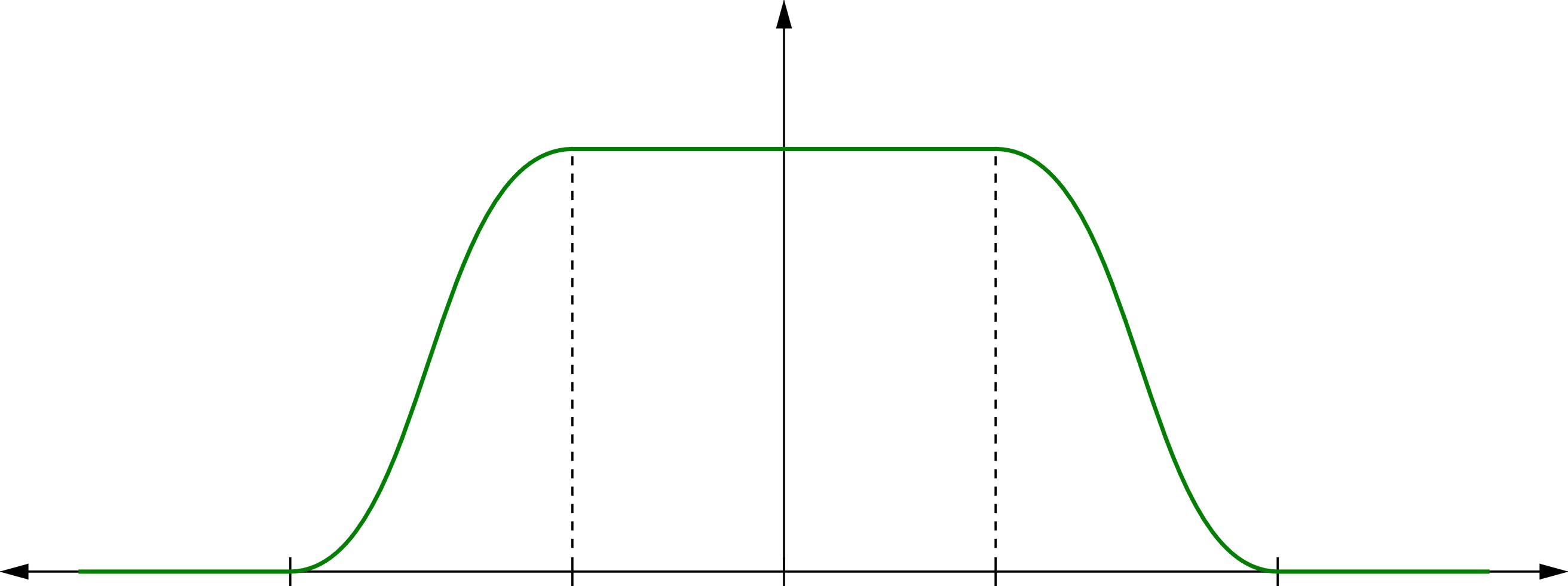}}
\put(0.2,3.9){\small $\overrightarrow{k}$}
\put(-5.6,-.05){\small $p_\mathrm{c}$}
\put(-3.8,-.05){\small $p_\mathrm{lim}$}
\put(-1.8,-.05){\small $p_\mathrm{opt}$}
\put(0,-.05){\small 0}
\put(1.4,-.05){\small $g_\mathrm{opt}$}
\put(3.4,-.05){\small $g_\mathrm{lim}$}
\put(5.5,-.05){\small $g_\mathrm{n}$}
\end{picture}
\caption{Mechano-sensitive osseointegration: 
Dependency of the bonding reaction rate $\protect\overrightarrow{k} = c_\mrr\,\protect\overrightarrow{\unde{K_c}}/\unde{n_c}$ 
on the contact pressure and contact gap according to Eqs.~\eqref{e:crosseo}  and \eqref{e:Kc}.}
\label{f:fm2}
\end{center}
\end{figure}

The final example proposes a model for bone-implant bonding, so-called \textit{osseointegration}, where the bonding state depends on the mechanical contact state but not vice versa.
The model is thus a one-way coupled model, which allows for a simple sequential solution scheme as is demonstrated in Sec.~\ref{s:ex2}. 
This can be achieved by making $c_\mrr$ instead of $\overrightarrow{\unde{K_c}} = \overrightarrow{\unde{K_0}}\,\overrightarrow{\unde{f_m}}\,\overrightarrow{\unde{f_T}}$ a function of the mechanical contact state.
From \eqref{e:ptMS} then follows $\partial\undetc/\partial\phi = \mathbf{0}$ (as long as there are no debonding forces).
The effect on the bonding ODE \eqref{e:bondODE} is the same, but no mechanical forces are generated from this, which is in contrast to the cases in Secs.~\ref{s:con1}-\ref{s:con3}.
This one-way coupling is reasonable, as osseointegration proceeds at large time scales and hence should not have a significant effect on the contact forces.
The proposed osseointegration model is based on the pressure- and separation-dependent reaction rate function
\eqb{l}
c_\mrr(\bg_\mre) = \ds\frac{1}{2} \left\{\begin{array}{ll}
\cos\bigg(\ds\frac{g_\mrn-g_\mathrm{opt}}{g_\mathrm{lim}-g_\mathrm{opt}}\pi\bigg) + 1 ~ & 
	$for$~~g_\mathrm{opt} \leq g_\mrn \leq g_\mathrm{lim}\,, \\[4mm]	
2 & $for$~~g_\mrn \leq g_\mathrm{opt}~~\&~~p_\mrc \leq p_\mathrm{opt}\,, \\[1mm]
\cos\bigg(\ds\frac{p_\mrc-p_\mathrm{opt}}{p_\mathrm{lim}-p_\mathrm{opt}}\pi\bigg) + 1 ~ & 
	$for$~~p_\mathrm{opt} \leq p_\mrc \leq p_\mathrm{lim}\,, \\[3mm]	
0 & $else$\,,
\end{array}\right.
\label{e:crosseo}\eqe
where $p_\mrc = -E_\mrn\,g_\mrn$ is the contact pressure according to the penalty method, and
$g_\mathrm{opt}$, $g_\mathrm{lim}$, $p_\mathrm{opt}$ and $p_\mathrm{lim}$ are model constants.
Function \eqref{e:crosseo} is motivated by the observation that implant osseointegration is inhibited by large contact pressures and gaps \citep{kienapfel99,sotto10}.
Fig.~\ref{f:fm2} illustrates the bonding reaction rate resulting from \eqref{e:crosseo} for constant~$\overrightarrow{\unde{K_c}}$.
For a fixed mechanical contact state (i.e.~$c_\mrr$ is not a function of time), ODE \eqref{e:bondODE}
can then be solved analytically at every contact point, as is shown in Appendix \ref{s:anaODE}, see Eq.~\eqref{e:anaphi}. 
The bonding state can thus be easily evaluated, even for complicated mechanical contact states.
This is illustrated in the example of Sec.~\ref{s:ex2}.

An alternative to osseointegration model \eqref{e:crosseo}, is to keep $c_\mrr$ constant and instead pick $\overrightarrow{\unde{f_m}}$ to follow the function in \eqref{e:crosseo}. 
This leads to exactly the same bonding ODE, but now the coupling is two-way and the mechanical contact state is affected, as Appendix \ref{s:con4a} shows.

\section{Finite element formulation}\label{s:FE}

This section presents the finite element (FE) formulation of the preceding large deformation contact model.
The focus is placed on the discretization of the mechanical, thermal and chemical contact terms, and their coupling.
The bulk terms, on the other hand, are taken from known thermoelasticity models and therefore not discussed in detail.
The novelty of the formulation lies in chemical contact and its coupling to the mechanical and thermal contact conditions.
The thermo-mechanical contact formulation itself is not a focus, and it is therefore described here by a simple penalty formulation.
However, the present formulation is general, and can in principle be combined with more advanced mechanical contact formulations in future work.
The nonlinear system of equations is solved iteratively using the Newton-Raphson method. 
This requires the linearization of the FE equations, which is therefore also presented here.

\subsection{Discretized field variables}

The two contacting bodies are distinguished into \textit{master} and \textit{slave} (see Fig.~\ref{f:cocobo})
and discretized into $n_\mathrm{el}$ finite elements using $n_\mathrm{no}$ nodes.
The total set of elements contains bulk elements and surface elements.
Scalar fields $s$ and vector fields $\bv$ are discretized on all elements by the interpolations
\eqb{l}
s \approx \bar\mN_e\,\mss_e\,,\qquad
\bv \approx \mN_e\,\mv_e\,,
\eqe
where $\mss_e = [s_1,\,s_2,\,...,\,s_{n_e}]^\mrT$ and $\mv_e = [\bv_1,\,\bv_2,\,...,\,\bv_{n_e}]^\mrT$ contain the $n_e$ nodal values of fields $s$ and $\bv$, and
$\bar\mN_e = [N_1,\,N_2,\,...,\,N_{n_e}]$ and $\mN_e = [\bone N_1,\,\bone N_2,\,...,\,\bone N_{n_e} ]$ are the elemental shape function arrays of dimension $1\times n_e$ and $3\times 3n_e$, respectively.

The elemental contact integrals are evaluated on the slave surface using Gaussian quadrature, where every quadrature point $\bx_2$ is projected onto the neighboring master surface to obtain the closest projection point $\bx_\mrp$.
The two interaction cases of Eq.~\eqref{e:ccasesi} are considered here: 
In case of frictionless contact (Case (ii)), $\bx_2$ directly interacts with the current $\bx_\mrp$.
In case of sticking contact (and subsequent bonding; Case (i)), $\bx_2$ interacts with the initial projection point.
Therefore, the initial projection point $\bX_\mrp = \bx_\mrp|_{t=0}$ on the master surface is marked and updated over time according to the master deformation to obtain its current position $\bx_1$, required for Eq.~\eqref{e:ccasesi}.
The discretization of the slave and master deformation is given by
\eqb{rll}
\bx_\mrs = \bx_2 \ais \mN_e\,\mx^e_\mrs\,, \\[1mm]
\bx_\mrm \ais \mN_{\bar e}\,\mx^{\bar e}_\mrm\,,
\label{e:mx}\eqe
where $e$ denotes a slave surface element, and $\bar e$ the corresponding master element that contains the interaction point, which is either $\bx_\mrm = \bx_1$, in case of sticking, or $\bx_\mrm = \bx_\mrp$, in case of frictionless contact.
The integration is taken over all active slave surface elements, evaluating $\mN_e$ at the quadrature points and $\mN_{\bar e}$ at their corresponding interaction points.
From \eqref{e:mx} follow the elastic gap vector of Eq.~\eqref{e:ccasesi},
\eqb{l}
\bg_\mre \approx  \mN_e\,\mx^e_\mrs - \mN_{\bar e}\,\mx^{\bar e}_\mrm\,,
\label{e:geh}\eqe
and the tangent vectors
\eqb{l}
\ba^\mrm_\alpha := \ds\pa{\bx_\mrm}{\xi^\alpha} \approx \mN_{\bar e,\alpha}\,\mx^{\bar e}_\mrm\,,\qquad
\mN_{\bar e,\alpha} := \ds\pa{\mN_{\bar e}}{\xi^\alpha}\,,
\label{e:bah}\eqe
at the interaction point on the master surface along the two element coordinates $\xi^\alpha$, $\alpha =1,\,2$. 

For the linearization of the FE formulation, the increments $\Delta\bg_\mre^\mathrm{(i)}$ and $\Delta\ba^\mrp_\alpha$ are required, see Appendix~\ref{s:lin-ii}.
Their discretization follows from \eqref{e:geh}, (\ref{e:Dge0}.1) and \eqref{e:bah} as
\eqb{lll}
\Delta\bg_\mre^\mathrm{(i)} \ais  \mN_e\,\Delta\mx^e_\mrs - \mN_{\bar e}\,\Delta\mx^{\bar e}_\mrm\,,\\[1mm] 
\Delta\ba^\mrp_\alpha \ais \mN_{\bar e,\alpha}\,\Delta\mx^{\bar e}_\mrm\,.
\label{e:Dbah}\eqe
In Case (ii), $\mN_{\bar e}$ is evaluated at the projection point, which depends on master and slave surface deformation.
This has to be accounted for in the linearization of Case (ii).
It results in the increment
\eqb{l}
\Delta\mN_{\bar e} = \mN_{\bar e,\alpha}\,\Delta\xi^\alpha_\mrp\,,
\label{e:DNm}\eqe
where $\Delta\xi^\alpha_\mrp$, provided in Appendix~\ref{s:Dxi}, comes from changes in the projection due to changes in the deformation of master and slave surfaces. 

The slave surface discretization is used to discretize the bonding state $\phi$. 
Thus,
\eqb{l}
\phi \approx \bar\mN_e\,\bphi^e\,.
\label{e:phih}\eqe
Its linearization then gives
\eqb{l}
\Delta\phi \approx \bar\mN_e\,\Delta\bphi^e\,.
\label{e:Dphih}\eqe
Since the bonding state follows from an ODE here, it can be solved independently at each quadrature point.
It can thus be treated as an internal variable at the quadrature point level, instead of interpolating it from nodal values.
In this case discretizations \eqref{e:phih} and \eqref{e:Dphih} are not needed.

The temperatures of master, slave and interface, $T_1$, $T_2$ and $T_\mrc$, are discretized by
\eqb{llllll}
T_1 \ais \bar\mN_{\bar e} \,\mT_1^{\bar e}\,,~~ & \Delta T_1 \ais \bar\mN_{\bar e}\,\Delta\mT_1^{\bar e}\,, \\[1mm]
T_2 \ais \bar\mN_e \,\mT_2^e\,, & \Delta T_2 \ais \bar\mN_e\,\Delta\mT_2^e\,, \\[1mm]
T_\mrc \ais \bar\mN_e \,\mT_\mrc^e\,, & \Delta T_\mrc \ais \bar\mN_e\,\Delta\mT_\mrc^e\,,
\label{e:Tch}\eqe
analogously to \eqref{e:mx}, \eqref{e:phih} and \eqref{e:Dphih}.
Note that the master body temperature, $T_1$, uses the master body discretization, while $T_2$ and $T_\mrc$ use the slave body discretization, cf.~\eqref{e:mx}.

\subsection{Discrete equation of motion}\label{s:FEx}

Discretization of the mechanical field equations of both bodies leads to the well-known ODE system 
\eqb{l}
\mM\ddot\muu + \mf_\mathrm{int} +  \mf_\mrc - \mf_\mathrm{ext} = \mathbf{0}\,.
\label{e:ODEx}\eqe
In the following examples inertia is considered zero ($\mM\ddot\muu=\mathbf{0}$) and external forces $\mf_\mathrm{ext}$ constant, while the internal finite element force vector $\mf_\mathrm{int}$ is taken from known elasticity models: 3D linear elasticity in the first example, and Neo-Hookean membrane elasticity in the remaining examples. 
The vector $\mf_\mrc$ contains the elemental contact forces from the slave surface,
\eqb{l}
\mf^e_\mrs = -\ds\int_{\Omega^e_0} \mN^\mrT_e\,\undet^\unde{c}_2\,\dif A
= \ds\int_{\Omega^e_0} \mN^\mrT_\mre\,\undetc\,\dif A\,,
\eqe
and the corresponding elemental contact forces from the master surface,
\eqb{l}
\mf^e_\mrm = -\ds\int_{\Omega^e_0} \mN^\mrT_{\bar e}\,\undet^\unde{c}_1\,\dif A
= - \ds\int_{\Omega^e_0} \mN^\mrT_{\bar e}\,\undetc\,\dif A\,,
\eqe
that derive from the contact tractions of Eqs.~\eqref{e:sf_moc} and \eqref{e:tMS}.
Linearization then gives
\eqb{lllllllllll}
\Delta\mf^e_\mrs \is \ds\int_{\Omega^e_0} \mN^\mrT_e\, \Delta\undetc\,\dif A 
	\is \mk^e_\mathrm{ss}\,\Delta\mx^e_\mrs \plus \mk^e_\mathrm{sm}\,\Delta\mx^{\bar e}_\mrm 
	\plus \mk^e_\mathrm{s\phi}\,\Delta\bphi^e \plus \mk^e_\mathrm{si}\,\Delta\mT^e_\mrc \,, \\[4mm]
\Delta\mf^e_\mrm \is - \ds\int_{\Omega^e_0} \big(\mN^\mrT_{\bar e}\, \Delta\undetc + \Delta\mN^\mrT_{\bar e}\, \undetc\big)\,\dif A
	\is \mk^e_\mathrm{ms}\,\Delta\mx^e_\mrs \plus \mk^e_\mathrm{mm}\,\Delta\mx^{\bar e}_\mrm 
	\plus \mk^e_\mathrm{m\phi}\,\Delta\bphi^e \plus \mk^e_\mathrm{mi}\,\Delta\mT^e_\mrc \,,
\label{e:Dfe}\eqe
with the eight tangent blocks
\eqb{llllll}
\mk^e_\mathrm{ss} \dis \ds\pa{\mf^e_\mrs}{\mx^e_\mrs} 
	\is \ds\int_{\Omega^e_0} \mN^\mrT_e\pa{\undetc}{\bg_\mre^\mathrm{(i)}}\,\mN_e\,\dif A\,, \\[4mm] 
\mk^e_\mathrm{sm} \dis \ds\pa{\mf^e_\mrs}{\mx^{\bar e}_\mrm} 
	\is -\ds\int_{\Omega^e_0} \mN^\mrT_e\pa{\undetc}{\bg_\mre^\mathrm{(i)}}\,\mN_{\bar e}\,\dif A
	+ \ds\int_{\Omega^e_0} \mN^\mrT_e\pa{\undetc}{\ba^\mrp_\alpha}\,\mN_{\bar e,\alpha}\,\dif A \,, \\[4mm]
\mk^e_\mathrm{ms} \dis \ds\pa{\mf^e_\mrm}{\mx^e_\mrs} 
	\is {\mk^e_\mathrm{sm}}^{\!\!\!\mrT}\,, \\[3.5mm]
\mk^e_\mathrm{mm} \dis \ds\pa{\mf^e_\mrm}{\bx^{\bar e}_\mrm}  
	\is \ds\int_{\Omega^e_0} \mN^\mrT_{\bar e}\pa{\undetc}{\bg_\mre^\mathrm{(i)}}\,\mN_{\bar e}\,\dif A
	- 2\!\ds\int_{\Omega^e_0}\!\!\bigg(\!\mN^\mrT_{\bar e}\pa{\undetc}{\ba^\mrp_\alpha}\,\mN_{\bar e,\alpha}\!\bigg)^{\!\!\mrs}\dif A \\[4mm]
	& & \plus \ds\int_{\Omega^e_0}\mN^\mrT_{\bar e,\alpha}\,E_\mrn^\mathrm{eff} 
	c^{\alpha\beta}_\mrp g_\mrn^2\,\bn_\mrp\otimes\bn_\mrp\,\mN^\mrT_{\bar e,\beta}\,\dif A\,, \\[4mm]
\mk^e_\mathrm{s\phi} \dis \ds\pa{\mf^e_\mrs}{\bphi^e}
	 \is  \ds\int_{\Omega^e_0} \mN^\mrT_e\pa{\undetc}{\phi}\,\bar\mN_e\,\dif A\,,  \\[4mm]
\mk^e_\mathrm{m\phi}  \dis \ds\pa{\mf^e_\mrm}{\bphi^e}
	\is -\ds\int_{\Omega^e_0} \mN^\mrT_{\bar e}\pa{\undetc}{\phi}\,\bar\mN_e\,\dif A\,, \\[4mm]
\mk^e_\mathrm{si} 
	\dis \ds\pa{\mf^e_\mrs}{\mT^e_\mrc} \is  \ds\int_{\Omega^e_0} \mN^\mrT_e\pa{\undetc}{T_\mrc}\,\bar\mN_e\,\dif A\,,  \\[4mm]	 
\mk^e_\mathrm{mi} 
	\dis \ds\pa{\mf^e_\mrm}{\mT^e_\mrc} \is -\ds\int_{\Omega^e_0} \mN^\mrT_{\bar e}\pa{\undetc}{T_\mrc}\,\bar\mN_e\,\dif A\,,
\label{e:mkx}\eqe
that follow from inserting \eqref{e:Dtc}, \eqref{e:Dbah}, \eqref{e:Dphih}, (\ref{e:Tch}.2), \eqref{e:DNm} and \eqref{e:Dxi0} into \eqref{e:Dfe}, and then using~\eqref{e:tcxi}. 
Here, $(...)^\mrs := (... + ...^\mrT)/2$.
The second and third integrals in tangent blocks $\mk^e_\mathrm{sm}$, $\mk^e_\mathrm{ms}$ and $\mk^e_\mathrm{mm}$ are only present in Case (ii), but not in Case (i).
The derivatives appearing in \eqref{e:mkx} were derived in the previous section for various constitutive examples.
In particular, the expressions for $\partial\undetc/\partial\bg_\mre^\mathrm{(i)}$ and $\partial\undetc/\partial\ba^\mrp_\alpha$ can be found in \eqref{e:tigi} and \eqref{e:tiigi}, for the two examples from Sec.~\ref{s:con1} and \ref{s:con3}, respectively.
The submatrix
\eqb{l}
\mk^e_\mrx := \begin{bmatrix}
\mk^e_\mathrm{ss} & \mk^e_\mathrm{sm} \\[1mm]
\mk^e_\mathrm{ms} & \mk^e_\mathrm{mm}
\end{bmatrix},
\eqe
that characterizes pure mechanical coupling (between Eqs.~I and III in Fig.~\ref{f:cocobo}), is symmetric and consistent with the tangent given in \citet{spbf}.

\begin{remark}\label{r:NFEM}
The elemental force vectors and tangent matrices from above and Secs.~\ref{s:FEphi}-\ref{s:FET} are assembled into a global force vector and tangent matrix accounting for all the active contact elements on the current contact surface. 
This dependency on the current contact surface, as well as its underlying large deformation kinematics, makes the present formulation fully nonlinear, even when quadratic contact potentials are considered, such as those from Sec.~\ref{s:consti}.
\end{remark}

\subsection{Discrete reaction equation}\label{s:FEphi}

The reaction equation can be discretized in two ways.
One can use the usual nodal FE discretization, which is discussed first.
Alternatively, the bonding field can be treated as an internal variable and eliminated locally through static condensation, which is presented second.  

\subsubsection{Nodal formulation}

Discretization of the bonding ODE \eqref{e:bondODE} based on \eqref{e:phih} leads to the ODE system
\eqb{l}
\bar\mM\dot\bphi + \bar\mf_\mrc = \mathbf{0}\,,
\label{e:bondODEh}\eqe
where $\bar\mM$ and $\bar\mf_\mrc$ are assembled from
\eqb{l}
\bar\mm^e = \ds\int_{\Omega^e_0} \bar\mN_e^\mrT\bar\mN_e\frac{\unde{n_c}}{c_\mrr}\,\dif A
\eqe
and
\eqb{l}
\bar\mf_\mrc^e = \ds\int_{\Omega^e_0} \bar\mN_e^\mrT\,\unde{M_c}\,\dif A\,,
\eqe
respectively.
Linearization then leads to
\eqb{l}
\Delta\bar\mf_\mrc^e = \ds\int_{\Omega^e_0} \bar\mN_e^\mrT\,\Delta\unde{M_c}\,\dif A
	= \mk^e_\mathrm{\phi s}\,\Delta\mx^e_\mrs + \mk^e_\mathrm{\phi m}\,\Delta\mx^{\bar e}_\mrm 
	+ \mk^e_\mathrm{\phi\phi}\,\Delta\bphi^e + \mk^e_\mathrm{\phi i}\,\Delta\mT^e_\mrc \,,
\label{e:Dfbc}\eqe
with
\eqb{lllll}
\mk^e_\mathrm{\phi s} \dis \ds\pa{\bar\mf^e_\mrc}{\mx^e_\mrs} 
	\is \ds\int_{\Omega^e_0} \bar\mN^\mrT_e\pa{\unde{M_c}}{\bg_\mre}\,\mN_e\,\dif A = {\mk^e_\mathrm{s\phi}}^{\!\!\mrT}, \\[4mm]
\mk^e_\mathrm{\phi m} \dis \ds\pa{\bar\mf^e_\mrc}{\mx^{\bar e}_\mrm} 
	\is -\ds\int_{\Omega^e_0}\bar\mN^\mrT_e\pa{\unde{M_c}}{\bg_\mre}\,\mN_{\bar e}\,\dif A = {\mk^e_\mathrm{m\phi}}^{\!\!\mrT},\\[4mm]
\mk^e_\mathrm{\phi\phi} \dis \ds\pa{\bar\mf^e_\mrc}{\bphi^e} 
	\is \ds\int_{\Omega^e_0} \bar\mN^\mrT_e\pa{\unde{M_c}}{\phi}\,\bar\mN_e\,\dif A\,, \\[4mm]
\mk^e_\mathrm{\phi i}
	\dis \ds\pa{\bar\mf^e_\mrc}{\mT^e_\mrc} \is \ds\int_{\Omega^e_0} \bar\mN^\mrT_e\pa{\unde{M_c}}{T_\mrc}\,\bar\mN_e\,\dif A\,,
\label{e:kphi}\eqe
due to (\ref{e:DMSii}.1), (\ref{e:Dbah}.1), \eqref{e:Dphih} and (\ref{e:Tch}.2).
The second equalities in (\ref{e:kphi}.1) and (\ref{e:kphi}.2) follow from \eqref{e:ptMS} and \eqref{e:mkx}.
This makes the submatrix
\eqb{l}
\begin{bmatrix}
\mk^e_\mathrm{ss} & \mk^e_\mathrm{sm} & \mk^e_\mathrm{s\phi} \\[1mm]
\mk^e_\mathrm{ms} & \mk^e_\mathrm{mm} & \mk^e_\mathrm{m\phi} \\[1mm]
\mk^e_\mathrm{\phi s} & \mk^e_\mathrm{\phi m} & \mk^e_\mathrm{\phi\phi}
\end{bmatrix},
\eqe
that characterizes the chemo-mechanical contact coupling (between Eqs.~I, III and V, in Fig.~\ref{f:cocobo}), symmetric.
ODE system \eqref{e:bondODEh} can be solved by a time stepping algorithm such as the trapezoidal rule presented in the following subsection.

\subsubsection{Quadrature point formulation}\label{s:QPform}

Since the bonding ODE \eqref{e:bondODE} contains no spatial derivatives, it can be integrated in time at each FE quadrature point.
This is done here with the trapezoidal rule
\eqb{l}
\phi_{n+1} \approx \phi_n + \Delta t_{n+1}\big((1-\gamma)\,\dot\phi_n + \gamma\,\dot\phi_{n+1} \big)\,,
\label{e:trap}\eqe
which is second order accurate in time for $\gamma=1/2$ and includes the explicit and implicit Euler schemes for $\gamma=0$ and $\gamma=1$, respectively.
Inserting \eqref{e:trap} into \eqref{e:bondODE} leads to the time discretized ODE
\eqb{l}
\unde{\bar m}\,\phi_{n+1} + \gamma\,\Delta t_{n+1}\,\unde{M_c}(\phi_{n+1}) = \unde{\bar m}\,\phi_n + (1-\gamma)\,\unde{\bar m}\,\Delta t_{n+1}\,\dot\phi_n\,,
\label{e:dtODE}\eqe
that can be solved for $\phi_{n+1}$ at each quadrature point separately; here $\unde{\bar m} := \unde{n_c}/c_\mrr$.
The bonding state $\phi$ thus becomes a history variable that is stored at each quadrature point.
It can be eliminated from the global system of equations as follows:
Linearizing \eqref{e:dtODE} at the current time step gives (skipping index $n+1$)
\eqb{l}
\unde{\bar m}\,\Delta\phi + \gamma\,\Delta t\,\Delta\unde{M_c} = 0\,,
\eqe
which can be used to eliminate $\Delta\unde{M_c}$ from (\ref{e:DMSii}.1) to get 
\eqb{l}
\Delta\phi = - \ds\bigg(\frac{\unde{\bar m}}{\gamma\,\Delta t} + \pa{\unde{M_c}}{\phi} \bigg)^{\!\!-1}
	\bigg(\pa{\unde{M_c}}{\bg_\mre}\cdot\Delta\bg_\mre^\mathrm{(i)} + \pa{\unde{M_c}}{T_\mrc}\,\Delta T_\mrc\bigg).
\label{e:dphi_qp}\eqe
That is, the update $\Delta\phi$ is fully determined by $\Delta\bg_\mre^\mathrm{(i)}$ and $\Delta T_\mrc$.
Inserting \eqref{e:dphi_qp} into \eqref{e:Dtc} and (\ref{e:DMSii}.2), and using some of the identities of \eqref{e:ptMS} then gives
\eqb{rll}
\Delta\undetc \is \ds\widehat{\pa{\undetc}{\bg_\mre^\mathrm{(i)}}}\,\Delta\bg_\mre^\mathrm{(i)} 
	+ \ds\widehat{\pa{\undetc}{T_\mrc}}\Delta T_\mrc + \pa{\undetc}{\ba^\mrp_\beta} \,\Delta\ba_\beta^\mrp \,, \\[4.5mm]
\Delta\unde{S_c} \is \ds\widehat{\pa{\unde{S_c}}{\bg_\mre^\mathrm{(i)}}}\,\Delta\bg_\mre^\mathrm{(i)} 
	+ \ds\widehat{\pa{\unde{S_c}}{T_\mrc}}\,\Delta T_\mrc\,,
\eqe
with
\eqb{rll}
\ds\widehat{\pa{\undetc}{\bg_\mre^\mathrm{(i)}}} \dis \ds\pa{\undetc}{\bg_\mre^\mathrm{(i)}} 
	- \bigg(\frac{\unde{\bar m}}{\gamma\,\Delta t} + \pa{\unde{M_c}}{\phi} \bigg)^{\!\!-1}\,\pa{\undetc}{\phi}\otimes\pa{\undetc}{\phi}\,, \\[4.5mm]
\ds\widehat{\pa{\undetc}{T_\mrc}} \dis \ds\pa{\undetc}{T_\mrc} 
	+ \bigg(\frac{\unde{\bar m}}{\gamma\,\Delta t} + \pa{\unde{M_c}}{\phi} \bigg)^{\!\!-1}\,\pa{\undetc}{\phi}\,\pa{\unde{S_c}}{\phi} 
	= - \ds\widehat{\pa{\unde{S_c}}{\bg_\mre}}\,, \\[4mm]
\ds\widehat{\pa{\unde{S_c}}{T_\mrc}} \dis \ds\pa{\unde{S_c}}{T_\mrc} 
	+ \bigg(\frac{\unde{\bar m}}{\gamma\,\Delta t} + \pa{\unde{M_c}}{\phi} \bigg)^{\!\!-1}\bigg(\pa{\unde{S_c}}{\phi}\bigg)^{\!\!2}\,,
\eqe
which retains the original symmetry of the formulation.\\
In case there is no temperature dependency, $\Delta T_\mrc = 0$ and $\Delta\unde{S_c} = 0$.
In this case the FE system, which now only consists of the mechanical ODE \eqref{e:ODEx},
thus only requires the stiffness matrices $\mk^e_\mathrm{ss}$, $\mk^e_\mathrm{ms} = {\mk^e_\mathrm{sm}}^{\!\!\!\mrT}$ and $\mk^e_\mathrm{mm}$ that are as given in \eqref{e:mkx}, but with $\partial\undetc/\partial\bg_\mre^\mathrm{(i)}$ replaced by $\widehat{\partial\undetc}/\partial\bg_\mre^\mathrm{(i)}$.
For the chemical potentials of (\ref{e:tMS1}.2) and (\ref{e:fttMS}.2), Eq.~\eqref{e:dtODE} is a linear function of $\phi_{n+1}$ that is easily solved in closed form.

\subsection{Discrete heat equation}\label{s:FET}

After FE discretization the heat equation for each body (master body $\sB_\mrm=\sB_1$, slave body $\sB_\mrs=\sB_2$ and interfacial medium $\sB_\mri$) takes the form
\eqb{l}
\mC\dot\mT + \tilde\mf_\mathrm{int} + \tilde\mf_\mrc - \tilde\mf_\mathrm{ext} = \mathbf{0}\,.
\label{e:FET}\eqe
Here, $\mC$ is the heat capacity matrix assembled from the elemental contributions
\eqb{l}
\mcc_e = \ds\int_{\Omega^e_0} \bar\mN_e^\mrT\bar\mN_e\,\unde{C}\,\dif V\,.
\eqe
This volume integration is replaced by a surface integration in case the corresponding body is very thin and discretized by surface FE. 
This is always considered for the interface, here.
In the examples of Sec.~\ref{s:ex3a}-\ref{s:ex3d}, it is also considered for the slave body.
The heat capacities of the three bodies, $\unde{C_1}$, $\unde{C_2}$ and $\unde{C_c}$, 
are taken as constant here, see Remark~\ref{r:C}.\\
The internal heat flux vector follows classical heat conduction, $\tilde\mf_\mathrm{int} = \tilde\mK\mT$,
where the conductivity matrix $\tilde\mK$ is considered constant and isotropic, here.
It is assembled from the elemental contributions
\eqb{l}
\tilde\mk_e = \ds\int_{\Omega^e_0} \bar\mN_{e,\alpha}^\mrT\,\unde{k}\,A^{\alpha\beta}\,\bar\mN_{e,\beta}\,\dif A\,,
\eqe
for thin-walled bodies.
Here $\unde{k}$ is the thermal surface conductivity (units Watt/Kelvin), which follows from its volumetric counterpart (Watt/(meter Kelvin)) multiplied by the body's thickness. 
The present formulation assumes the interfacial body so thin that its $\unde{k}$ becomes negligible, and hence there is no $\tilde\mf_\mathrm{int}$ considered in the discrete interfacial heat equation. 
\\
The external heat flux vector $\tilde\mf_\mathrm{ext}$ captures heat sources in or on the bodies, which are considered zero for master and slave body.
On the interface the chemical (de)bonding reactions produce the heat $-\mu_\mrc\,R_\mrc$ (per current area, or $-\mu_\mrc\,\unde{R_c}$ per reference area, cf.~Eq.~(\ref{e:tMS2}.2)). 
The external heat source for a FE of the interface thus is\footnote{Friction, which also generates heat on the interface, is not considered in the FE formulation here.}
\eqb{l}
\tilde\mf_\mathrm{ext}^e 
= -\!\ds\int_{\Omega^e_0} \bar\mN_e^\mrT \mu_\mrc\,\unde{R_c}\,\dif A
= \ds\int_{\Omega^e_0} \bar\mN_e^\mrT \unde{\bar m}\,\dot\phi^2\,\dif A\,.
\label{e:ftext}\eqe
Here the latter identity follows from \eqref{e:bondODE0}, $\unde{M_c}=\unde{n_c}\mu_c$ and \eqref{e:bondODE} for $\unde{\bar m} := \unde{n_c}/c_\mrr$.
Since $\unde{\bar m}\,\dot\phi^2>0$, any changes in the bonding state generate heat. 
\\
The remaining term in \eqref{e:FET}, $\tilde\mf_\mrc$ (defined positive as an outflow), is assembled from the elemental heat fluxes due to thermal contact.
These are the inflows on the master and slave surfaces,
\eqb{lll}
\tilde\mf^e_1 \is -\!\ds\int_{\Omega^e} \bar\mN_{\bar e}^\mrT\,q^\mrc_1\,\dif a = -\!\ds\int_{\Omega^e_0} \bar\mN_{\bar e}^\mrT\,\unde{q^c_1}\,\dif A \,,\\[4mm]
\tilde\mf^e_2 \is -\!\ds\int_{\Omega^e} \bar\mN_e^\mrT\,q^\mrc_2\,\dif a = -\!\ds\int_{\Omega^e_0} \bar\mN_e^\mrT\,\unde{q^c_2}\,\dif A\,,
\label{e:ftq1}\eqe
which equilibrate the outflow from the interface,
\eqb{l}
\tilde\mf^e_\mri = \ds\int_{\Omega^e_0} \bar\mN_e^\mrT\,\big(\unde{q^c_1}+\unde{q^c_2}\big)\,\dif A
\label{e:ftq2}\eqe
according to Fig.~\ref{f:qc}a.
Here, the fluxes $q^\mrc_k$ are given by the simple heat transfer model \eqref{e:qck}.
The equations above complete the general \textit{six-field formulation} of Fig.~\ref{f:cocobo}:

\fbox{\begin{minipage}{158.7mm}
\textit{Six-field formulation:} Solve \eqref{e:FET} together with \eqref{e:ODEx} and \eqref{e:bondODEh} for $\bx_k$, $\phi$, $T_k$ and $T_\mrc$ ($k=1,2$) using the heat flux vectors in \eqref{e:ftext}-\eqref{e:ftq2}.
Those flux vectors use the viewpoint of Fig.~\ref{f:qc}a.
\end{minipage}}

Using \eqref{e:qct} and \eqref{e:qcm0} the contact heat fluxes can be reformulated into 
\eqb{lll}
\tilde\mf^e_1 \is -\!\ds\int_{\Omega^e_0} \bar\mN_{\bar e}^\mrT\,\unde{q^c_m}\,\dif A - \!\int_{\Omega^e_0} \bar\mN_{\bar e}^\mrT\,\unde{q^c_t}\,\dif A \,, \\[4mm]
\tilde\mf^e_2 \is -\!\ds\int_{\Omega^e_0} \bar\mN_e^\mrT\,\unde{q^c_m}\,\dif A + \!\int_{\Omega^e_0} \bar\mN_e^\mrT\,\unde{q^c_t}\,\dif A \,, \\[4mm]
\tilde\mf^e_\mri \is 2\!\ds\int_{\Omega^e_0} \bar\mN_e^\mrT\,\unde{q^c_m}\,\dif A\,,
\label{e:ftq3}\eqe
see Fig.~\ref{f:qc}b.
For the special heat transfer coefficients $h_1 = h_2 =: 2h$, one for instance finds
\eqb{l}
\unde{q^c_m} = \unde{h}\big(2T_\mrc - T_1 - T_2\big)\,,\qquad
\unde{q^c_t} = \unde{h}\big(T_2 - T_1\big)\,.
\label{e:qmth}\eqe
Given Eqs.~\eqref{e:ftext} and (\ref{e:ftq3}.3), the discrete heat equation for the interface simplifies to
\eqb{l}
\mC_\mrc\dot\mT_\mrc + \tilde\mf_\mathrm{eff} = \mathbf{0}\,,\qquad \tilde\mf^e_\mathrm{eff} := \tilde\mf^e_\mri - \tilde\mf^e_\mathrm{ext} 
= \ds\int_{\Omega^e_0} \bar\mN_e^\mrT\Big(2\unde{q^c_m} + \mu_\mrc\,\unde{R_c}\Big)\dif A\,.
\label{e:FET6}\eqe
If $\mC_\mrc = \mathbf{0}$, this is solved by $\unde{q^c_m} = -\mu_\mrc\,\unde{R_c}/2 = \unde{\bar m}\,\dot\phi^2/2$,
and temperature field $T_\mrc$ is not needed anymore. 
This case is referred to as the \textit{five-field formulation}:

\fbox{\begin{minipage}{158.7mm}
\textit{Five-field formulation:} 
Solve \eqref{e:FET} together with \eqref{e:ODEx} and \eqref{e:bondODEh} for $\bx_k$, $\phi$ and $T_k$ ($k=1,2$) using the heat fluxes in (\ref{e:ftq3}.1) and (\ref{e:ftq3}.2) for $\unde{q^c_m} = \unde{\bar m}\,\dot\phi^2/2$.
Those fluxes use the viewpoint of Fig.~\ref{f:qc}b.
\end{minipage}}

If one body, e.g.~$\sB_1$, is thermally fixed, temperature field $T_1$ is not needed anymore, which results in a \textit{four-field formulation} for $\bx_k$, $\phi$ and $T_2$.
If deformation $\bx_1$ is also fixed, a \textit{three-field formulation} for $\bx_2$, $\phi$ and $T_2$ follows.
The numerical examples in Secs.~\ref{s:ex3a}-\ref{s:ex3d} illustrate these chemo-thermo-mechanical three-, four-, five- and six-field formulations one by one.

\begin{remark}\label{r:C}
In finite thermoelasticity, the heat capacity is defined by
\eqb{l}
\unde{C} := T\ds\pa{\unde{S}}{T}\,,
\eqe
e.g.~see \citet{holzapfel}.
This $\unde{C}$ is only constant if the free energy has the thermal part
\eqb{l}
\unde{\Psi} = \unde{C}\bigg(T-T_0-T\ln\ds\frac{T}{T_0}\bigg)\,.
\label{e:PsiT}\eqe
The entropy in this case is
\eqb{l}
\unde{S} := -\ds\pa{\unde{\Psi}}{T} = \unde{C}\ln\ds\frac{T}{T_0}\,.
\label{e:ST}\eqe
The fourth term in \eqref{e:Psi0} is the 2nd order Taylor expansion of \eqref{e:PsiT}, and the leading term in \eqref{e:Sc} is the linearization of \eqref{e:ST} (for $T_\mrc = T$ and $\unde{C_\mrc} = \unde{C}$).
In this linearized setting, the heat capacity becomes $\unde{C} = T_0\partial\unde{S}/\partial T$,
which is again constant.
For a generalization of \eqref{e:Psi0}, its fourth term can be replaced by \eqref{e:PsiT}.
\end{remark}

\begin{remark}
No thermoelastic heating is considered in \eqref{e:FET}.
Such a term comes from a temperature dependence of the stresses, e.g.~see \citet{holzapfel}.
If the contact traction $\undetc$ is temperature-dependent, it also contributes to thermoelastic heating; 
likewise $\unde{M_c}$.
This is not the case when $\overrightarrow{\unde{f_T}}$ and $\overleftarrow{\unde{f_T}}$ are const., which is the case considered in the following examples.
This means that $\partial\undetc/\partial T_\mrc$ and $\partial\unde{M_c}/\partial T_\mrc$ are zero and the corresponding tangent matrices vanish.
\end{remark}

Next the linearization of the thermal problem in \eqref{e:FET} is presented.
Contributions $\mC\dot\mT$ and $\tilde\mf_\mathrm{int}$ are linear in $\dot\mT$ and $\mT$, respectively, and considered independent of deformation and bonding state.
This then leaves us with the linearization of $\tilde\mf_\mathrm{ext}$ and $\tilde\mf_\mrc$.

\subsubsection{Linearization of $\tilde\mf^e_\mathrm{ext}$}

The dependency of $\tilde\mf_\mathrm{ext}$ on the bonding state in \eqref{e:ftext} leads to the linearization 
\eqb{l}
\Delta\tilde\mf^e_\mathrm{ext} = 2\!\ds\int_{\Omega^e_0} \bar\mN_e^\mrT \unde{\bar m}\,\dot\phi\,\pa{\dot\phi}{\phi}\Delta\phi\,\dif A 
\label{e:Dtfext}\eqe
for constant $\unde{\bar m}$.
For the trapezoidal rule, used for the time discretization of $\phi$ in \eqref{e:trap},
\eqb{l}
\ds\pa{\dot\phi}{\phi} = \frac{1}{\gamma\,\Delta t}
\eqe
at the current time step.
In case $\phi$ is interpolated from nodal values, inserting Eq.~\eqref{e:Dphih} leads to
\eqb{l}
\Delta\tilde\mf^e_\mathrm{ext} = \mk^e_{\mri\phi}\,\Delta\bphi^e,
\label{e:Dftext}\eqe
with
\eqb{l}
\mk^e_{\mri\phi} := \ds\pa{\tilde\mf^e_\mathrm{ext}}{\bphi^e} =  
	2\!\ds\int_{\Omega^e_0} \bar\mN_e^\mrT\,\bar\mN_e\,\frac{\unde{\bar m}\,\dot\phi}{\gamma\,\Delta t}\,\dif A\,.
\eqe
In case $\phi$ is a history variable at the quadrature point, inserting Eq.~\eqref{e:dphi_qp} into \eqref{e:Dtfext} leads to
\eqb{l}
\Delta\tilde\mf^e_\mathrm{ext} = \widehat\mk^e_{\mri\mrs}\,\Delta\mx^e_\mrs 
	+ \widehat\mk^e_{\mri\mrm}\,\Delta\mx^{\bar e}_\mrm 
	+ \widehat\mk^e_\mathrm{ii}\,\Delta\mT^e_\mrc\,,
\label{e:Dftext2}\eqe
with
\eqb{lll}
\widehat\mk^e_{\mri\mrs} \dis -2\!\ds\int_{\Omega^e_0} \bar\mN_e^\mrT \dot\phi
	\bigg(1+ \frac{\gamma\,\Delta t}{\unde{\bar m}}\pa{\unde{M_c}}{\phi} \bigg)^{\!\!-1}
	\pa{\unde{M_c}}{\bg_\mre}\cdot\mN_e\,\dif A\,, \\[4mm]
\widehat\mk^e_{\mri\mrm} \dis ~~2\!\ds\int_{\Omega^e_0} \bar\mN_e^\mrT \dot\phi
	\bigg(1+ \frac{\gamma\,\Delta t}{\unde{\bar m}}\pa{\unde{M_c}}{\phi} \bigg)^{\!\!-1}
	\pa{\unde{M_c}}{\bg_\mre}\cdot\mN_{\bar e}\,\dif A\,,	\\[4mm]
\widehat\mk^e_\mathrm{ii} \dis -2\!\ds\int_{\Omega^e_0} \bar\mN_e^\mrT \dot\phi
	\bigg(1+ \frac{\gamma\,\Delta t}{\unde{\bar m}}\pa{\unde{M_c}}{\phi} \bigg)^{\!\!-1}
	\pa{\unde{M_c}}{T_\mrc}\,\bar\mN_e\,\dif A\,.
\eqe
Note that $\widehat\mk^e_{\mri\mrm} \neq -\widehat\mk^e_{\mri\mrs}$ due to the different shape function evaluation of slave and master surface.

\subsubsection{Linearization of the six-field formulation}\label{s:6f-lin}

Next the linearization of $\tilde\mf^e_\mrc$ is presented -- first according to the six-field formulation, where $\tilde\mf^e_\mrc$ contains the elemental contributions of \eqref{e:ftq1} and \eqref{e:ftq2}.
Their linearization requires the linearization of heat transfer model \eqref{e:qck}, which yields
\eqb{lll}
\Delta\unde{q^c_1} \is \unde{h_1} \big(\Delta T_\mrc - \Delta T_1 - T_{1,\alpha}\,\Delta\xi^\alpha_\mrp \big)\,, \\[1mm]
\Delta\unde{q^c_2} \is \unde{h_2} \big(\Delta T_\mrc - \Delta T_2 \big)
\eqe
for constant $\unde{h}_k$ and accounting for the fact that $T_1$ is evaluated at the closest projection point defined by coordinate $\xi^\alpha_\mrp$.
In the discrete case, the temperature increments follow from \eqref{e:Tch}, while $\Delta\xi^\alpha_\mrp$ is given in Appendix~\ref{s:Dxi}.
The derivative $T_{1,\alpha} := \partial T_1/\partial\xi^\alpha$ vanishes if the temperature distribution across the master surface is uniform, which is the case in the examples of Sec.~\ref{s:ex}.
The linearization of \eqref{e:ftq1} and \eqref{e:ftq2} thus lead to
\eqb{lll}
\Delta\tilde\mf^e_1 \is -\!\ds\int_{\Omega^e_0} \!\bar\mN_{\bar e}^\mrT\,\Delta\unde{q^c_1}\,\dif A 	
	-\!\ds\int_{\Omega^e_0} \!\Delta\bar\mN_{\bar e}^\mrT\,\unde{q^c_1}\,\dif A 
	= \mk^e_{11}\,\Delta\mT_1^{\bar e} + \mk^e_{1\mri}\,\Delta\mT_\mrc^e 
	+ \mk^e_{1\mrs}\,\Delta\mx_\mrs^e + \mk^e_{1\mrm}\,\Delta\mx_\mrm^{\bar e} \,,\\[4mm]
\Delta\tilde\mf^e_2 \is -\!\ds\int_{\Omega^e_0} \!\bar\mN_e^\mrT\,\Delta\unde{q^c_2}\,\dif A
	\hspace{20.5mm} = \mk^e_{22}\,\Delta\mT_2^e + \mk^e_{2\mri}\,\Delta\mT_\mrc^e \,, \\[4mm] 
\Delta\tilde\mf^e_\mri \is \ds\int_{\Omega^e_0} \!\bar\mN_e^\mrT\big(\Delta\unde{q^c_1}+\Delta\unde{q^c_2}\big)\,\dif A
	= \mk^e_{\mri1}\,\Delta\mT_1^{\bar e}  + \mk^e_{\mri2}\,\Delta\mT_2^e + \mk^e_{\mri\mri}\,\Delta\mT_\mrc^e 
	+ \mk^e_{\mri\mrs}\,\Delta\mx_\mrs^e + \mk^e_{\mri\mrm}\,\Delta\mx_\mrm^{\bar e}\,, 
\label{e:Dftc}\eqe
with
\eqb{llllll}
\mk^e_{11} \dis \ds\pa{\tilde\mf^e_1}{\mT_1^{\bar e}} = \ds\int_{\Omega^e_0} \!\unde{h_1}\,\bar\mN_{\bar e}^\mrT\,\bar\mN_{\bar e}\,\dif A \,, &
	\mk^e_{1\mri} \dis \ds\pa{\tilde\mf^e_1}{\mT_\mrc^e} 
	= -\!\ds\int_{\Omega^e_0} \!\unde{h_1}\,\bar\mN_{\bar e}^\mrT\,\bar\mN_e\,\dif A \neq - \mk^e_{11}\,, \\[4mm]
\mk^e_{22} \dis \ds\pa{\tilde\mf^e_2}{\mT_2^e} = \ds\int_{\Omega^e_0} \!\unde{h_2}\,\bar\mN_e^\mrT\,\bar\mN_e\,\dif A \,, &
	\mk^e_{2\mri} \dis \ds\pa{\tilde\mf^e_2}{\mT_\mrc^e} = - \mk^e_{22}\,, \\[4mm]	
\mk^e_{\mri\mri} \dis \ds\pa{\tilde\mf^e_\mri}{\mT_\mrc^e} = \ds\int_{\Omega^e_0} \!(\unde{h_1}+\unde{h_2})\,\bar\mN_e^\mrT\,\bar\mN_e\,\dif A \,,~~ &
	\mk^e_{\mri1} \dis \ds\pa{\tilde\mf^e_\mri}{\mT_1^{\bar e}} = \mk_{1\mri}^{e\,\mrT}\,, ~~~~~ 
	\mk^e_{\mri2} := \ds\pa{\tilde\mf^e_\mri}{\mT_2^e} = \mk_{2\mri}^{e\,\mrT}\,, 	
\eqe
and
\eqb{lll}
\mk^e_{1\bullet} \is \ds\pa{\tilde\mf^e_1}{\mx_\bullet^e} 
	= \ds\int_{\Omega^e_0} \!\Big(\unde{h_1}\,T_{1,\alpha}\,\bar\mN_{\bar e}^\mrT - \unde{q_1^\mrc}\,\bar\mN_{\bar e,\alpha}^\mrT\Big)
	\pa{\xi^\alpha_\mrp}{\mx_\bullet^e}\,\dif A \,, \\[4mm] 
\mk^e_{\mri\bullet} \is \ds\pa{\tilde\mf^e_\mri}{\mx_\bullet^e} 
	= -\!\ds\int_{\Omega^e_0} \!\unde{h_1}\,T_{1,\alpha}\,\bar\mN_e^\mrT\,\pa{\xi^\alpha_\mrp}{\mx_\bullet^e}\,\dif A \,,	
\eqe
for $\bullet = \mrm,\,\mrs$.
The corresponding vectors $\partial\xi^\alpha_\mrp/\partial\mx_\mrs^e$ and $\partial\xi^\alpha_\mrp/\partial\mx_\mrm^{\bar e}$ are given in \eqref{e:Dxi}. 
Combining the linearizations from \eqref{e:Dfe}, \eqref{e:Dfbc}, \eqref{e:Dftext} and \eqref{e:Dftc} leads to the six-field tangent matrix equation
\eqb{l}
\begin{bmatrix}
\Delta\mf^e_\mrm  \\[1mm]
\Delta\mf^e_\mrs  \\[1mm]
\Delta\bar\mf^e_\mrc  \\[1mm]
\Delta\tilde\mf^e_1  \\[1mm]
\Delta\tilde\mf^e_2  \\[1mm]
\Delta\tilde\mf^e_\mri - \Delta\tilde\mf^e_\mathrm{ext}
\end{bmatrix}
=
\begin{bmatrix}
\mk^e_\mathrm{mm} & \mk^e_\mathrm{ms} & \mk^e_\mathrm{m\phi} & \mathbf{0} & \mathbf{0} & \mk^e_\mathrm{mi} \\[1mm]
\mk^e_\mathrm{sm} & \mk^e_\mathrm{ss} & \mk^e_\mathrm{s\phi} & \mathbf{0} & \mathbf{0} & \mk^e_\mathrm{si} \\[1mm]
\mk^e_\mathrm{\phi m} & \mk^e_\mathrm{\phi s} & \mk^e_\mathrm{\phi\phi} & \mathbf{0} & \mathbf{0} & \mk^e_\mathrm{\phi i} \\[1mm]
\mk^e_\mathrm{1m} & \mk^e_\mathrm{1s} & \mathbf{0} & \mk^e_{11} & \mathbf{0} & \mk^e_\mathrm{1i} \\[1mm]
\mathbf{0} & \mathbf{0} & \mathbf{0} & \mathbf{0} & \mk^e_{22} & \mk^e_\mathrm{2i} \\[1mm]
\mk^e_\mathrm{im} & \mk^e_\mathrm{is} & -\mk^e_\mathrm{i\phi} & \mk^e_\mathrm{i1} & \mk^e_\mathrm{i2} & \mk^e_\mathrm{ii}
\end{bmatrix}
\begin{bmatrix}
\Delta\mx^{\bar e}_\mrm  \\[1mm]
\Delta\mx^e_\mrs  \\[1mm]
\Delta\bphi^e  \\[1mm]
\Delta\mT^{\bar e}_1  \\[1mm]
\Delta\mT^e_2  \\[1mm]
\Delta\mT^e_\mrc
\end{bmatrix}.
\label{e:kc6}\eqe
It is emphasized that the coupling matrix appearing here only contains the contributions from contact and (de)bonding reactions. 
Additional contributions from the internal forces $\mf^e_\mathrm{int}$ and $\tilde\mf^e_\mathrm{int}$ will appear in the blocks associated with master and slave body. 
In the last example of Sec.~\ref{s:ex}, the master body has only a single displacement and temperature node, hence $\mN_{\bar e,\alpha}$, $\bar\mN_{\bar e,\alpha}$ and $T_{1,\alpha}$ vanish. 
Further, $\undetc$ and $\unde{M_c}$ are not considered temperature dependent. 
Then Eq.~\eqref{e:kc6} simplifies to
\eqb{l}
\begin{bmatrix}
\Delta\mf^e_\mrm  \\[1mm]
\Delta\mf^e_\mrs  \\[1mm]
\Delta\bar\mf^e_\mrc  \\[1mm]
\Delta\tilde\mf^e_1  \\[1mm]
\Delta\tilde\mf^e_2  \\[1mm]
\Delta\tilde\mf^e_\mri - \Delta\tilde\mf^e_\mathrm{ext}
\end{bmatrix}
=
\begin{bmatrix}
\mk^e_\mathrm{mm} & \mk^e_\mathrm{ms} & \mk^e_\mathrm{m\phi} & \mathbf{0} & \mathbf{0} & \mathbf{0} \\[1mm]
\mk^e_\mathrm{sm} & \mk^e_\mathrm{ss} & \mk^e_\mathrm{s\phi} & \mathbf{0} & \mathbf{0} & \mathbf{0}  \\[1mm]
\mk^e_\mathrm{\phi m} & \mk^e_\mathrm{\phi s} & \mk^e_\mathrm{\phi\phi} & \mathbf{0} & \mathbf{0} & \mathbf{0} \\[1mm]
\mathbf{0}  & \mathbf{0}  & \mathbf{0} & \mk^e_{11} & \mathbf{0} & \mk^e_\mathrm{1i} \\[1mm]
\mathbf{0} & \mathbf{0} & \mathbf{0} & \mathbf{0} & \mk^e_{22} & \mk^e_\mathrm{2i} \\[1mm]
\mathbf{0}  & \mathbf{0} & -\mk^e_\mathrm{i\phi} & \mk^e_\mathrm{i1} & \mk^e_\mathrm{i2} & \mk^e_\mathrm{ii}
\end{bmatrix}
\begin{bmatrix}
\Delta\mx^{\bar e}_\mrm  \\[1mm]
\Delta\mx^e_\mrs  \\[1mm]
\Delta\bphi^e  \\[1mm]
\Delta\mT^{\bar e}_1  \\[1mm]
\Delta\mT^e_2  \\[1mm]
\Delta\mT^e_\mrc
\end{bmatrix}.
\label{e:kc6f}\eqe
If the bonding field $\phi$ is eliminated according to Sec.~\eqref{s:QPform}, this reduces to
\eqb{l}
\begin{bmatrix}
\Delta\mf^e_\mrm  \\[1mm]
\Delta\mf^e_\mrs  \\[1mm]
\Delta\tilde\mf^e_1  \\[1mm]
\Delta\tilde\mf^e_2  \\[1mm]
\Delta\tilde\mf^e_\mri - \Delta\tilde\mf^e_\mathrm{ext}
\end{bmatrix}
=
\begin{bmatrix}
\widehat\mk^e_\mathrm{mm} & \widehat\mk^e_\mathrm{ms} & \mathbf{0} & \mathbf{0} & \mathbf{0} \\[1mm]
\widehat\mk^e_\mathrm{sm} & \widehat\mk^e_\mathrm{ss} & \mathbf{0} & \mathbf{0} & \mathbf{0} \\[1mm]
\mathbf{0} & \mathbf{0} & \mk^e_{11} & \mathbf{0} & \mk^e_\mathrm{1i} \\[1mm]
\mathbf{0} & \mathbf{0} & \mathbf{0} & \mk^e_{22} & \mk^e_\mathrm{2i} \\[1mm]
-\widehat\mk^e_\mathrm{im} & -\widehat\mk^e_\mathrm{is} & \mk^e_\mathrm{i1} & \mk^e_\mathrm{i2} & \mk^e_\mathrm{ii} - \widehat\mk^e_\mathrm{ii}
\end{bmatrix}
\begin{bmatrix}
\Delta\mx^{\bar e}_\mrm  \\[1mm]
\Delta\mx^e_\mrs  \\[1mm]
\Delta\mT^{\bar e}_1  \\[1mm]
\Delta\mT^e_2  \\[1mm]
\Delta\mT^e_\mrc
\end{bmatrix}.
\eqe
according to \eqref{e:Dftext2}.
Here, the tangent matrices $\widehat\mk^e_\mathrm{ss}$, $\widehat\mk^e_\mathrm{ms} = \widehat\mk^{e~\mrT}_\mathrm{sm}$ and $\widehat\mk^e_\mathrm{mm}$ are as given in \eqref{e:mkx}, but with $\partial\undetc/\partial\bg_\mre^\mathrm{(i)}$ replaced by $\widehat{\partial\undetc}/\partial\bg_\mre^\mathrm{(i)}$.
An analogous reduction applies to \eqref{e:kc6}.

\subsubsection{Linearization of the five-field formulation}\label{s:5f-lin}

Next the linearization of $\tilde\mf^e_\mrc$ according to the five-field formulation is presented.
Now, $\tilde\mf^e_\mrc$ only requires the elemental contributions of (\ref{e:ftq3}.1) and (\ref{e:ftq3}.2). 
Their linearization yields
\eqb{lll}
\Delta\tilde\mf^e_1 \is -\!\ds\int_{\Omega^e_0} \!\bar\mN_{\bar e}^\mrT\,\big(\Delta\unde{q^c_t} + \Delta\unde{q^c_m} \big)\,\dif A 	
	-\!\ds\int_{\Omega^e_0} \!\Delta\bar\mN_{\bar e}^\mrT\,\big(\unde{q^c_t} + \unde{q^c_m} \big)\,\dif A\,,  \\[4mm]
\Delta\tilde\mf^e_2 \is \ds\int_{\Omega^e_0} \!\bar\mN_e^\mrT\,\big(\Delta\unde{q^c_t} - \Delta\unde{q^c_m} \big)\,\dif A\,.
\label{e:Dftc5}\eqe
As shown above, the mean flux is now given by $\unde{q^c_m} = \unde{\bar m}\dot\phi^2/2$.
Considering this and the case of (\ref{e:qmth}.2) leads to 
\eqb{lll}
\Delta\unde{q^c_m} \is \ds\unde{\bar m}\,\dot\phi\,\pa{\dot\phi}{\phi}\Delta\phi \,, \\[3mm]
\Delta\unde{q^c_t} \is \unde{h} \big(\Delta T_2 - \Delta T_1 - T_{1,\alpha}\,\Delta\xi^\alpha_\mrp \big)\,, 
\label{e:Dqmt5}\eqe
for constant $\unde{\bar m}$ and $\unde{h}$.
Inserting this into \eqref{e:Dftc5} then gives
\eqb{lll}
\Delta\tilde\mf^e_1 \is \mk^e_{11}\,\Delta\mT_1^{\bar e} + \mk^e_{12}\,\Delta\mT_2^e + \mk^e_{1\phi}\,\Delta\bphi^e
	+ \mk^e_{1\mrs}\,\Delta\mx_\mrs^e + \mk^e_{1\mrm}\,\Delta\mx_\mrm^{\bar e} \,,\\[4mm]
\Delta\tilde\mf^e_2 \is \mk^e_{21}\,\Delta\mT_1^{\bar e} + \mk^e_{22}\,\Delta\mT_2^e + \mk^e_{2\phi}\,\Delta\bphi^e 
	+ \mk^e_{2\mrs}\,\Delta\mx_\mrs^e + \mk^e_{2\mrm}\,\Delta\mx_\mrm^{\bar e} \,,
\label{e:Dftc6}\eqe
with
\eqb{llllll}
\mk^e_{11} \dis \ds\pa{\tilde\mf^e_1}{\mT_1^{\bar e}} = \ds\int_{\Omega^e_0} \!\unde{h}\,\bar\mN_{\bar e}^\mrT\,\bar\mN_{\bar e}\,\dif A \,, &
	\mk^e_{21} \dis \ds\pa{\tilde\mf^e_2}{\mT_1^{\bar e}} = \mk^{e~\mrT}_{12}  \,, \\[4mm]
\mk^e_{12} \dis \ds\pa{\tilde\mf^e_1}{\mT_2^e} = -\!\ds\int_{\Omega^e_0} \!\unde{h}\,\bar\mN_{\bar e}^\mrT\,\bar\mN_e\,\dif A \,, &
	\mk^e_{22} \dis \ds\pa{\tilde\mf^e_2}{\mT_2^e} = \ds\int_{\Omega^e_0} \!\unde{h}\,\bar\mN_e^\mrT\,\bar\mN_e\,\dif A \neq \mk^e_{11}  \,, \\[4mm]	
\mk^e_{1\phi} \dis \ds\pa{\tilde\mf^e_1}{\bphi^e} 
	= -\!\ds\int_{\Omega^e_0} \!\bar\mN_{\bar e}^\mrT\,\bar\mN_e\,\frac{\unde{\bar m}\dot\phi}{\gamma\,\Delta t}\,\dif A \,,~~ &
	\mk^e_{2\phi} \dis \ds\pa{\tilde\mf^e_2}{\bphi^e} 
	= -\!\ds\int_{\Omega^e_0} \!\bar\mN_e^\mrT\,\bar\mN_e\,\frac{\unde{\bar m}\dot\phi}{\gamma\,\Delta t}\,\dif A \neq \mk^e_{1\phi} \,,
\eqe
and
\eqb{lll}
\mk^e_{1\bullet} \is \ds\pa{\tilde\mf^e_1}{\mx_\bullet^e} 
	= \ds\int_{\Omega^e_0} \!\Big(\unde{h}\,T_{1,\alpha}\,\bar\mN_{\bar e}^\mrT - \big(\unde{q_t^c}+\unde{q_m^c}\big)\,\bar\mN_{\bar e,\alpha}^\mrT\Big)
	\pa{\xi^\alpha_\mrp}{\mx_\bullet^e}\,\dif A \,, \\[4mm] 
\mk^e_{2\bullet} \is \ds\pa{\tilde\mf^e_2}{\mx_\bullet^e} 
	= -\!\ds\int_{\Omega^e_0} \!\unde{h}\,T_{1,\alpha}\,\bar\mN_e^\mrT\,\pa{\xi^\alpha_\mrp}{\mx_\bullet^e}\,\dif A \,,	
\eqe
for $\bullet = \mrm,\,\mrs$, and $\partial\xi^\alpha_\mrp/\partial\mx_\mrs^e$ and $\partial\xi^\alpha_\mrp/\partial\mx_\mrm^{\bar e}$ given in \eqref{e:Dxi}. 
Combining the linearizations from \eqref{e:Dfe}, \eqref{e:Dfbc} and \eqref{e:Dftc6} then leads to the five-field elemental tangent matrix equation
\eqb{l}
\begin{bmatrix}
\Delta\mf^e_\mrm  \\[1mm]
\Delta\mf^e_\mrs  \\[1mm]
\Delta\bar\mf^e_\mrc  \\[1mm]
\Delta\tilde\mf^e_1  \\[1mm]
\Delta\tilde\mf^e_2 
\end{bmatrix}
=
\begin{bmatrix}
\mk^e_\mathrm{mm} & \mk^e_\mathrm{ms} & \mk^e_\mathrm{m\phi} & \mathbf{0} & \mathbf{0} \\[1mm]
\mk^e_\mathrm{sm} & \mk^e_\mathrm{ss} & \mk^e_\mathrm{s\phi} & \mathbf{0} & \mathbf{0} \\[1mm]
\mk^e_\mathrm{\phi m} & \mk^e_\mathrm{\phi s} & \mk^e_\mathrm{\phi\phi} & \mathbf{0} & \mathbf{0} \\[1mm]
\mk^e_\mathrm{1m} & \mk^e_\mathrm{1s} & \mk^e_\mathrm{1\phi} & \mk^e_{11} & \mk^e_{12} \\[1mm]
\mk^e_\mathrm{2m} & \mk^e_\mathrm{2s} & \mk^e_\mathrm{2\phi} & \mk^e_{21} & \mk^e_{22} 
\end{bmatrix}
\begin{bmatrix}
\Delta\mx^{\bar e}_\mrm  \\[1mm]
\Delta\mx^e_\mrs  \\[1mm]
\Delta\bphi^e  \\[1mm]
\Delta\mT^{\bar e}_1  \\[1mm]
\Delta\mT^e_2
\end{bmatrix},
\label{e:kc5}\eqe
for the case that $\undetc$ and $\unde{M_c}$ do not depend on $T_\mrc$.
Otherwise, see Remark~\ref{r:Tc}.
As above, the coupling matrix appearing in \eqref{e:kc5} only contains the contributions from contact and (de)bonding reactions. 
The complete coupling matrix additionally has contributions from $\mf^e_\mathrm{int}$ and $\tilde\mf^e_\mathrm{int}$ in the master and slave blocks.
For a single master temperature dof, $T_{1,\alpha}$ and $\bar\mN_{\bar e,\alpha}$ vanish, such that \eqref{e:kc5} becomes
\eqb{l}
\begin{bmatrix}
\Delta\mf^e_\mrm  \\[1mm]
\Delta\mf^e_\mrs  \\[1mm]
\Delta\bar\mf^e_\mrc  \\[1mm]
\Delta\tilde\mf^e_1  \\[1mm]
\Delta\tilde\mf^e_2 
\end{bmatrix}
=
\begin{bmatrix}
\mk^e_\mathrm{mm} & \mk^e_\mathrm{ms} & \mk^e_\mathrm{m\phi} & \mathbf{0} & \mathbf{0} \\[1mm]
\mk^e_\mathrm{sm} & \mk^e_\mathrm{ss} & \mk^e_\mathrm{s\phi} & \mathbf{0} & \mathbf{0} \\[1mm]
\mk^e_\mathrm{\phi m} & \mk^e_\mathrm{\phi s} & \mk^e_\mathrm{\phi\phi} & \mathbf{0} & \mathbf{0} \\[1mm]
\mathbf{0} & \mathbf{0} & \mk^e_\mathrm{1\phi} & \mk^e_{11} & \mk^e_{12} \\[1mm]
\mathbf{0} & \mathbf{0} & \mk^e_\mathrm{2\phi} & \mk^e_{21} & \mk^e_{22} 
\end{bmatrix}
\begin{bmatrix}
\Delta\mx^{\bar e}_\mrm  \\[1mm]
\Delta\mx^e_\mrs  \\[1mm]
\Delta\bphi^e  \\[1mm]
\Delta\mT^{\bar e}_1  \\[1mm]
\Delta\mT^e_2
\end{bmatrix}.
\label{e:kc5a}\eqe
In case $\phi$ is a history variable at the quadrature point, $\Delta\phi$ is given by Eq.~\eqref{e:dphi_qp}. 
Inserting this into (\ref{e:Dqmt5}.1) for $\unde{M_c}\neq\unde{M_c}(T_\mrc)$ leads to
\eqb{l}
\mk^e_{k\phi}\,\Delta\bphi^e = \widehat\mk^e_{k\mrm}\,\Delta\mx_\mrm^{\bar e} + \widehat\mk^e_{k\mrs}\,\Delta\mx_\mrs^e \,,\quad~  k = 1,\,2\,,	
\label{e:Dftc7}\eqe
with
\eqb{lll}
\widehat\mk^e_{1\mrm} \dis -\!\ds\int_{\Omega^e_0} \bar\mN_{\bar e}^\mrT \dot\phi
	\bigg(1+ \frac{\gamma\,\Delta t}{\unde{\bar m}}\pa{\unde{M_c}}{\phi} \bigg)^{\!\!-1} \pa{\unde{M_c}}{\bg_\mre}\cdot\mN_{\bar e}\,\dif A\,, \\[4mm]
\widehat\mk^e_{1\mrs} \dis \ds\int_{\Omega^e_0} \bar\mN_{\bar e}^\mrT \dot\phi
	\bigg(1+ \frac{\gamma\,\Delta t}{\unde{\bar m}}\pa{\unde{M_c}}{\phi} \bigg)^{\!\!-1} \pa{\unde{M_c}}{\bg_\mre}\cdot\mN_e\,\dif A\,, \\[4mm] 
\widehat\mk^e_{2\mrm} \dis -\!\ds\int_{\Omega^e_0} \bar\mN_e^\mrT \dot\phi
	\bigg(1+ \frac{\gamma\,\Delta t}{\unde{\bar m}}\pa{\unde{M_c}}{\phi} \bigg)^{\!\!-1} \pa{\unde{M_c}}{\bg_\mre}\cdot\mN_{\bar e}\,\dif A\,, \\[4mm] 
\widehat\mk^e_{2\mrs} \dis \ds\int_{\Omega^e_0} \bar\mN_e^\mrT \dot\phi
	\bigg(1+ \frac{\gamma\,\Delta t}{\unde{\bar m}}\pa{\unde{M_c}}{\phi} \bigg)^{\!\!-1} \pa{\unde{M_c}}{\bg_\mre}\cdot\mN_e\,\dif A\,, 
\eqe
such that \eqref{e:kc5a} becomes
\eqb{l}
\begin{bmatrix}
\Delta\mf^e_\mrm  \\[1mm]
\Delta\mf^e_\mrs  \\[1mm]
\Delta\tilde\mf^e_1  \\[1mm]
\Delta\tilde\mf^e_2 
\end{bmatrix}
=
\begin{bmatrix}
\widehat\mk^e_\mathrm{mm} & \widehat\mk^e_\mathrm{ms} & \mathbf{0} & \mathbf{0} \\[1mm]
\widehat\mk^e_\mathrm{sm} & \widehat\mk^e_\mathrm{ss} & \mathbf{0} & \mathbf{0} \\[1mm]
\widehat\mk^e_\mathrm{1m} & \widehat\mk^e_\mathrm{1s} & \mk^e_{11} & \mk^e_{12} \\[1mm]
\widehat\mk^e_\mathrm{2m} & \widehat\mk^e_\mathrm{2s} & \mk^e_{21} & \mk^e_{22} 
\end{bmatrix}
\begin{bmatrix}
\Delta\mx^{\bar e}_\mrm  \\[1mm]
\Delta\mx^e_\mrs  \\[1mm]
\Delta\mT^{\bar e}_1  \\[1mm]
\Delta\mT^e_2
\end{bmatrix}.
\label{e:kc5b}\eqe

\begin{remark}\label{r:Tc}
In the five-field formulation the interface temperature $T_\mrc$ is not needed.
If desired it can be recovered from (\ref{e:qmth}.1) and $\unde{q^c_m} = \unde{\bar m}\dot\phi^2/2$, giving
\eqb{l}
T_\mrc = \ds\frac{T_1 + T_2}{2} + \frac{\unde{\bar m}\dot\phi^2}{4\unde{h}}\,.
\label{e:5fTc}\eqe
The linearization of this expression yields
\eqb{l}
\Delta T_\mrc = \ds\frac{\Delta T_1 + \Delta T_2}{2} + \frac{\unde{\bar m}\dot\phi}{2\unde{h}}\pa{\dot\phi}{\phi}\Delta\phi 
	+ \frac{T_{1,\alpha}}{2}\Delta\xi^\alpha_\mrp\,.
\eqe
This can be used to eliminate $\Delta\mT_\mrc^e$ from \eqref{e:kc6} and thus derive a more general five-field formulation.
\end{remark}

The discussion closes by noting that in the tangent expressions from Secs.~\ref{s:6f-lin} and \ref{s:5f-lin} further rows and columns disappear when the master temperature or master deformation is fixed.
This is considered in some of the following examples.

\section{Numerical examples}\label{s:ex}

\begin{figure}[h]
\begin{center} \unitlength1cm
\begin{picture}(0,6.7)
\put(-8,3.5){\includegraphics[height=32mm]{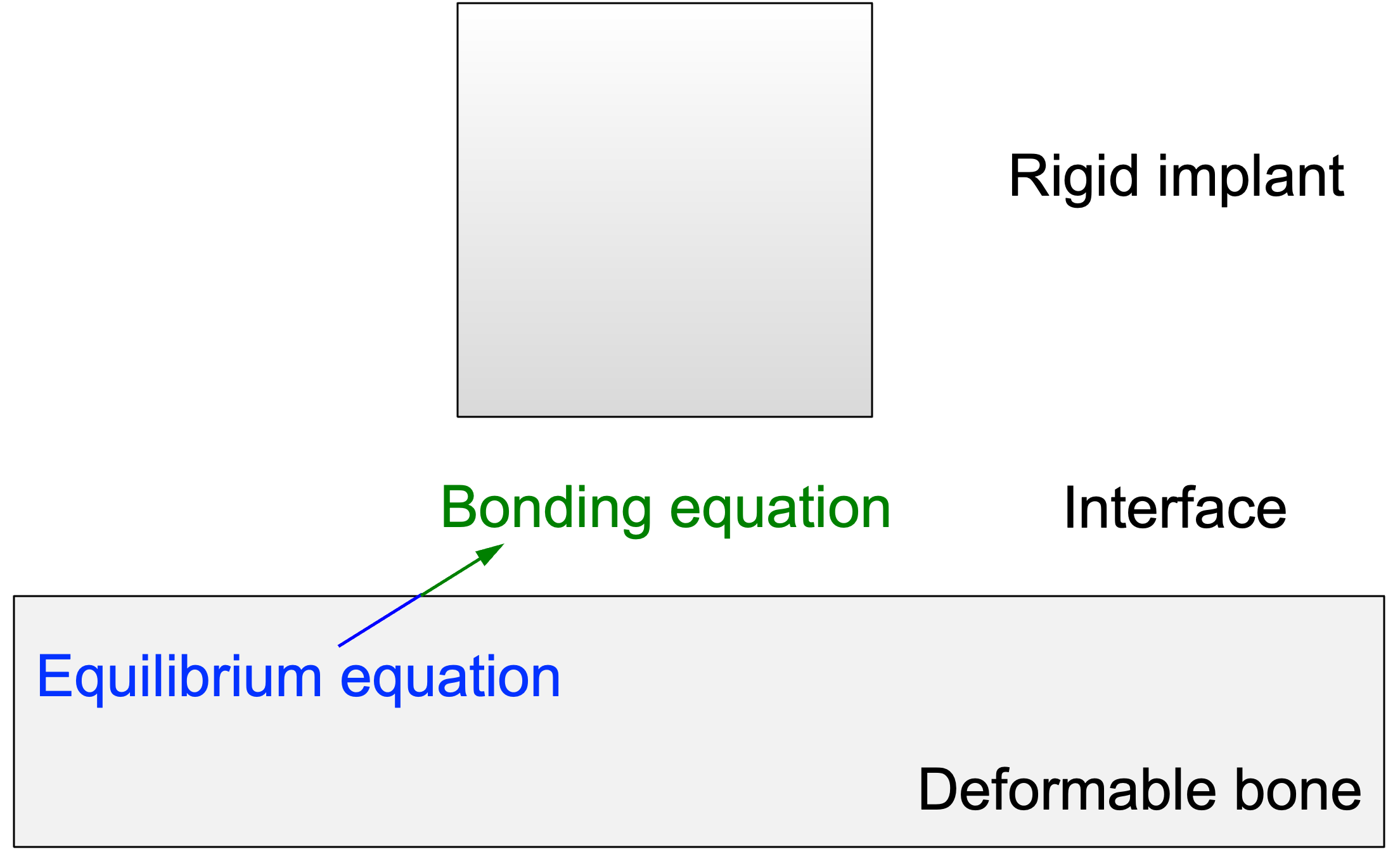}}
\put(-2.56,3.5){\includegraphics[height=32mm]{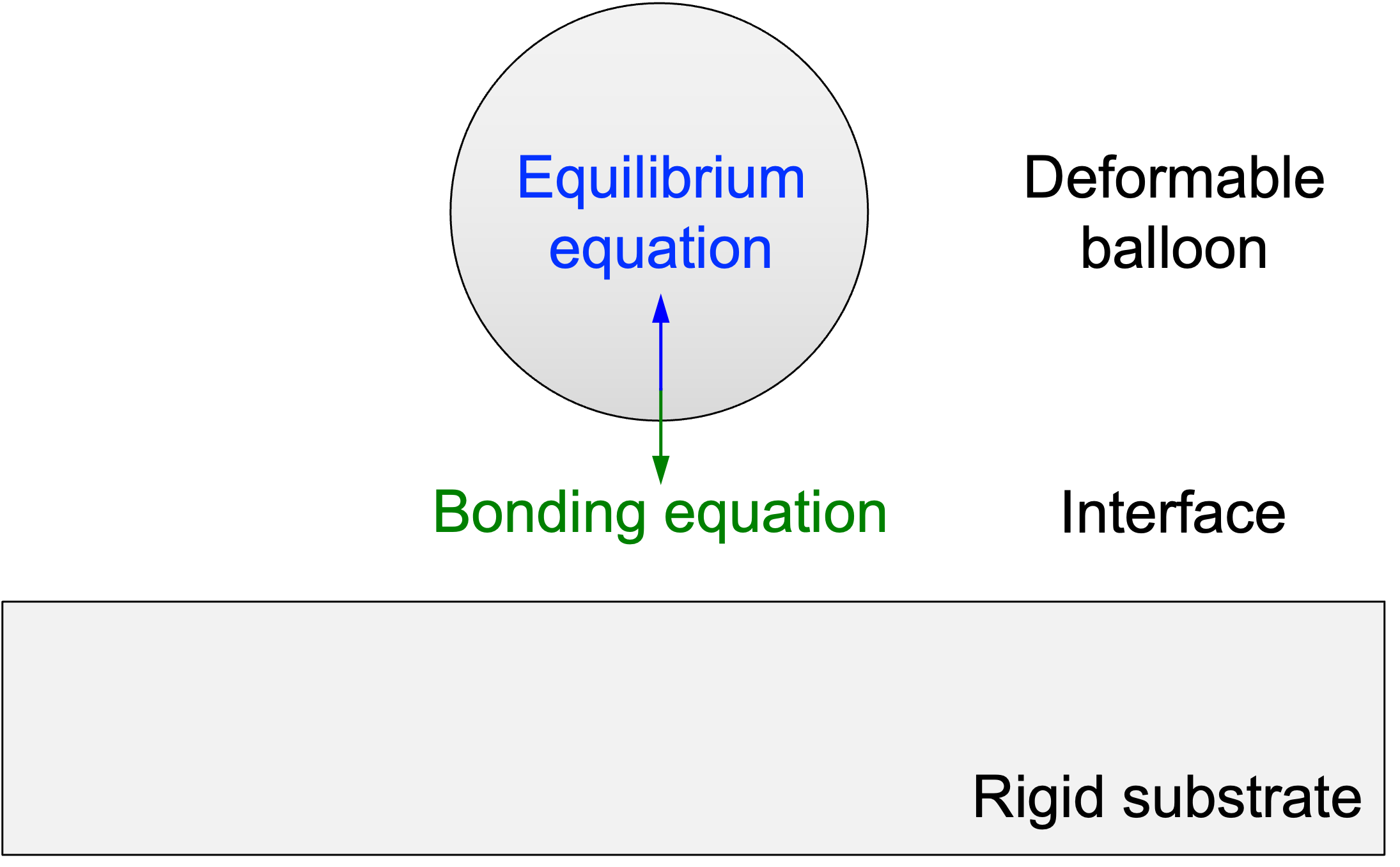}}
\put(2.8,3.5){\includegraphics[height=32mm]{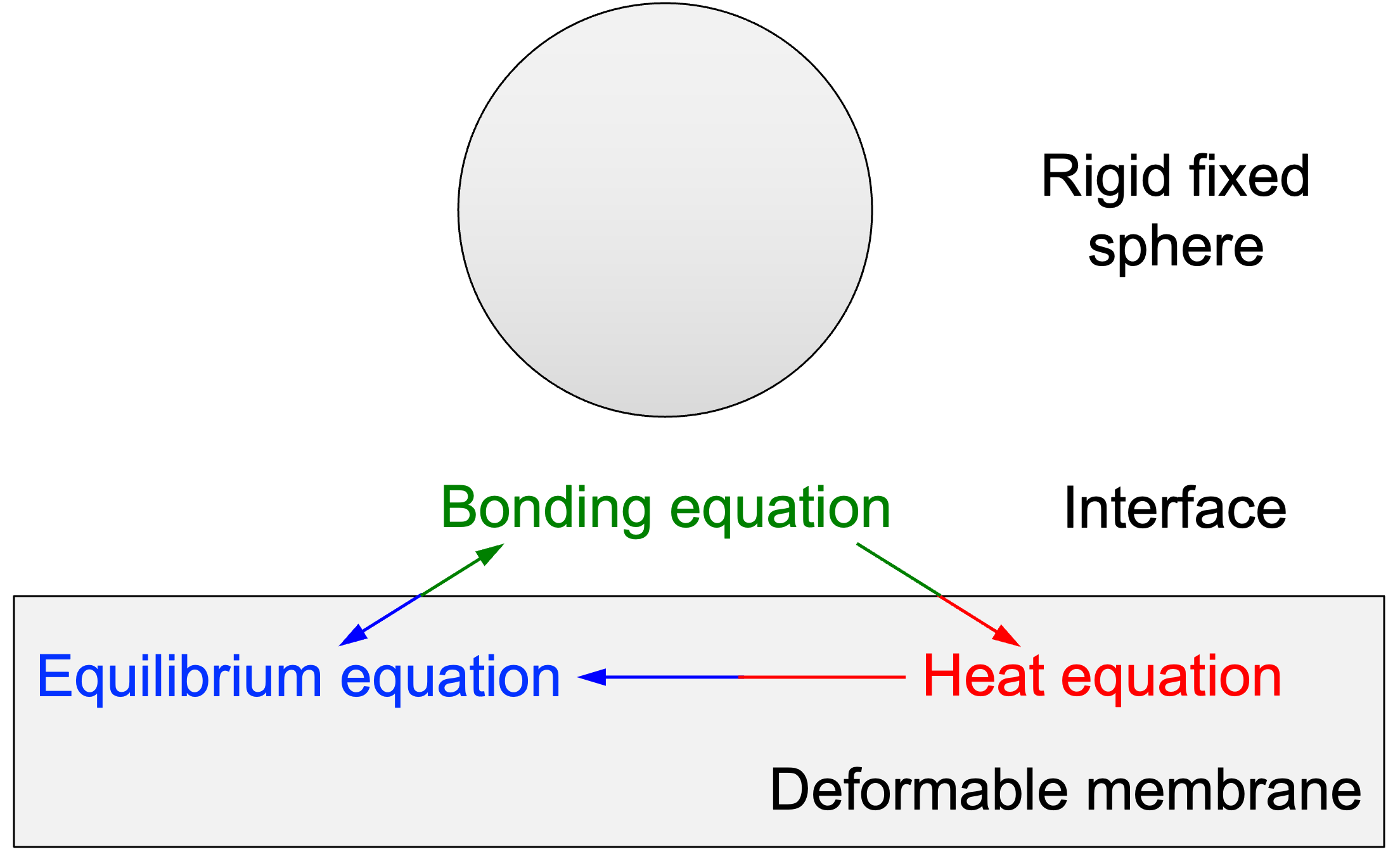}}
\put(-8,-.15){\includegraphics[height=32mm]{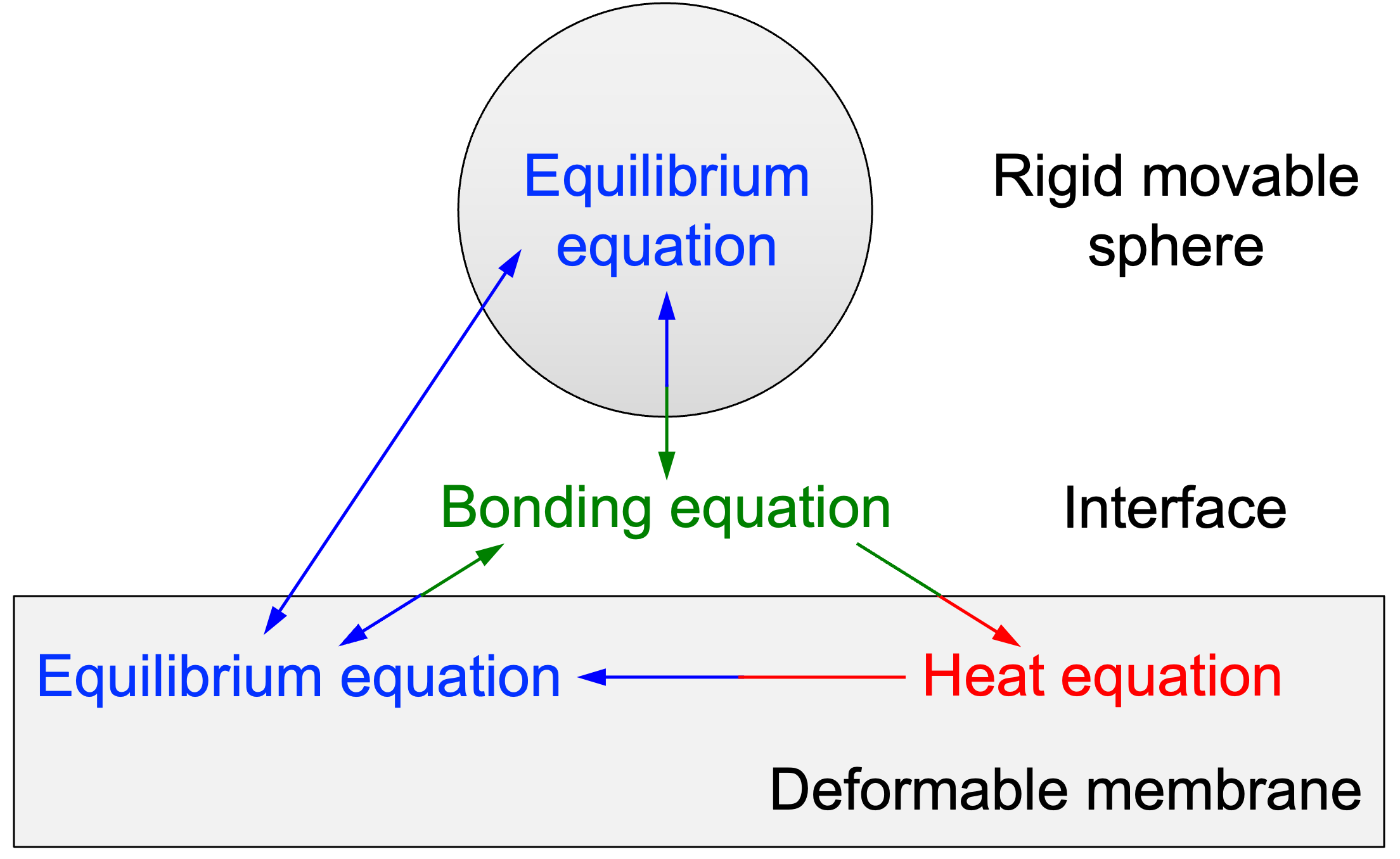}}
\put(-2.6,-.15){\includegraphics[height=32mm]{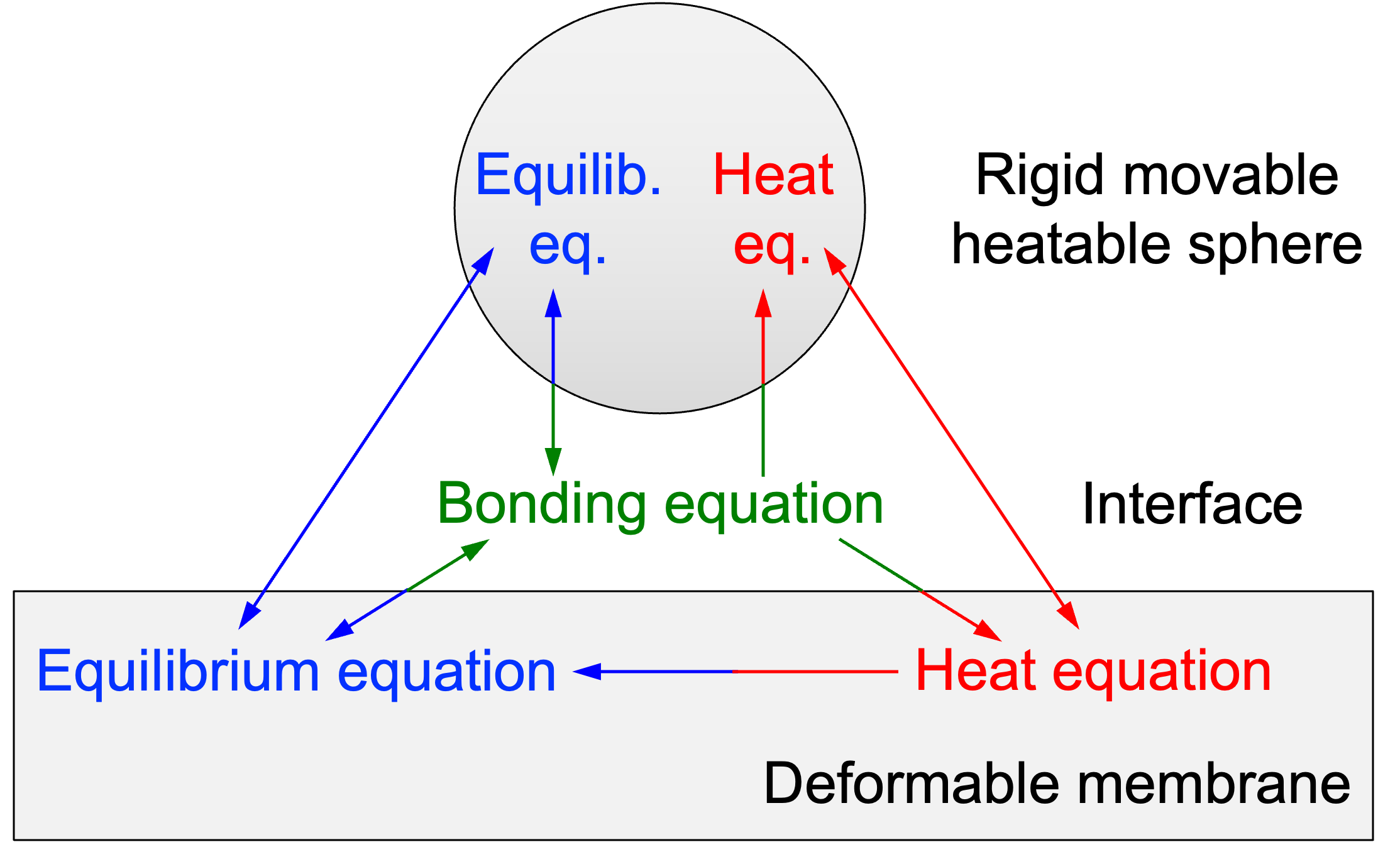}}
\put(2.8,-.15){\includegraphics[height=32mm]{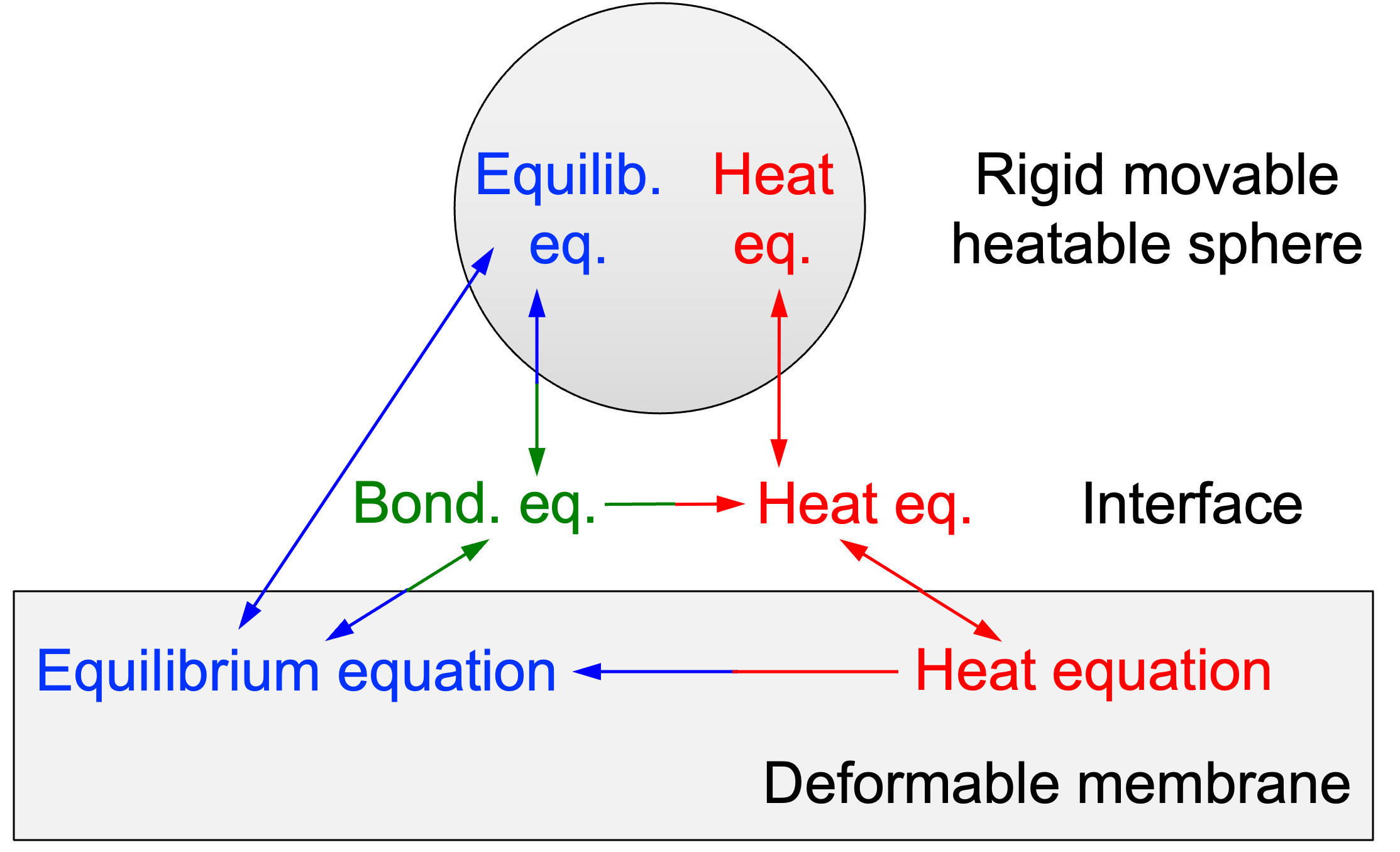}}
\put(-7.9,3.68){\footnotesize (a)}
\put(-2.5,3.68){\footnotesize (b)}
\put(2.9,3.68){\footnotesize (c)}
\put(-7.9,.03){\footnotesize (d)}
\put(-2.5,.03){\footnotesize (e)}
\put(2.9,.03){\footnotesize (f)}
\end{picture}
\caption{Numerical coupling examples: 
(a) one-way (Sec.~\ref{s:ex2}), (b) two-way (Sec.~\ref{s:ex1}), (c) three-field (Sec.~\ref{s:ex3a}), (d) four-field (Sec.~\ref{s:ex3b}), (e) five-field (Sec.~\ref{s:ex3c}) and (f) six-field chemo-thermo-mechanical coupling (Sec.~\ref{s:ex3d}).
The arrows in (e) and (f) represent the coupling in Eqs.~\eqref{e:kc5a} and \eqref{e:kc6f}, respectively, plus the influence of thermal expansion of the membrane.
Beyond this coupling, all heat and bonding equations depend on the current contact domain governed by the two equilibrium equations.}
\label{f:CoboEx}
\end{center}
\end{figure}

This section presents six different numerical examples that highlight different coupling aspects of the proposed chemo-mechano-thermodynamical contact model, as illustrated in Fig.~\ref{f:CoboEx}.
The first example in Sec.~\ref{s:ex2} considers bonding with one-way chemo-mechanical coupling during implant osseointegration (Fig.~\ref{f:CoboEx}a).
The second example in Sec.~\ref{s:ex1} considers debonding with two-way chemo-mechanical coupling for a simple model system (Fig.~\ref{f:CoboEx}b).
The remaining examples in Secs.~\ref{s:ex3a}-\ref{s:ex3d} consider bonding with full chemo-thermo-mechanical coupling for three-field (Fig.~\ref{f:CoboEx}c), four-field (Fig.~\ref{f:CoboEx}d), five-field (Fig.~\ref{f:CoboEx}e) and six-field model systems (Fig.~\ref{f:CoboEx}f).
The examples involve large mechanical and thermal deformations, large temperature changes, simultaneous bonding and debonding, and the strong coupling of up to six different fields.
Further, all the coupling is nonlinear as noted in Remark~\ref{r:NFEM}.
The proper convergence of the Newton-Raphson iteration used to solve the examples monolithically is confirmed in Appendix~\ref{s:NRconv}.

\subsection{One-way coupling: Mechanically dependent implant osseointegration}\label{s:ex2}

The first example illustrates pressure- and gap-dependent bonding for the case of implant osseointegration.
For bone, the chemical contact forces are expected to be so small that they do not affect the mechanical behavior.
The problem therefore only exhibits one-way chemo-mechanical contact bonding according to Fig.~\ref{f:CoboEx}a,  which allows for a sequential solution scheme:
First the mechanical contact state is computed, and then this is used to determine the chemical bonding state.
The dependency of bonding on the mechanical contact state is taken from the bonding model of Sec.~\ref{s:con4}.
According to this model, bonding is optimal for $p_\mrc \leq p_\mathrm{opt}$ and $g_\mrn \leq g_\mathrm{opt}$.

The considered bonding parameters are listed in Table~\ref{t:RCSI-para}, together with all other model parameters used in this example.
\begin{table}[h]
\centering
\begin{tabular}{|r|r|r|r|}
  \hline
   symbol & material parameter & value & unit \\[0mm] \hline 
   & & & \\[-3.5mm]   
   $R$ & implant radius & 2.5 & mm \\ [.5mm] 
   $H$ & implant height & 3 & mm \\ [.5mm] 
   $\unde{E_b}$ & Young's modulus of bone & 18 & GPa \\[.5mm] 
   $\nu$ & Poisson's ratio of bone & 0.3 & -- \\[.5mm]
   $\unde{E_n}$ & contact penalty stiffness & 200 & $\unde{E_b}/R$ \\[0mm]
   $\overrightarrow{\unde{K_0}}$ & implant-bone bond energy & 90 & J/m$^2$  \\ [.5mm] 
   $\unde{\bar m}$ & bonding ``mass" & 1000 & Nd/m \\[.5mm]
   $\tau$ & bonding time scale & 11.11 & d \\[.5mm]
   $p_\mathrm{lim}$ & contact pressure limit & 12 & MPa \\ [.5mm]   
   $p_\mathrm{opt}$ & optimal contact pressure & 0& MPa \\ [.5mm] 
   $g_\mathrm{opt}$ & optimal contact gap & 0& mm \\ [.5mm] 
   $g_\mathrm{lim}$ & contact gap limit & 0.3 & mm \\ [0mm]    
    \hline
\end{tabular}
\vspace*{-2mm}
\caption{Osseointegration example: Considered model parameters. (d = days, $\tau = \protect\unde{\bar m}/\protect\overrightarrow{\unde{K_0}}$.)}
\label{t:RCSI-para}
\end{table}
The bonding energy $\overrightarrow{\unde{K_0}}$ is taken equal to the fracture energy of the bone-implant bond, since it is the energy that needs to be overcome during its separation.
The listed value for this is taken from the mode III fracture experiments of \citet{mathieu12}.
The bonding time scale thus becomes $\tau = \unde{\bar m}/\overrightarrow{\unde{K_0}} = 11.11$ days.
In this case $\overrightarrow{\unde{K_0}}/g_0^2 \ll \unde{E_n}$ (for $g_0 = 1$mm), which justifies the one-way coupling.

\begin{figure}[h!]
\begin{center} \unitlength1cm
\begin{picture}(0,5.6)
\put(-7.95,.2){\includegraphics[height=53mm]{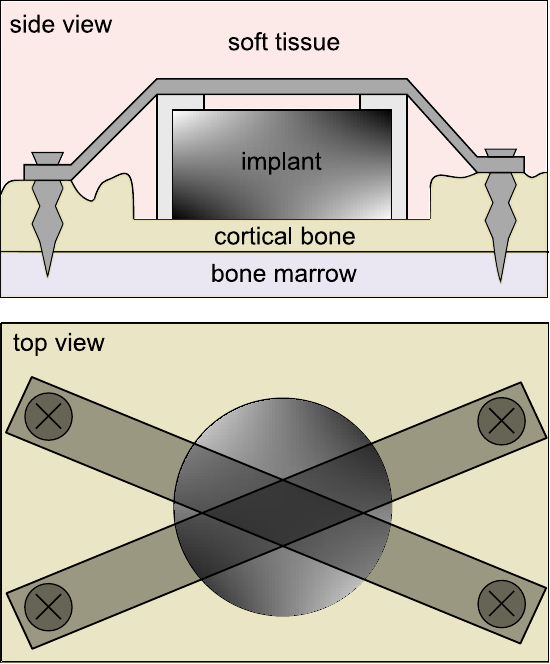}}
\put(-3.0,-.1){\includegraphics[height=56mm]{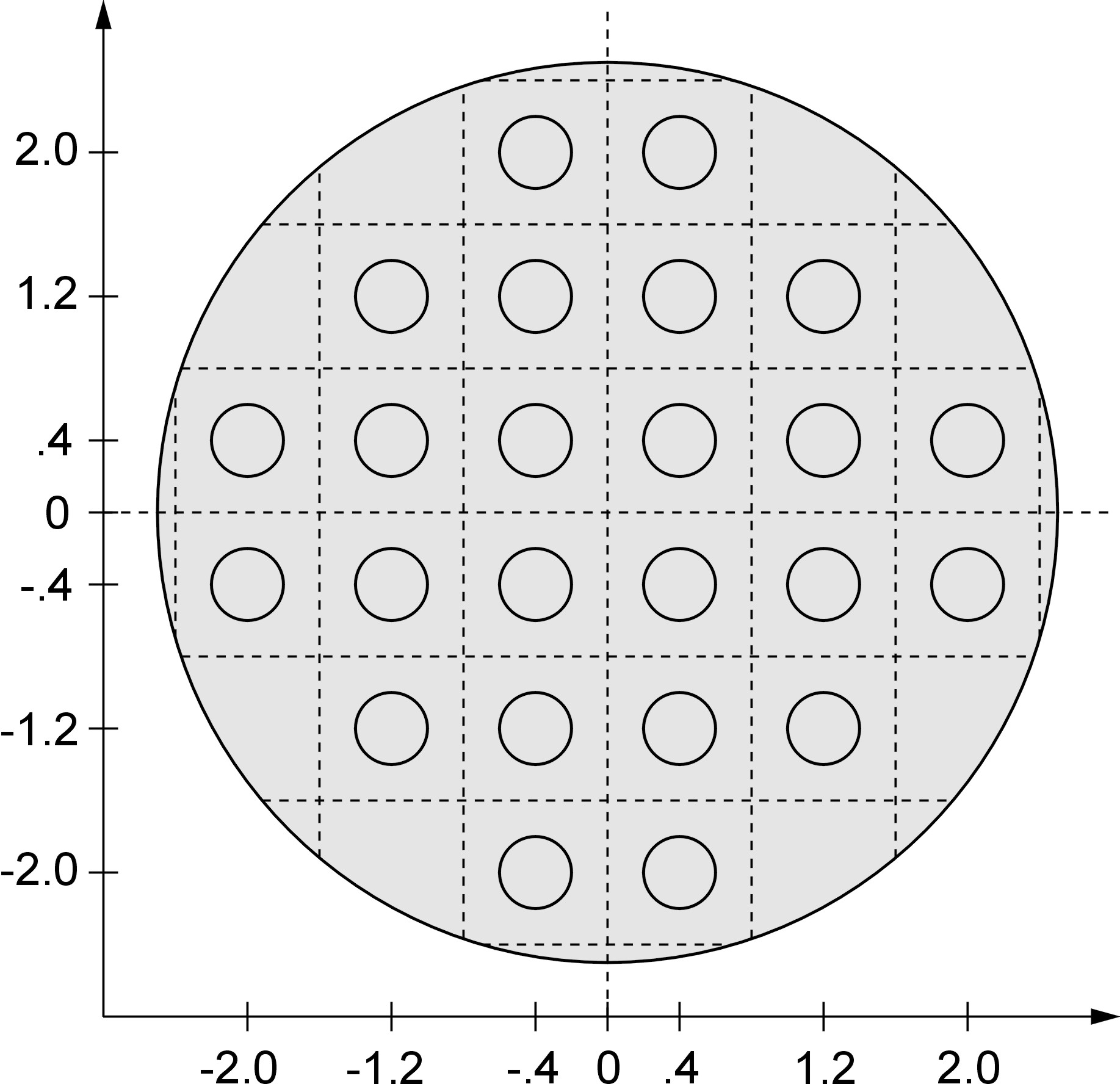}}
\put(3.25,0){\includegraphics[height=53mm]{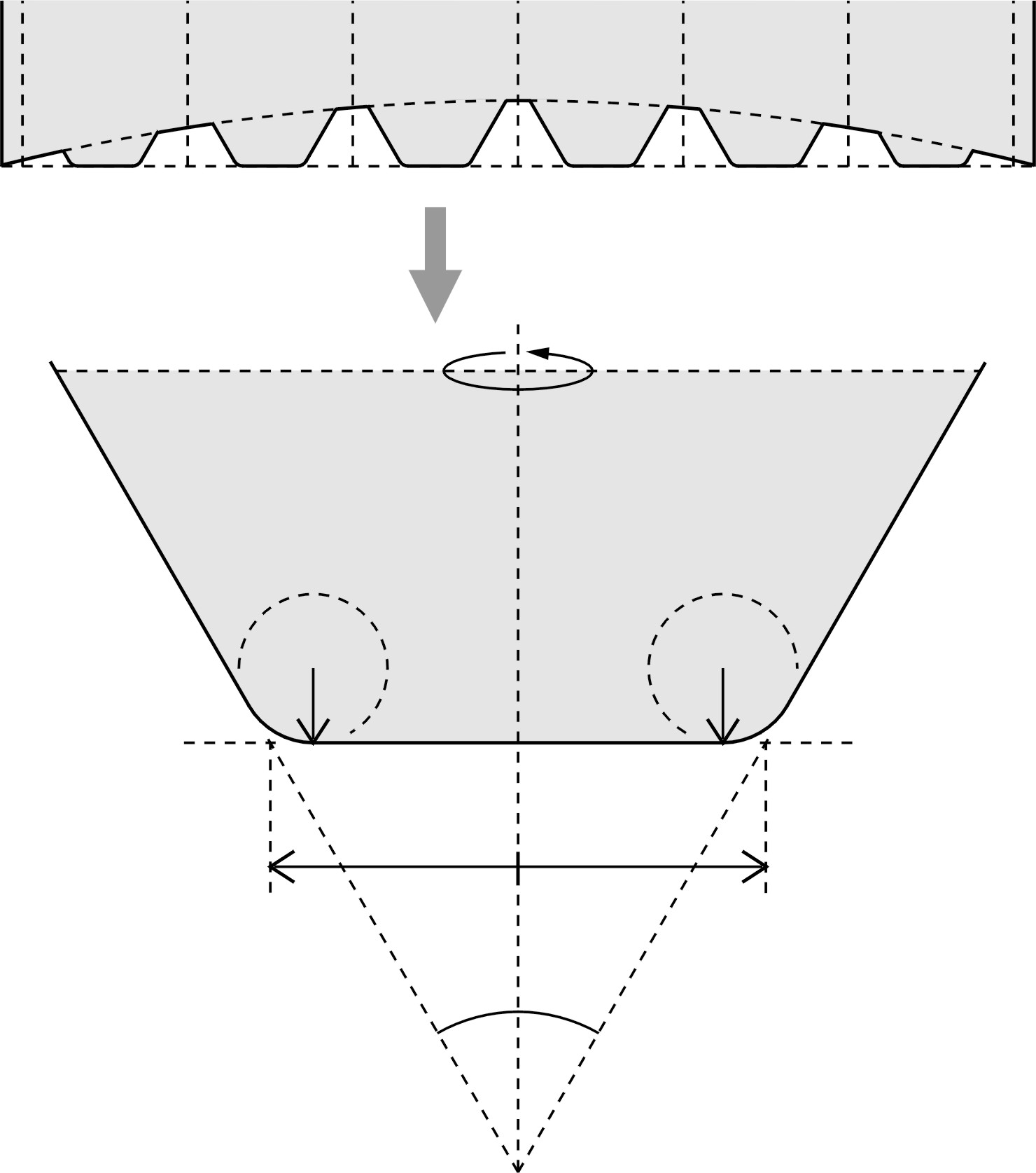}}
\put(2.1,.45){\scriptsize [mm]}
\put(-2.4,5.2){\scriptsize [mm]}
\put(5.25,0.8){\footnotesize $\alpha$}
\put(5.75,0.8){\footnotesize $\alpha$}
\put(5,1.5){\footnotesize $r_\mra$}
\put(5.9,1.5){\footnotesize $r_\mra$}
\put(4.57,2.38){\footnotesize $r_\mrf$}
\put(6.42,2.38){\footnotesize $r_\mrf$}
\put(-7.97,-.1){\footnotesize (a)}
\put(-3.0,-.1){\footnotesize (b)}
\put(3.3,-.1){\footnotesize (c)}
\end{picture}
\caption{Osseointegration example: (a) Experimental setup of \citet{mathieu12}, 
(b) implant bottom view with asperity distribution and (c) implant side view with asperity geometry.}
\label{f:RCSI-setup}
\end{center}
\vspace{-10mm}
\end{figure}

\begin{figure}[h]
\begin{center} \unitlength1cm
\begin{picture}(0,3.7)
\put(-6.8,1.2){\line(0,1){1.2}}
\put(-6.2,0.6){\line(0,1){2.4}}
\put(-5.6,0){\line(0,1){3.6}}
\put(-5,0){\line(0,1){3.6}}
\put(-4.4,0){\line(0,1){3.6}}
\put(-3.8,0.6){\line(0,1){2.4}}
\put(-3.2,1.2){\line(0,1){1.2}}
\put(-5.6,3.6){\line(1,0){1.2}}
\put(-6.2,3.0){\line(1,0){2.4}}
\put(-6.8,2.4){\line(1,0){3.6}}
\put(-6.8,1.8){\line(1,0){3.6}}
\put(-6.8,1.2){\line(1,0){3.6}}
\put(-6.2,0.6){\line(1,0){2.4}}
\put(-5.6,0.0){\line(1,0){1.2}}
\put(-5.52,3.18){\footnotesize $45^\circ$}
\put(-4.92,3.18){\footnotesize $45^\circ$}
\put(-6.12,2.58){\footnotesize $45^\circ$}
\put(-5.52,2.58){\footnotesize $30^\circ$}
\put(-4.92,2.58){\footnotesize $30^\circ$}
\put(-4.32,2.58){\footnotesize $45^\circ$}
\put(-6.72,1.98){\footnotesize $45^\circ$}
\put(-6.12,1.98){\footnotesize $30^\circ$}
\put(-5.52,1.98){\footnotesize $30^\circ$}
\put(-4.92,1.98){\footnotesize $30^\circ$}
\put(-4.32,1.98){\footnotesize $30^\circ$}
\put(-3.72,1.98){\footnotesize $45^\circ$}
\put(-6.72,1.38){\footnotesize $45^\circ$}
\put(-6.12,1.38){\footnotesize $30^\circ$}
\put(-5.52,1.38){\footnotesize $30^\circ$}
\put(-4.92,1.38){\footnotesize $30^\circ$}
\put(-4.32,1.38){\footnotesize $30^\circ$}
\put(-3.72,1.38){\footnotesize $45^\circ$}
\put(-6.12,0.78){\footnotesize $45^\circ$}
\put(-5.52,0.78){\footnotesize $30^\circ$}
\put(-4.92,0.78){\footnotesize $30^\circ$}
\put(-4.32,0.78){\footnotesize $45^\circ$}
\put(-5.52,0.18){\footnotesize $45^\circ$}
\put(-4.92,0.18){\footnotesize $45^\circ$}
\put(-1.8,1.2){\line(0,1){1.2}}
\put(-1.2,0.6){\line(0,1){2.4}}
\put(-.6,0){\line(0,1){3.6}}
\put(0,0){\line(0,1){3.6}}
\put(.6,0){\line(0,1){3.6}}
\put(1.2,0.6){\line(0,1){2.4}}
\put(1.8,1.2){\line(0,1){1.2}}
\put(-0.6,3.6){\line(1,0){1.2}}
\put(-1.2,3.0){\line(1,0){2.4}}
\put(-1.8,2.4){\line(1,0){3.6}}
\put(-1.8,1.8){\line(1,0){3.6}}
\put(-1.8,1.2){\line(1,0){3.6}}
\put(-1.2,0.6){\line(1,0){2.4}}
\put(-0.6,0.0){\line(1,0){1.2}}
\put(-.52,3.18){\footnotesize $.12$}
\put(0.08,3.18){\footnotesize $.12$}
\put(-1.12,2.58){\footnotesize $.16$}
\put(-.52,2.58){\footnotesize $.12$}
\put(0.08,2.58){\footnotesize $.12$}
\put(0.68,2.58){\footnotesize $.16$}
\put(-1.72,1.98){\footnotesize $.20$}
\put(-1.12,1.98){\footnotesize $.20$}
\put(-.52,1.98){\footnotesize $.16$}
\put(0.08,1.98){\footnotesize $.16$}
\put(0.68,1.98){\footnotesize $.20$}
\put(1.28,1.98){\footnotesize $.20$}
\put(-1.72,1.38){\footnotesize $.20$}
\put(-1.12,1.38){\footnotesize $.20$}
\put(-.52,1.38){\footnotesize $.16$}
\put(0.08,1.38){\footnotesize $.16$}
\put(0.68,1.38){\footnotesize $.20$}
\put(1.28,1.38){\footnotesize $.20$}
\put(-1.12,0.78){\footnotesize $.16$}
\put(-.52,0.78){\footnotesize $.12$}
\put(0.08,0.78){\footnotesize $.12$}
\put(0.68,0.78){\footnotesize $.16$}
\put(-.52,0.18){\footnotesize $.12$}
\put(0.08,0.18){\footnotesize $.12$}
\put(3.2,1.2){\line(0,1){1.2}}
\put(3.8,0.6){\line(0,1){2.4}}
\put(4.4,0){\line(0,1){3.6}}
\put(5,0){\line(0,1){3.6}}
\put(5.6,0){\line(0,1){3.6}}
\put(6.2,0.6){\line(0,1){2.4}}
\put(6.8,1.2){\line(0,1){1.2}}
\put(4.4,3.6){\line(1,0){1.2}}
\put(3.8,3.0){\line(1,0){2.4}}
\put(3.2,2.4){\line(1,0){3.6}}
\put(3.2,1.8){\line(1,0){3.6}}
\put(3.2,1.2){\line(1,0){3.6}}
\put(3.8,0.6){\line(1,0){2.4}}
\put(4.4,0.0){\line(1,0){1.2}}
\put(4.48,3.18){\footnotesize $.05$}
\put(5.08,3.18){\footnotesize $.00$}
\put(3.88,2.58){\footnotesize $.05$}
\put(4.48,2.58){\footnotesize $.05$}
\put(5.08,2.58){\footnotesize $.00$}
\put(5.68,2.58){\footnotesize $.00$}
\put(3.28,1.98){\footnotesize $.05$}
\put(3.88,1.98){\footnotesize $.05$}
\put(4.48,1.98){\footnotesize $.05$}
\put(5.08,1.98){\footnotesize $.00$}
\put(5.68,1.98){\footnotesize $.00$}
\put(6.28,1.98){\footnotesize $.00$}
\put(3.28,1.38){\footnotesize $.10$}
\put(3.88,1.38){\footnotesize $.10$}
\put(4.48,1.38){\footnotesize $.10$}
\put(5.08,1.38){\footnotesize $.20$}
\put(5.68,1.38){\footnotesize $.20$}
\put(6.28,1.38){\footnotesize $.20$}
\put(3.88,0.78){\footnotesize $.10$}
\put(4.48,0.78){\footnotesize $.10$}
\put(5.08,0.78){\footnotesize $.20$}
\put(5.68,0.78){\footnotesize $.20$}
\put(4.48,0.18){\footnotesize $.10$}
\put(5.08,0.18){\footnotesize $.20$}
\put(-6.85,0.07){\footnotesize (a)}
\put(-1.85,0.07){\footnotesize (b)}
\put(3.15,0.07){\footnotesize (c)}
\end{picture}
\vspace{-1mm}
\caption{Osseointegration example: Variation of the asperities on the implant: (a) asperity angle $\alpha$, (b) asperity radius $r_\mra$ in [mm], (c) initial gap below the asperity in [$\mu$m].
For all asperities, $r_\mrf = 50\mu$m.}
\label{f:RCSI-para}
\end{center}
\end{figure}
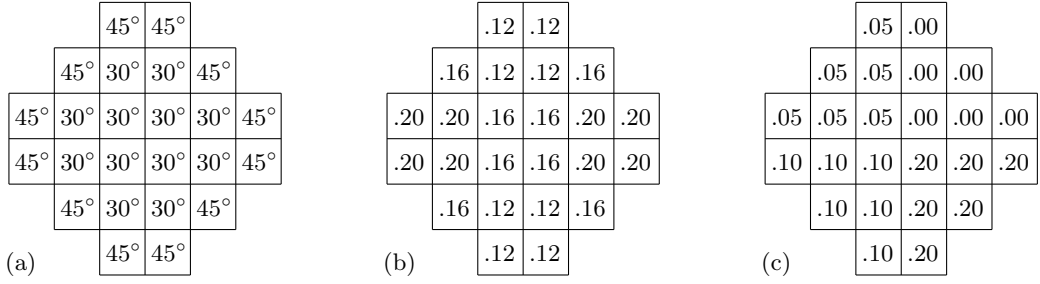

Following the experimental setup of \citet{mathieu12}, a cylindrical implant specimen (radius $R=2.5$mm, height $H = 3$mm) on a reamed bone surface is considered, as shown in Fig.~\ref{f:RCSI-setup}a.
\vspace{10mm}
Since the bone deformation is small and localized, the bone is modeled as a block with dimension 5.6mm $\times$ 5.6mm $\times$ 0.6mm that is supported in normal direction at the bottom and on all four sides.
The finite element mesh consists of 852,992 quadratic B-spline elements ($224 \times 224 \times 16$ brick elements of size $25\mu$m $\times~25\mu$m $\times~37.5\mu$m and $224 \times 224$ contact elements of size $25\mu$m $\times~25\mu$m) resulting in 919,368 FE nodes ($=$ control points) and 2,758,104 dofs.
To illustrate how the implant osseointegration depends on the mechanical contact conditions, the bottom surface of the implant specimen is modeled as a rough surface with 24 well-defined asperities as shown in Fig.~\ref{f:RCSI-setup}b-c. 
The asperity parameters vary as specified in Fig.~\ref{f:RCSI-para}, so that the contact conditions differ at each asperity. 

\begin{figure}[h!]
\begin{center} \unitlength1cm
\begin{picture}(0,5.1)
\put(-6.5,-.7){\includegraphics[height=60mm]{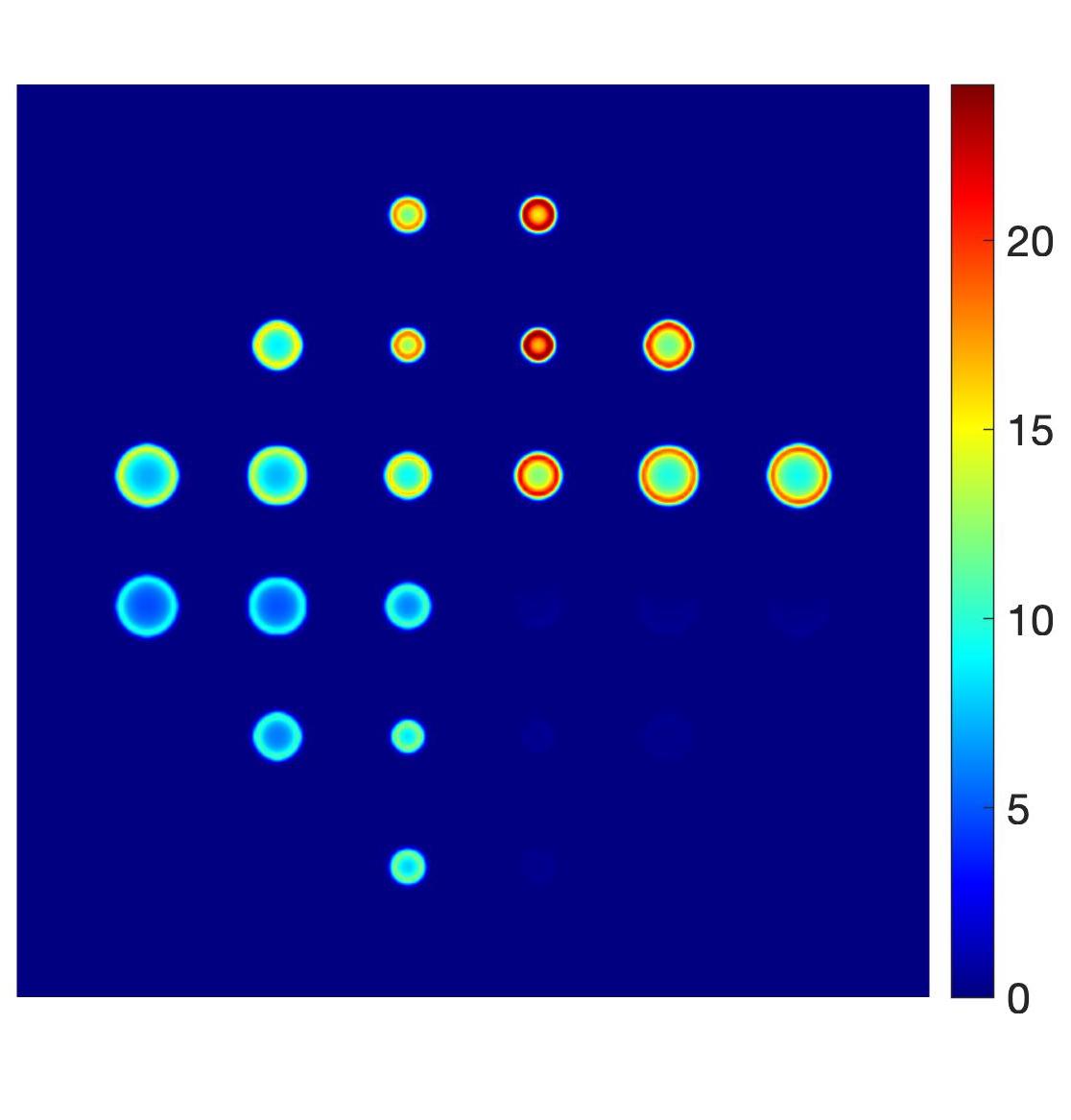}}
\put(-.3,-.7){\includegraphics[height=60mm]{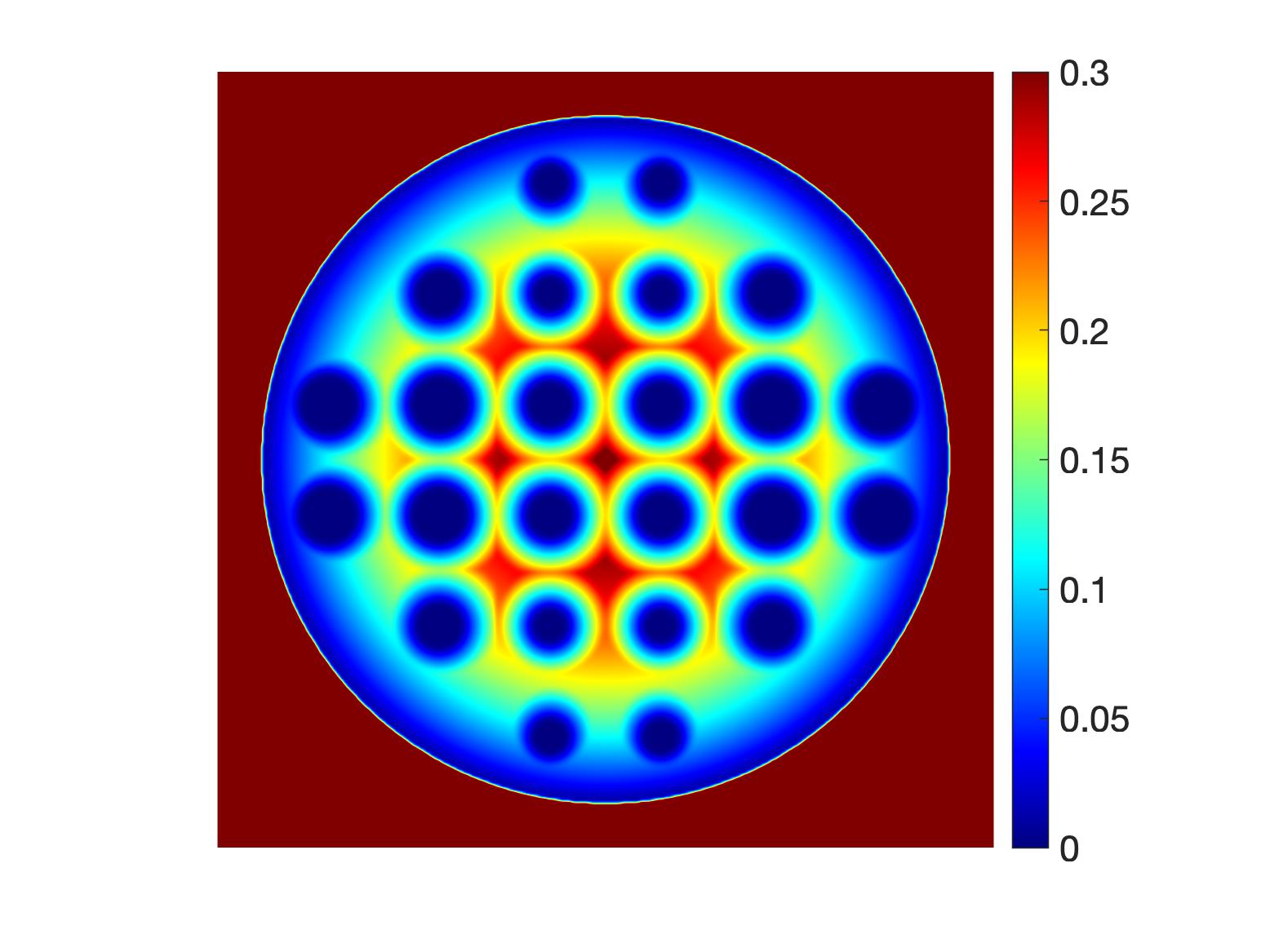}}
\end{picture}
\caption{Osseointegration example: Normal contact pressure $p_\mrc$ (left; units [MPa]) and contact gap $g_\mrn$ (right; units [mm]) due to the implant asperities prior to osseointegration.}
\label{f:RCSI-pg}
\end{center}
\end{figure}

\begin{figure}[h!]
\begin{center} \unitlength1cm
\begin{picture}(0,3.3)
\put(-8.02,-.5){\includegraphics[height=37mm]{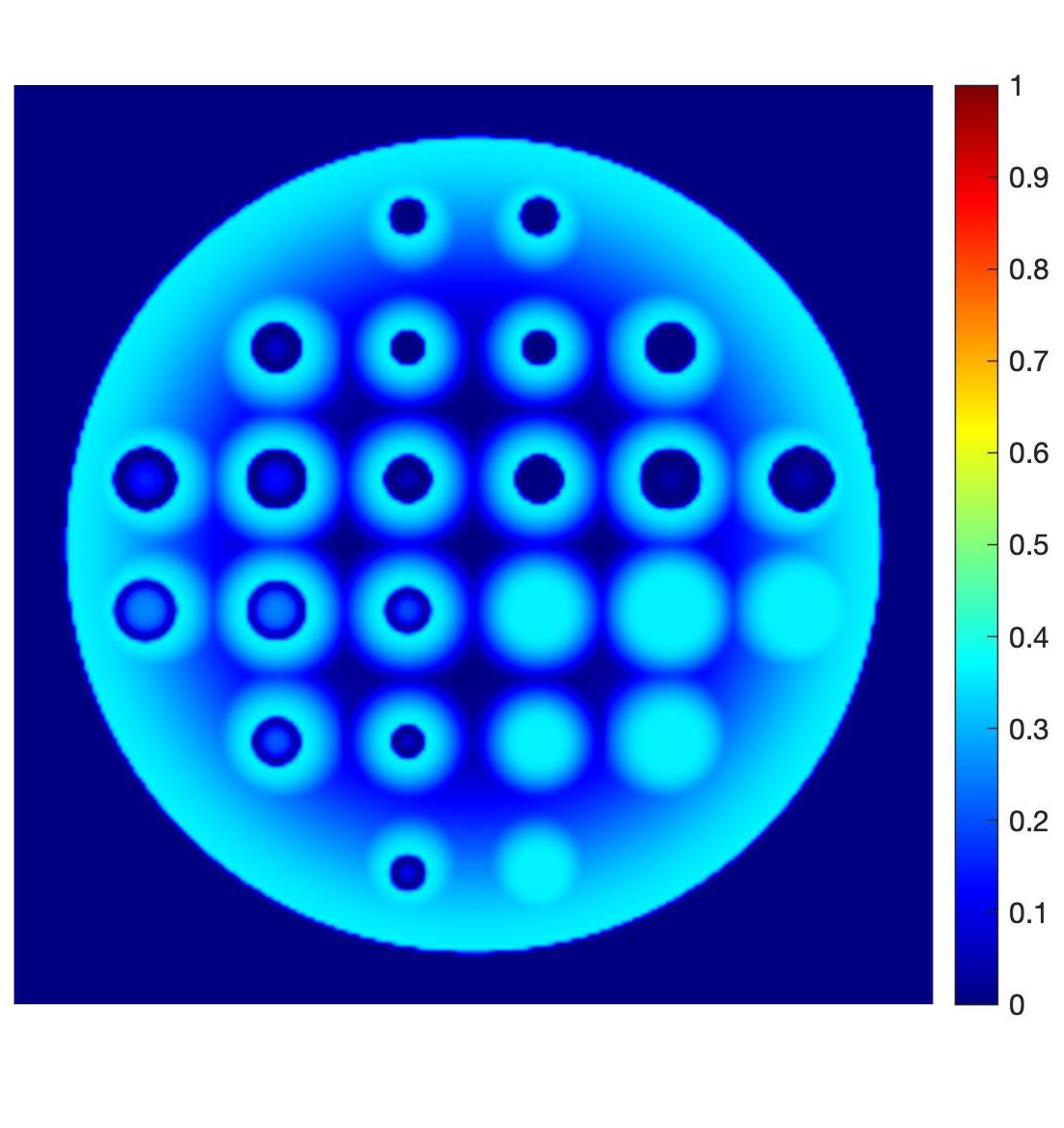}}
\put(-4.92,-.5){\includegraphics[height=37mm]{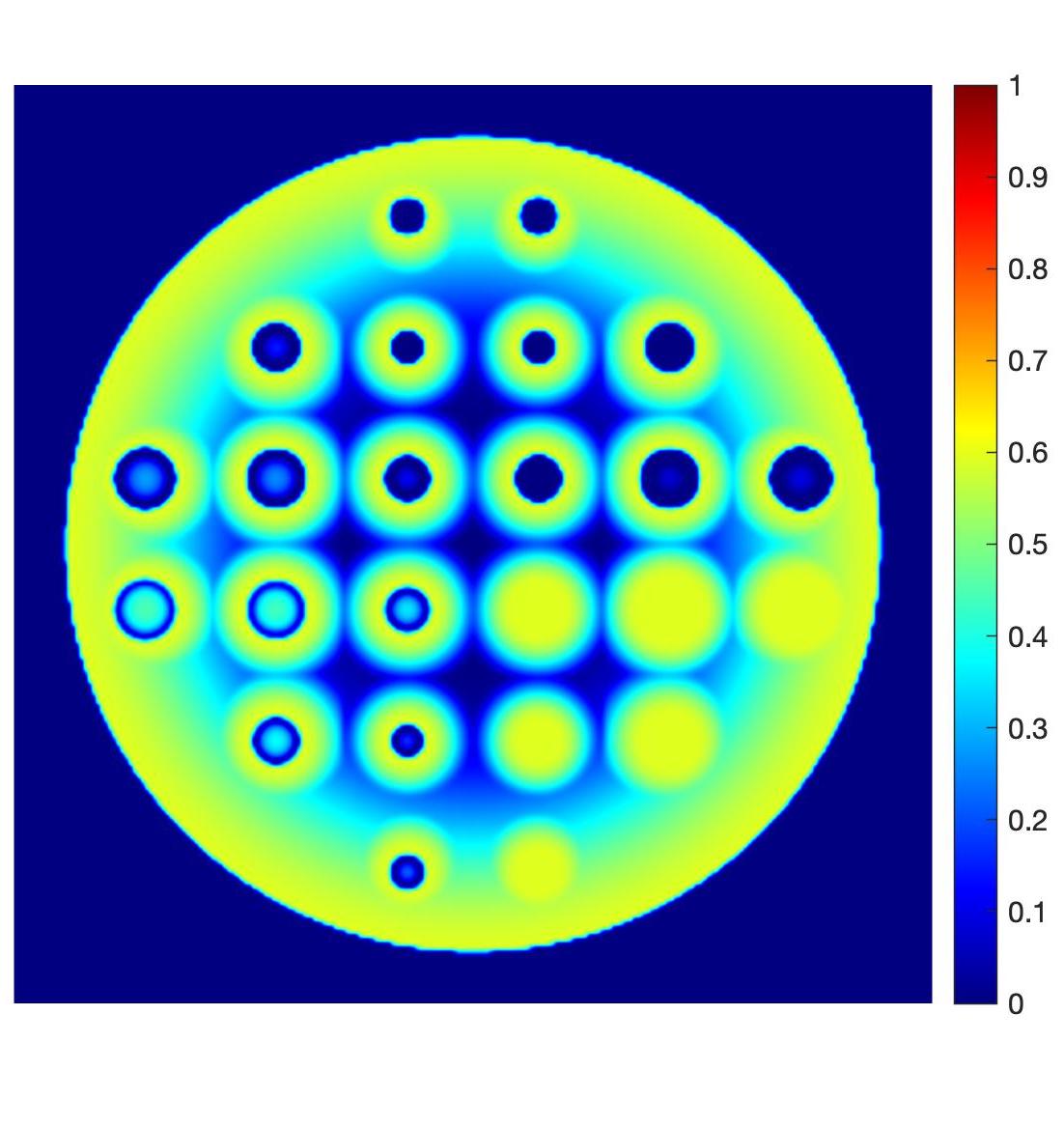}}
\put(-1.82,-.5){\includegraphics[height=37mm]{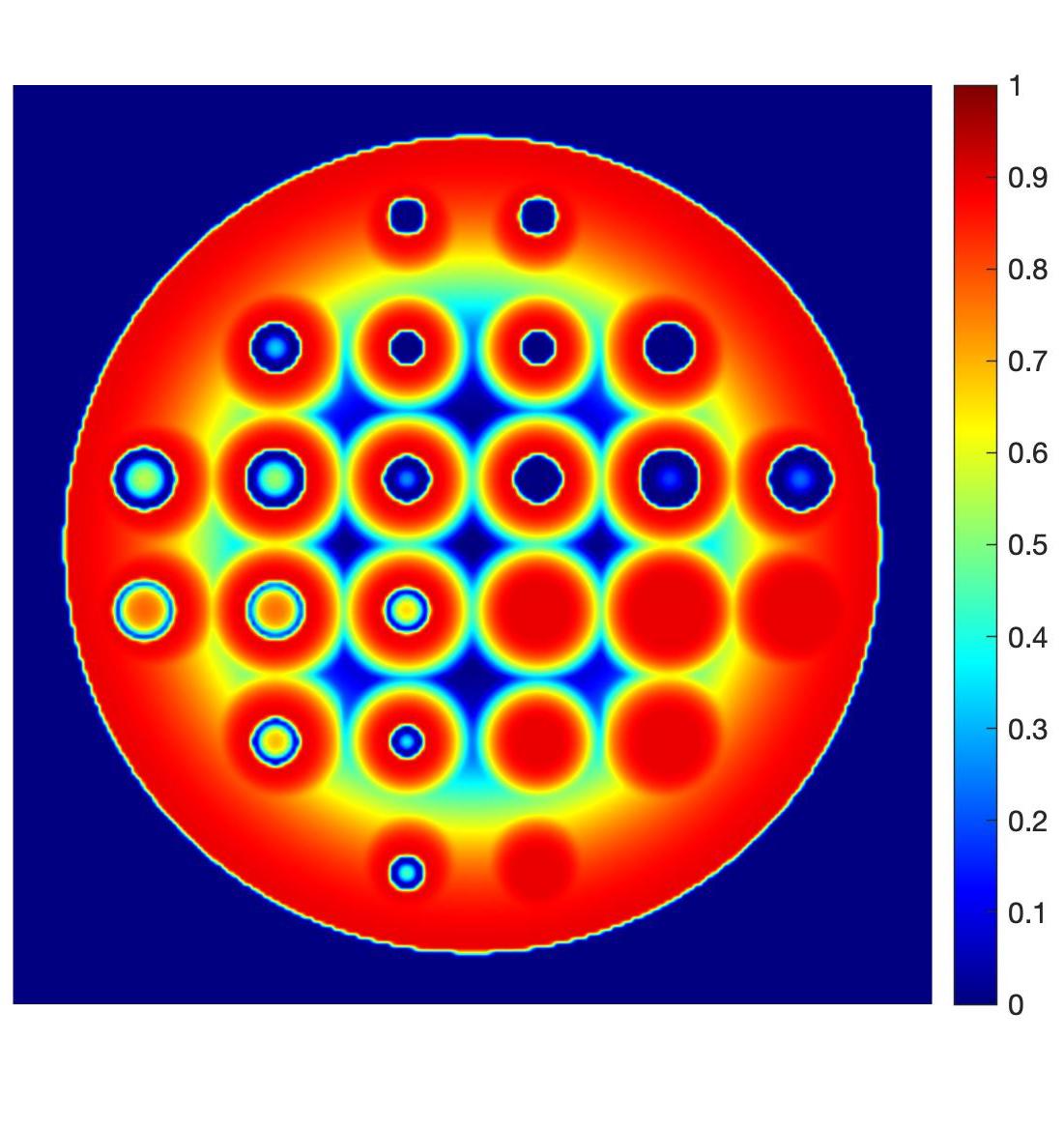}}
\put(1.28,-.5){\includegraphics[height=37mm]{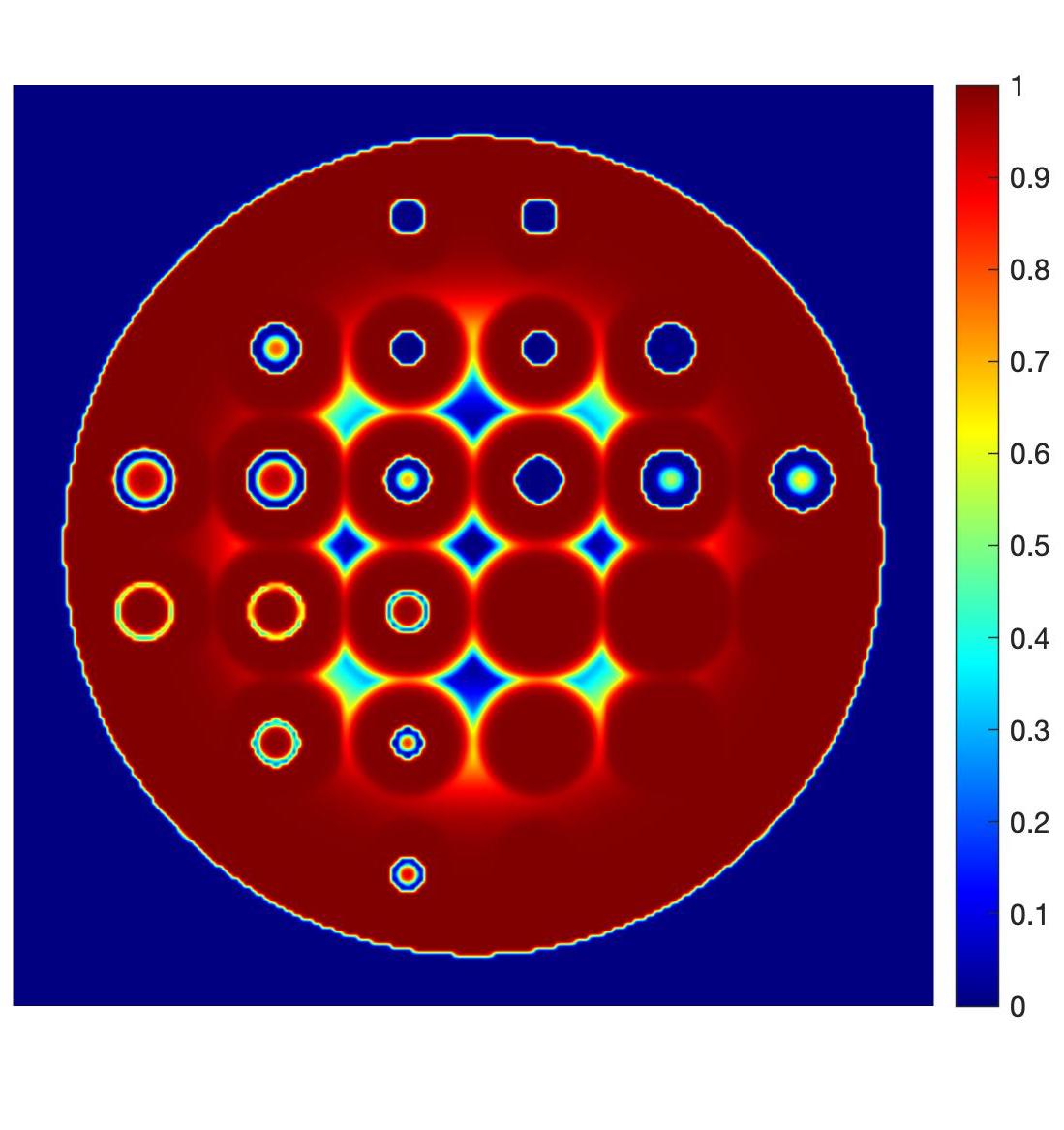}}
\put(4.38,-.5){\includegraphics[height=37mm]{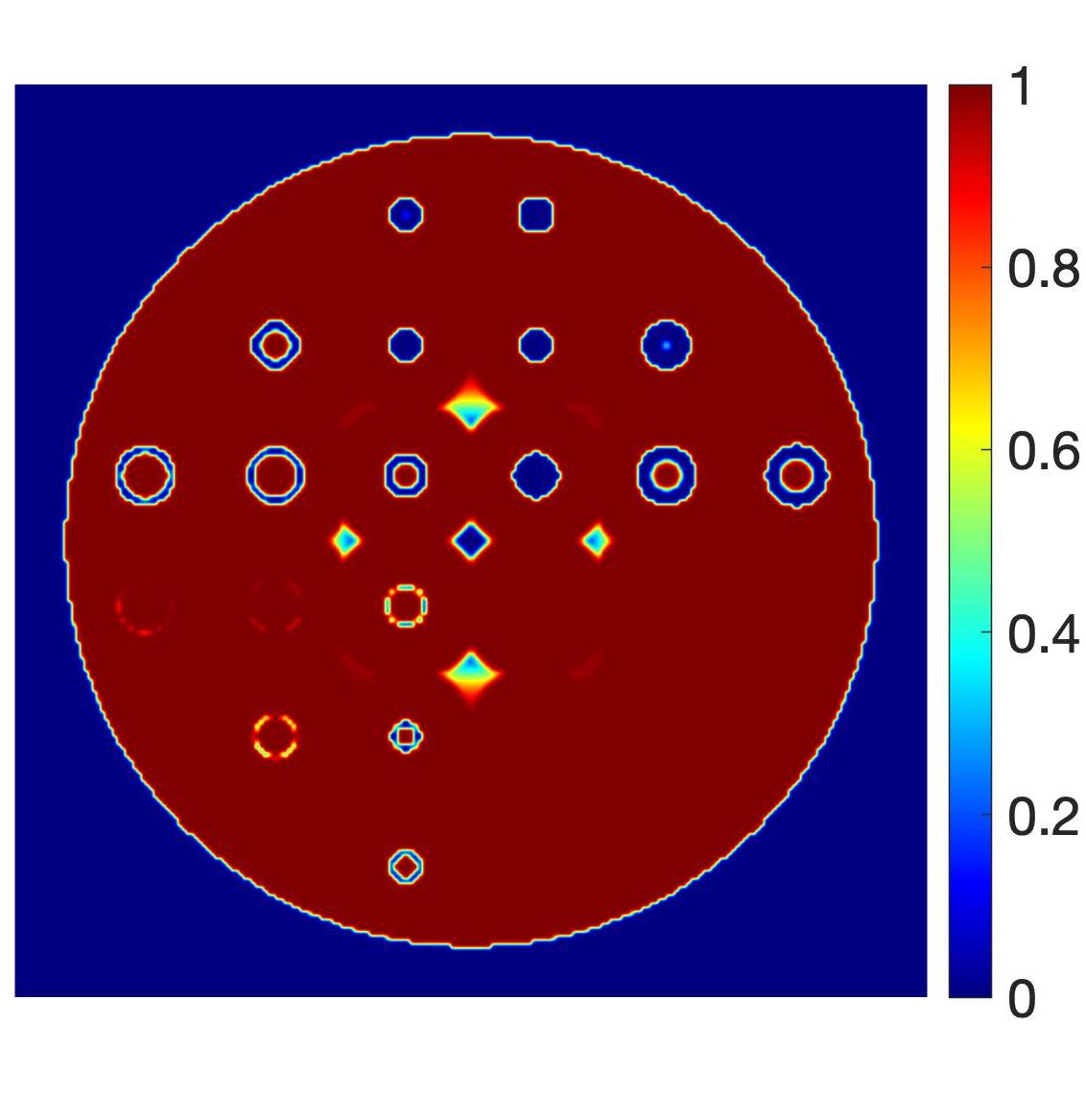}}
\put(-7.92,.0){\textcolor{white}{\scriptsize{5d}}}
\put(-4.82,.0){\textcolor{white}{\scriptsize{10d}}}
\put(-1.72,.0){\textcolor{white}{\scriptsize{25d}}}
\put(1.38,.0){\textcolor{white}{\scriptsize{100d}}}
\put(4.48,.0){\textcolor{white}{\scriptsize{1000d}}}
\end{picture}
\caption{Osseointegration example: Evolution of the bonding state $\phi(t)$ due to the varying contact gap and pressure shown in Fig.~\ref{f:RCSI-pg}.
(d = days.)}
\label{f:RCSI-res1}
\end{center}
\end{figure}

Fig.~\ref{f:RCSI-pg} shows the resulting contact pressure and normal gap for a normal contact load of 13.46N.
With these, osseointegration model \eqref{e:crosseo} can be evaluated, and then bonding ODE \eqref{e:bondODE} can be integrated analytically, see Appendix~\ref{s:anaODE}.
The resulting bonding evolution is shown in Fig.~\ref{f:RCSI-res1}.
As seen, bonding is fastest at locations where $p_\mrc \leq p_\mathrm{opt}$ and $g_\mrn \leq g_\mathrm{opt}$, while no bonding occurs at locations where either $p_\mrc \geq p_\mathrm{lim}$ or $g_\mrn \geq g_\mathrm{lim}$.
The model is thus able to capture the mechanically-sensitive bonding behavior of implant osseointegration.
The results serve as a proof-of-concept.
Further studies are needed to calibrate and refine the model properties for real osseointegration scenarios.

\subsection{Two-way coupling: Chemo-mechanical debonding of a balloon}\label{s:ex1}

\begin{figure}[h]
\begin{center} \unitlength1cm
\begin{picture}(0,4.0)
\put(-8.9,-.6){\includegraphics[height=48mm]{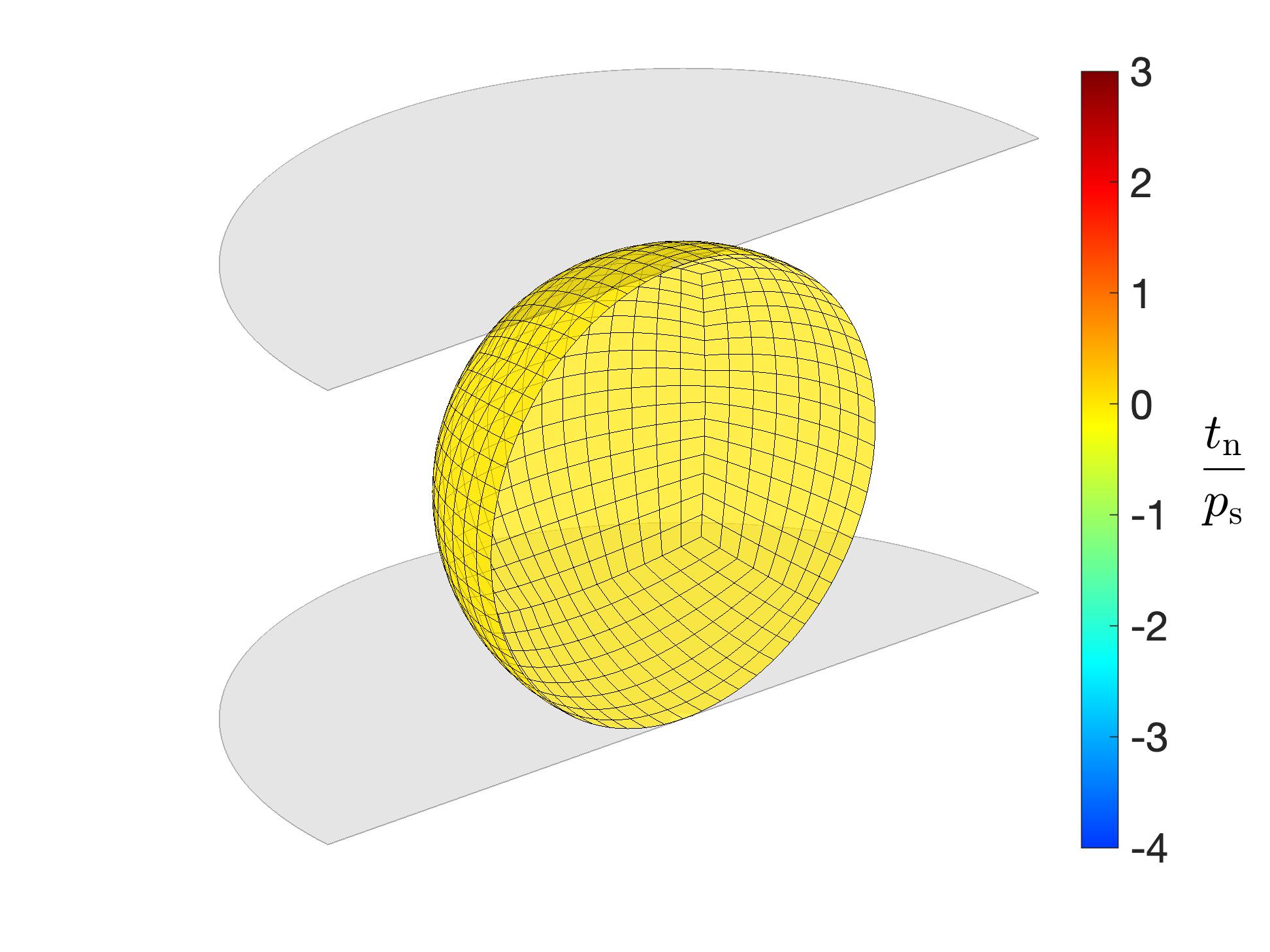}}
\put(-3.65,-.6){\includegraphics[height=48mm]{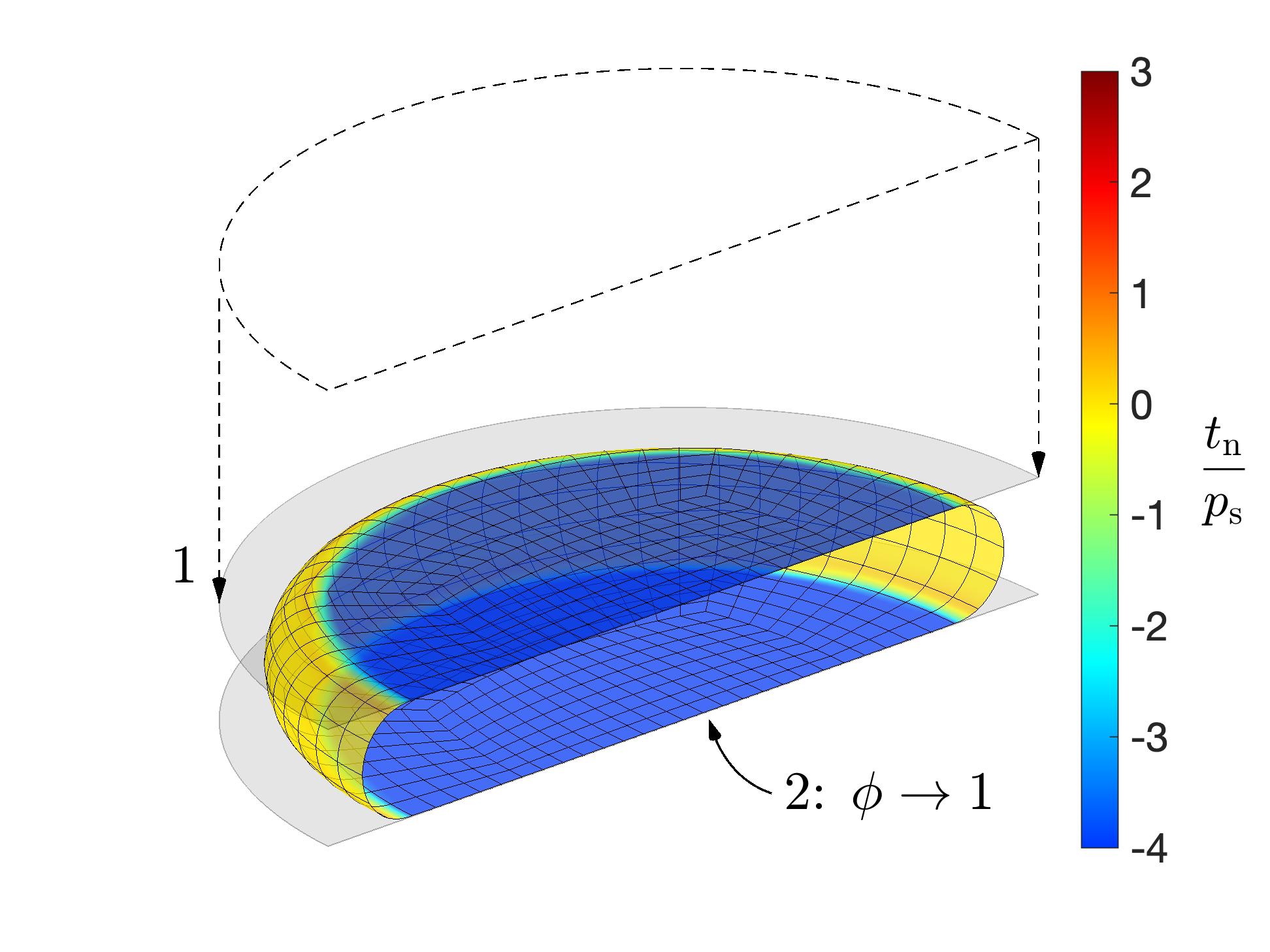}}
\put(1.65,-.6){\includegraphics[height=48mm]{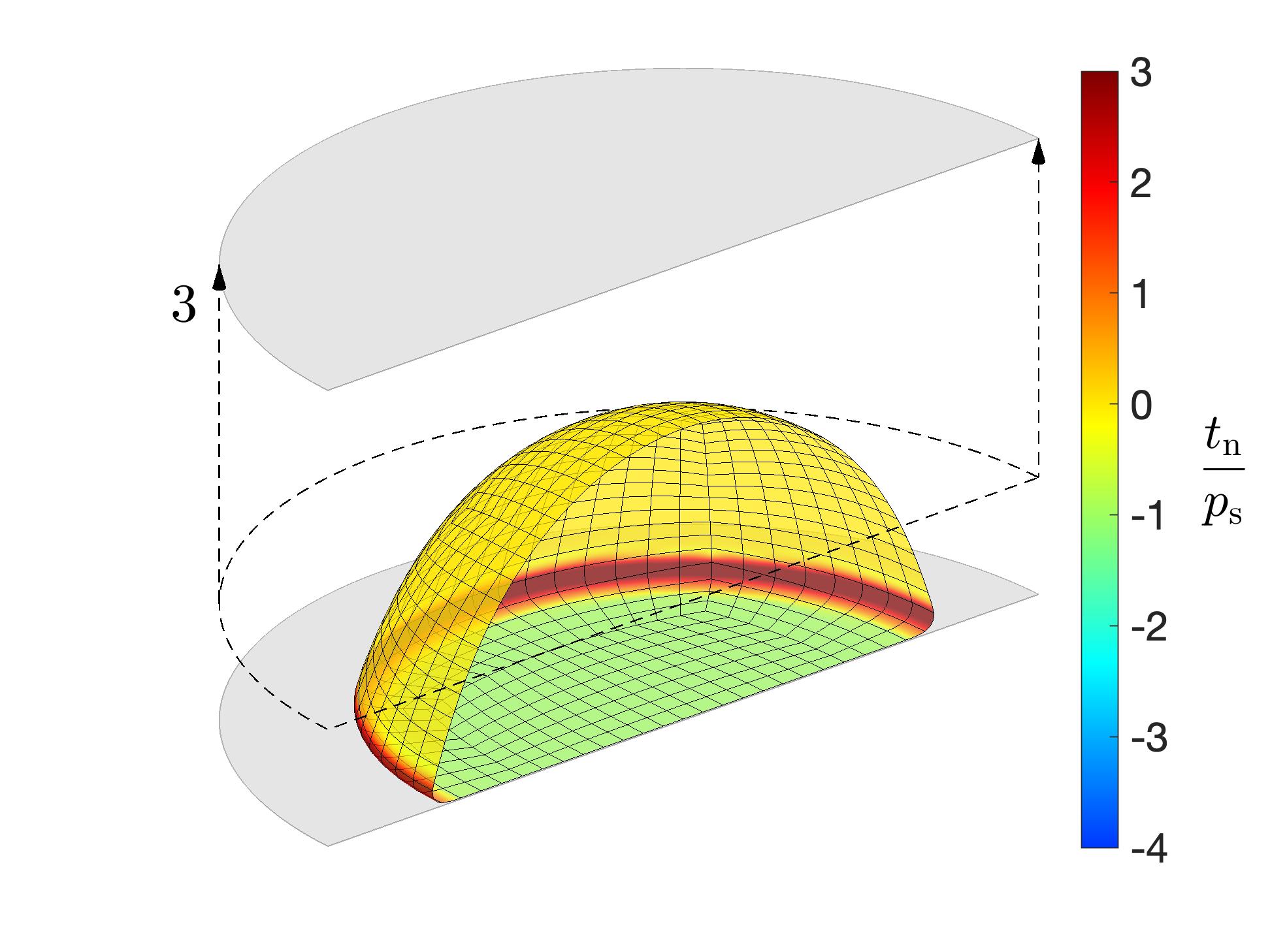}}
\put(-7.95,-.15){\small (a)}
\put(-2.75,-.15){\small (b)}
\put(2.6,-.15){\small (c)}
\end{picture}
\caption{Balloon example -- squeezing \& unloading: An initially spherical balloon (a) is squeezed between two rigid plates (step 1, b). 
Perfect bonding is then assumed to occur at the bottom surface (step 2) such that the removal of the top plate (step 3) produces an initially deformed and bonded configuration (c).
The colors show the normal contact traction $t_\mrn$ (= negative contact pressure $p_\mrc$). 
The contact pressure during maximum squeezing is $p_\mrc = 4.41\,p_\mrs$, the maximum tension after unloading is $t_\mrn = 3.90\,p_\mrs$. 
The figure shows half of the symmetric balloon using mesh $m=8$.}
\label{f:CoboBalloon1}
\end{center}
\end{figure}

The second example considers the debonding of a rubber balloon that is sticking to a rigid surface.
The initial sticking configuration is obtained in three steps:
An initially spherical balloon is squeezed between two rigid plates (step 1).
Bonding is then assumed to occur at the bottom surface (step 2).
Once full bonding occurs, the loading is released by removing the top plate (step 3).
The process is illustrated in Fig.~\ref{f:CoboBalloon1}.

The balloon is described by the finite element membrane model of \citet{membrane}, which is based on an incompressible Neo-Hookean material model.
The interior of the balloon is also assumed incompressible, using a Lagrange multiplier model to enforce the incompressibility constraint.
Initially the balloon is a sphere with radius $R$ that has the internal pressure $p_\mrs$ and surface tension $\gamma = p_\mrs R/2$, 
which arise due to an isotropic pre-stretch of the membrane by $\lambda = 1.1$.
All material parameters used in the example are given in Table~\ref{t:CoboBalloon1}.
\begin{table}[h]
\centering
\begin{tabular}{|r|r|r|r|}
  \hline
   symbol & material parameter & value & unit \\[0mm] \hline 
   & & & \\[-3.5mm]   
   $R$ &  initial balloon radius & 1 & $R$ \\ [.5mm] 
   $p_\mrs$ & initial balloon pressure & 1 & $p_\mrs$ \\ [.5mm] 
   $\unde{G}$ & membrane shear stiffness & 1 & $p_\mrs R$ \\[0mm] 
   $\overleftarrow{\unde{K_0}}/g_0^2$ & debonding energy density & 100 & $p_\mrs/R$  \\ [.5mm] 
   $\unde{\bar m}$ & debonding ``mass" & 0.04 \& 1  & $Tp_\mrs R$ \\[.5mm]
   $\unde{E_c}$ & contact stiffness & 1000 & $p_\mrs/R$ \\[0mm]
    \hline
\end{tabular}
\vspace{-1mm}
\caption{Balloon debonding: Considered material parameters. 
$R$, $T$ and $p_\mrs$ are a length, time and pressure scale that can be arbitrary.}
\label{t:CoboBalloon1}
\end{table}
They are based on the arbitrary length, time and pressure scales $R$, $T$ and $p_\mrs$.
The finite element meshes shown in Table~\ref{t:CoboBalloon2} are used to discretize the balloon surface.
They are for a quarter sphere model composed of six patches with $m^2$ elements each.
\begin{table}[h]
\centering
\begin{tabular}{|r|r|r|r|r|r|r|r|}
  \hline
   $m$ & $n_\mathrm{el}$ & $n_\mathrm{no}$ & 3-dof & 4-dof & $n_{t3}$ & $n_{t4a}$ & $n_{t4c}$ \\[0mm] \hline 
   & & & & & & & \\[-4mm]   
   1 &  6 & 33 & 100 & 133 & 10 & 25 & 40 \\ [0mm] 
   2 & 24 & 113 & 340 & 453 & 20 & 50 & 80 \\ [0mm] 
   4 & 96 & 417 & 1,252 & 1,669 & 40 & 100 & 160 \\[0mm] 
   8 &  384 & 1,601 & 4,804 & 6,405 & 80 & 200 & 320 \\[0mm] 
   16 & 1,536 & 6,273 & 18,820 & 25,093 & 160 & 400 & 640 \\[0mm] 
   32 & 6,144 & 24,833 & 74,500 & 99,333 & 320 & 800 & 1,280 \\[0mm]    
   64 & 24,576 & 98,817 & 296,452 & 395,269 & 640 & 1,600 & 2,560\\[0mm]
   \hline
\end{tabular}
\vspace{-1mm}
\caption{Balloon debonding: FE meshes of a quarter balloon based on quadratic Lagrange elements. 
Condensing-out the bonding field according to Sec.~\ref{s:QPform} leaves three of the original four nodal dofs.}
\label{t:CoboBalloon2}
\end{table}

\begin{figure}[h]
\begin{center} \unitlength1cm
\begin{picture}(0,5.7)
\put(-8,-.12){\includegraphics[height=58mm]{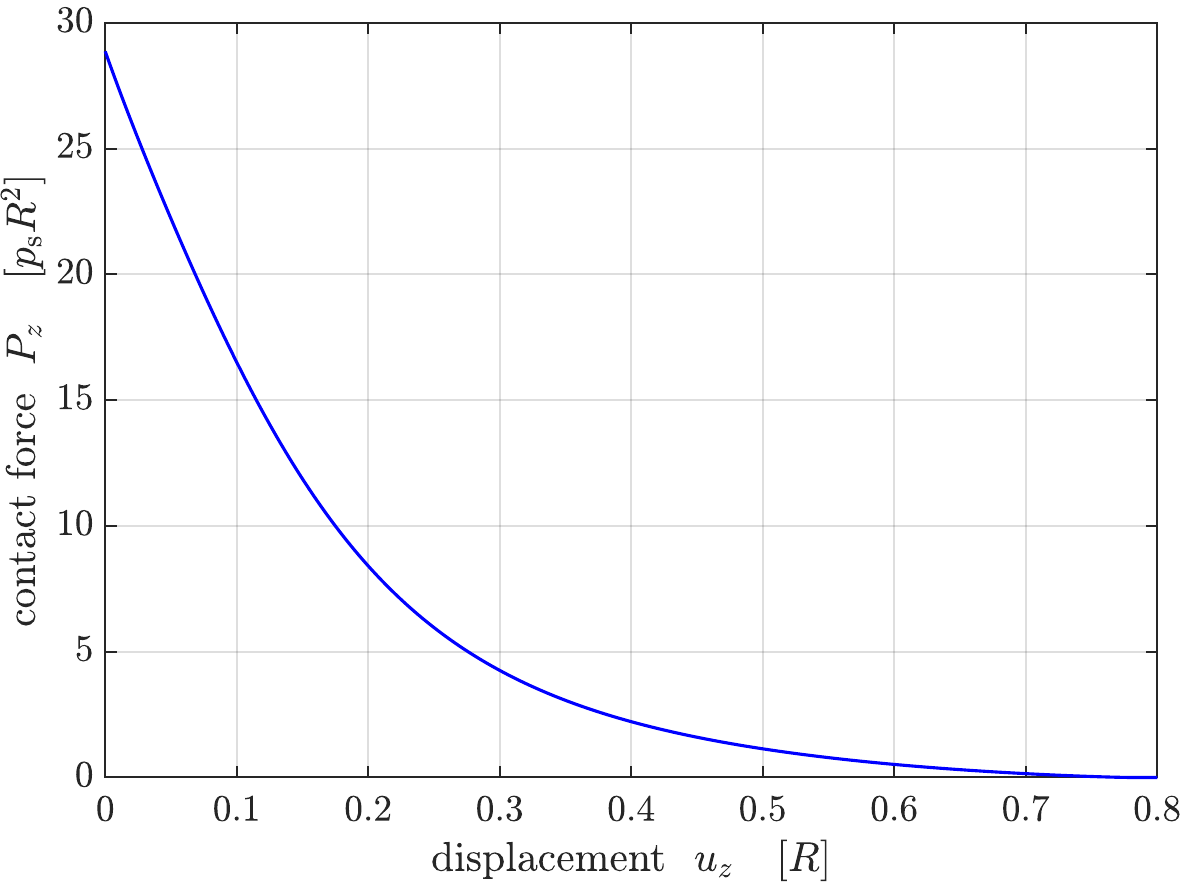}}
\put(0.2,-.2){\includegraphics[height=58mm]{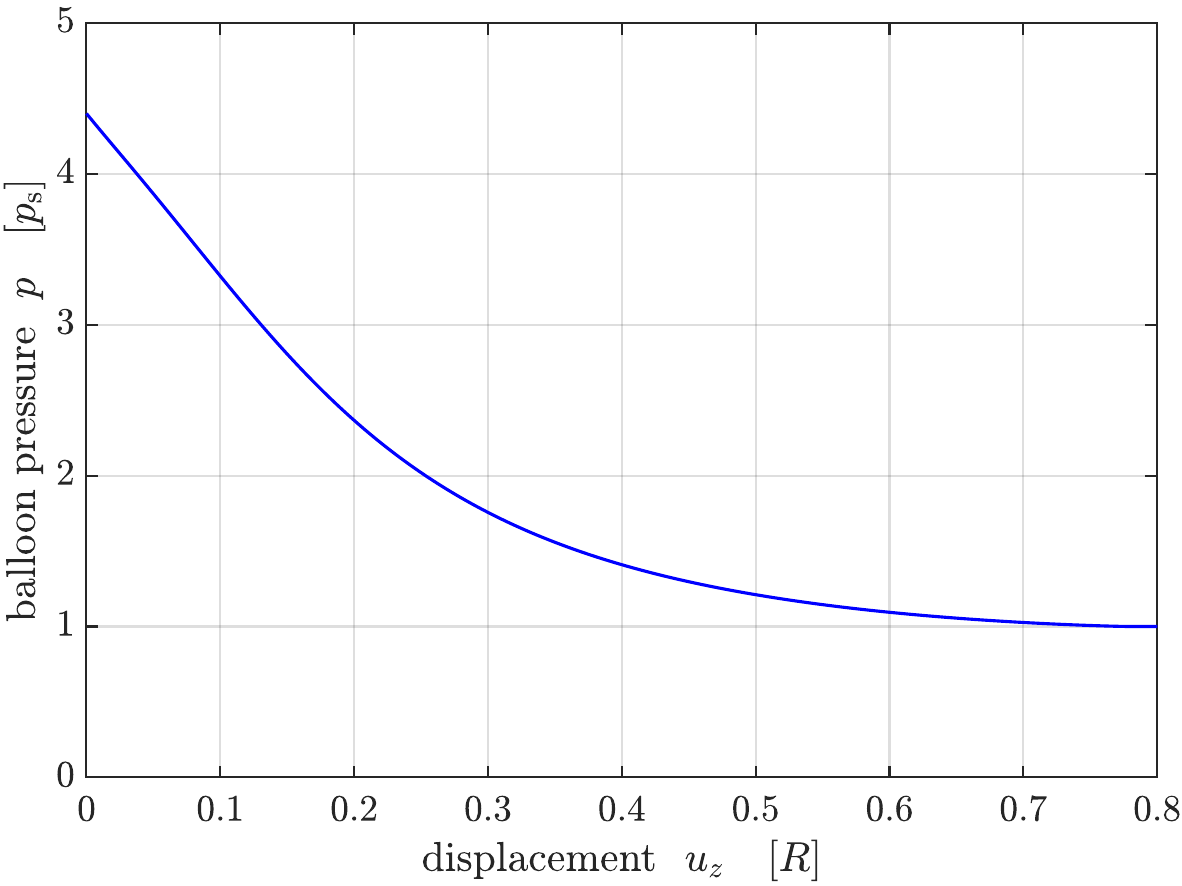}}
\put(-7.95,-.15){\footnotesize (a)}
\put(0.2,-.15){\footnotesize (b)}
\end{picture}
\caption{Balloon example: Decrease of (a) contact force and (b) balloon pressure during unloading (step 3).}
\label{f:CoboBalloon2}
\end{center}
\end{figure}

During step 1 the balloon is squeezed to the height $R/2$ as is shown in Fig.~\ref{f:CoboBalloon1}b, which results in the constant contact pressure $p_\mrc = 4.41\,p_\mrs$.
Bonding is then considered to occur on the lower contact surface during step 2.
Since the contact pressure is constant, bonding can be expected to be uniform, and the initial bonding state is thus set to $\phi = 1$ on the entire lower contact surface.
When the load imposed by the two plates is released in step 3, the contact pressure redistributes itself as shown in Fig.~\ref{f:CoboBalloon1}c.
The change in the maximum contact pressure (= balloon pressure) and contact force (= integrated pressure) during unloading are shown in Fig.~\ref{f:CoboBalloon2}.

\begin{figure}[h]
\begin{center} \unitlength1cm
\begin{picture}(0,6.0)
\put(-8,2.7){\includegraphics[height=34mm]{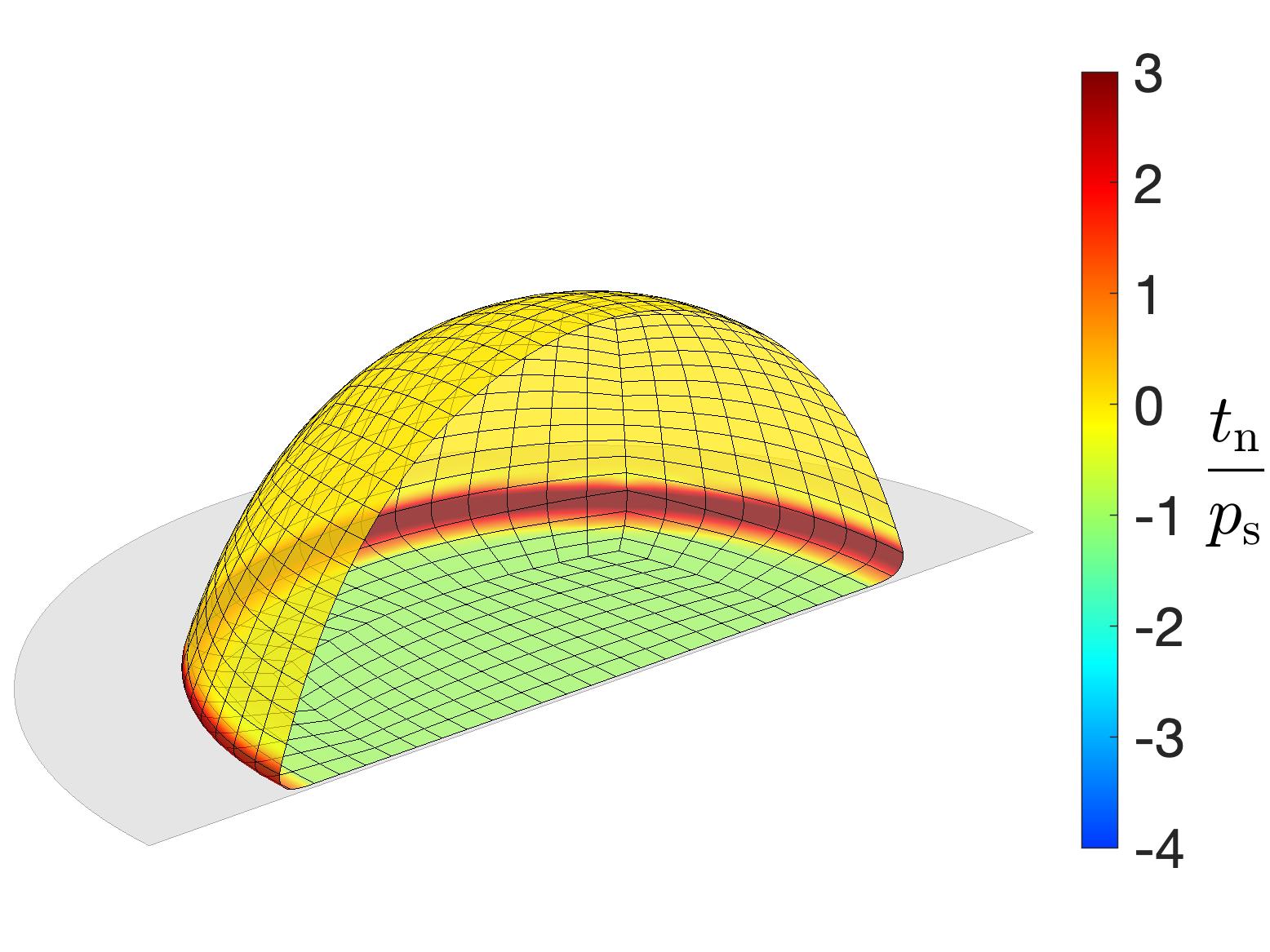}}
\put(-4.2,2.7){\includegraphics[height=34mm]{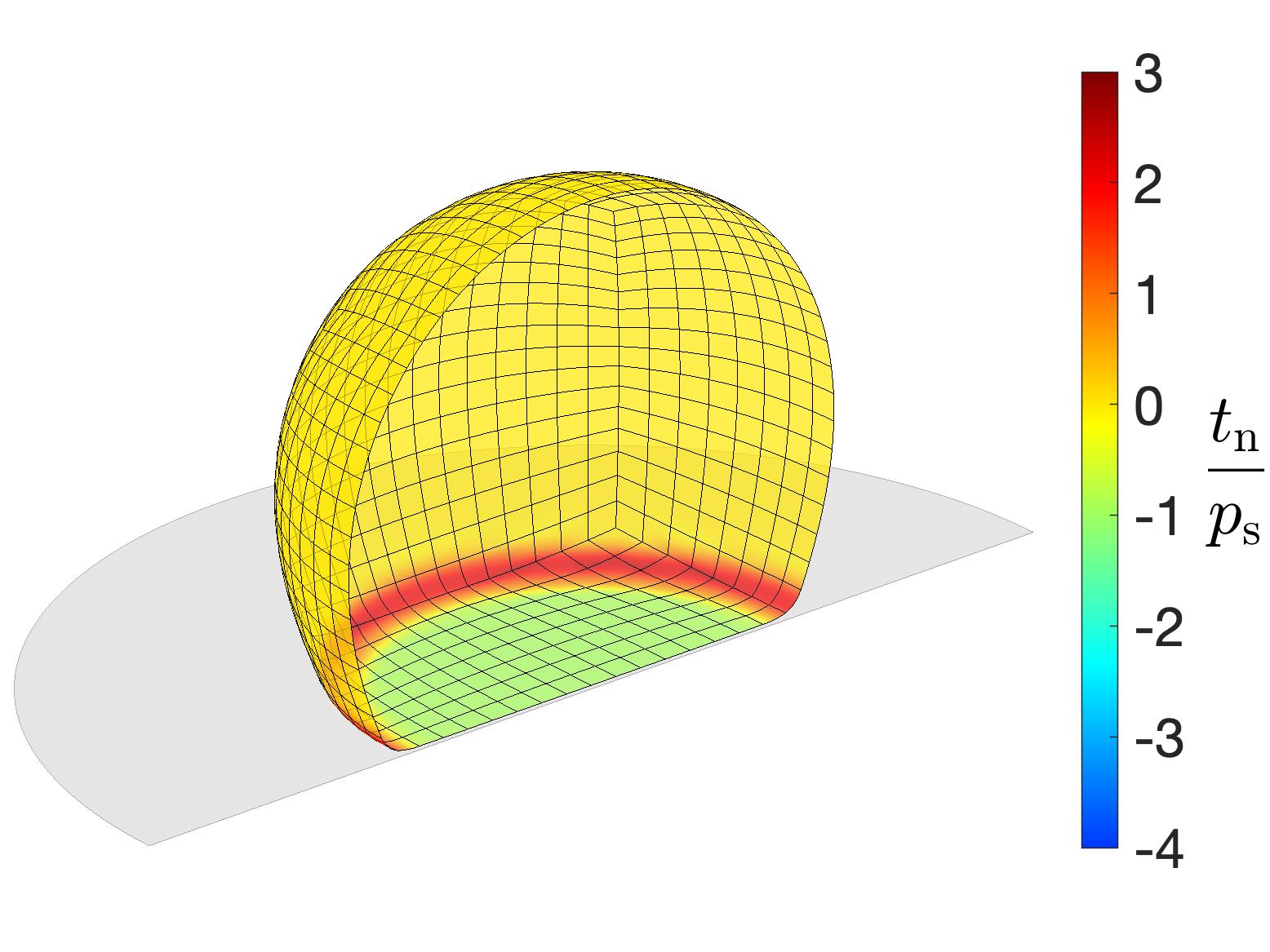}}
\put(-0.4,2.7){\includegraphics[height=34mm]{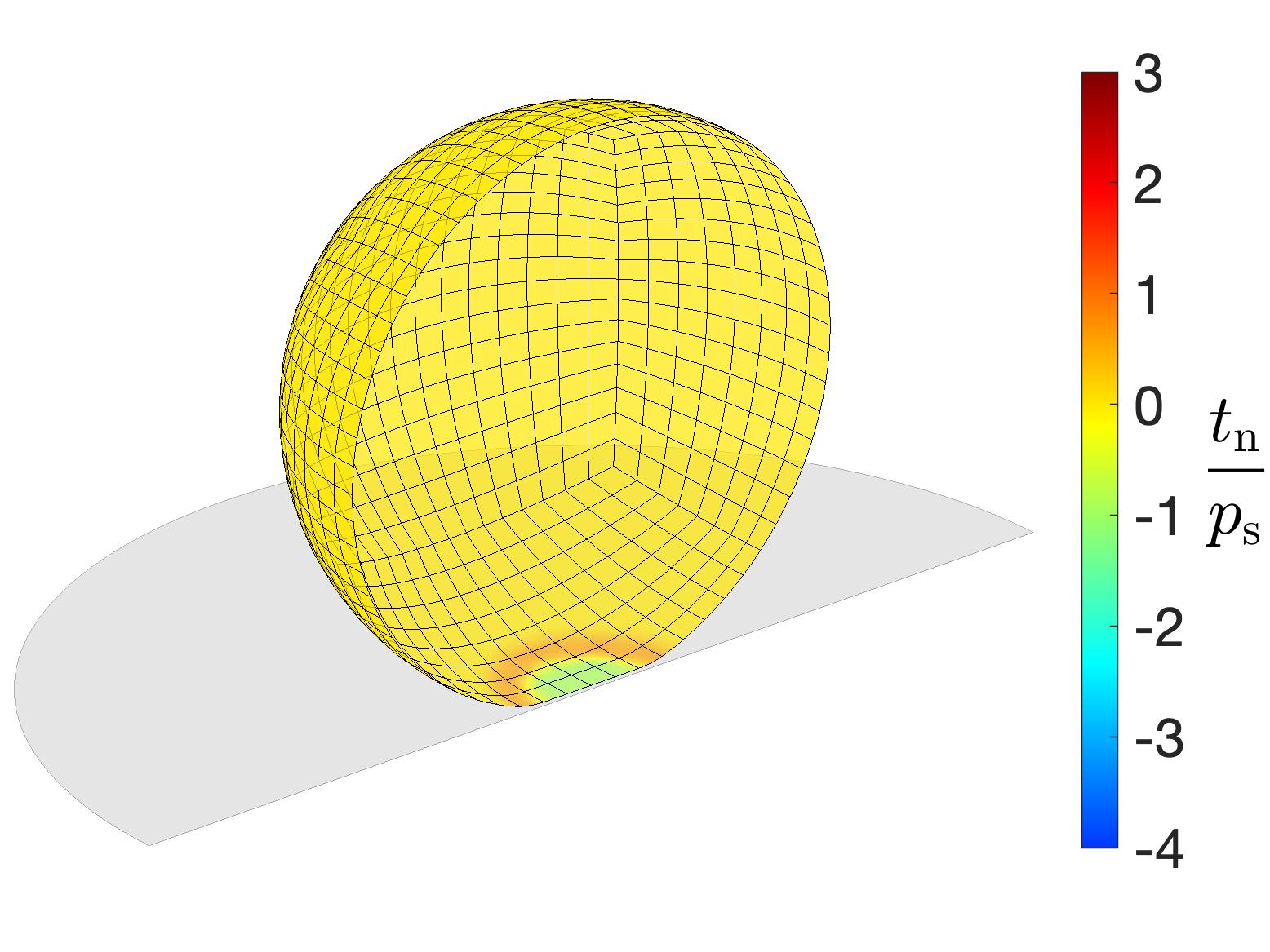}}
\put(3.4,2.7){\includegraphics[height=34mm]{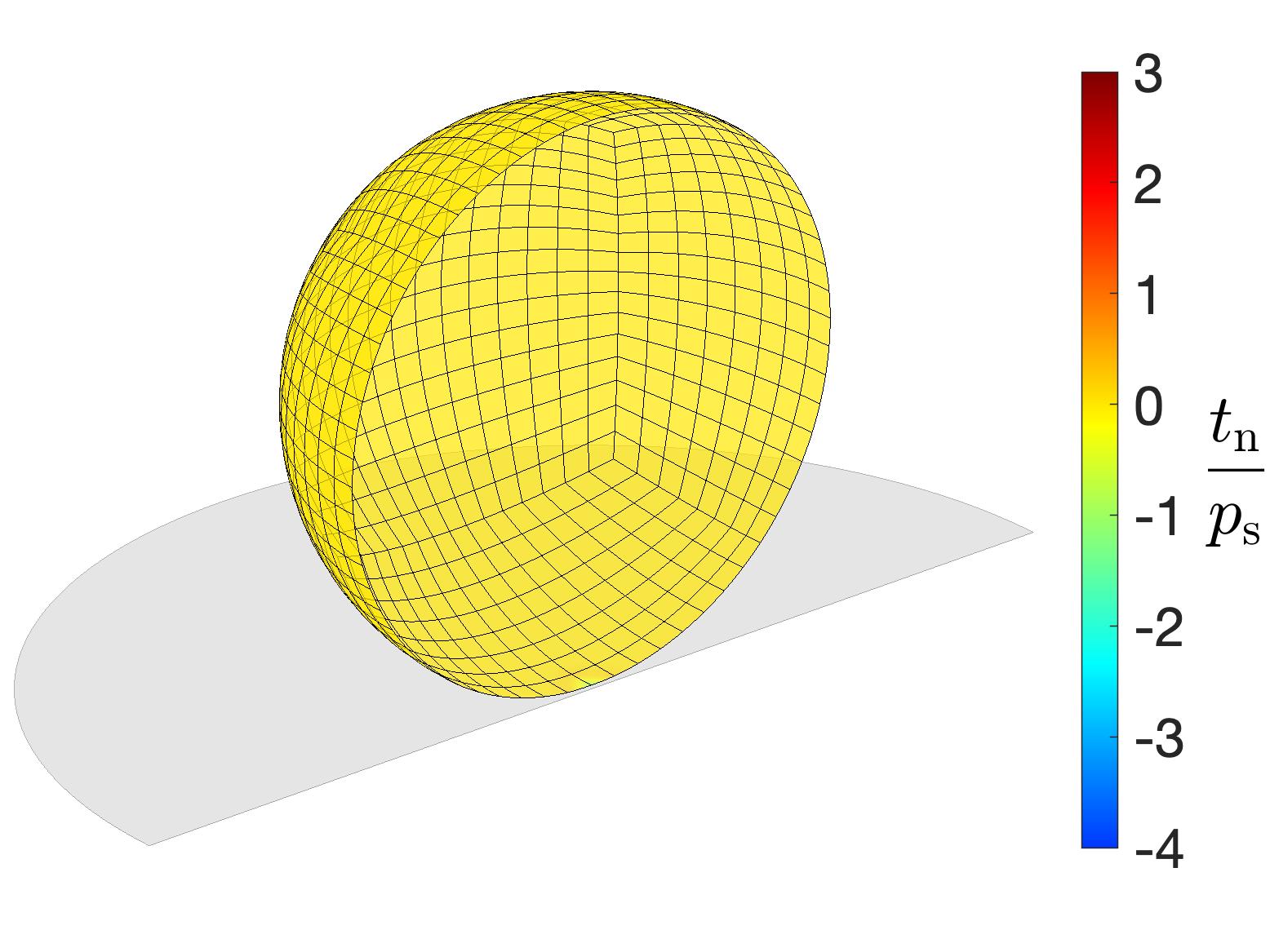}}
\put(-8,-.4){\includegraphics[height=34mm]{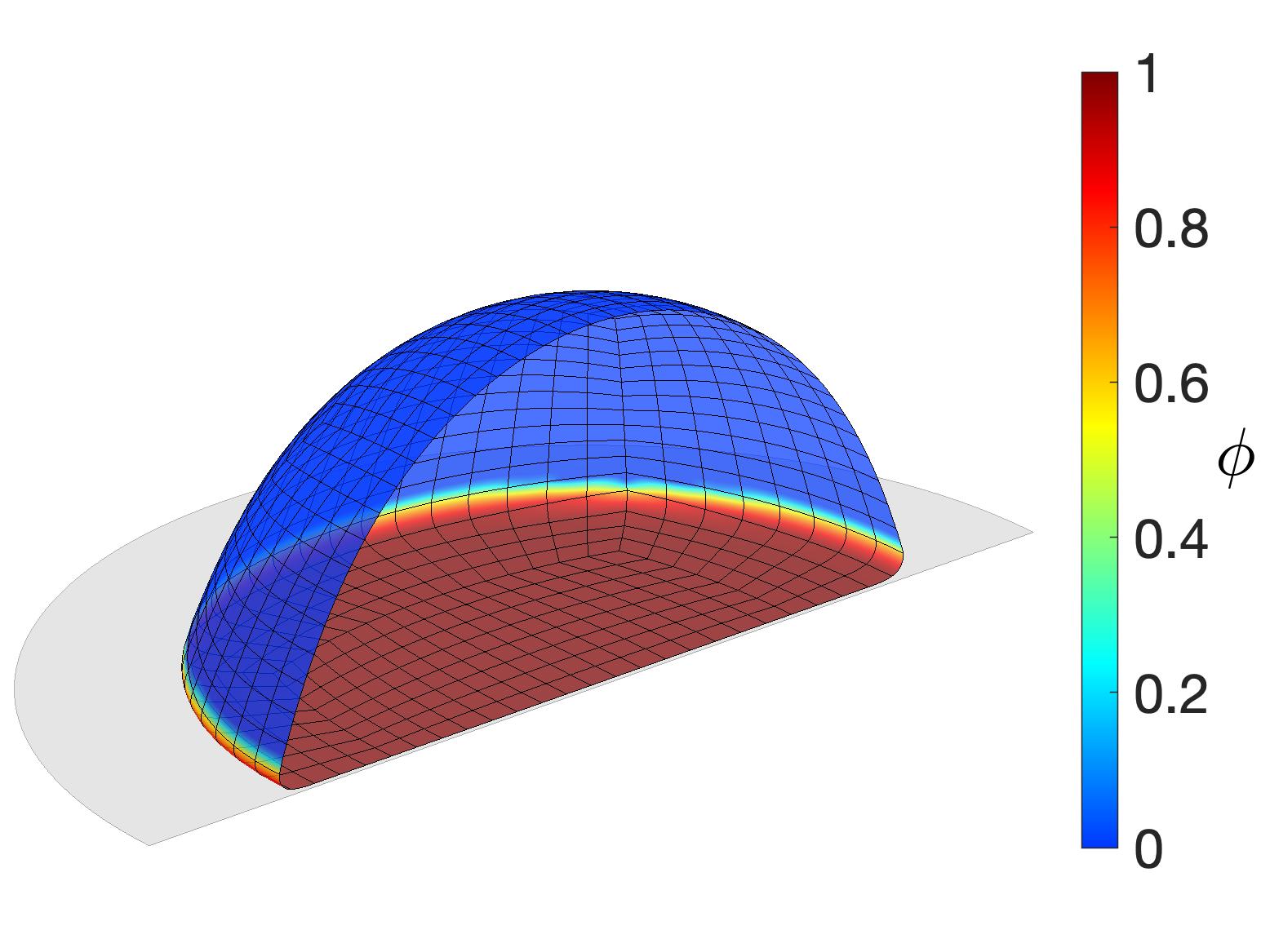}}
\put(-4.2,-.4){\includegraphics[height=34mm]{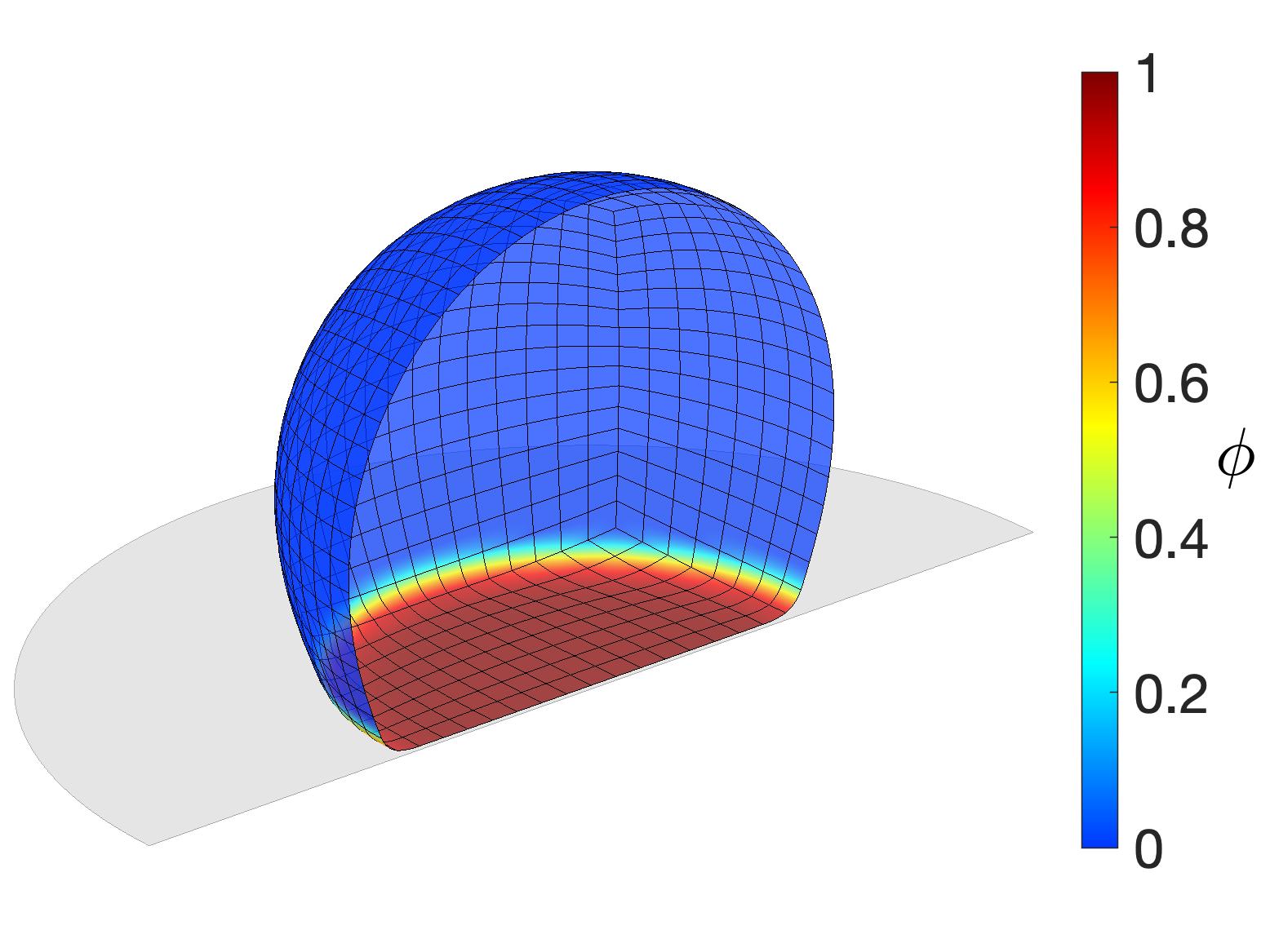}}
\put(-0.4,-.4){\includegraphics[height=34mm]{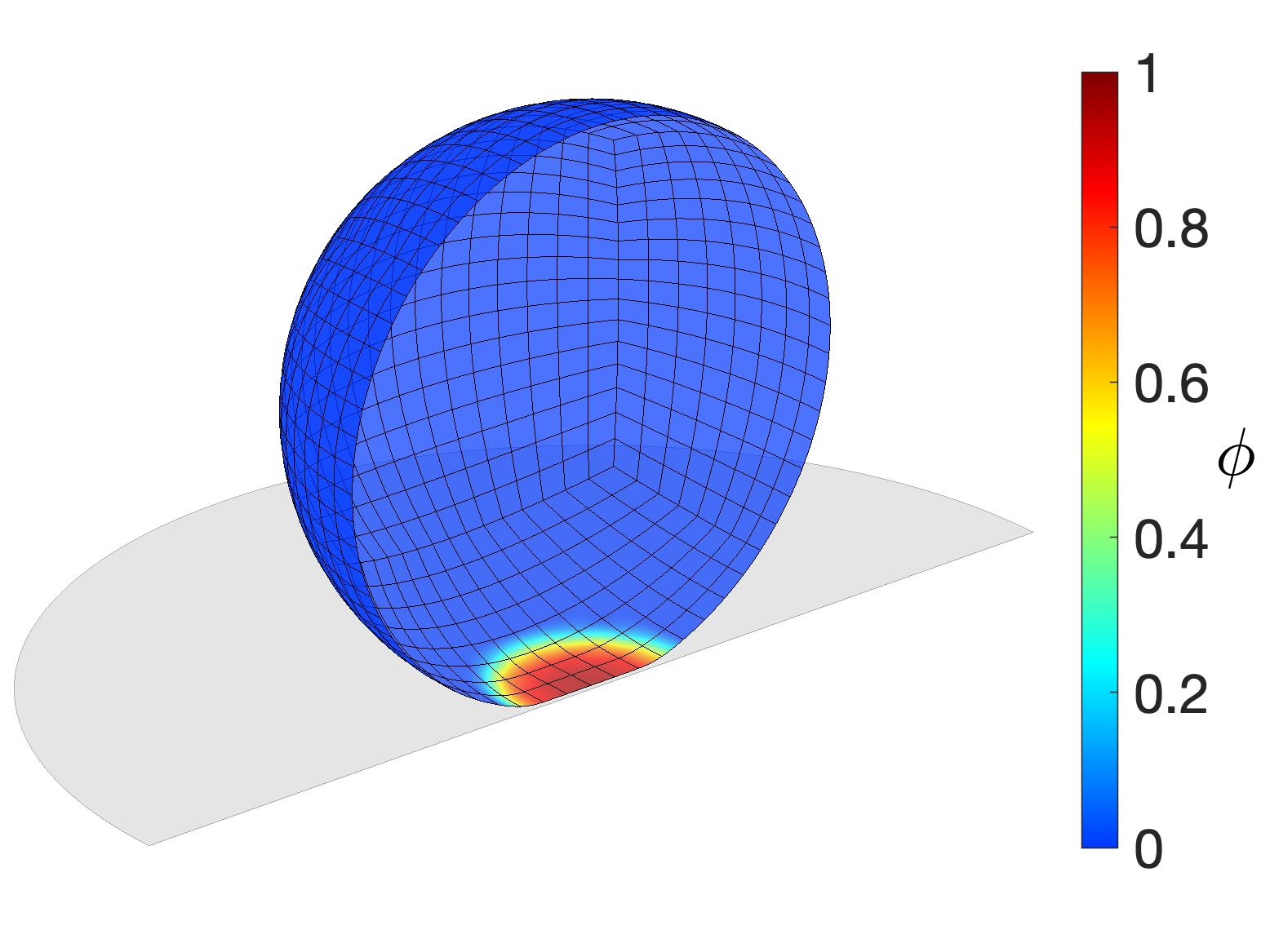}}
\put(3.4,-.4){\includegraphics[height=34mm]{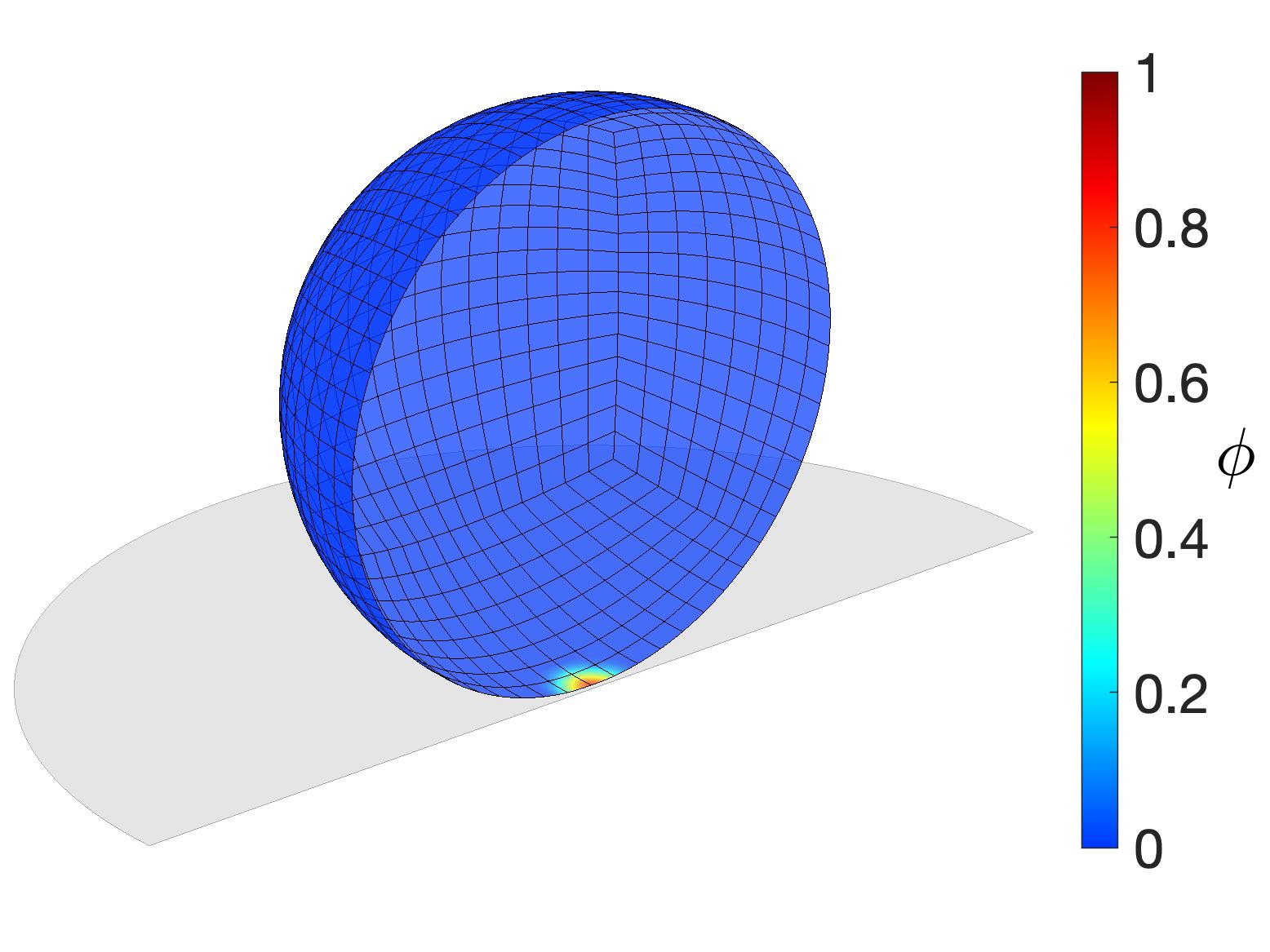}}
\end{picture}
\caption{Balloon example -- debonding (step 4): Evolution of the contact traction $t_\mrn$ (top row) and bonding state $\phi$ (bottom row) during gradual debonding (at $t_3+[0,\,0.25,\,2,\,10]T$; from left to right).
The results show half of the balloon using  mesh $m=8$.}	
\label{f:CoboBalloon3}
\end{center}
\end{figure}

The final, 4.~step then simulates debonding, using the quadrature point formulation of Sec.~\ref{s:QPform} and the debonding model of Sec.~\ref{s:con1}.
Debonding starts at the boundary of the contact surface, due to the high tensile forces there, and then propagates inward, leading to a gradual release of the balloon from the lower plate.
This process is shown in Fig.~\ref{f:CoboBalloon3}.

\begin{figure}[h]
\begin{center} \unitlength1cm
\begin{picture}(0,5.8)
\put(-7.95,-.12){\includegraphics[height=58mm]{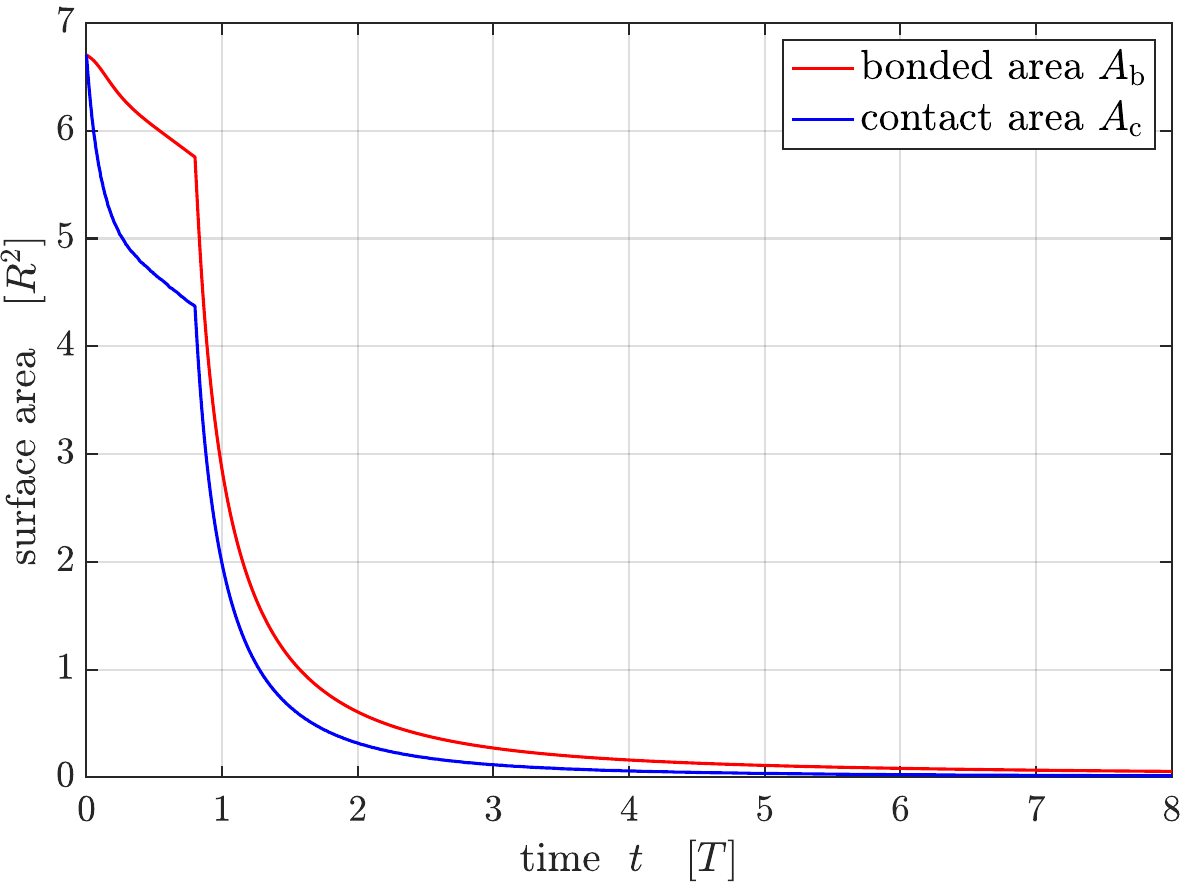}}
\put(0.2,-.2){\includegraphics[height=58mm]{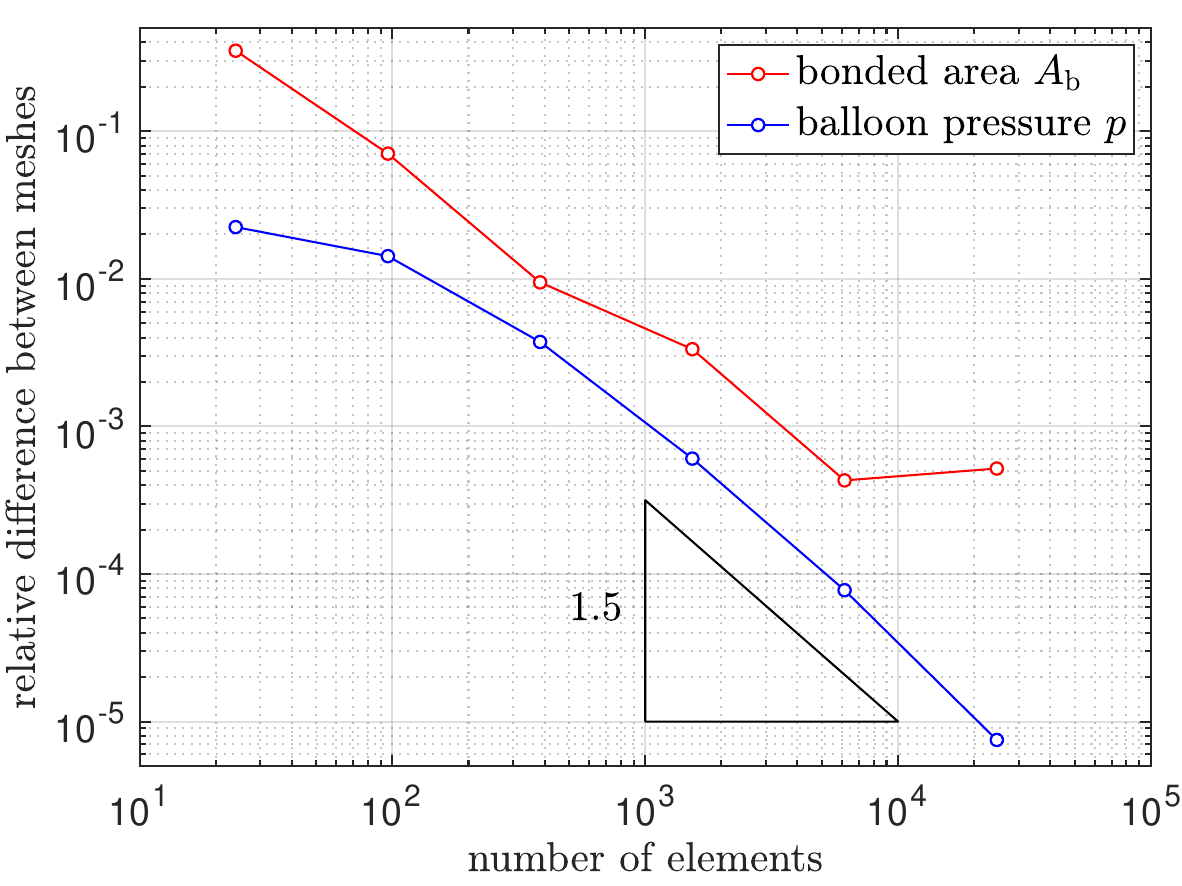}}
\put(-7.95,-.15){\footnotesize (a)}
\put(0.2,-.15){\footnotesize (b)}
\end{picture}
\caption{Balloon example: 
(a) Decrease of the bonded area and the contact area during unloading and debonding (step 3 \& 4). 
(b) FEM convergence of the balloon pressure $p$ (= max.~contact pressure) and bonded area $A_\mrb$ during debonding.
The convergence rates are $O(n_\mathrm{el}^{-1.5})=O(h^3)$ for $p$ and around $O(n_\mathrm{el}^{-1})=O(h^2)$ for $A_\mrb$; (the rise at the last point is caused by a sign change in the relative difference).}
\label{f:CoboBalloon4}
\end{center}
\end{figure}

The decrease of the bonded surface area, together with the contact area, is shown in Fig.~\ref{f:CoboBalloon4}a.
The contact area is the portion of the bonded area that is in compression.
As seen, debonding occurs during both step 3 and 4, but much slower in the former.
This is achieved here by setting the debonding time scale to $\unde{\bar m}=1$ and $\unde{\bar m}=0.04$ during step 3 and 4, respectively.
Consequently a much finer time step size is needed for step 4 than for step 3.

Step 3 is run for $n_{t3}=10m$ constant time steps $\Delta t_3 = 0.08 T/m$ (until time $t_3 = 0.8T$), during which the top plate is moved upward by the velocity $v=R/T$ (such that it is located at $u_z = 0.8 R$ at the end of step 3).
Step 4 uses adaptive time stepping to speed-up computations:
Starting from the initial time step $\Delta t_4 = 0.016 T/m$ for mesh $m$, the step size is increased by a constant factor such that the final time $t_{4\mra} = t_3 + 10T$ is reached in $n_{t4\mra}=25m$ adaptive steps.
For the purpose of studying FE convergence, also the constant time step $\Delta t_4 = 0.0125 T/m$ has been used for $n_{t4\mrc}=40m$ steps (i.e. up to $t_{4\mrc} = t_3 + T/2$).
The convergence behavior for constant time steps is shown in Fig.~\ref{f:CoboBalloon4}b.
Here, the relative differences
\eqb{l}
\Delta\bar p^{\,m} := \ds\frac{\big|\bar p^{\,m} - \bar p^{\,m/2}\big|}{\bar p^{\,64}}\,,\quad 
\Delta\bar A_\mrb^m := \ds\frac{\big |\bar A_\mrb^m - \bar A_\mrb^{m/2}\big|}{\bar A_\mrb^{64}}\,,\quad 
m\in\{2,\,4,\,8,\,16,\,32,\,64\}\,,
\eqe
of the time averaged pressure and area
\eqb{lllll}
\bar p^{\,m} \dis \ds\frac{1}{T_\mre}\int_0^{T_\mre} p^m(t)\,\dif t 
	\ais \ds\frac{1}{n_t^m}\bigg( \frac{p^m_1}{2} + \sum_{n\,=\,2}^{n_t^m-1}p_n^m + \frac{p_{n_t}^m}{2} \bigg)\,, \\[5mm]
\bar A_\mrb^m \dis \ds\frac{1}{T_\mre}\int_0^{T_\mre} A_\mrb^m(t)\,\dif t 
	\ais \ds\frac{1}{n_t^m}\bigg( \frac{A_{\mrb\,1}^m}{2} + \sum_{n\,=\,2}^{n_t^m-1}A_{\mrb\,n}^m + \frac{A_{\mrb\,n_t}^m}{2} \bigg)
\eqe
between the FE results $p^m_n$ and $A_{\mrb\,n}^m$ at time step $t_n$ of subsequent meshes $m$ are examined.
The time average is taken over $n_t^m = n_{t4\mrc}=40m$ steps (i.e.~for the time span $T_\mre = T/2$).
Convergence rates between $O(h^2)$ and $O(h^3)$ are observed.

\begin{figure}[h]
\begin{center} \unitlength1cm
\begin{picture}(0,5.8)
\put(-8,2.7){\includegraphics[height=34mm]{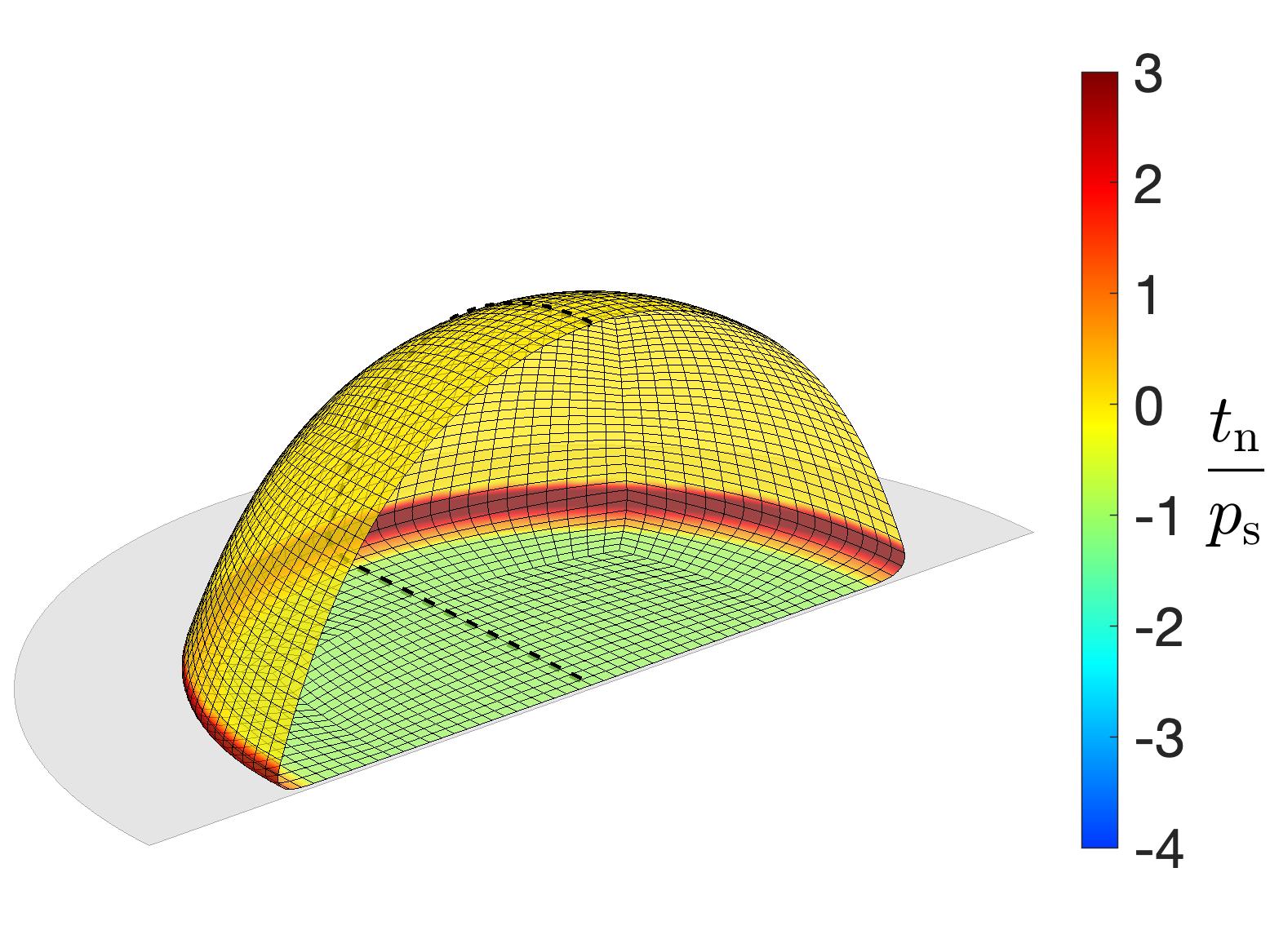}}
\put(-4.2,2.7){\includegraphics[height=34mm]{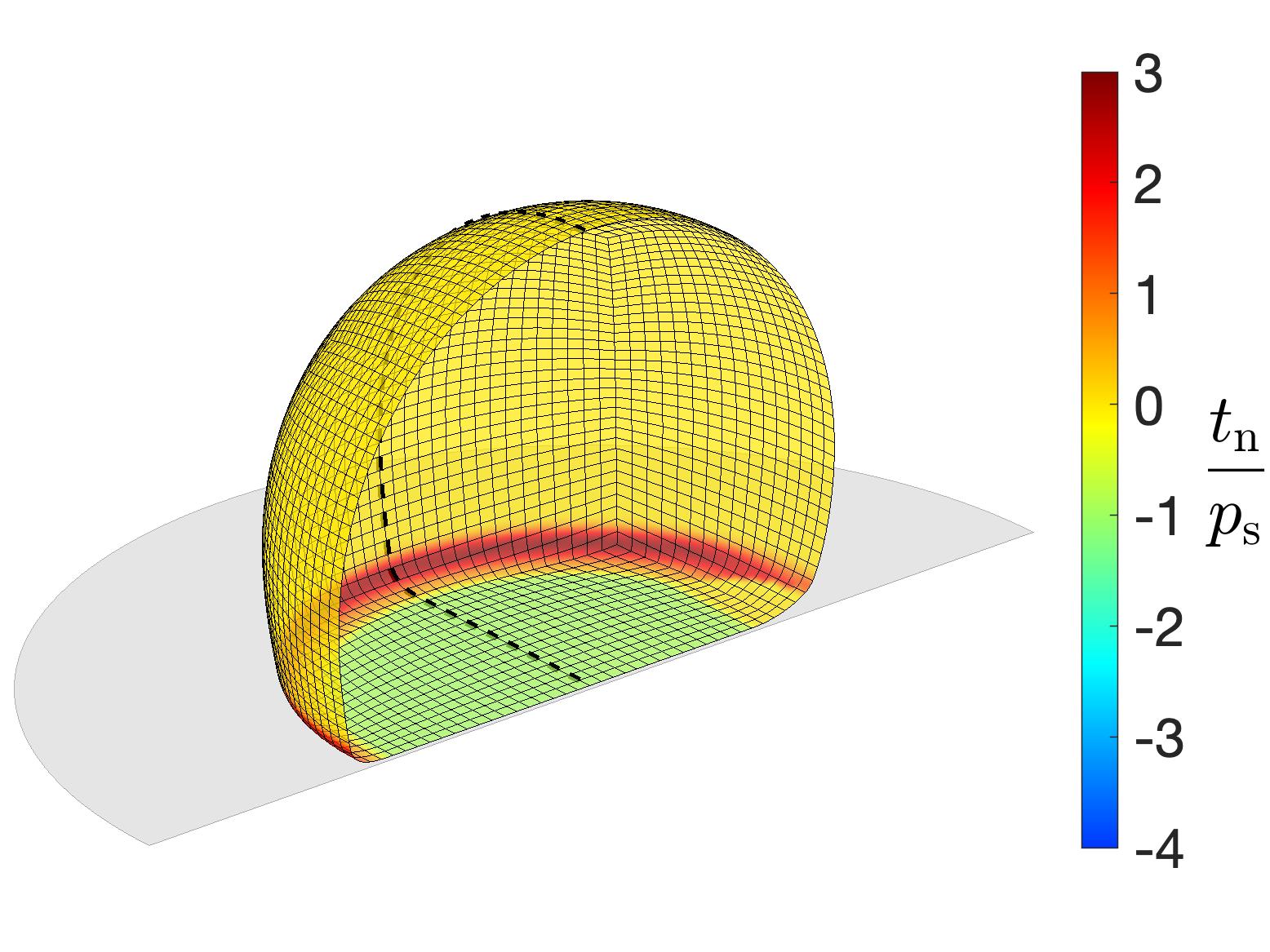}}
\put(-0.4,2.7){\includegraphics[height=34mm]{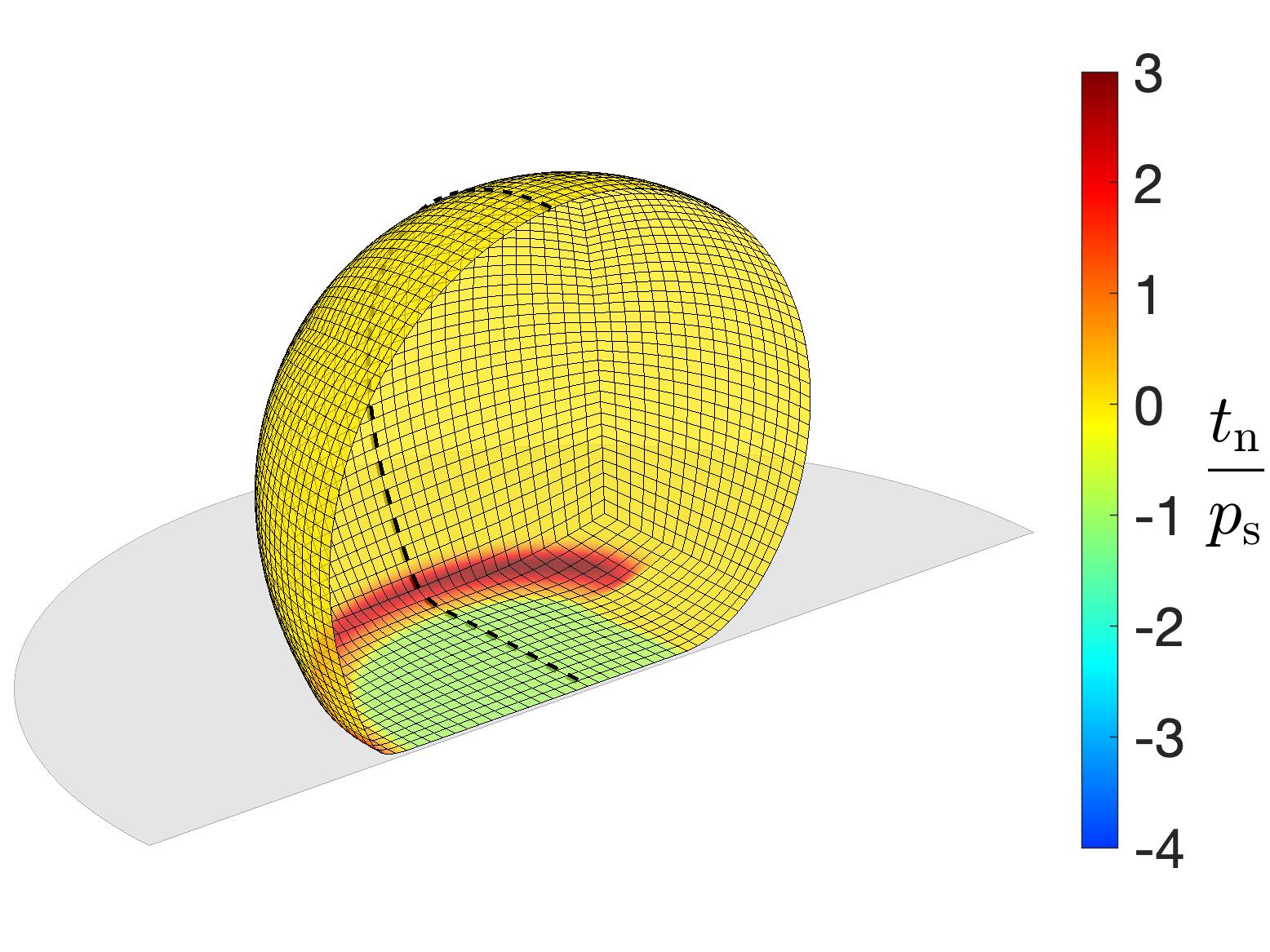}}
\put(3.4,2.7){\includegraphics[height=34mm]{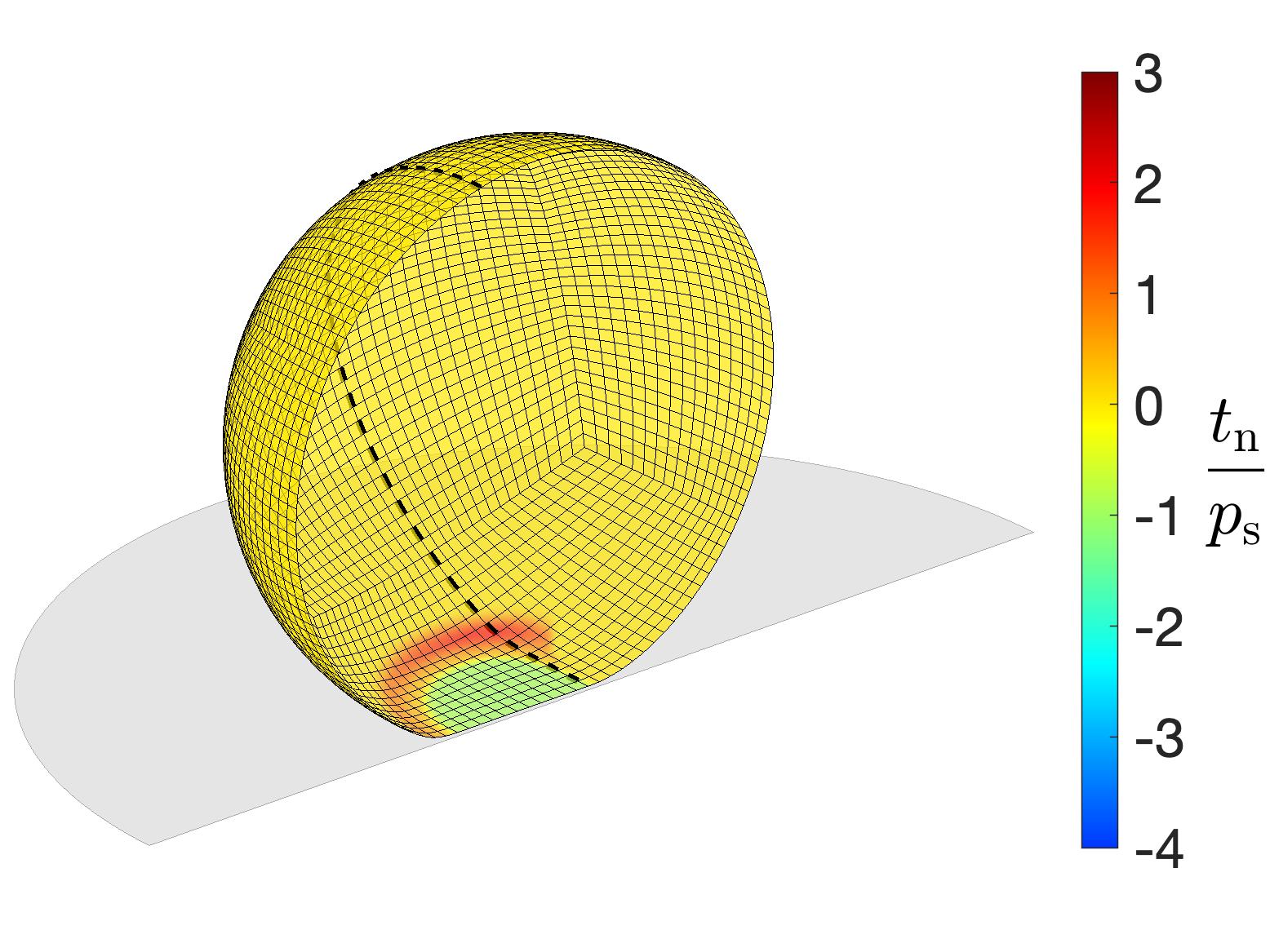}}
\put(-8,-.4){\includegraphics[height=34mm]{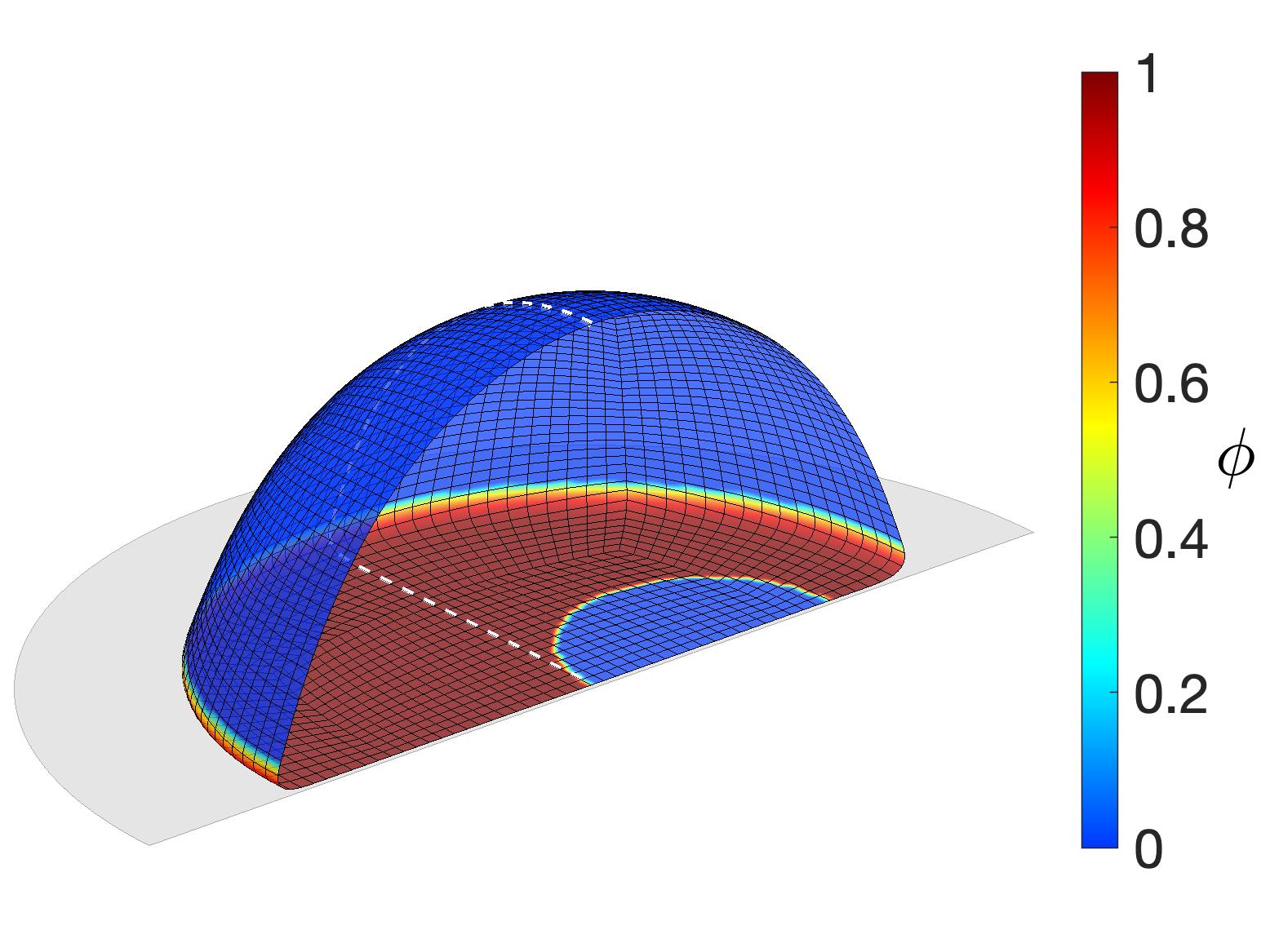}}
\put(-4.2,-.4){\includegraphics[height=34mm]{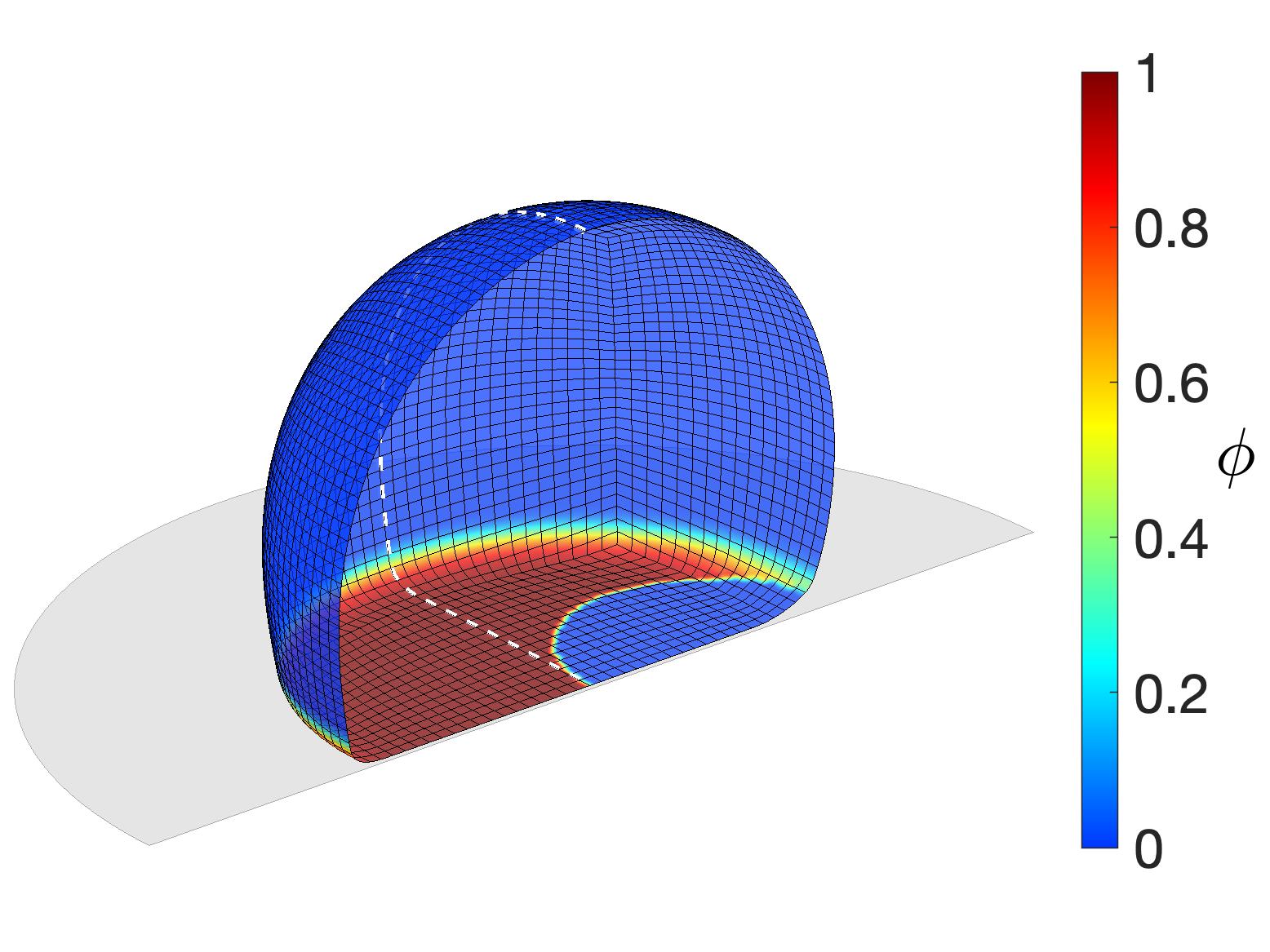}}
\put(-0.4,-.4){\includegraphics[height=34mm]{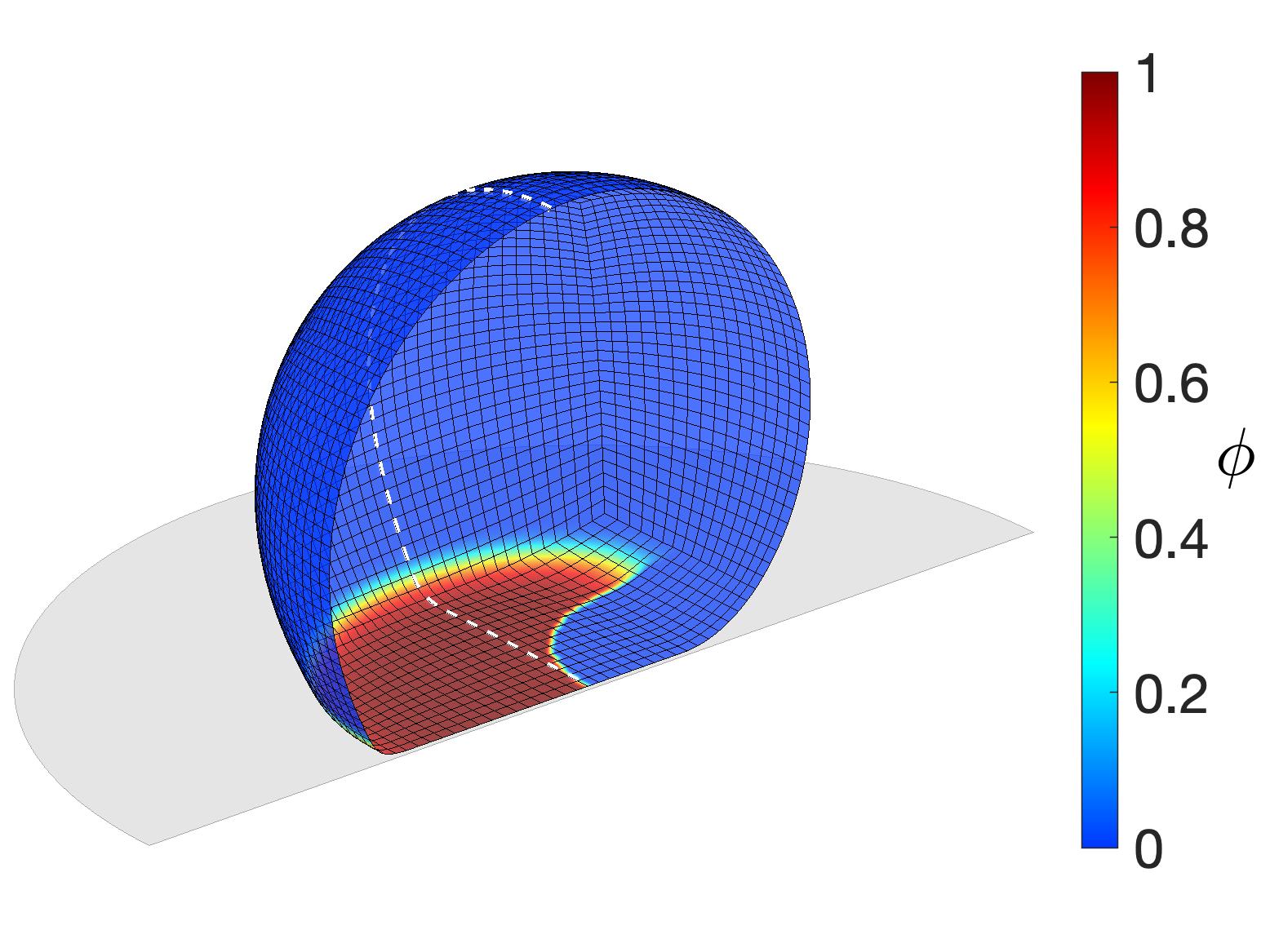}}
\put(3.4,-.4){\includegraphics[height=34mm]{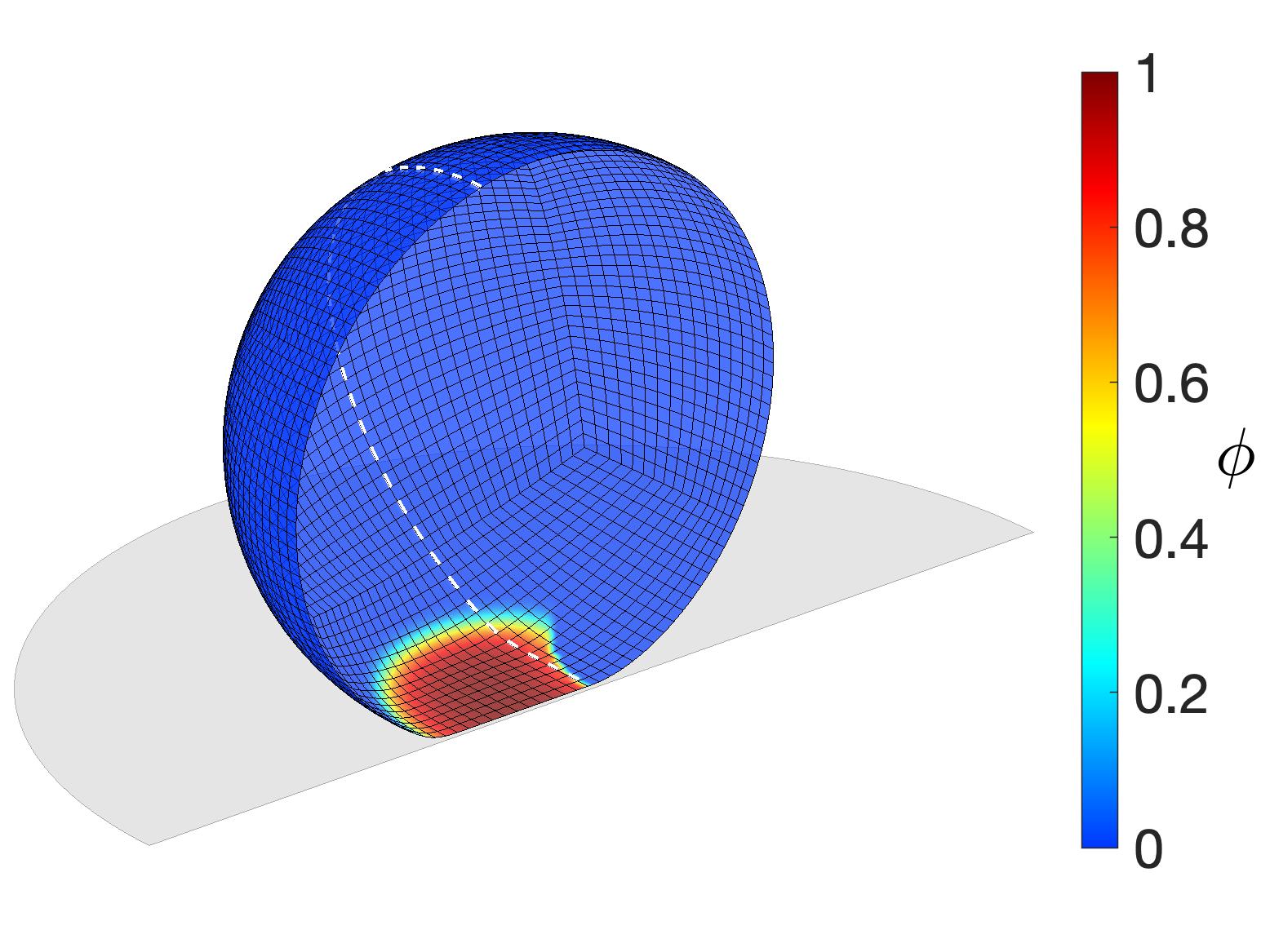}}
\end{picture}
\cprotect\caption{Balloon example -- debonding (step 4): Evolution of the contact traction $t_\mrn$ (top) and bonding state $\phi$ (bottom) for the case of an initial circular bonding defect (at $t_3+[0,\,0.15,\,0.25,\,0.75]T$; from left to right).
Due to the eccentricity of the defect, the balloon rotates during debonding, as the dashed line shows.
The results show half of the balloon using  mesh $m=16$. 
See also supplementary movie \verb|debo_balloon_phi.mpg| 
at \href{https://doi.org/10.5446/73320}{doi.org/10.5446/73320}.}	   
\label{f:CoboBalloon5}
\end{center}
\end{figure}

Finally, Fig.~\ref{f:CoboBalloon5} shows the influence of an initial bonding defect on the debonding behavior.
As seen, the balloon rotates during debonding due to the eccentricity of the defect.
The example thus illustrates the capability of the proposed model to resolve spatially and temporally varying debonding behavior. 

\subsection{Three-field chemo-thermo-mechanical bonding: Fixed sphere}\label{s:ex3a}

The third example extends the previous example to temperature evolution, and thus illustrates three-field coupled chemo-thermo-mechanical bonding.
The example of a rigid sphere in contact with a flexible membrane is used, as shown in Fig.~\ref{f:CoboEx}c.
The weight of the sphere presses on the membrane and deforms it.
Pressure-dependent bonding then ensues at the interface.
The bonding reaction causes heating which in turn affects the membrane deformation through isotropic thermal expansion and the Gough-Joule effect.
Therefore, the membrane deformation is decomposed multiplicatively into its elastic part and its thermal expansion part,
\eqb{l}
\bF = \bF_\mathrm{el}\,\bF_\mathrm{in}\,,\qquad \bF_\mathrm{in} = \lambda_\mathrm{in}\bI\,,\qquad \lambda_\mathrm{in} 
	= \exp\big(\alpha_\mrT(T-T_0)\big)\,,
\eqe
following the framework of \citet{FeFi}.
Now the compressible membrane model of \citet{shelltheo} is used for the elastic part.
To capture the Gough-Joule effect, a linear dependency of bulk and shear moduli on temperature is considered in the form $\unde{G} = \unde{G_0}\,T/T_0$ and $\unde{K} = \unde{K_0}\,T/T_0$ \citep{chadwick74}.
The membrane is discretized with quadratic NURBS elements and 4 dofs per node (temperature and three displacement components). 
As before, the bonding state is a history variable at each quadrature point. 
The membrane is prestretched before bonding, and the prestretched contact configuration is used as the initial configuration for bonding.
This means that, here, the stiffness constants $\unde{E_n}$ and $\overrightarrow{\unde{K_0}}/g_0^2$ are taken as constants w.r.t.~this prestretched configuration.

\begin{figure}[h!]
\begin{center} \unitlength1cm
\begin{picture}(0,9.8)
\put(-7.9,6.45){\includegraphics[height=34mm]{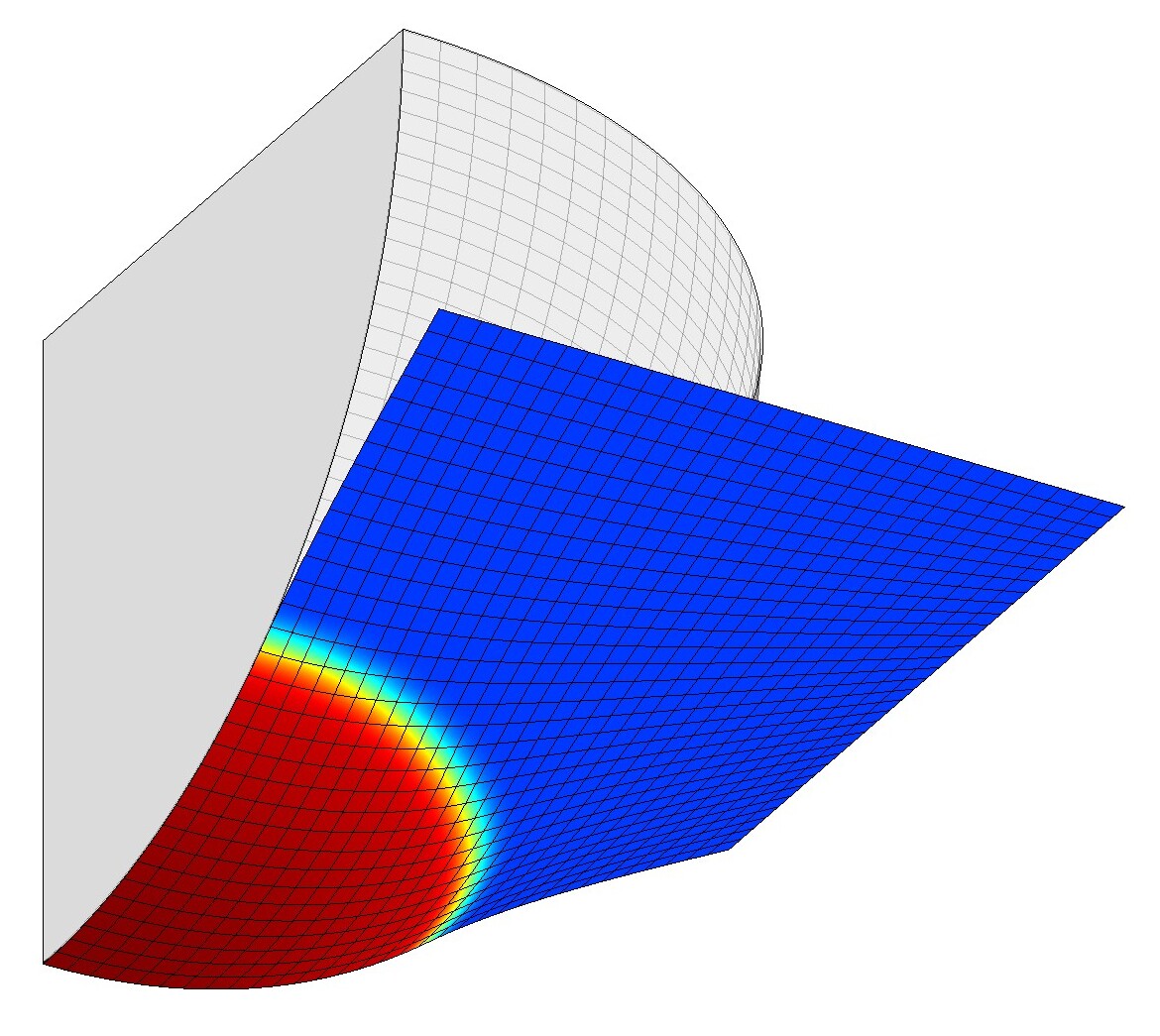}}
\put(-4.1,6.45){\includegraphics[height=34mm]{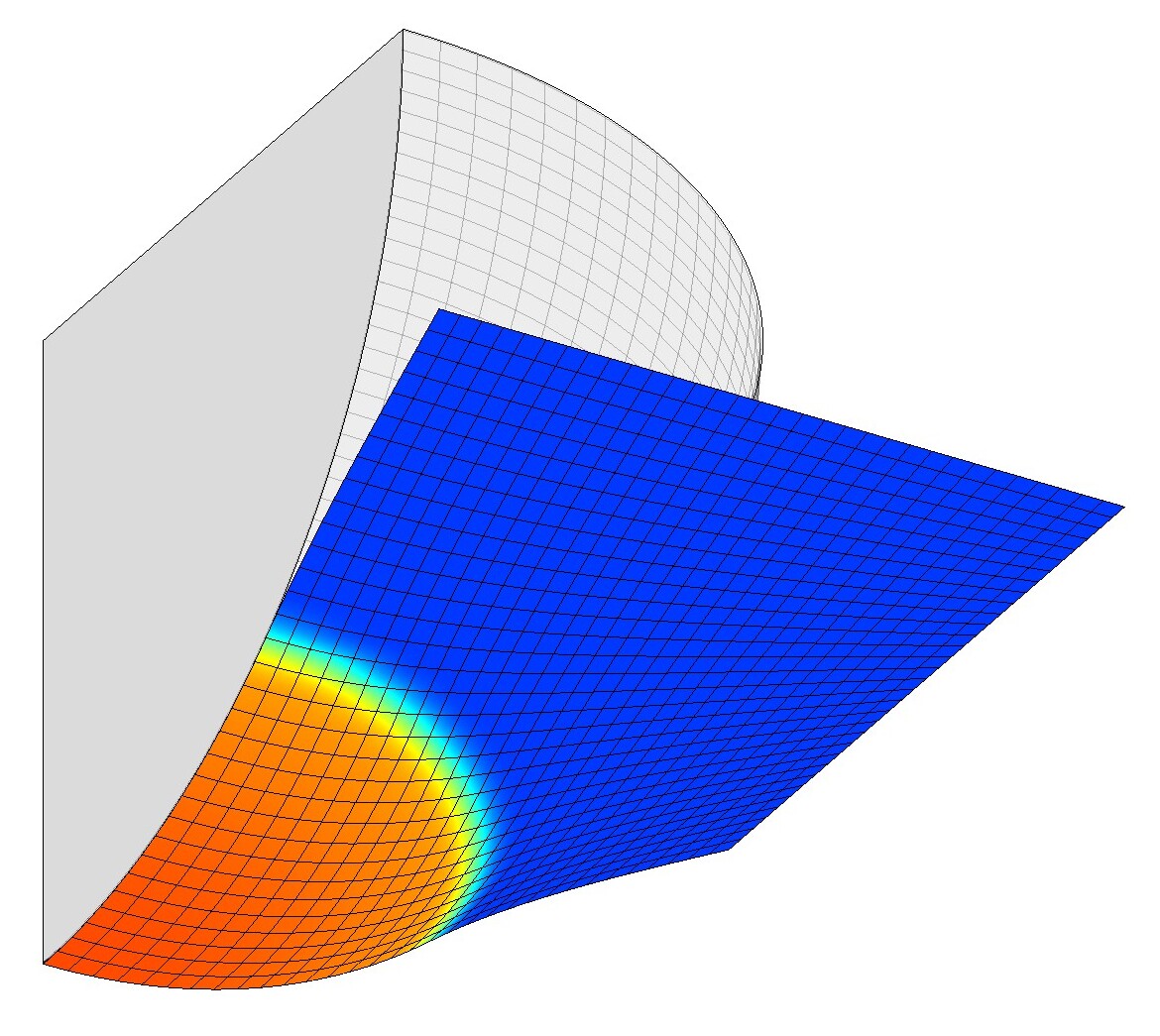}}
\put(-.3,6.45){\includegraphics[height=34mm]{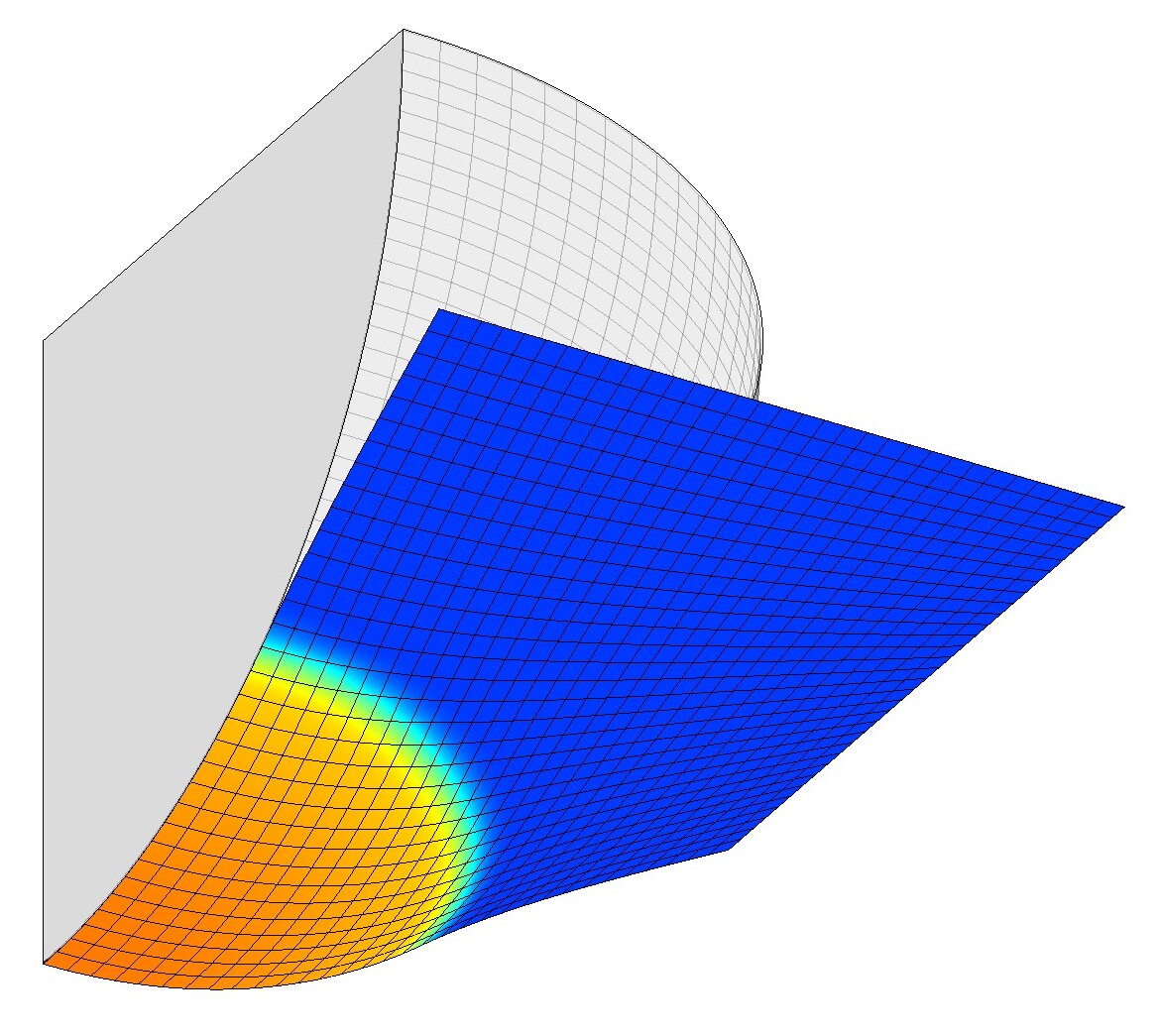}}
\put(3.5,6.45){\includegraphics[height=34mm]{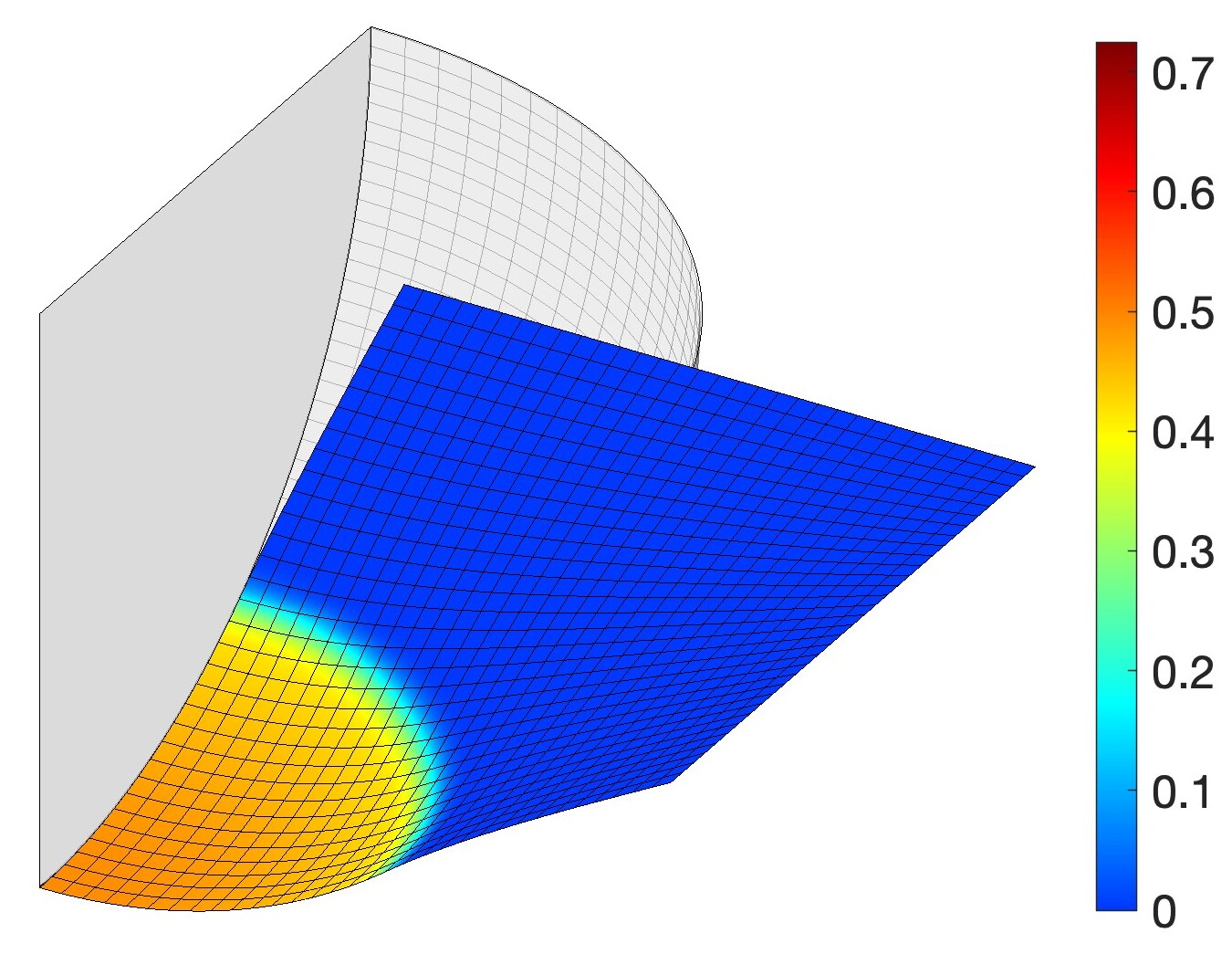}}
\put(-7.9,3.1){\includegraphics[height=34mm]{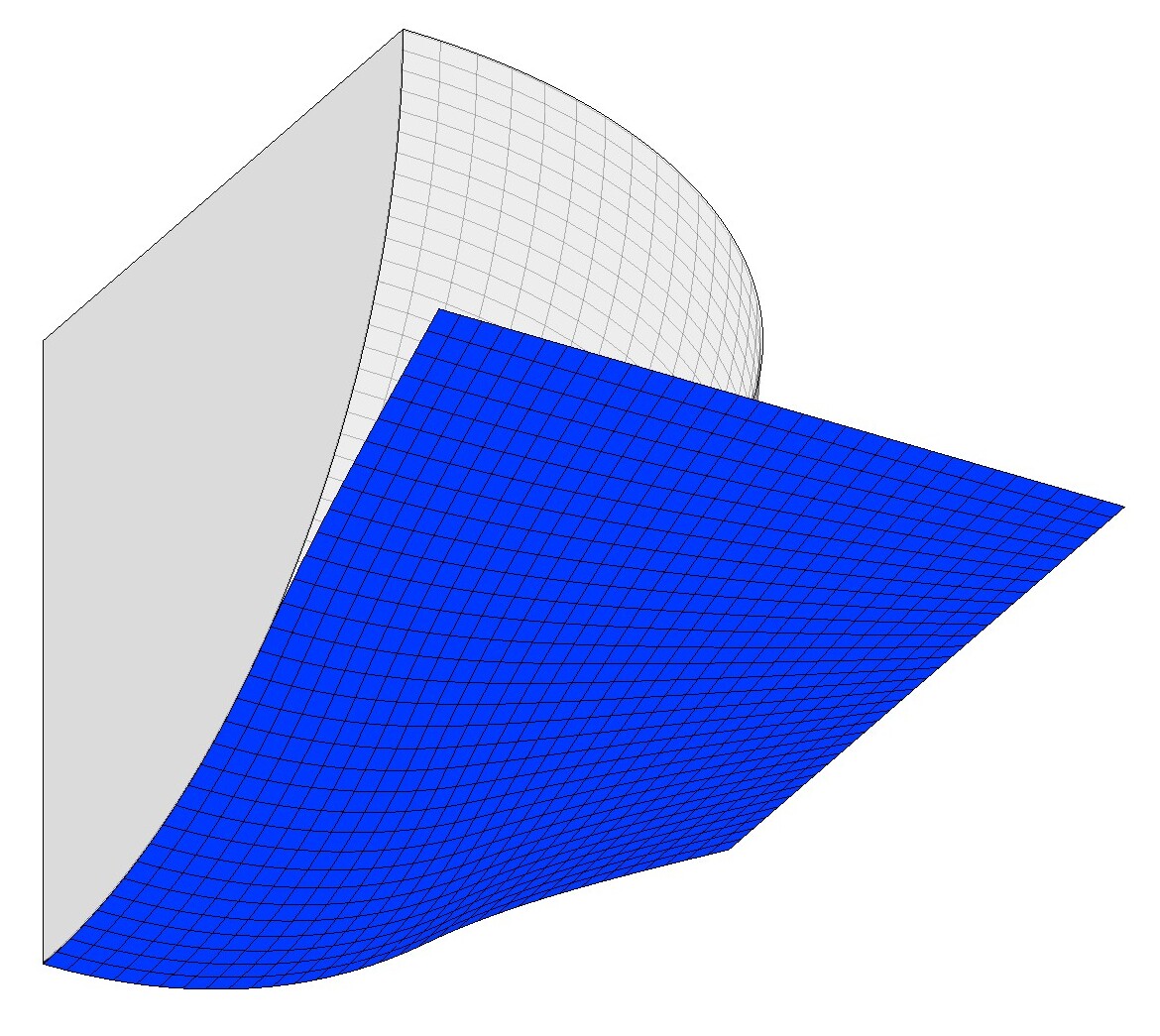}}
\put(-4.1,3.1){\includegraphics[height=34mm]{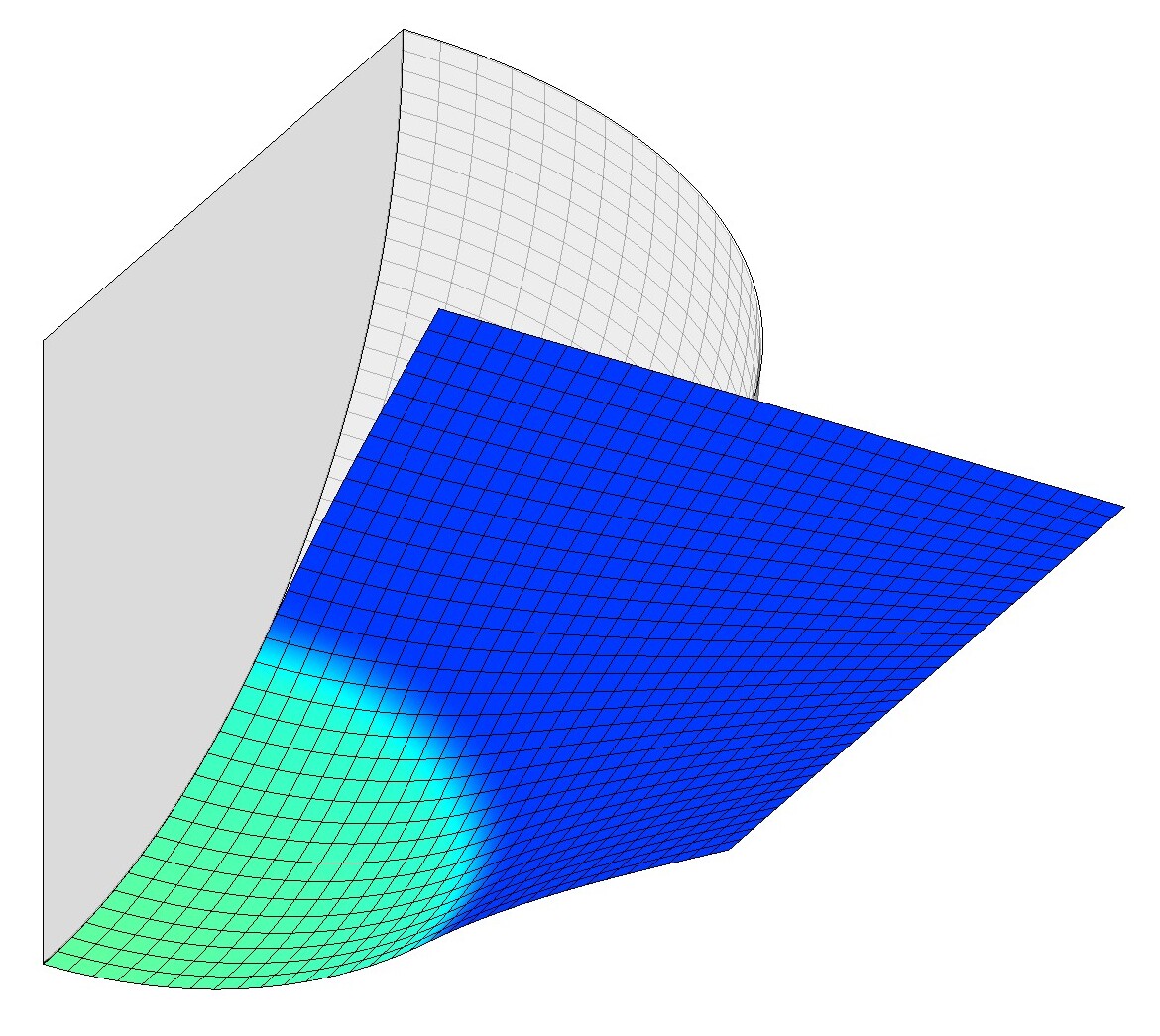}}
\put(-.3,3.1){\includegraphics[height=34mm]{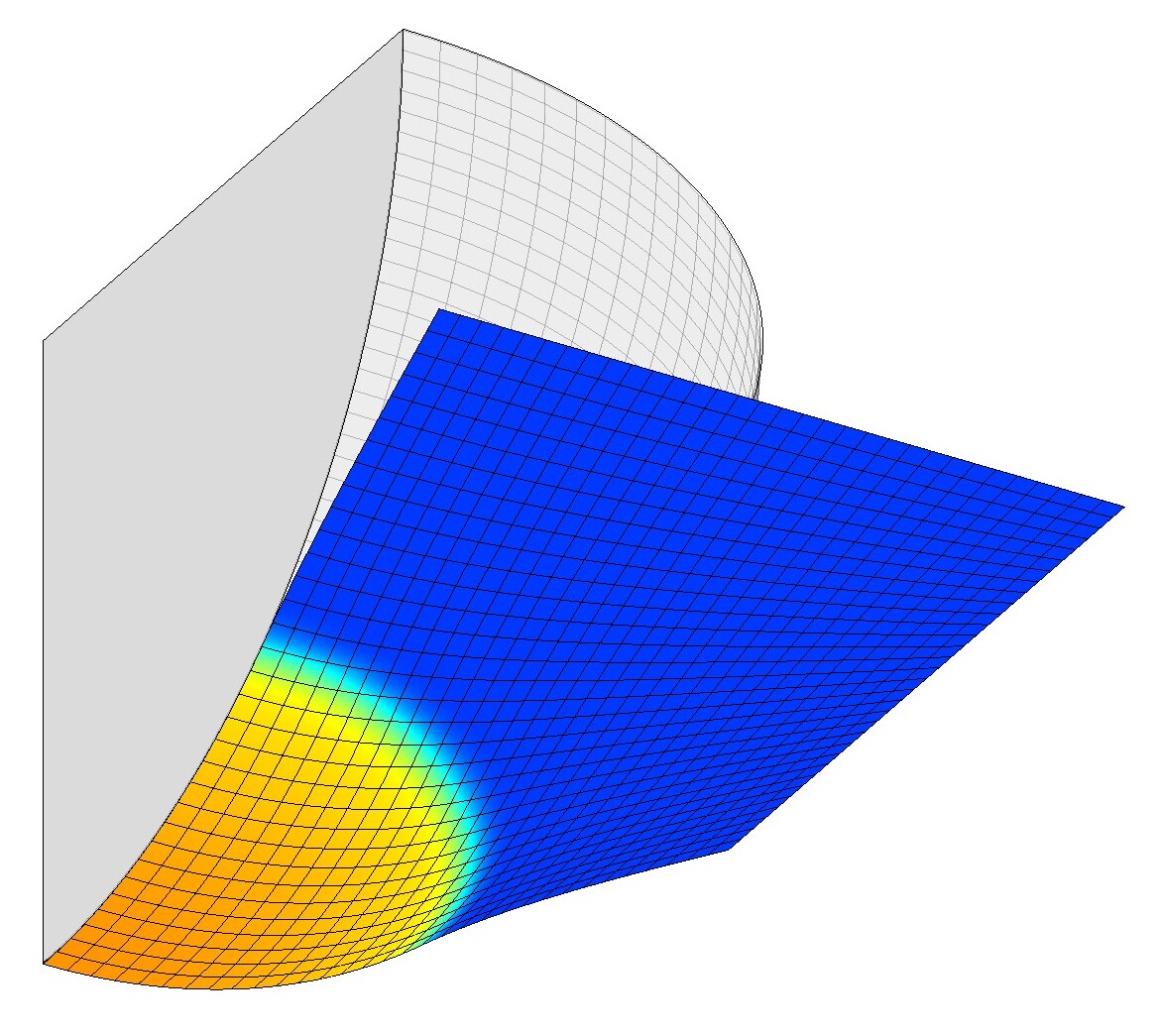}}
\put(3.5,3.1){\includegraphics[height=34mm]{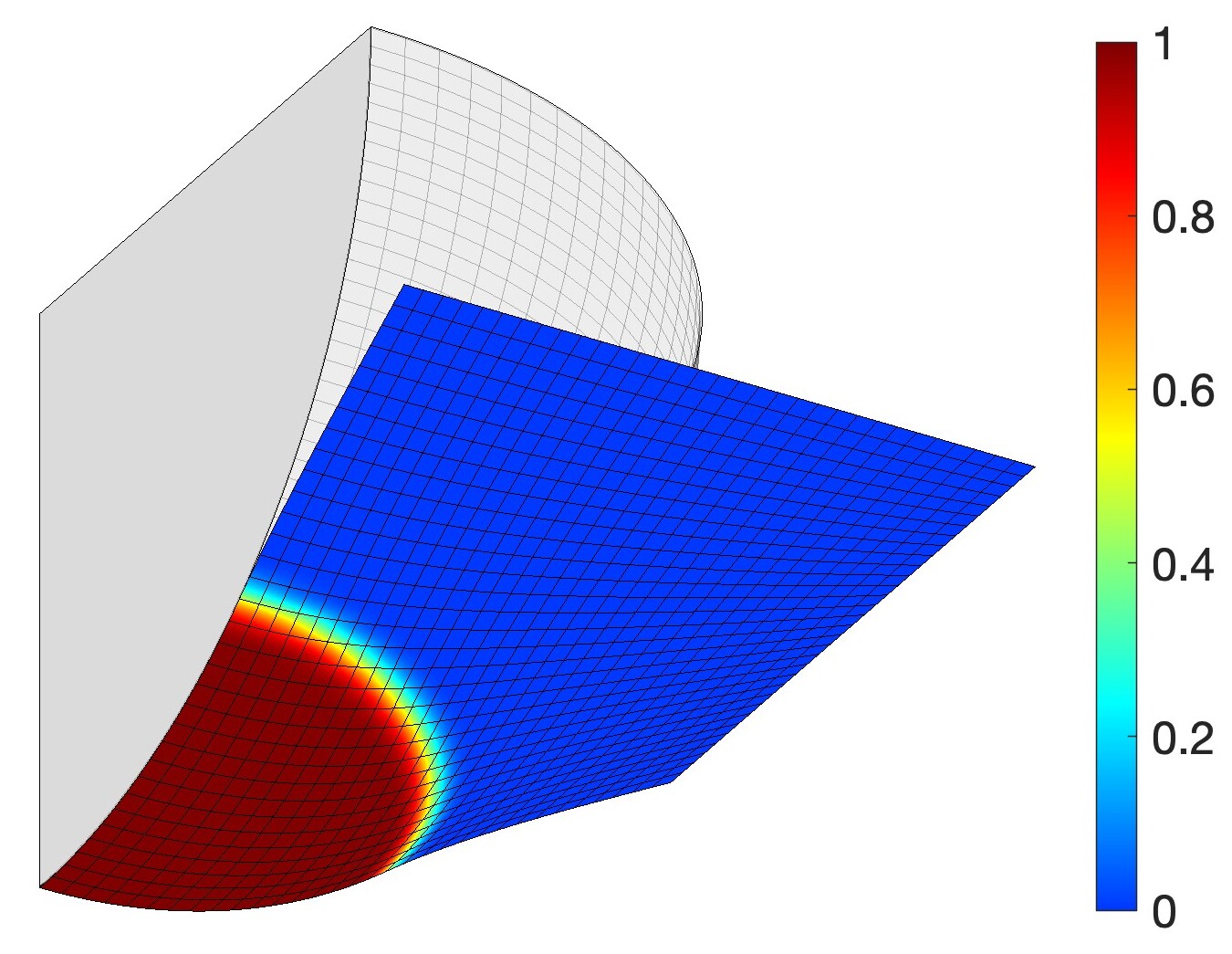}}
\put(-7.9,-.25){\includegraphics[height=34mm]{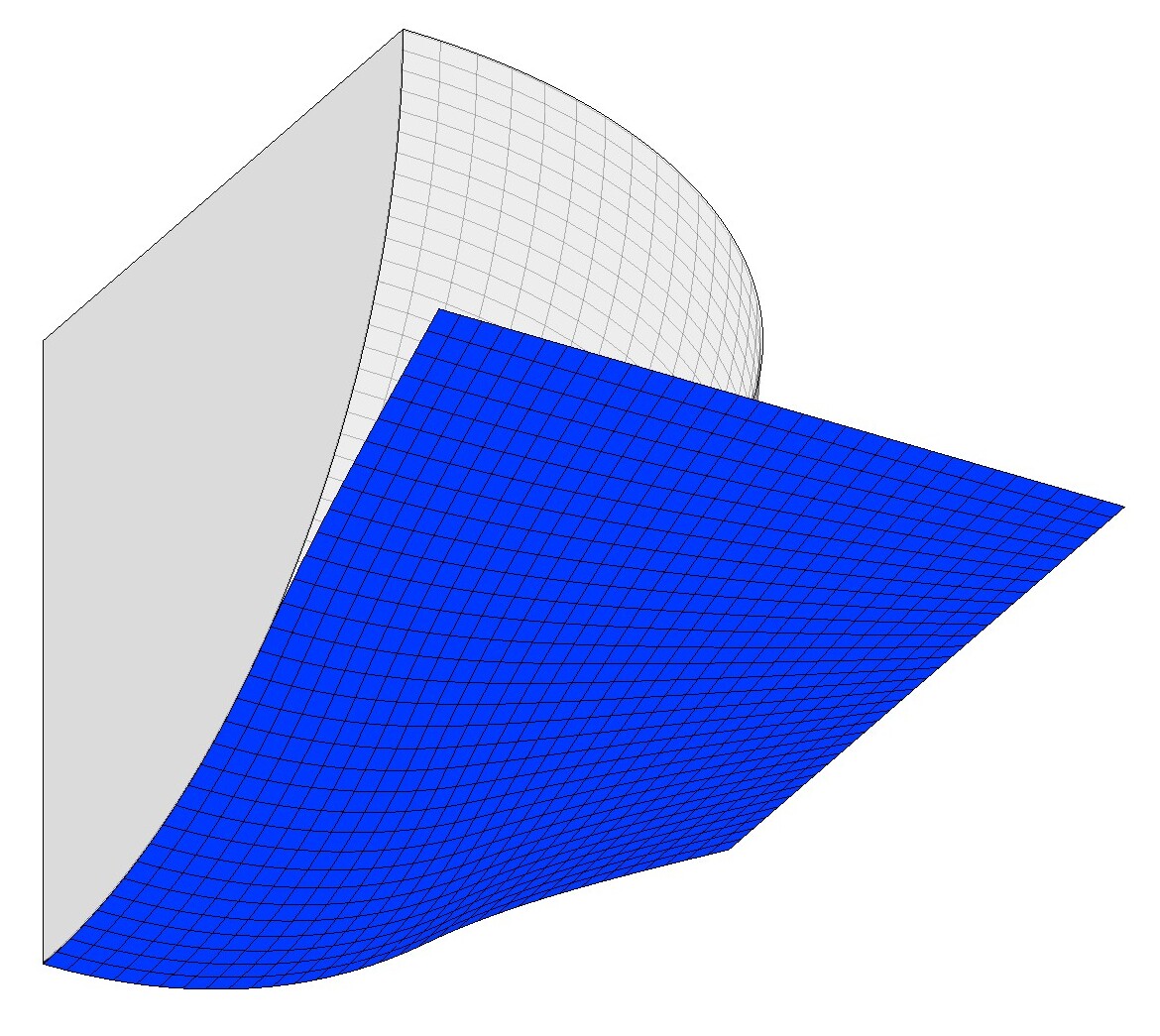}}
\put(-4.1,-.25){\includegraphics[height=34mm]{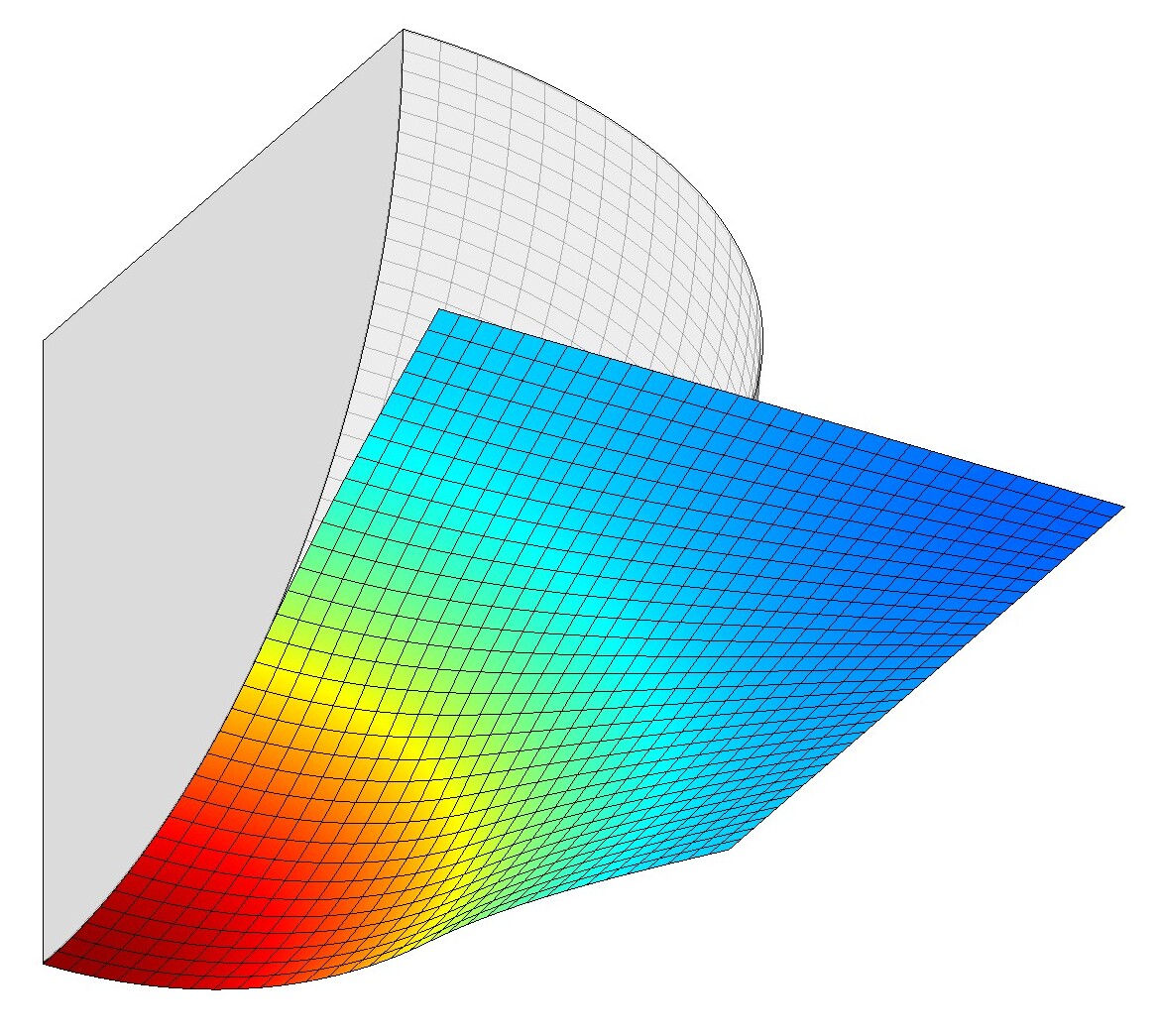}}
\put(-.3,-.25){\includegraphics[height=34mm]{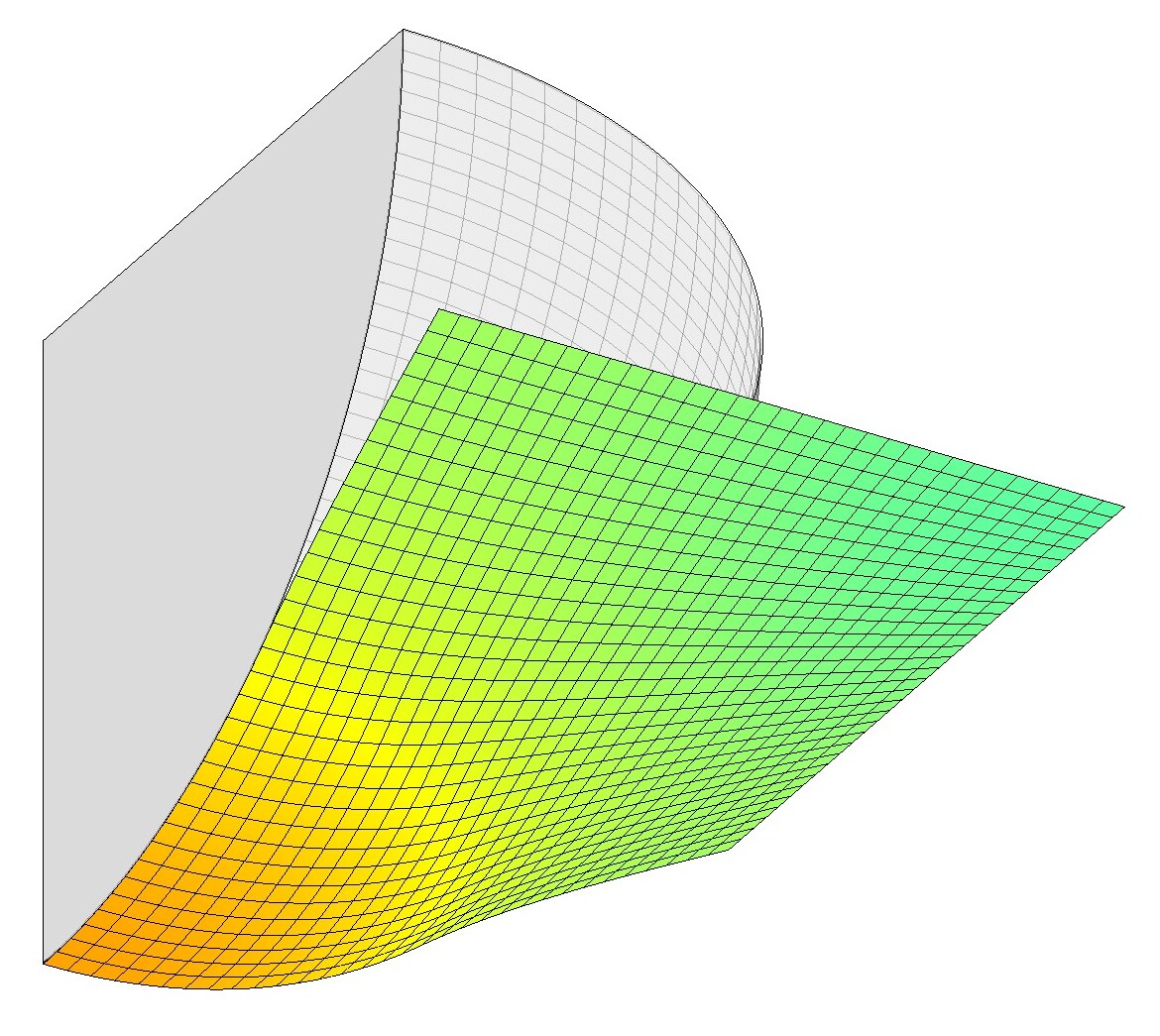}}
\put(3.5,-.25){\includegraphics[height=34mm]{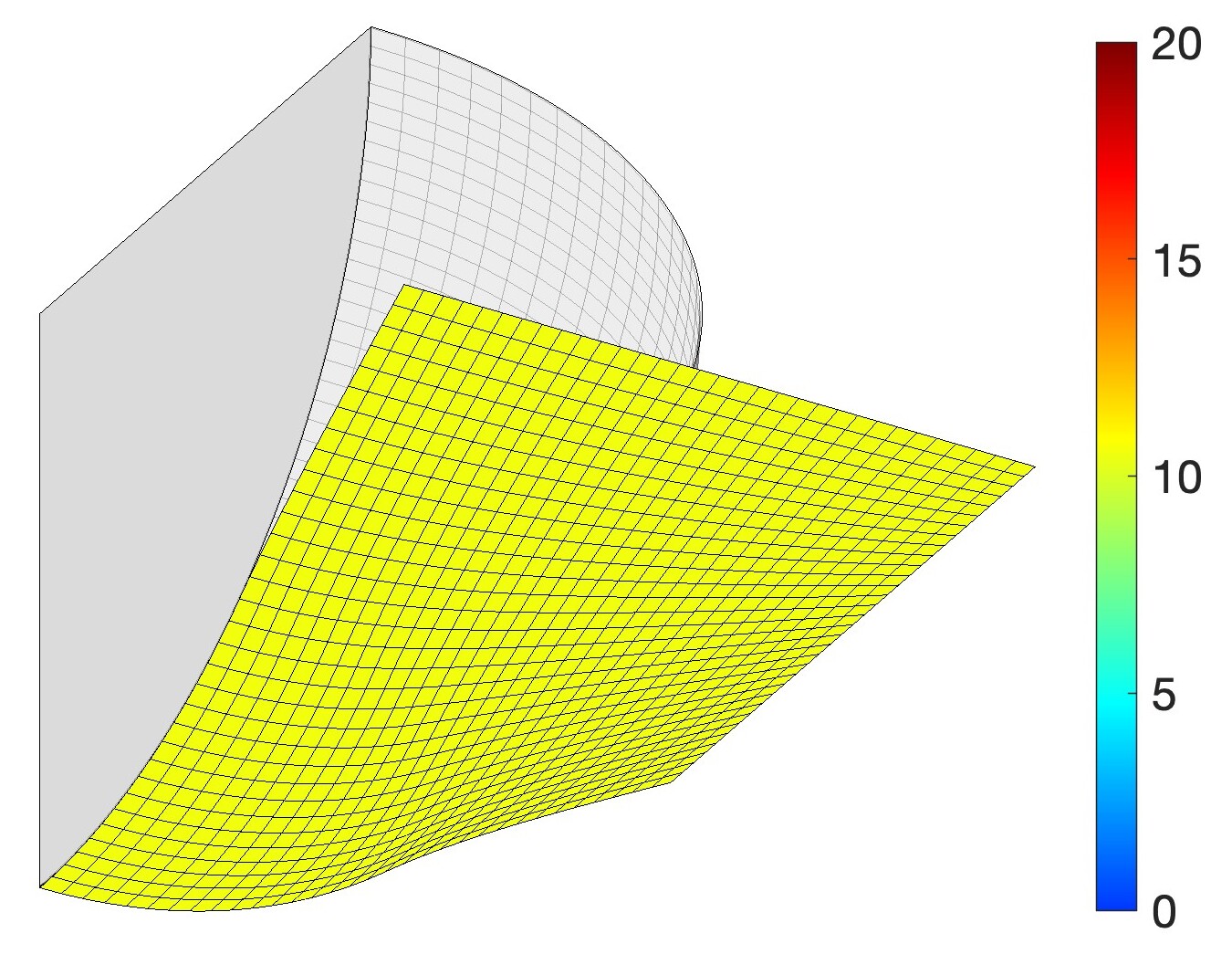}}
\put(7.03,9.35){\scriptsize $p_\mrc$}
\put(7.1,6.0){\scriptsize $\phi$}
\put(6.85,2.65){\scriptsize $\Delta T$}
\put(-5.8,6.65){\scriptsize $t=0$}
\put(-2.0,6.65){\scriptsize $t=0.5$}
\put(1.8,6.65){\scriptsize $t=1.5$}
\put(5.6,6.65){\scriptsize $t=10$}
\put(-5.8,3.3){\scriptsize $t=0$}
\put(-2.0,3.3){\scriptsize $t=0.5$}
\put(1.8,3.3){\scriptsize $t=1.5$}
\put(5.6,3.3){\scriptsize $t=10$}
\put(-5.8,-.05){\scriptsize $t=0$}
\put(-2.0,-.05){\scriptsize $t=0.5$}
\put(1.8,-.05){\scriptsize $t=1.5$}
\put(5.6,-.05){\scriptsize $t=10$}
\end{picture}
\cprotect\caption{Three-field bonding between membrane and fixed sphere: Evolution of the contact pressure $p_\mrc$ [$F_0/L_0^2$] (top row), contact bonding state $\phi$ (middle row) and relative membrane temperature $\Delta T$ [K] (bottom row).
The results show a quarter of the sheet using $32\times 32$ quadratic NURBS finite elements.
See also supplementary movie \verb|cobo_sphere_fixed.mp4| at \href{https://doi.org/10.5446/73728}{doi.org/10.5446/73728}.}   
\label{f:CoboT1}
\end{center}
\end{figure}

The interfacial heat capacity $\unde{C_c}$ is taken as zero and the sphere is considered fixed in space and temperature, thus eliminating position $\bx_1$ and temperatures $T_1$ and $T_\mrc$ as unknowns.
Additionally, transfer flux $\unde{q^c_t}$ is considered zero, i.e.~$\unde{h}=0$ here.
Fig.~\ref{f:CoboT1} shows the evolution of the contact pressure $p_\mrc$, bonding state $\phi$, and sheet temperature increase $\Delta T$ for this case.
The considered material parameters are listed in Table~\ref{t:CoboSheet}.
\begin{table}[h!]
\centering
\begin{tabular}{|r|l|r|l|}
  \hline
   symbol & material parameter & value & unit \\[0mm] \hline 
   & & & \\[-3.5mm]   
   $R$ & sphere radius & 2 & $L_0$ \\ [.5mm] 
   $L$ & initial length of the square sheet & 4 & $L_0$ \\ [.5mm]
   $\lambda$ & prestretch of the sheet before contact & 1.1 & -- \\ [.5mm]
   $F$ & initial contact force (on quarter sheet) & $0.8$ & $F_0$ \\[.5mm]
   $\unde{G_0}$ & shear modulus of the sheet at $T_0$ & 1 & $F_0/L_0$ \\[.5mm] 
   $\unde{K_0}$ & areal bulk modulus of the sheet at $T_0$ & 2 & $F_0/L_0$  \\[.5mm] 
   $\unde{E_n}$ & contact penalty stiffness & 1000 & $F_0/L_0^3$ \\[.5mm]
   $\overrightarrow{\unde{K_0}}/g_0^2$ & areal bond energy density &  50 & $F_0/L_0^3$  \\ [.5mm] 
   $\unde{\bar m}$ & bonding ``mass" & $2\cdot10^{-5}$  & $F_0t_0/L_0$ \\[.5mm]
   $T_0$ & reference temperature & 290 &  K  \\[.5mm]
   $\unde{C}$ & heat capacity of the sheet ($=\unde{C_2}$) &  $10^{-7}$ & $F_0/(L_0$K)   \\[.5mm] 
   $\unde{k}$ & thermal conductivity of the sheet & $10^{-7}$ & $F_0L_0/(t_0$K)  \\[.5mm]
   $\alpha_\mrT$ & coefficient of thermal expansion & 0.005 & 1/K  \\[.5mm]
   $\unde{h}$ & heat transfer coefficient & $10^{-7}$ & $F_0/(L_0t_0$K) \\[.5mm]
    \hline
\end{tabular}
\caption{Bonding between membrane and sphere: Considered material parameters. 
$L_0$, $F_0$ and $t_0$ are length, force and time scales that can be arbitrary.
No debonding is considered at first ($\protect\overleftarrow{\unde{K_0}}=0$). \\[-3mm]}
\label{t:CoboSheet}
\end{table}
For these, the sheet heats up and thus thermally expands substantially during bonding.\footnote{Larger values for $\unde{C}$ lead to less heating of the sheet, while larger $\unde{k}$ leads to faster heat conduction s.t.~there may be no initial temperature peak at the center.}
As the sphere is considered fixed, this reduces the contact pressure, which in turn slows down the bonding reaction.

\begin{figure}[h]
\begin{center} \unitlength1cm
\begin{picture}(0,11.6)
\put(-8,5.8){\includegraphics[height=58mm]{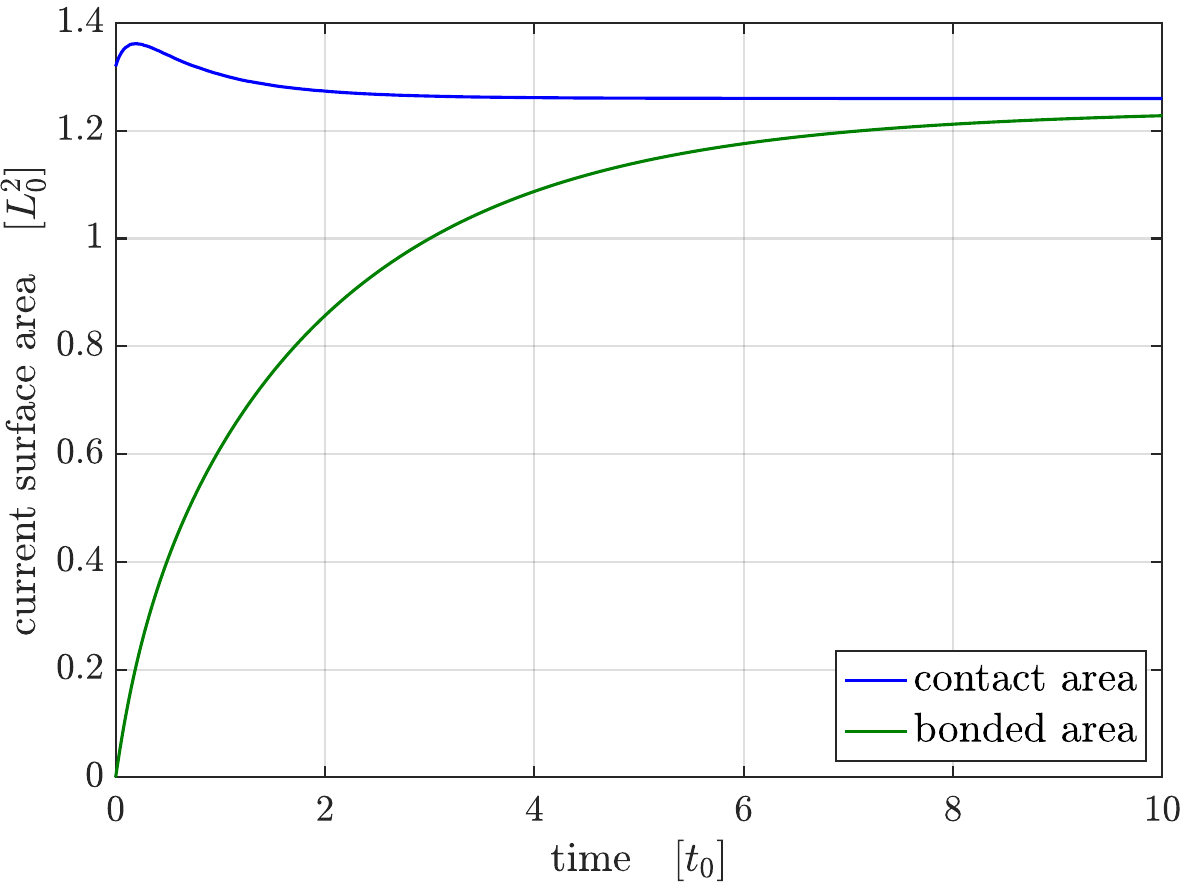}}
\put(0.2,5.8){\includegraphics[height=58mm]{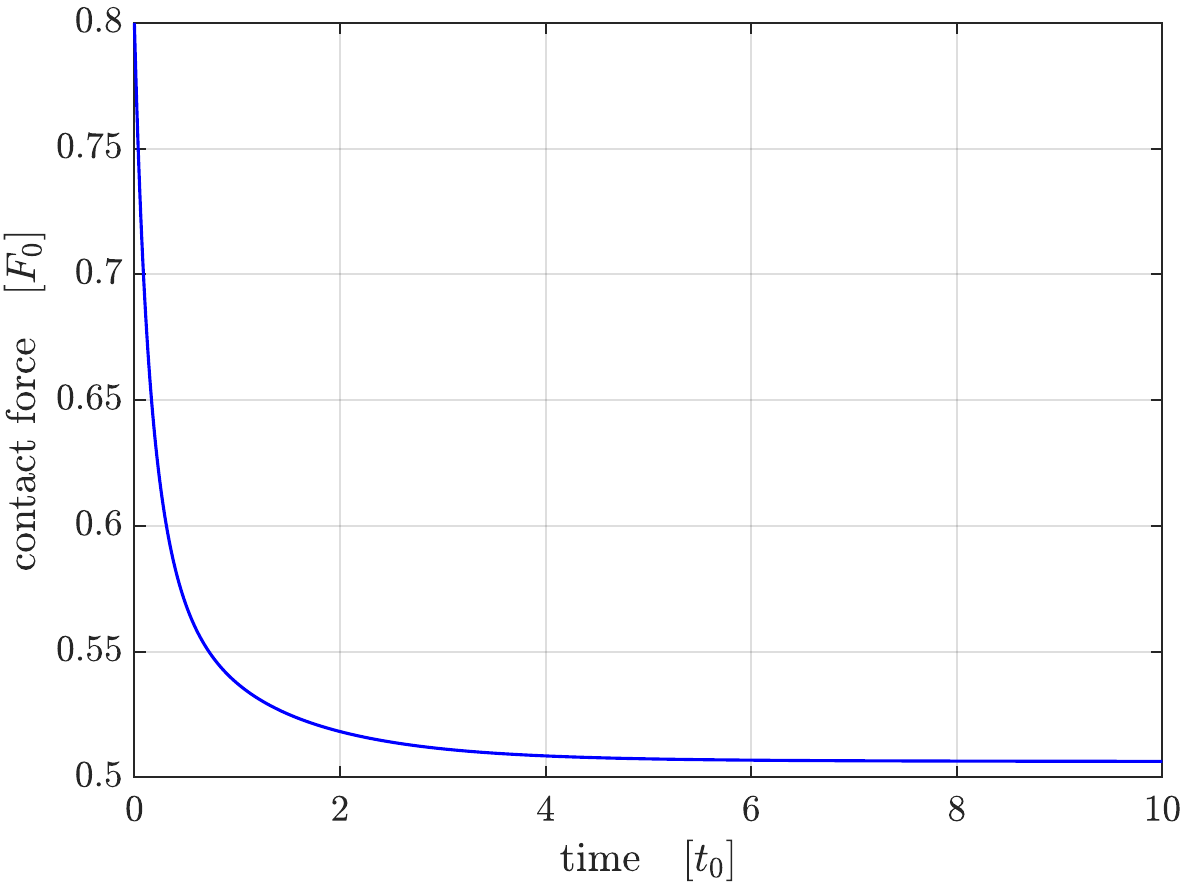}}
\put(-8,-.2){\includegraphics[height=58mm]{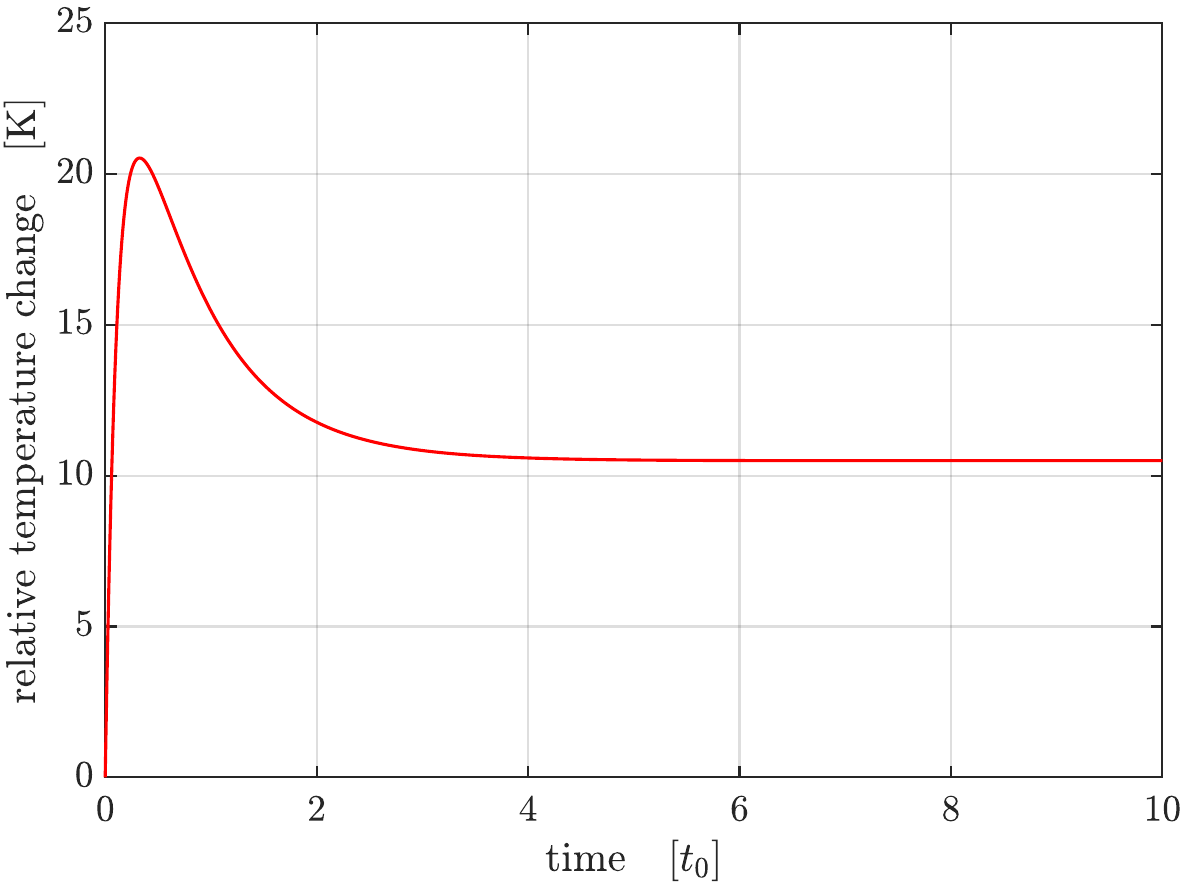}}
\put(.2,-.2){\includegraphics[height=58mm]{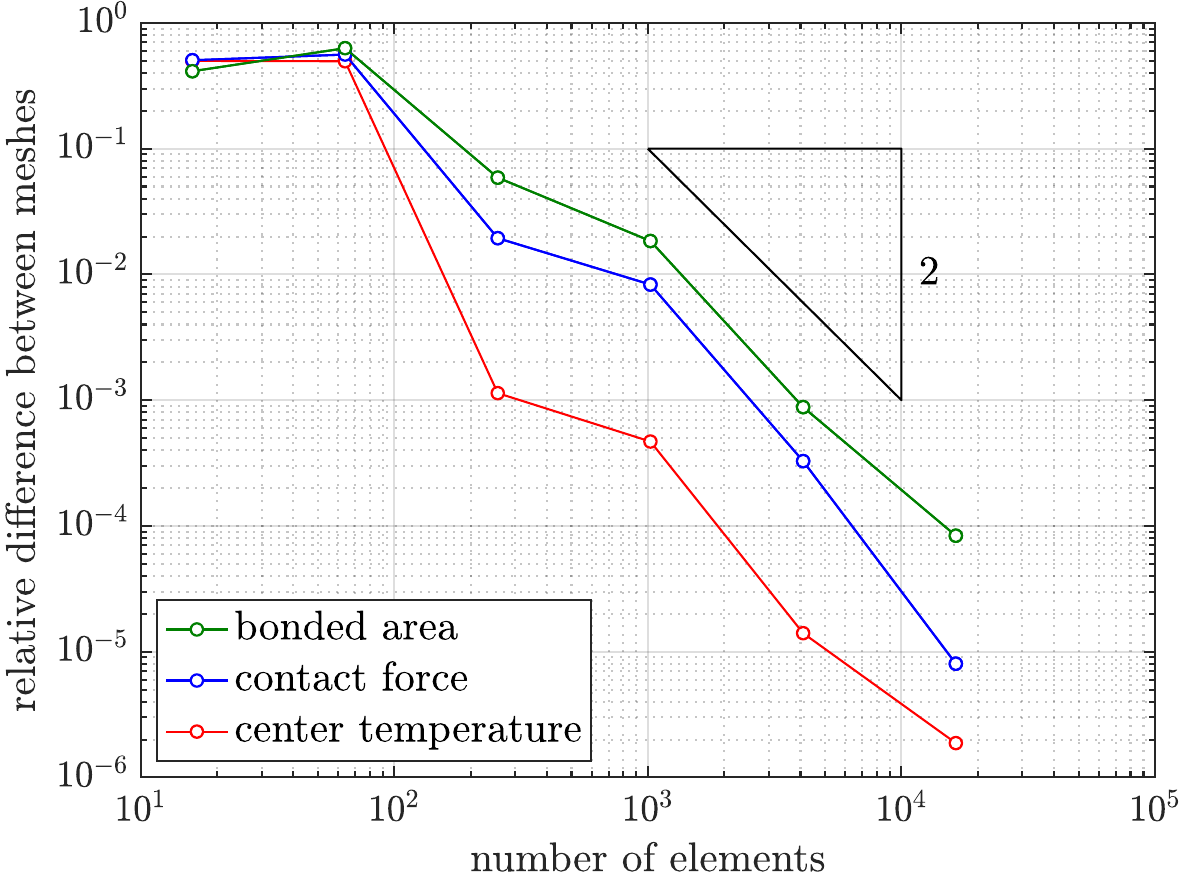}}
\put(-7.95,5.9){\footnotesize (a)}
\put(0.2,5.9){\footnotesize (b)}
\put(-7.95,-.1){\footnotesize (c)}
\put(0.2,-.1){\footnotesize (d)}
\end{picture}
\caption{Three-field bonding between membrane and fixed sphere: 
(a) Increase of the bonded area in comparison to the contact area, 
(b) decrease of the contact force, 
(c) evolution of the temperature at the center, and 
(d) their FE mesh convergence. 
The contact force decreases due to thermal expansion.
The convergence rates follow $O(n_\mathrm{el}^{-2})=O(h^4)$.
Panel (a) \& (b) are for the quarter sheet; (d) accounts for all time steps additively.}
\label{f:CoboT2}
\end{center}
\end{figure}

Fig.~\ref{f:CoboT2} shows the evolution of the bonded surface area, the sheet temperature at the center and the total contact force.
The center temperature peaks at the beginning when heat production from bonding exceeds heat loss due to conduction.
The sheet boundary is treated isolated such that all the heat remains in the membrane.
The energy balance can thus be verified:
The energy released by bonding (from $\phi=0$ until current $\phi$ at time $t$) follows from \eqref{e:Psi0} as
\eqb{l}
\Delta\Pi_\mrb = \ds\int_{\sS_0} \frac{1}{2} \overrightarrow{\unde{K_c}}\,\dif A - \int_{\sS_0} \frac{1}{2} \overrightarrow{\unde{K_c}}(\phi-1)^2\,\dif A\,. 
\label{e:dPi3}\eqe
For pressure-sensitive bonding according to \eqref{e:fm3} and temperature-insensitive bonding ($\overrightarrow{\unde{f_T}}=1$), Eqs. \eqref{e:Kc} and \eqref{e:dPi3} yield
\eqb{l}
\Delta\Pi_\mrb = \ds\frac{\overrightarrow{\unde{K_0}}}{2g_0^2}\int_{\sS_{\mrc0}} g_\mrn^2\,\dif A
\label{e:dPi3a}\eqe
at full bonding ($\phi = 1$) of the contact surface $\sS_{\mrc0}\subset\sS_0$.
Even though this expression looks straightforward, it is not easy to evaluate in the present example since the contact gap $g_\mrn$ varies across the contact surface.
Additionally, $g_\mrn$ varies in time, as the contact pressure changes during bonding, which is not even accounted for in \eqref{e:dPi3a} -- see Appendix~\ref{s:DPib} for an expression of $\Delta\Pi_\mrb$ that accounts for this.
However, Eq.~\eqref{e:dPi3a} can still be used to bound the expected energy release and thus assess the consistency of the temperature rise:
As $g_\mrn$ decreases monotonically over time, the bounding 
\eqb{l}
\ds\frac{\overrightarrow{\unde{K_0}}A_\mrc}{2g_0^2} \big\langle g_\mrn^2 \big\rangle_{t=10t_0} \leq \Delta\Pi_\mrb \leq
\ds\frac{\overrightarrow{\unde{K_0}}A_\mrc}{2g_0^2} \big\langle g_\mrn^2 \big\rangle_{t=0}
\label{e:dPi3b}\eqe
follows.
Here
\eqb{l}
\big\langle ... \big\rangle := \ds\frac{1}{A_\mrc}\int_{\sS_{\mrc0}} ...\,\dif A\,,
\eqe
defines the average over the contact surface, with $A_\mrc|_{t=0} = 5.277\,L_0^2$ and $A_\mrc|_{t=10t_0}=5.142L_0^2$ w.r.t.~the initial configuration.
The value of $\big\langle g_\mrn^2 \big\rangle$ at $t=0$ and $t=10t_0$ can be estimated from the average contact pressures, which are $\langle\unde{p_c}\rangle_{t=0} = 0.6064\,F_0/L_0^2$ and $\langle\unde{p_c}\rangle_{t=10t_0} = 0.3939\,F_0/L_0^2$ (per initial area $A_\mrc$) for the contact forces $F_\mrc|_{t=0} = 4\cdot0.8F_0$ and $F_\mrc|_{t=10t_0} = 4\cdot0.5064F_0$ from Fig.~\ref{f:CoboT2}b (for the full sheet).
This then leads to the average gap $\langle g_\mrn\rangle = -\langle\unde{p_c}\rangle/\unde{E_n}$ for $\phi=1$ according to Remark~\ref{r:31}. 
Taking $\big\langle g_\mrn^2 \big\rangle \approx \langle g_\mrn \rangle^2$ then leads to the bounding
\eqb{l}
1.995\cdot10^{-5}F_0L_0 \leq \Delta\Pi_\mrb \leq 4.400\cdot10^{-5}F_0L_0
\label{e:dPi3c}\eqe
for the parameters of Table~\ref{t:CoboSheet}.
Half of $\Delta\Pi_\mrb$ flows into sheet (and the other half into the sphere)
and is converted into the thermal energy\footnote{This is the thermal energy following from a quadratic free-energy expression, cf.~\eqref{e:Psi0}. If the logarithmic expression from \eqref{e:PsiT} is used, the thermal energy becomes $\Delta\Pi_\mrT = \unde{C}L^2\Delta T$, which leads to very similar $\Delta T$ bounds here.}
\eqb{l}
\Delta\Pi_\mrT = \ds\frac{\unde{C}L^2}{2T_0}(T_2^2-T_0^2)\,,
\eqe
once the sheet reaches uniform temperature \citep{cobo}.
From $\Delta\Pi_\mrT = \Delta\Pi_\mrb/2$ and the parameters of Table~\ref{t:CoboSheet} thus follows the temperature bounding
\eqb{l}
6.168\mathrm{K} \leq \Delta T = T_2 - T_0 \leq 13.440\mathrm{K}\,.
\label{e:dPi3d}\eqe
These bounds contain the FE result $\Delta T = 10.484$K, cf.~Fig.~\ref{f:CoboT2}c, which verifies the consistent energy conversion of the proposed formulation.
It is noted that the spread of the bounds solely comes from the difference of $\big\langle g_\mrn^2 \big\rangle$ at $t=0$ and $t=10t_0$, which shows the large influence of the contact pressure on bonding in the example.
It is also noted that this estimate, as well as the FE model used here, exclude temperature changes due to elastic energy changes that occur between $t = 0$ and $t = 10t_0$.

Fig.~\ref{f:CoboT2} also shows the convergence rates for the bonded area, contact force and center temperature.
They all follow $O(n_\mathrm{el}^{-2})=O(h^4)$. 
The compared meshes contain $m^2$ elements and $20m$ time steps, with $m=2,\,4,\,8,\,...,\,256$.

\begin{remark}
The results of Figs.~\ref{f:CoboT1} and \ref{f:CoboT2} are for contact integration over the initially deformed contact surface.
Integration over the undeformed reference configuration yields the same principal behavior but with faster bonding (higher $\dot\phi$) and thus larger sheet temperature $T_2$, larger thermal expansion and larger contact force reduction. 
This is due to the fact that the contact penetration increases when the elemental area reduces for constant FE force. (The FE size remains roughly the same, as the sheet stretch remains almost unchanged.)
This behavior can be argued to be an artifact of the proposed bonding formulation. 
But it can be compensated by adjusting the model parameters such as $\overrightarrow{\unde{K_0}}$.
\end{remark}

\subsection{Four-field chemo-thermo-mechanical bonding: Movable sphere}\label{s:ex3b}

\begin{figure}[h]
\begin{center} \unitlength1cm
\begin{picture}(0,11.6)
\put(-8,5.8){\includegraphics[height=58mm]{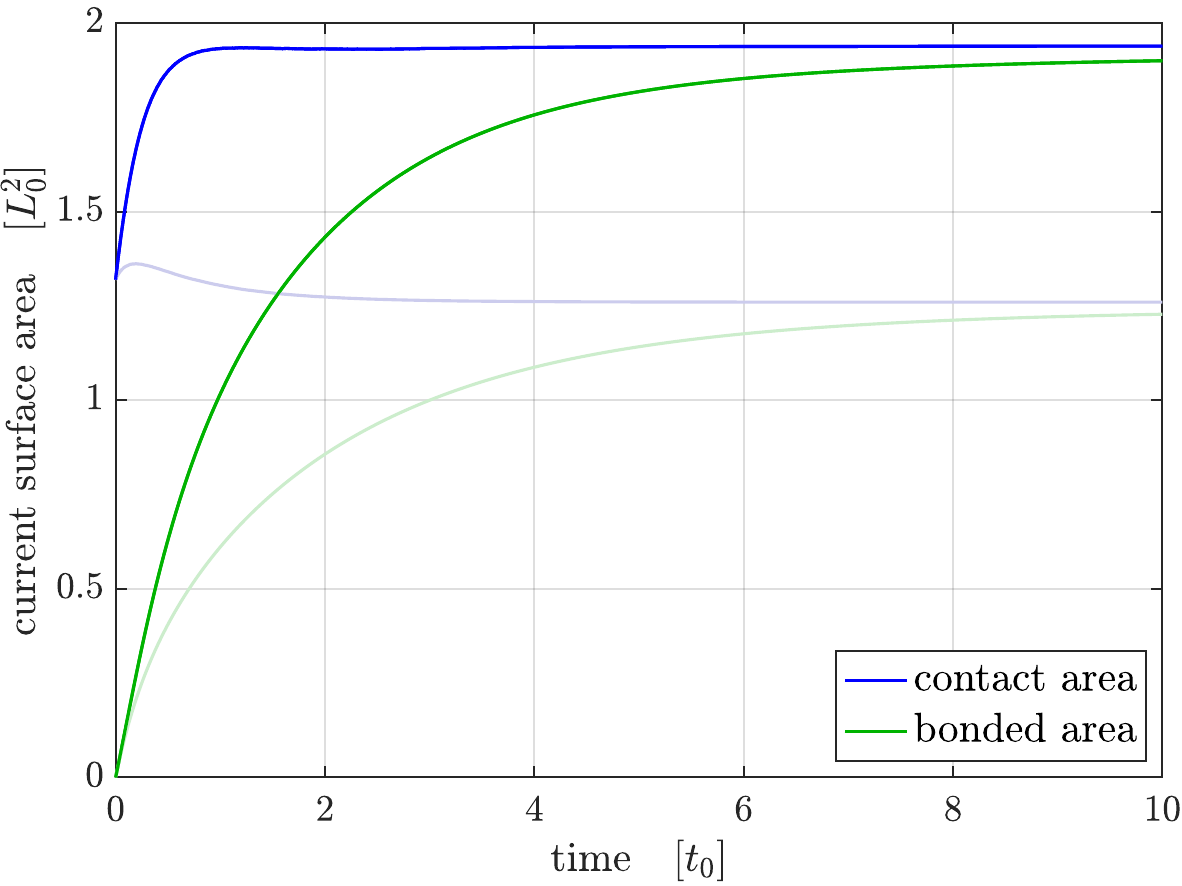}}
\put(0.2,5.8){\includegraphics[height=58mm]{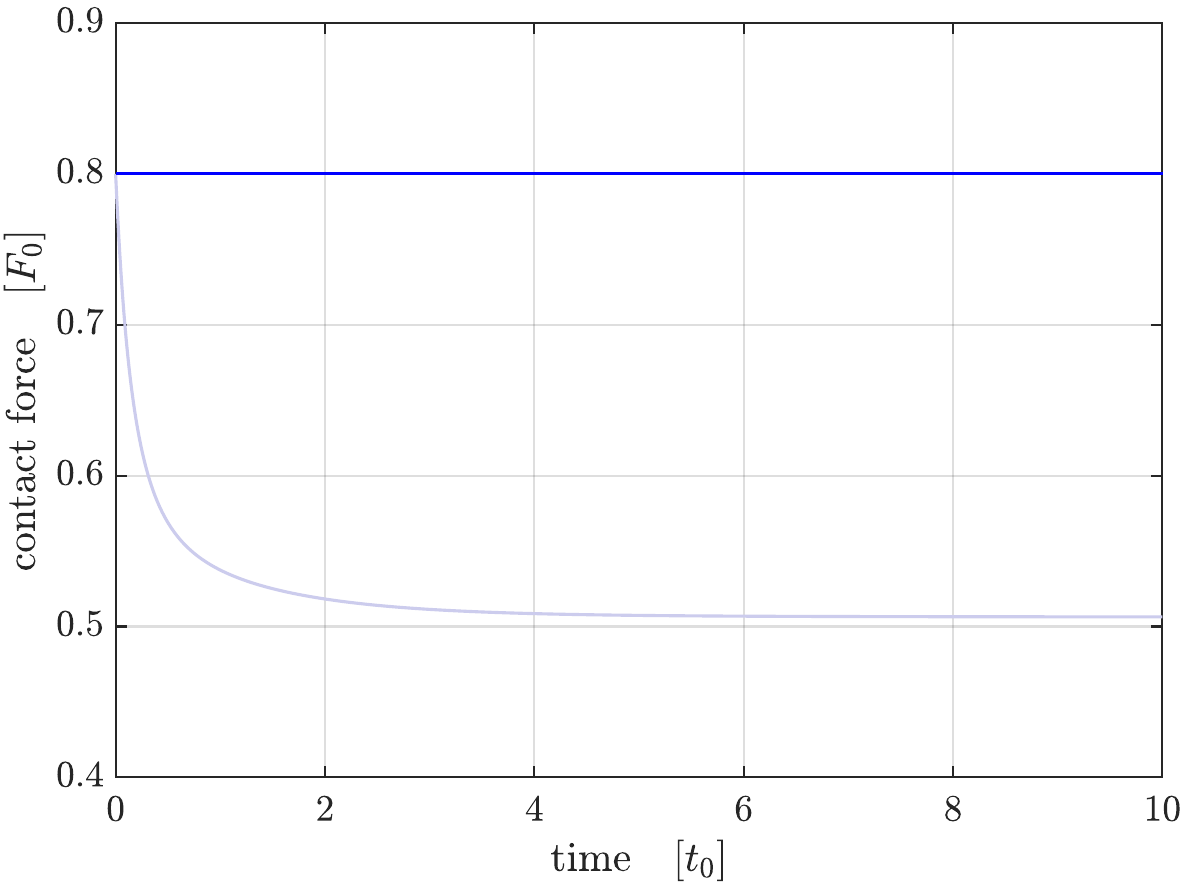}}
\put(-8,-.2){\includegraphics[height=58mm]{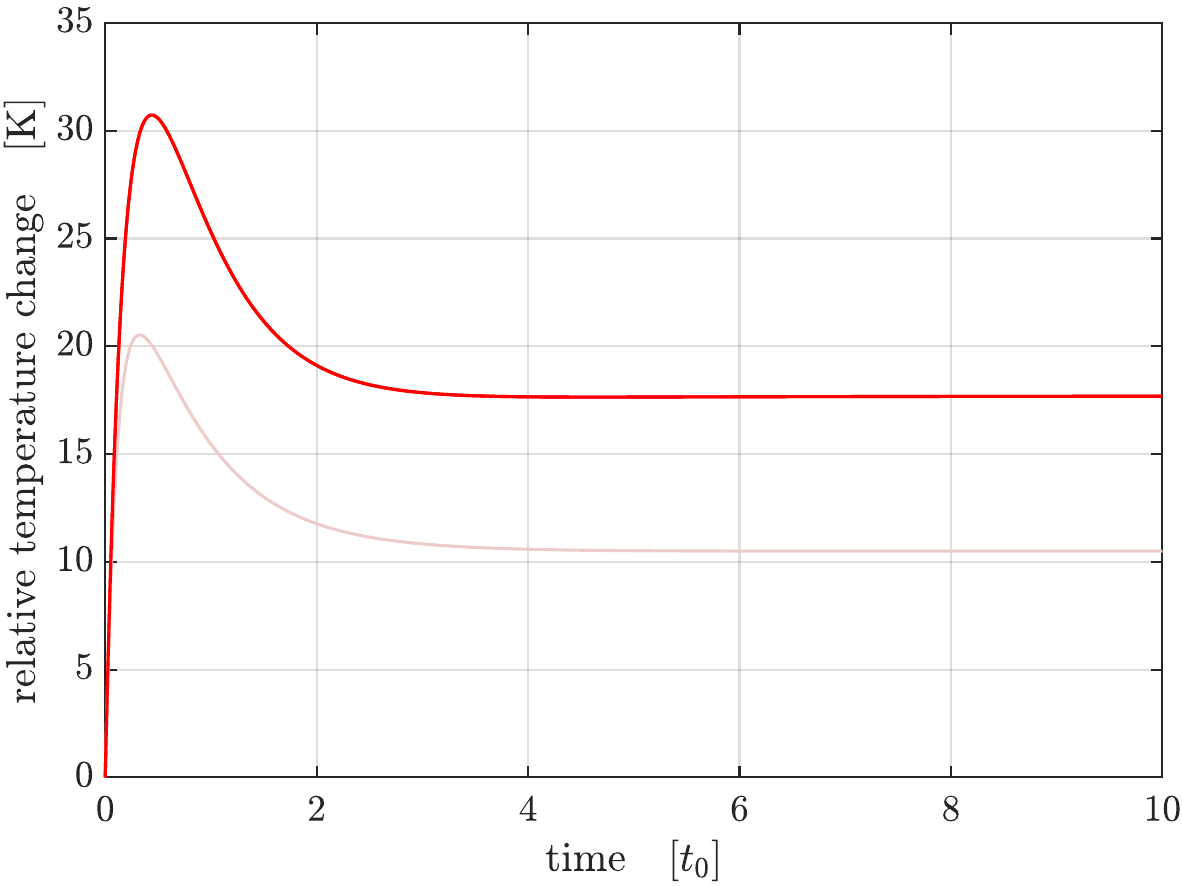}}
\put(.2,-.2){\includegraphics[height=58mm]{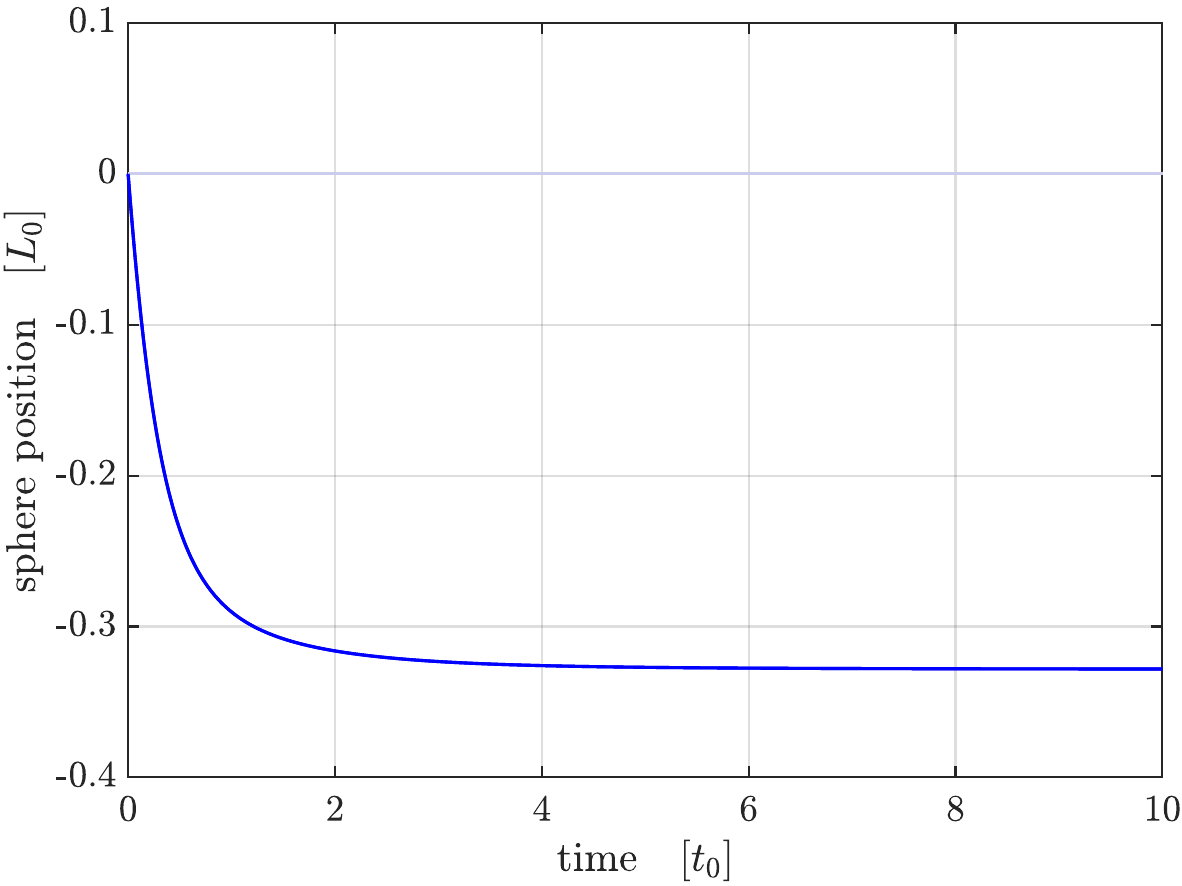}}
\put(-7.95,5.9){\footnotesize (a)}
\put(0.2,5.9){\footnotesize (b)}
\put(-7.95,-.1){\footnotesize (c)}
\put(0.2,-.1){\footnotesize (d)}
\end{picture}
\cprotect\caption{Four-field bonding between membrane and movable sphere for $\alpha_\mrT = 0.005$: 
(a) Evolution of the bonded area in comparison to the contact area, 
(b) evolution of the contact force, 
(c) evolution of the temperature at the center, and 
(d) evolution of the sphere position.
The pale lines show the fixed sphere results from Fig.~\ref{f:CoboT2}.
Compared to those, thermal expansion leads to downward sphere motion, which increases the contact area. 
The larger contact area leads to a larger bonding energy release and hence larger temperatures.
See also supplementary movie \verb|cobo_sphere_fall.mp4| at \href{https://doi.org/10.5446/73729}{doi.org/10.5446/73729}.}  	
\label{f:CoboT3}
\end{center}
\end{figure}

Next the sphere is considered movable.
The starting point is the same as before: same sphere position and same contact force, but now the contact force and not the position is kept constant.
Thus $\bx_1$ becomes an additional unknown, and so there are four unknown fields now, see Fig.~\ref{f:CoboEx}d.
For the same parameters as before (Table~\ref{t:CoboSheet}, $\unde{h}=0$), thermal expansion of the sheet leads to downward motion of the sphere.
This is illustrated in Fig.~\ref{f:CoboT3} together with the evolution of the bonded surface area and center sheet temperature.
In the figure, the pale lines show the results of the fixed sphere (from Sec.~\ref{s:ex3a}).
Compared to those, the bonded surface area and the temperature are now larger.
This is due to the fact that the contact area is now larger.
Therefore more surface can bond, the energy released by bonding is larger, and thus more heat is generated.

\begin{figure}[h]
\begin{center} \unitlength1cm
\begin{picture}(0,11.6)
\put(-8,5.8){\includegraphics[height=58mm]{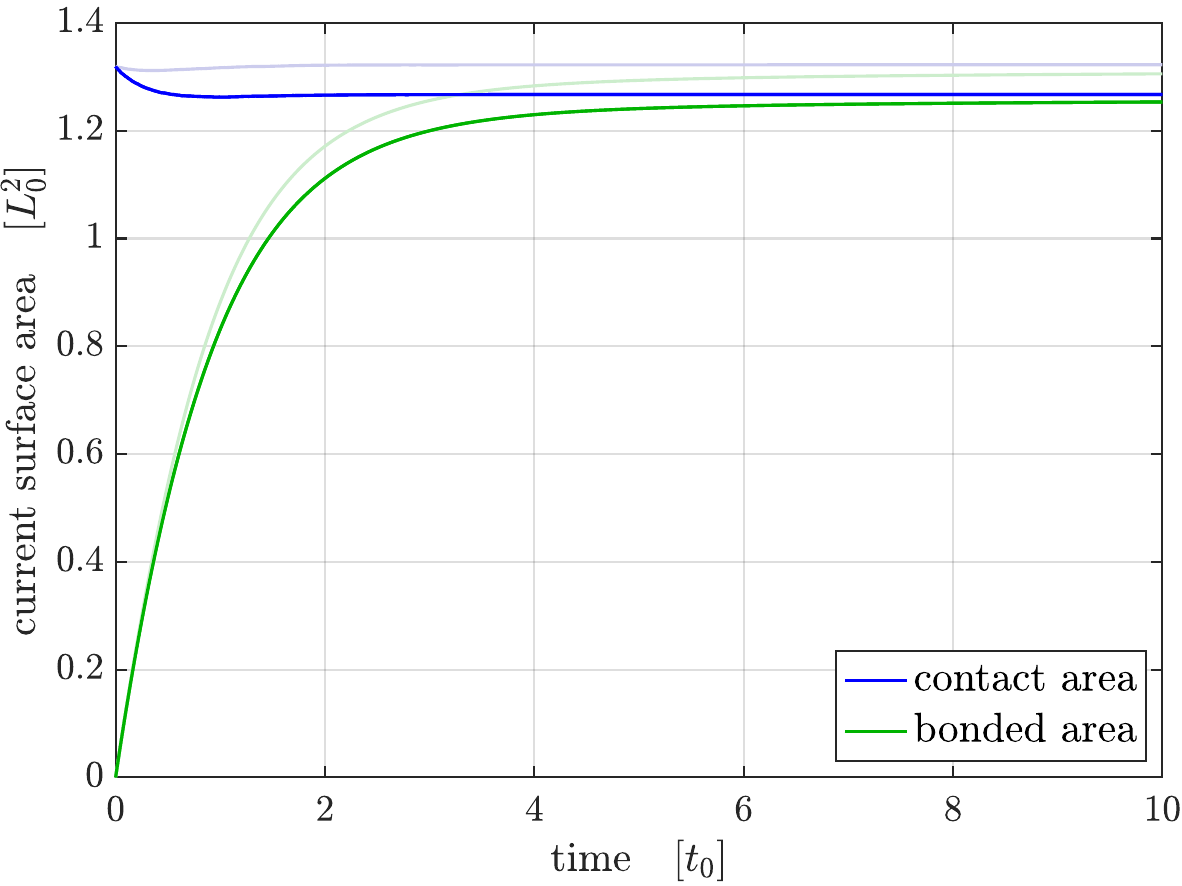}}
\put(0.2,5.8){\includegraphics[height=58mm]{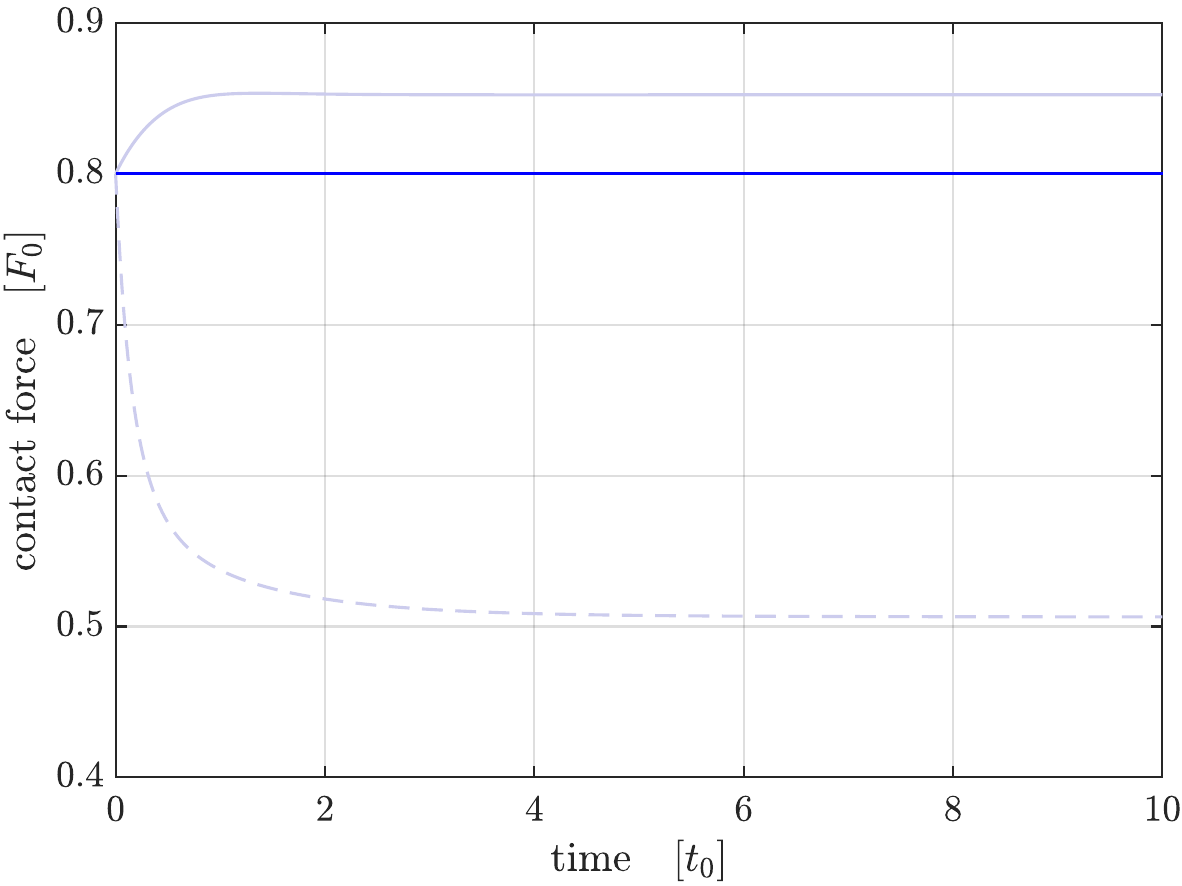}}
\put(-8,-.2){\includegraphics[height=58mm]{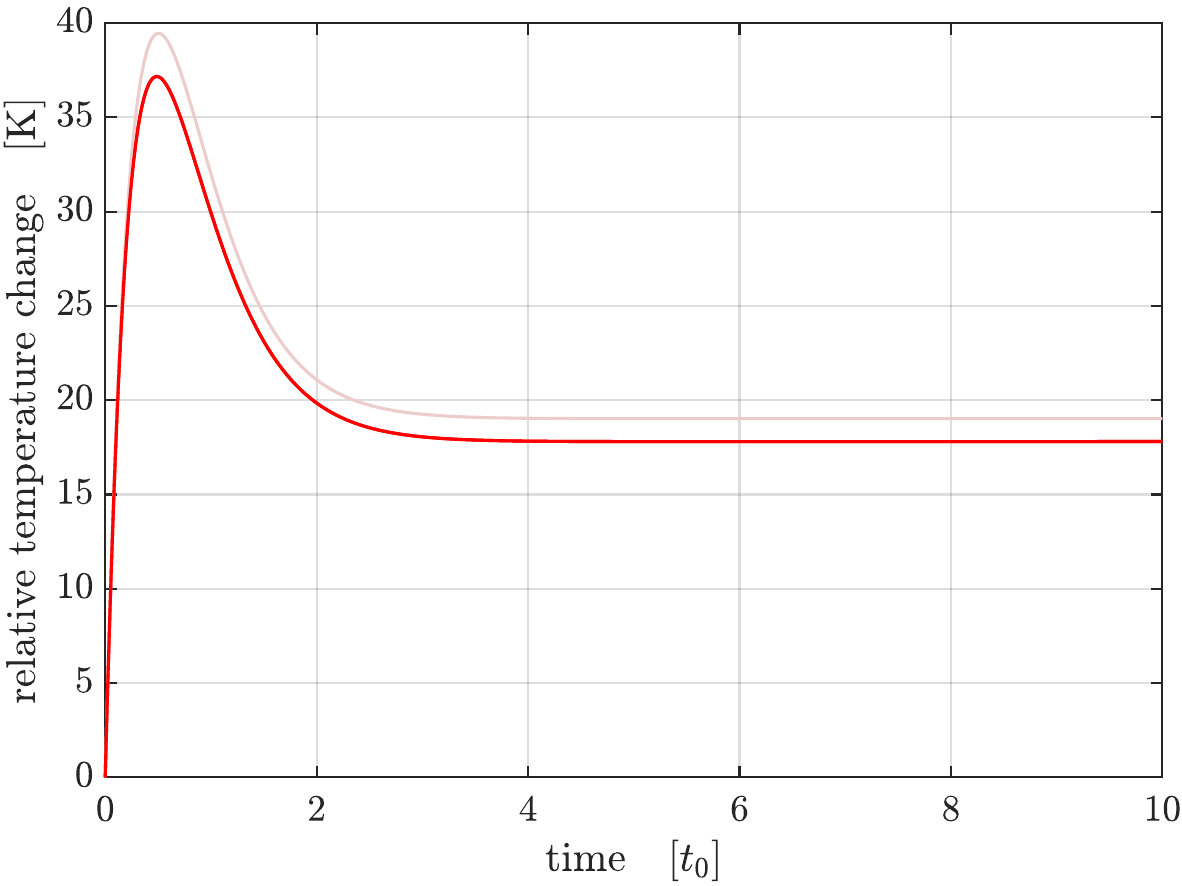}}
\put(.2,-.2){\includegraphics[height=58mm]{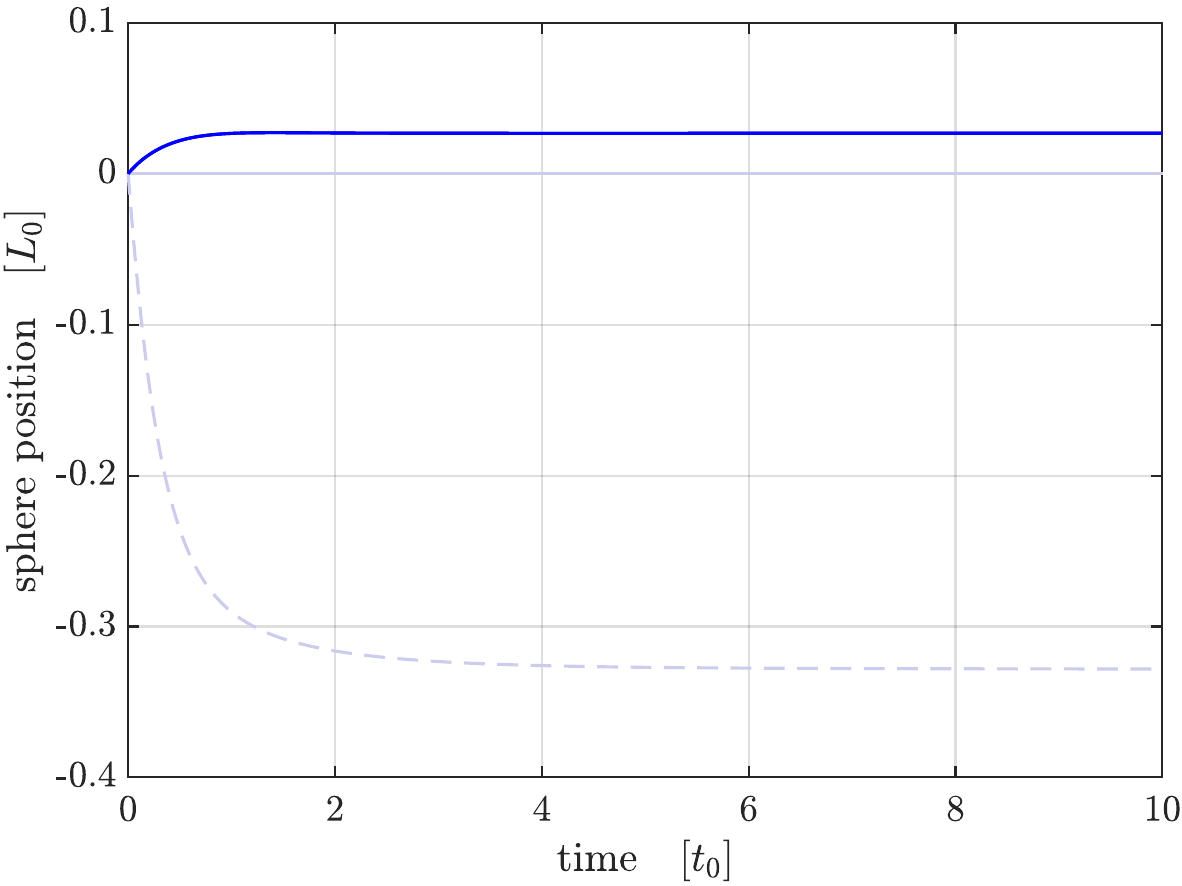}}
\put(-7.95,5.9){\footnotesize (a)}
\put(0.2,5.9){\footnotesize (b)}
\put(-7.95,-.1){\footnotesize (c)}
\put(0.2,-.1){\footnotesize (d)}
\end{picture}
\caption{Four-field bonding between membrane and movable sphere for $\alpha_\mrT = 0$: 
(a) Evolution of the bonded area in comparison to the contact area, 
(b) evolution of the contact force, 
(c) evolution of the temperature at the center, and 
(d) evolution of the sphere position.
The pale lines show the fixed sphere results for $\alpha_\mrT = 0$.
Compared to those, the Gough-Joule effect leads to upward sphere motion, which decreases the contact area. 
The decreased bonding area leads to smaller temperatures.
The pale dashed lines in (b) and (d) show the fixed and moving sphere solutions for $\alpha_\mrT = 0.005$, respectively.}
\label{f:CoboT4}
\end{center}
\end{figure}

\begin{figure}[h]
\begin{center} \unitlength1cm
\begin{picture}(0,11.6)
\put(-8,5.8){\includegraphics[height=58mm]{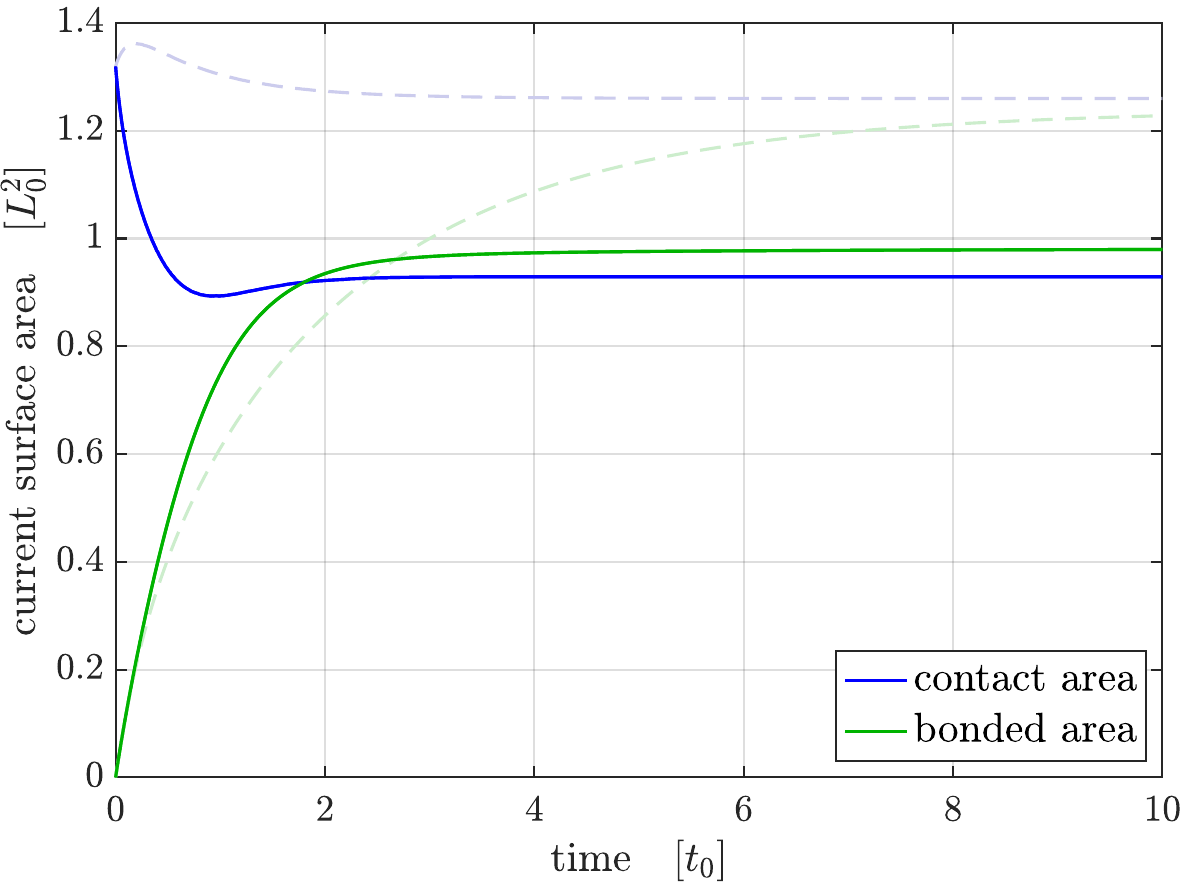}}
\put(0.2,5.8){\includegraphics[height=58mm]{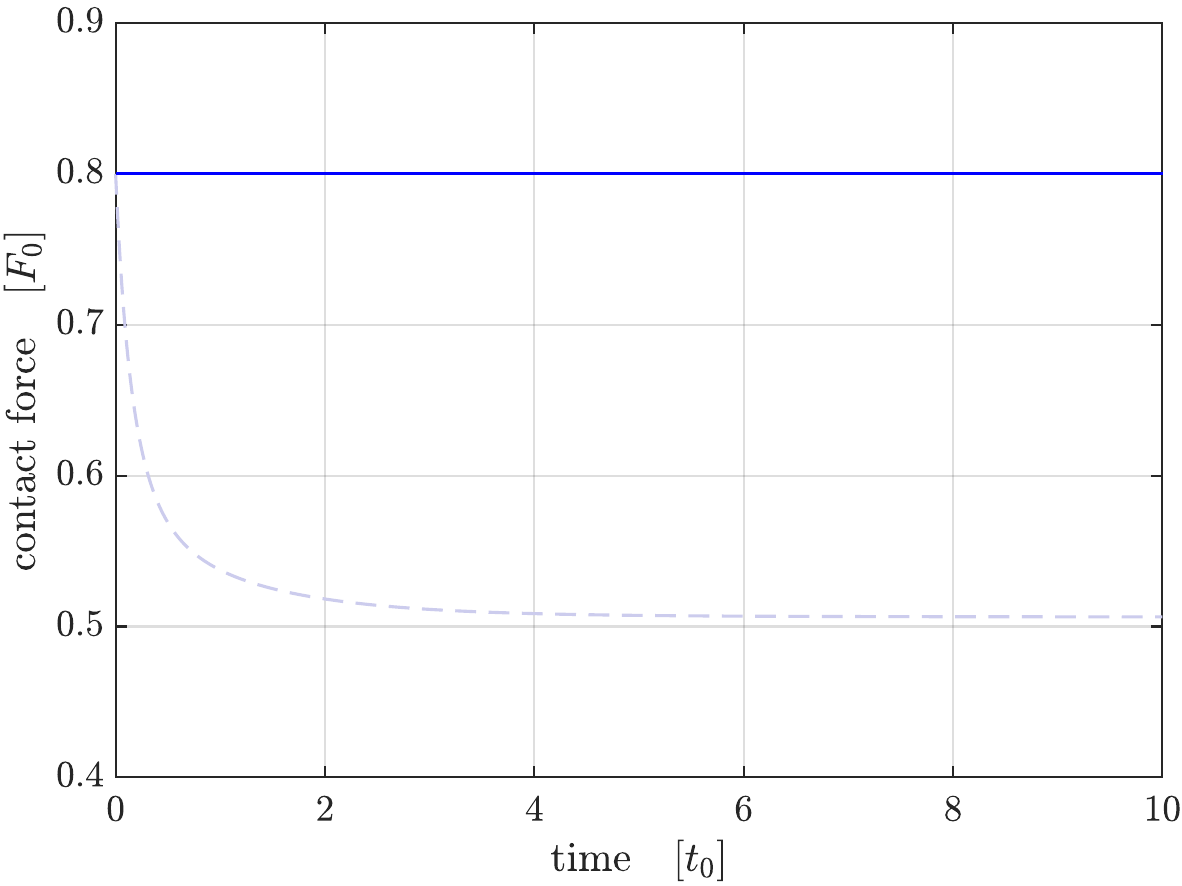}}
\put(-8,-.2){\includegraphics[height=58mm]{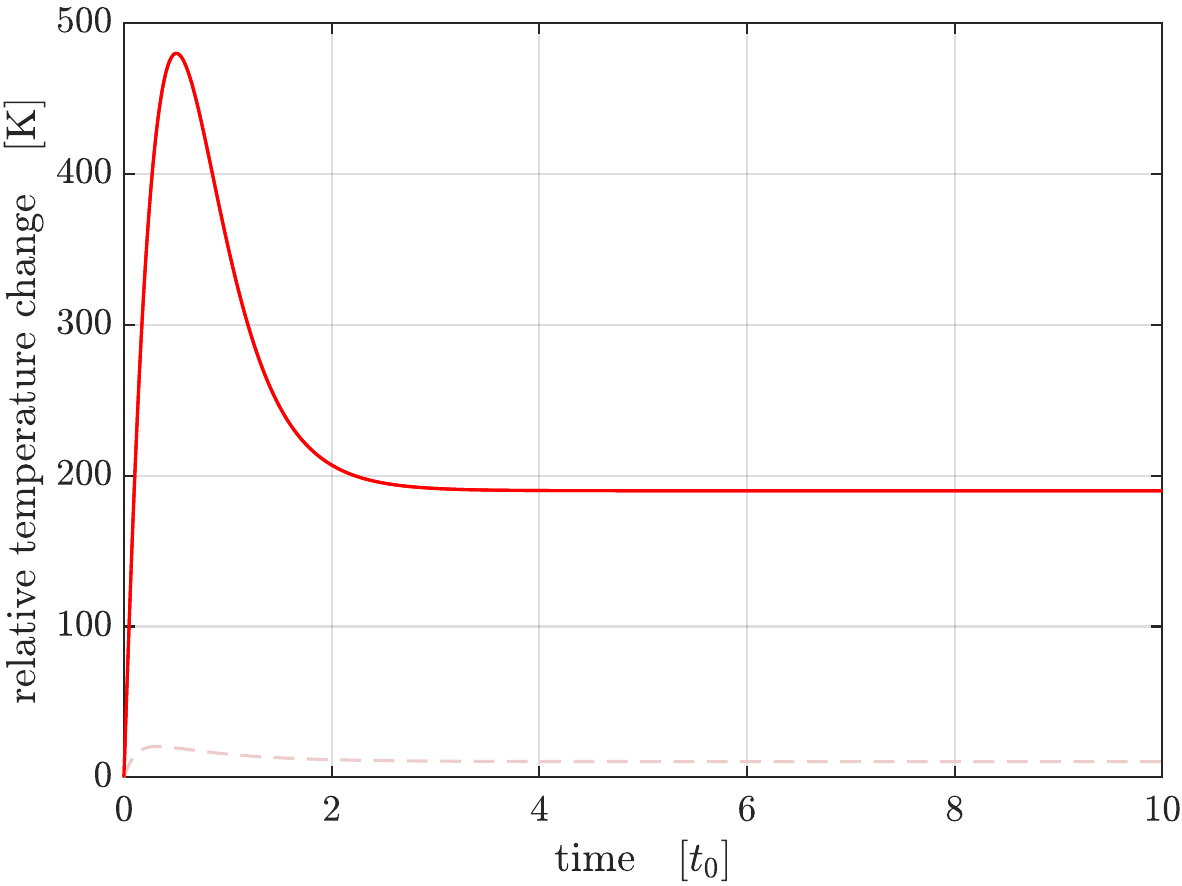}}
\put(.2,-.2){\includegraphics[height=58mm]{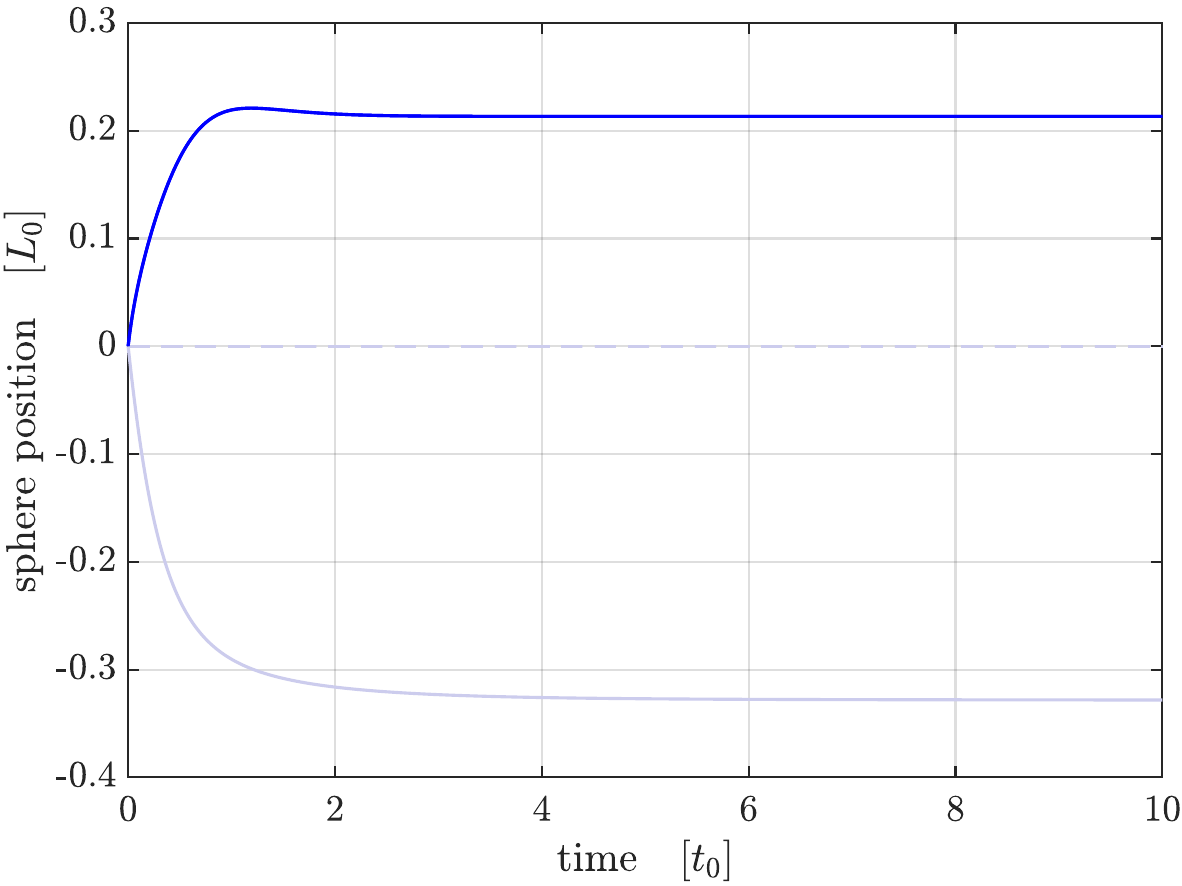}}
\put(-7.95,5.9){\footnotesize (a)}
\put(0.2,5.9){\footnotesize (b)}
\put(-7.95,-.1){\footnotesize (c)}
\put(0.2,-.1){\footnotesize (d)}
\end{picture}
\cprotect\caption{Four-field bonding between membrane and movable sphere for $\alpha_\mrT = 0$, $\protect\unde{C}=10^{-8}F_0/(L_0$K) and $\unde{k}=10^{-8}F_0L_0/(t_0$K) in comparison to the fixed sphere results of Fig.~\ref{f:CoboT2} (dashed line) and the moving sphere result from Fig.~\ref{f:CoboT3} (solid line in d).
For these parameters the sheet gets very hot and the Gough-Joule effect leads to major lifting of the sphere.
Due to this the contact area reduces substantially, and becomes less than the bonded area. 
This happens because there is no debonding considered here.
Thus initially bonded contact regions that lose contact during lifting stay bonded. 
Panels (a)-(d) as before.
See also supplementary movie \verb|cobo_sphere_rise.mp4| at \href{https://doi.org/10.5446/73730}{doi.org/10.5446/73730}.} 	
\label{f:CoboT5}
\end{center}
\end{figure}

\begin{figure}[H]
\begin{center} \unitlength1cm
\begin{picture}(0,5.6)
\put(-8,-.2){\includegraphics[height=58mm]{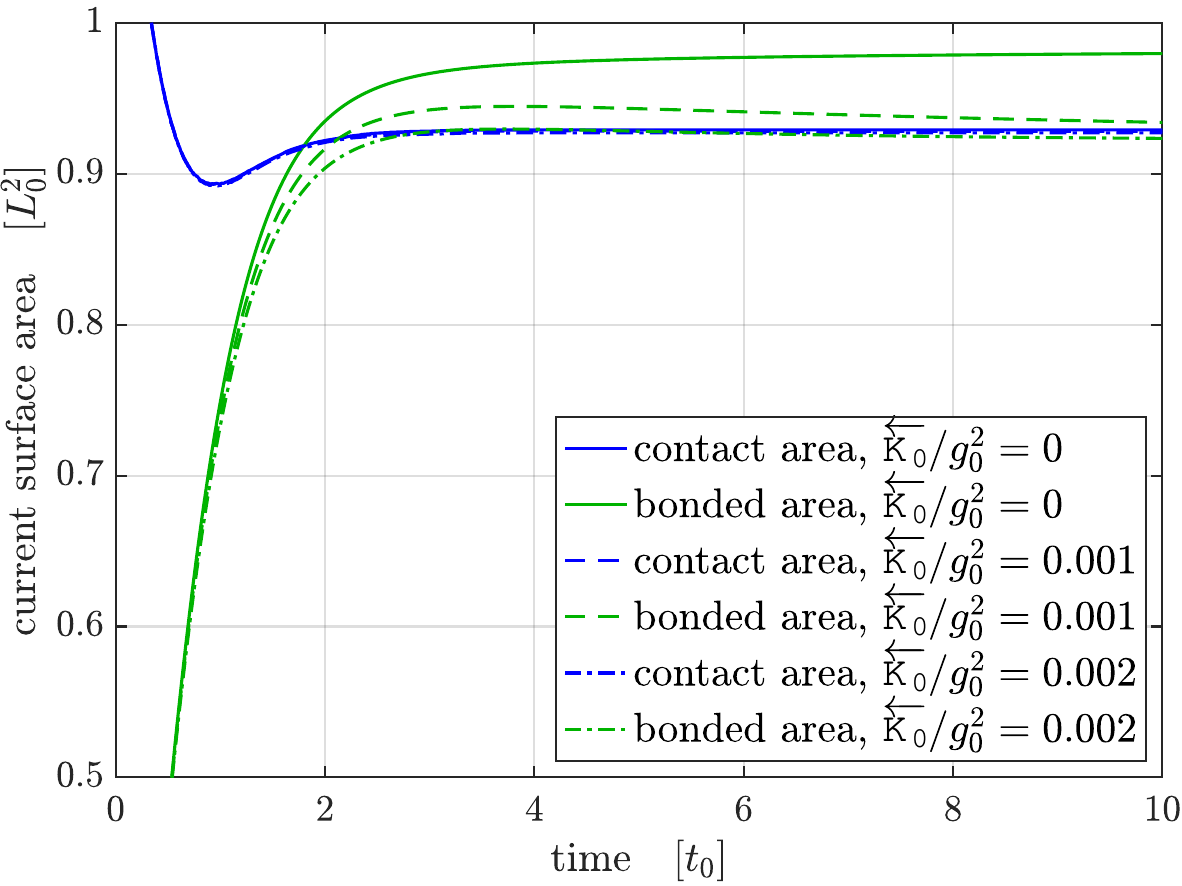}}
\put(.2,-.2){\includegraphics[height=58mm]{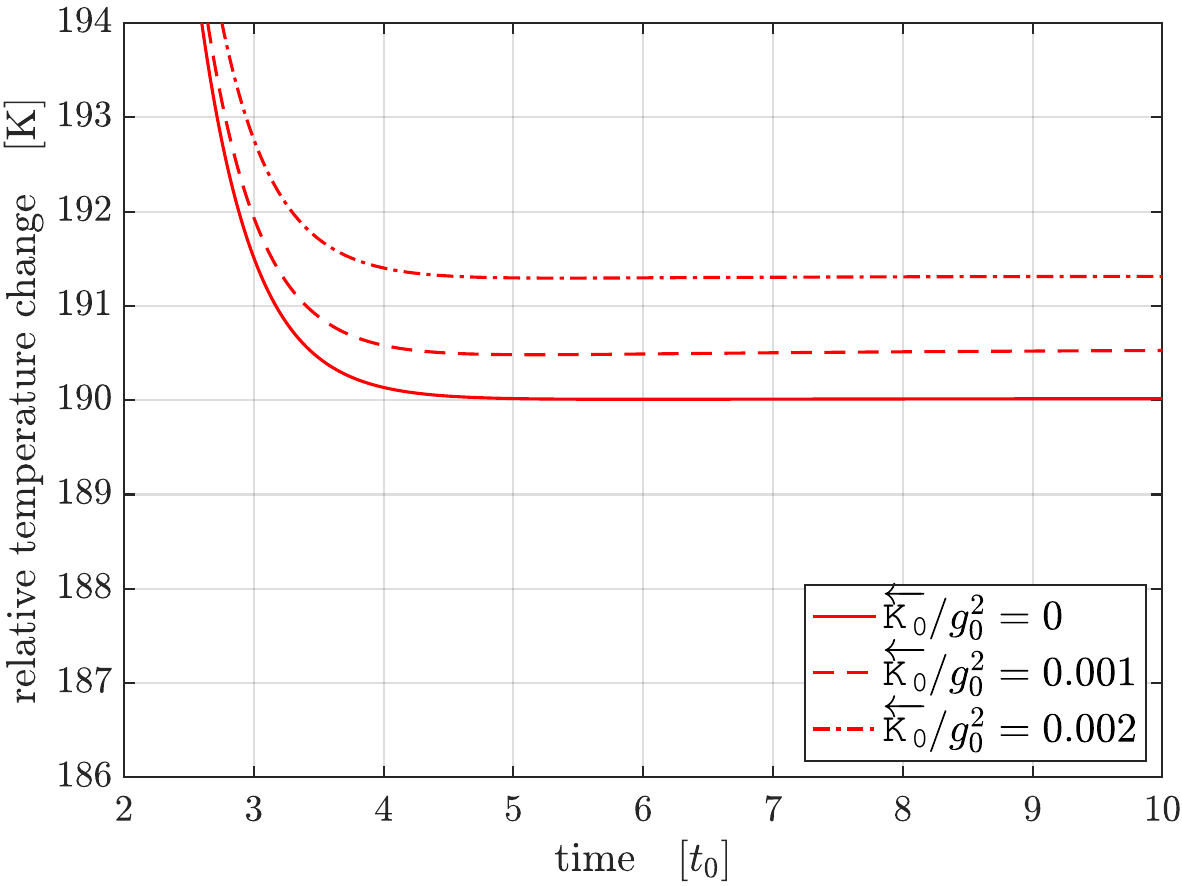}}
\put(-7.95,-.1){\footnotesize (a)}
\put(0.2,-.1){\footnotesize (b)}
\end{picture}
\caption{Four-field bonding between membrane and movable sphere for $\alpha_\mrT = 0$, $\unde{C}=10^{-8}F_0/(L_0$K), $\unde{k}=10^{-8}F_0L_0/(t_0$K) and various debonding parameters $\protect\overleftarrow{\unde{K_0}}/g_0^2$:
(a) Evolution of the bonded area in comparison to the contact area, and
(b) evolution of the center temperature.
$\protect\overleftarrow{\unde{K_0}}/g_0^2=0$ is the result from Fig.~\ref{f:CoboT5}. 
Compared to that, $\protect\overleftarrow{\unde{K_0}}/g_0^2>0$ ensures that initially bonded regions at the contact boundary can debond again.
The total bonded area then becomes equal to the contact area.
The debonding dissipates energy just as bonding does, as Eq.~\eqref{e:ftext} shows, which leads to an additional temperature increase.}	 
\label{f:CoboT6}
\end{center}
\end{figure}

The sphere can also move upward instead of downward, due to the Gough-Joule effect.
This is due to a stiffness increase with temperature, which is accounted for in the present sheet model.
This stiffness increase leads to a decrease in the sheet deformation (for fixed contact load) and hence a rising sphere.
The rise is most pronounced for $\alpha_\mrT = 0$, as thermal expansion otherwise counters the Gough-Joule effect.
Fig.~\ref{f:CoboT4} shows the evolution of the bonded surface area, center sheet temperature and sphere position for this case.
Again, the pale lines show the results of the fixed sphere (now for $\alpha_\mrT = 0$).
Compared to those, the bonded surface area and the temperature are now smaller.
The rise is not as large as the drop from Fig.~\ref{f:CoboT3}, as the stiffness increase, which is proportional to the temperature increase, is only 17.8K/290K = 6.1\%.
A more dramatic rise occurs when the heat capacity and conductivity are decreased by a factor of 10, which is shown in Fig.~\ref{f:CoboT5}.
Then, the temperature increases by 190K, leading to a 190/290 = 66\% stiffness increase.
This case is also interesting, as now the bonded area is larger than the contact area, as initially bonded contact regions lose contact but remain bonded during the rise of the sphere.
These bonds are only lost if the debonding process is accounted for simultaneously in the model, as Fig.~\ref{f:CoboT6}  shows. 

\subsection{Five-field chemo-thermo-mechanical bonding: Heatable sphere}\label{s:ex3c}

\begin{figure}[h]
\begin{center} \unitlength1cm
\begin{picture}(0,11.6)
\put(-8,5.8){\includegraphics[height=58mm]{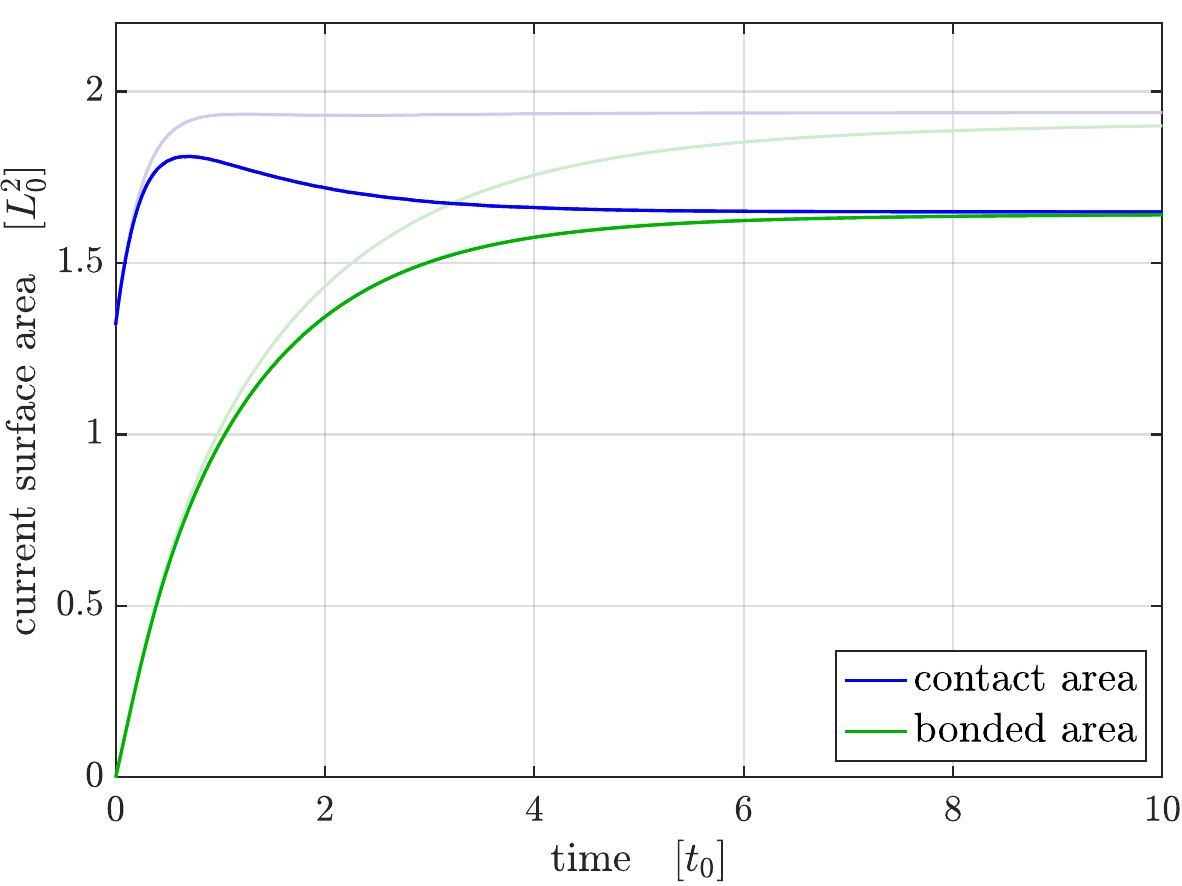}}	
\put(0.2,5.8){\includegraphics[height=58mm]{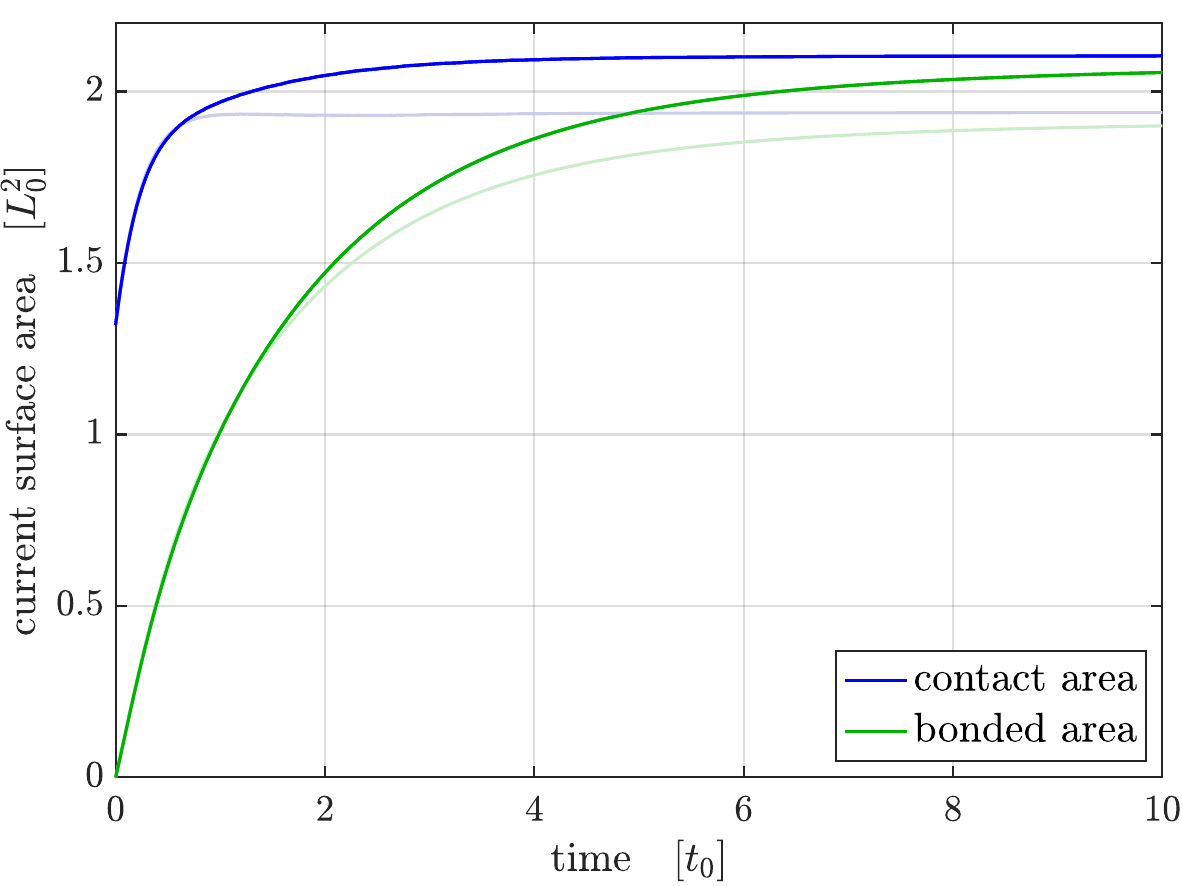}}
\put(-8,-.2){\includegraphics[height=58mm]{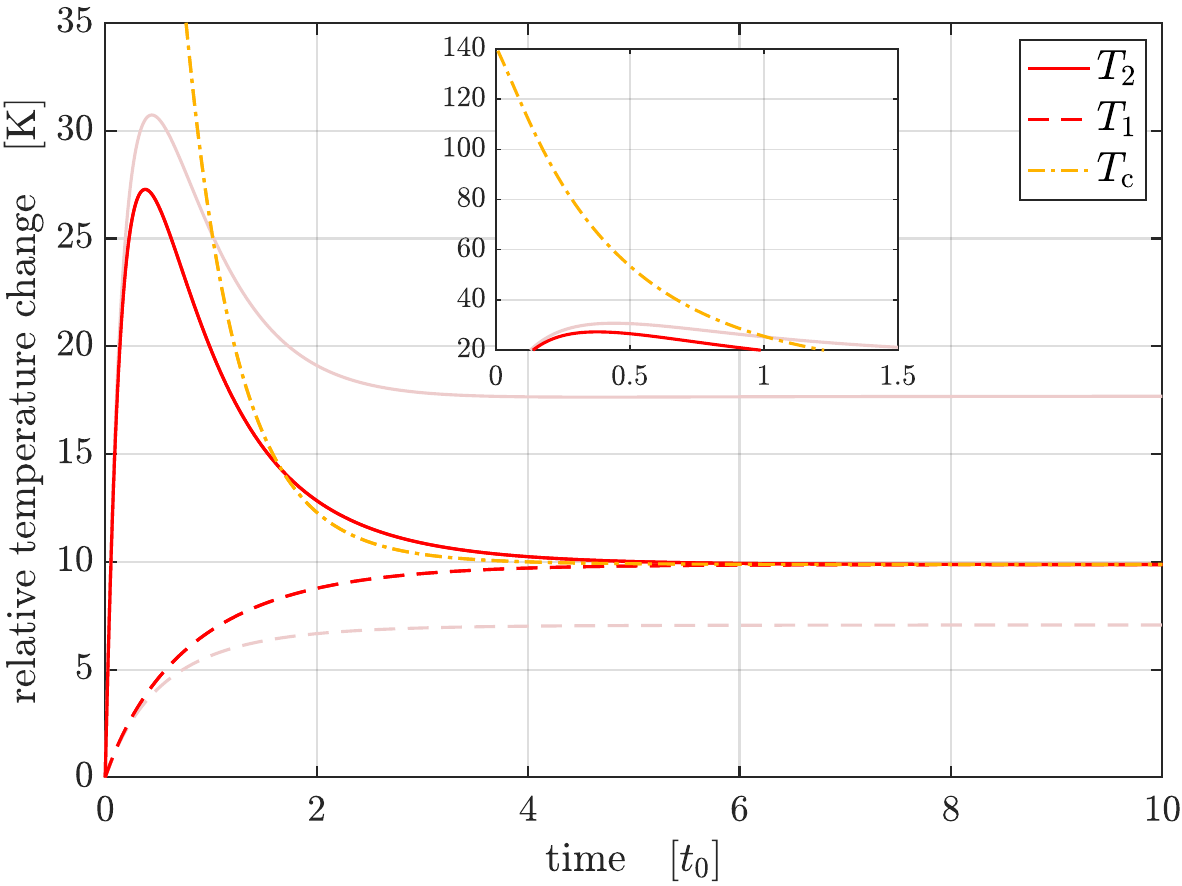}} 
\put(.2,-.2){\includegraphics[height=58mm]{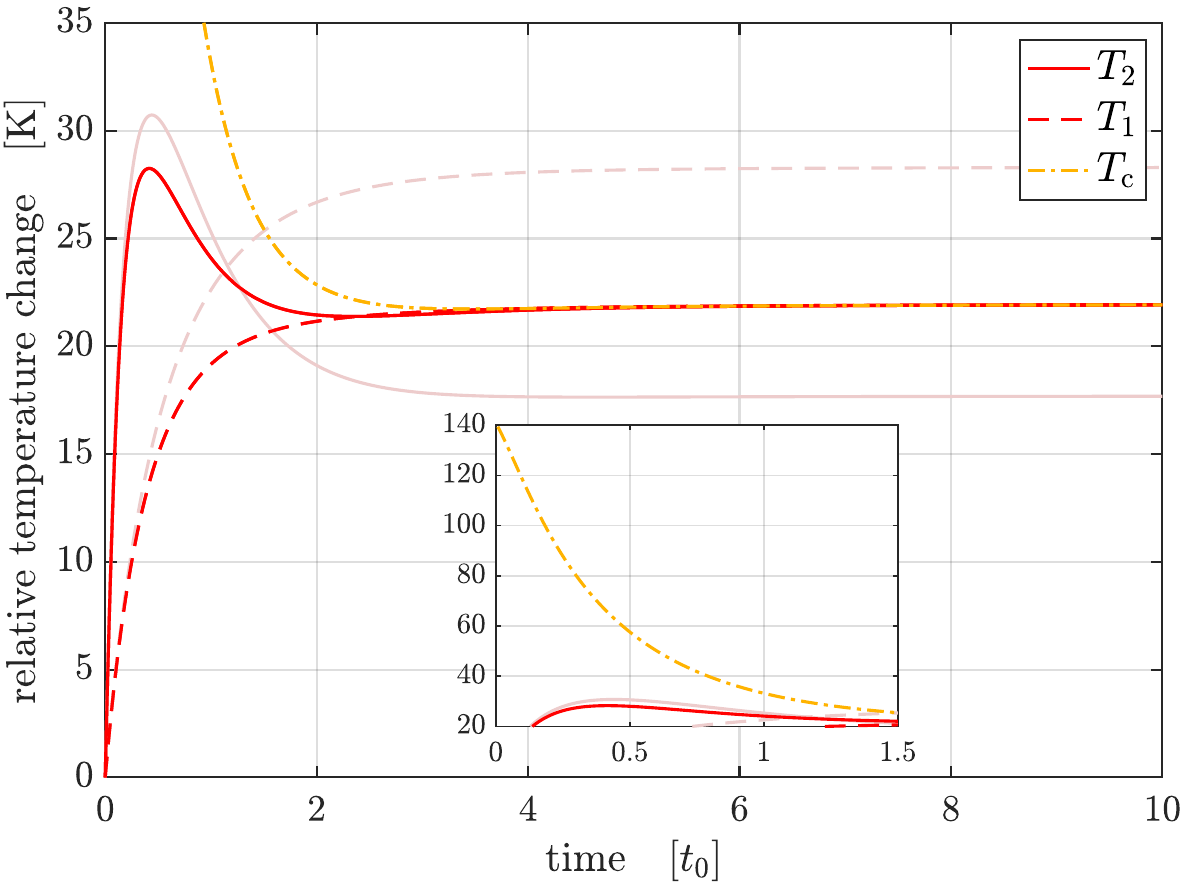}} 
\put(-7.95,5.9){\footnotesize (a)}
\put(0.2,5.9){\footnotesize (b)}
\put(-7.95,-.1){\footnotesize (c)}
\put(0.2,-.1){\footnotesize (d)}
\end{picture}
\caption{Five-field bonding between membrane and heatable sphere: 
Evolution of bonded area and center temperature for (a, c) $\unde{C_1} = 10^{-6}F_0L_0/$K and (b, d) $\unde{C_1} = 2.5\cdot10^{-7}F_0L_0/$K.
All other parameters as in Table \ref{t:CoboSheet}. 
The pale solid lines are for $\unde{h} = 0$ (which is the result of Fig.~\ref{f:CoboT3}).
The higher temperatures in (b, d) lead to larger thermal expansion and hence larger contact and bonding areas.
The interface temperature $T_\mrc$ is a post-processing result that follows from \eqref{e:5fTc}.}
\label{f:CoboT7}
\end{center}
\end{figure}

Next the sphere is allowed to heat up, such that there are five unknown fields, see Fig.~\ref{f:CoboEx}e.
Fig.~\ref{f:CoboT7} shows the evolution of the bonded area and temperatures for the two sphere heat capacities $\unde{C_1} = 10^{-6}F_0L_0/$K and $\unde{C_1} = 2.5\cdot10^{-7}F_0L_0/$K (for the quarter model), and all other parameters as in Table~\ref{t:CoboSheet}.
In case there is no transfer heat flux ($\unde{h}=0$), membrane and sphere heat up differently and end up having different temperatures: 
$T_1<T_2$ for the first heat capacity and  $T_1>T_2$ for the second.
This is indicated by the pale lines in the figure.
If a transfer occurs -- here for $\unde{h}=10^{-7}F_0/(L_0t_0$K) -- the temperatures equilibrate at the same level, as the figure shows.
For the five-field formulation, the interface temperature $T_\mrc$ can be obtained from \eqref{e:5fTc}.
It is highest at $t=0$, when $\dot\phi$ is largest.

\subsection{Six-field chemo-thermo-mechanical bonding: Interfacial heat capacity}\label{s:ex3d}

\begin{figure}[h]
\begin{center} \unitlength1cm
\begin{picture}(0,11.6)
\put(-8,5.8){\includegraphics[height=58mm]{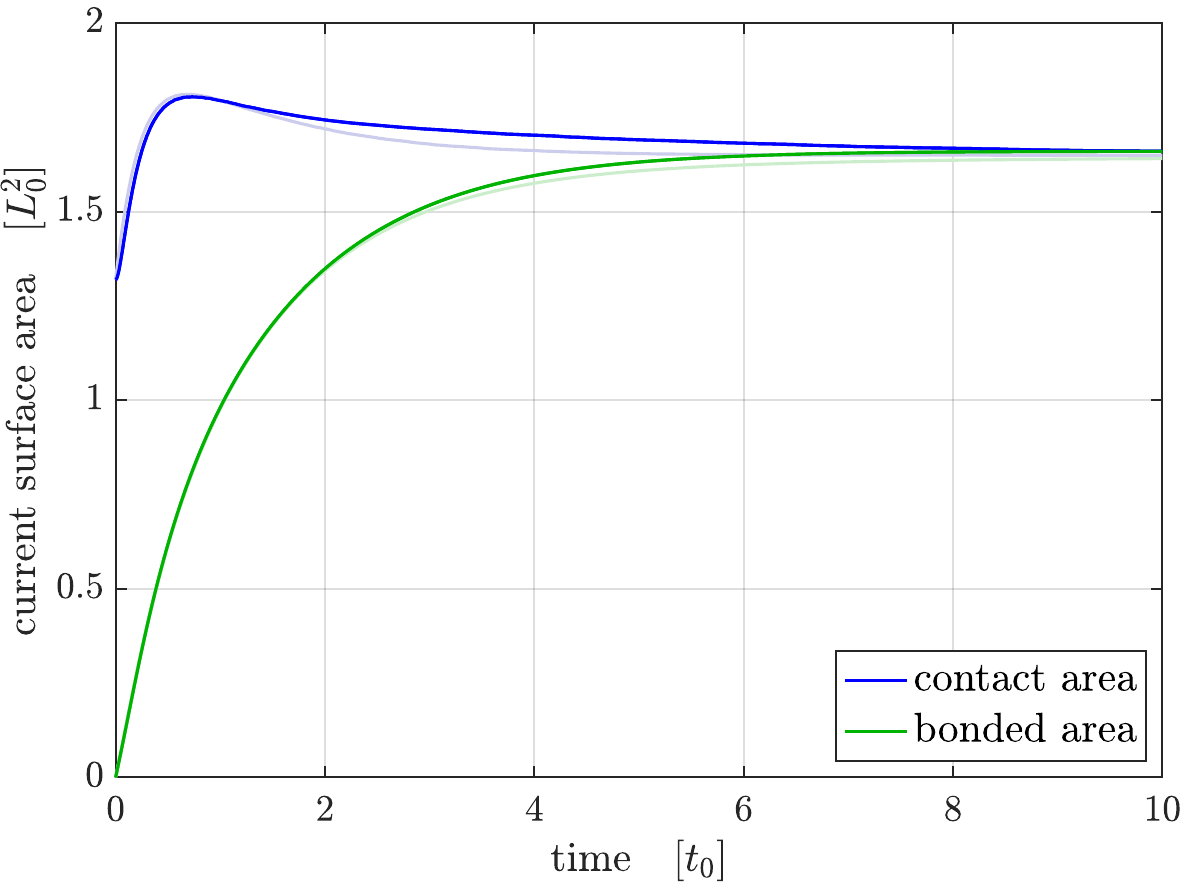}} 
\put(0.2,5.8){\includegraphics[height=58mm]{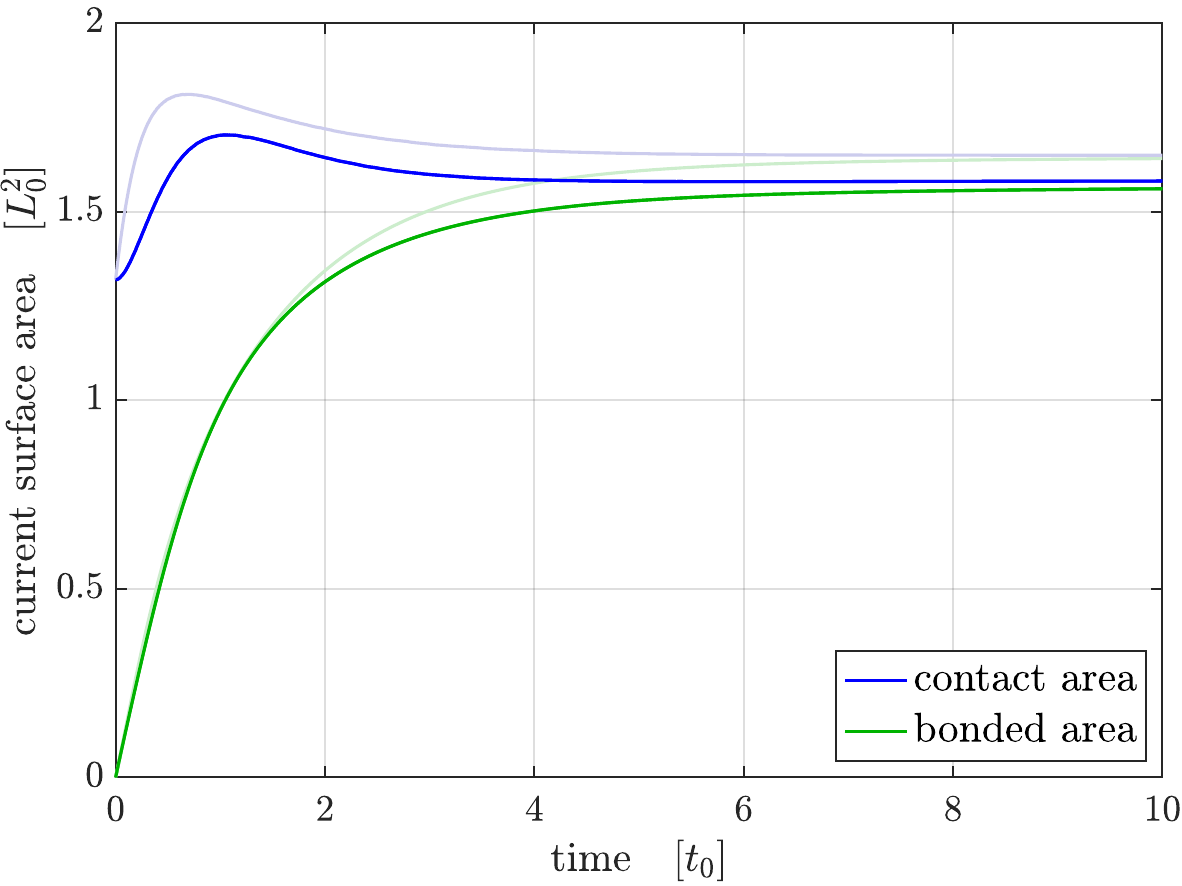}}	 
\put(-8,-.2){\includegraphics[height=58mm]{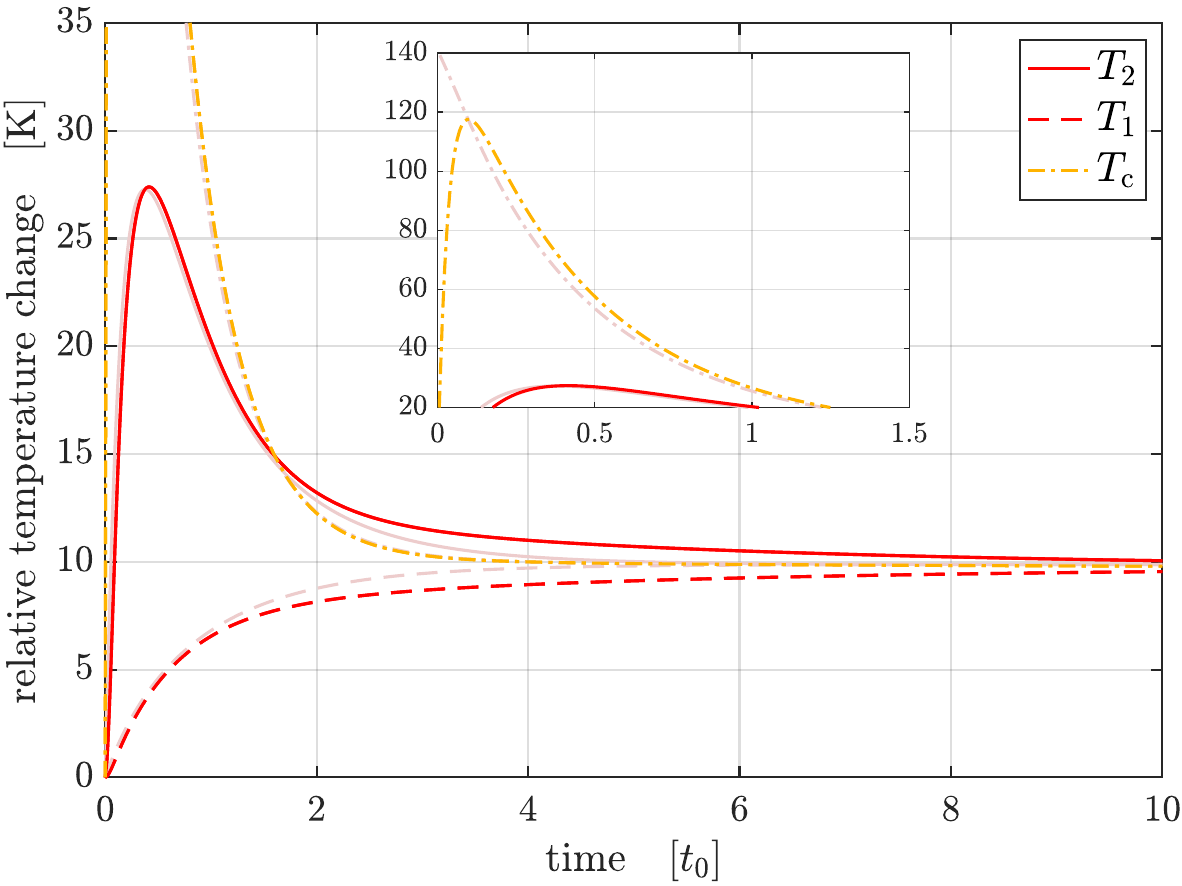}}       
\put(.2,-.2){\includegraphics[height=58mm]{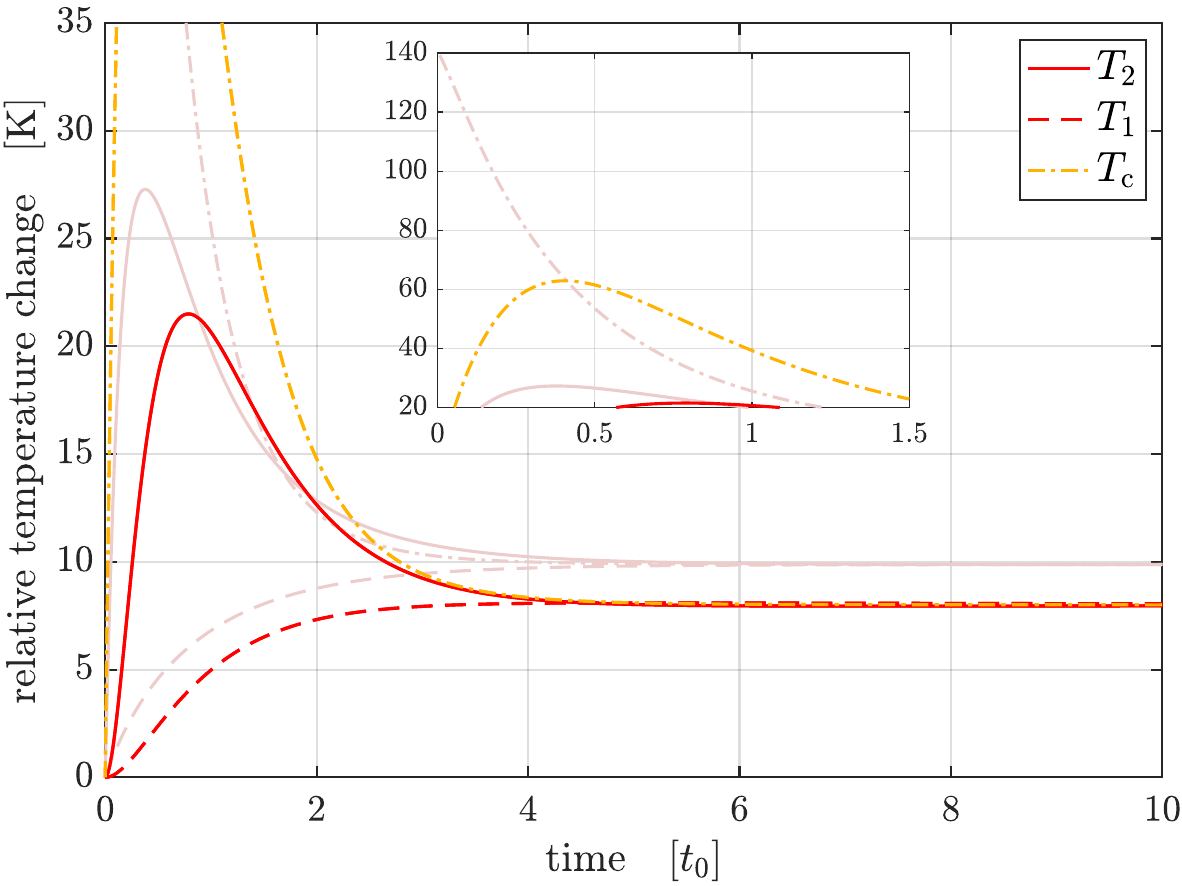}}    
\put(-7.95,5.9){\footnotesize (a)}
\put(0.2,5.9){\footnotesize (b)}
\put(-7.95,-.1){\footnotesize (c)}
\put(0.2,-.1){\footnotesize (d)}
\end{picture}
\caption{Six-field bonding between membrane and heatable sphere: 
Evolution of bonding and temperature for interfacial heat capacity (a, c) $\unde{C_c} = 10^{-8}F_0/(L_0$K) and
(b, d) $\unde{C_c} = 10^{-7}F_0/(L_0$K).
Both for $\unde{C_1} = 10^{-6}F_0L_0/$K and all other parameters as in Table \ref{t:CoboSheet}. 
The pale lines are for $\unde{C_c} = 0$ (which is the case of Fig.~\ref{f:CoboT7}a \& c).
As $\unde{C_c}\rightarrow0$ the five-field bonding results from Fig.~\ref{f:CoboT7}a \& c are approached.}
\label{f:CoboT8}
\end{center}
\end{figure}

\begin{figure}[h!]
\begin{center} \unitlength1cm
\begin{picture}(0,13.2)
\put(-7.9,9.86){\includegraphics[height=34mm]{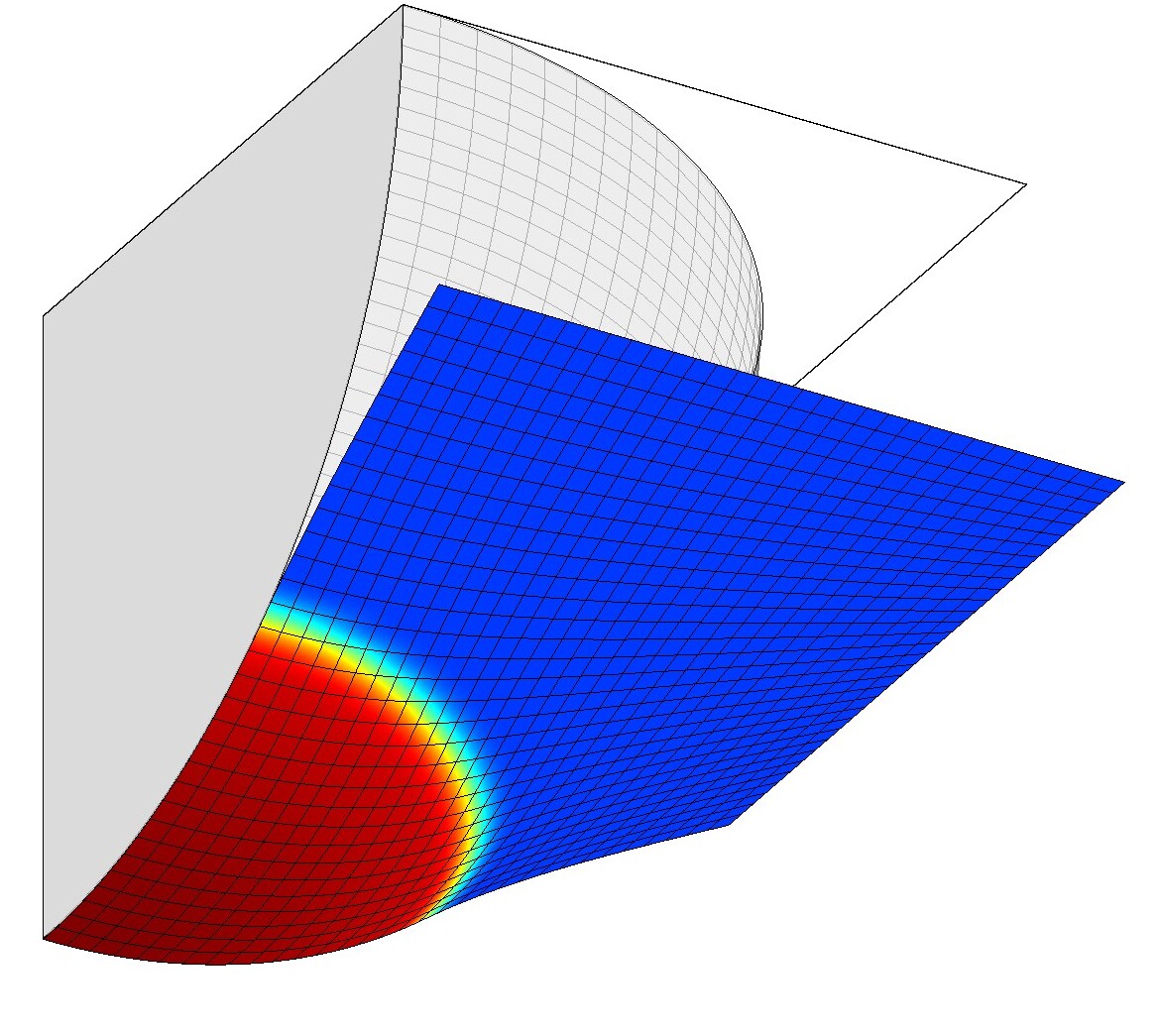}}
\put(-4.1,9.86){\includegraphics[height=34mm]{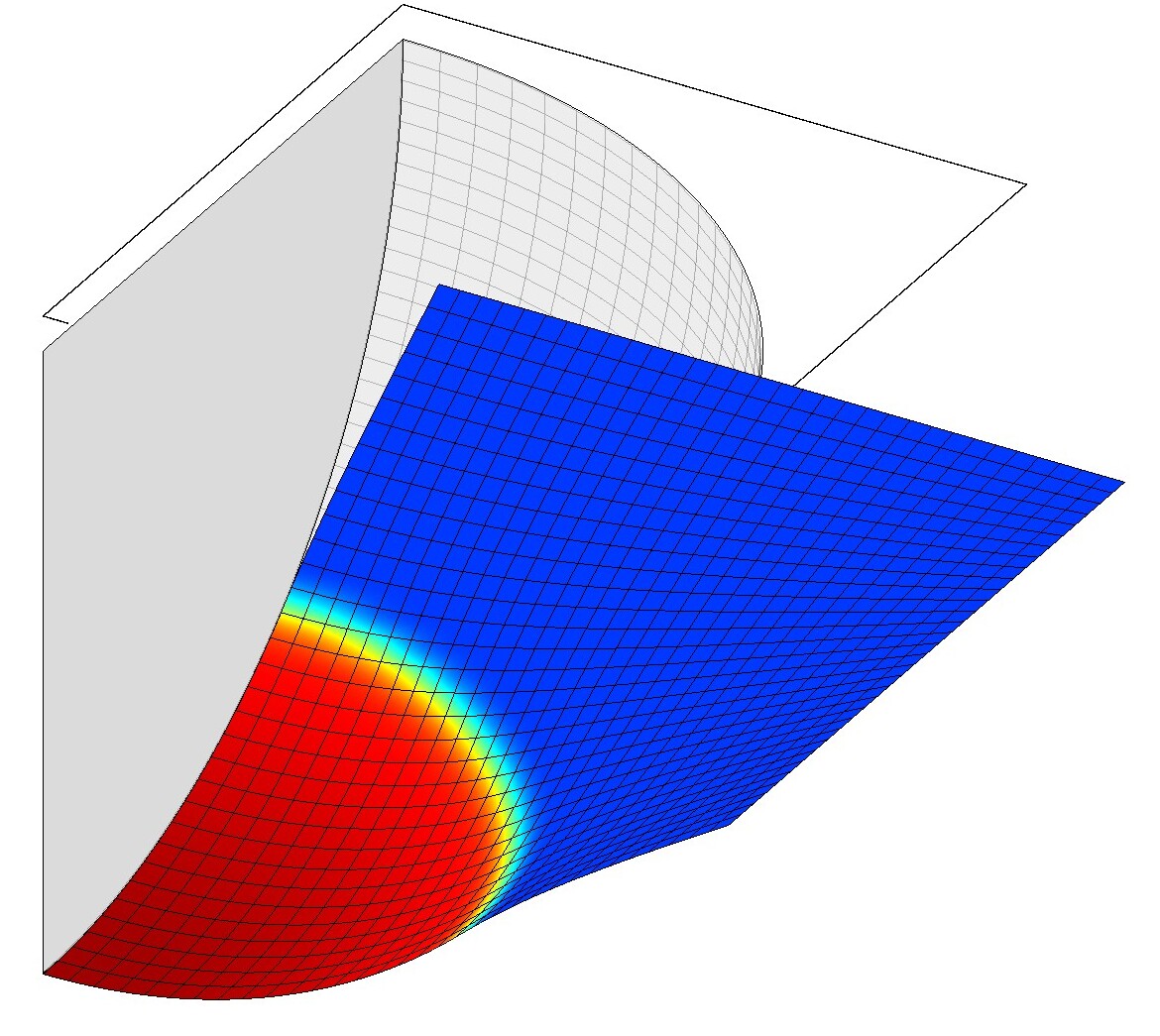}}
\put(-.3,9.86){\includegraphics[height=34mm]{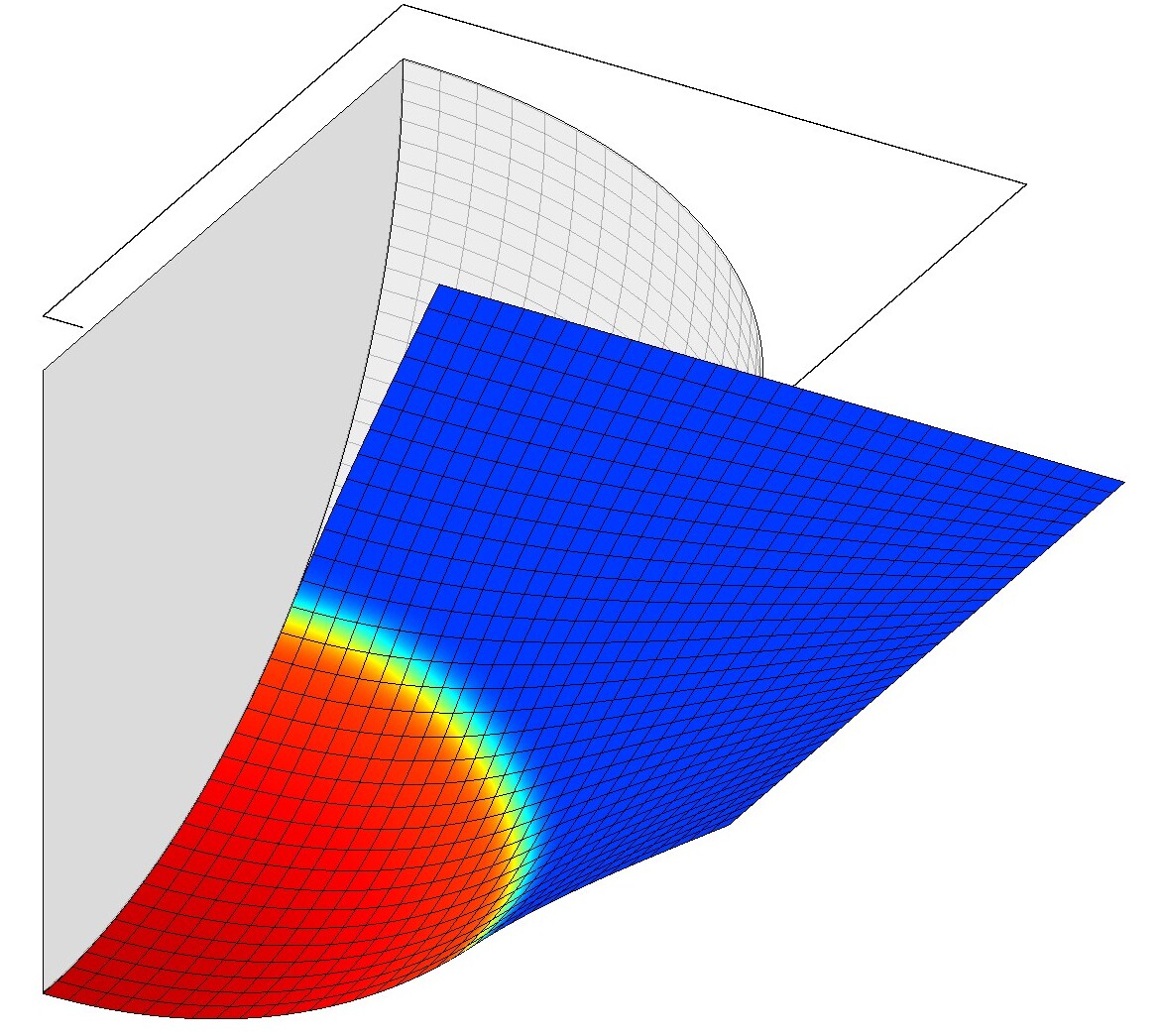}}
\put(3.5,9.86){\includegraphics[height=34mm]{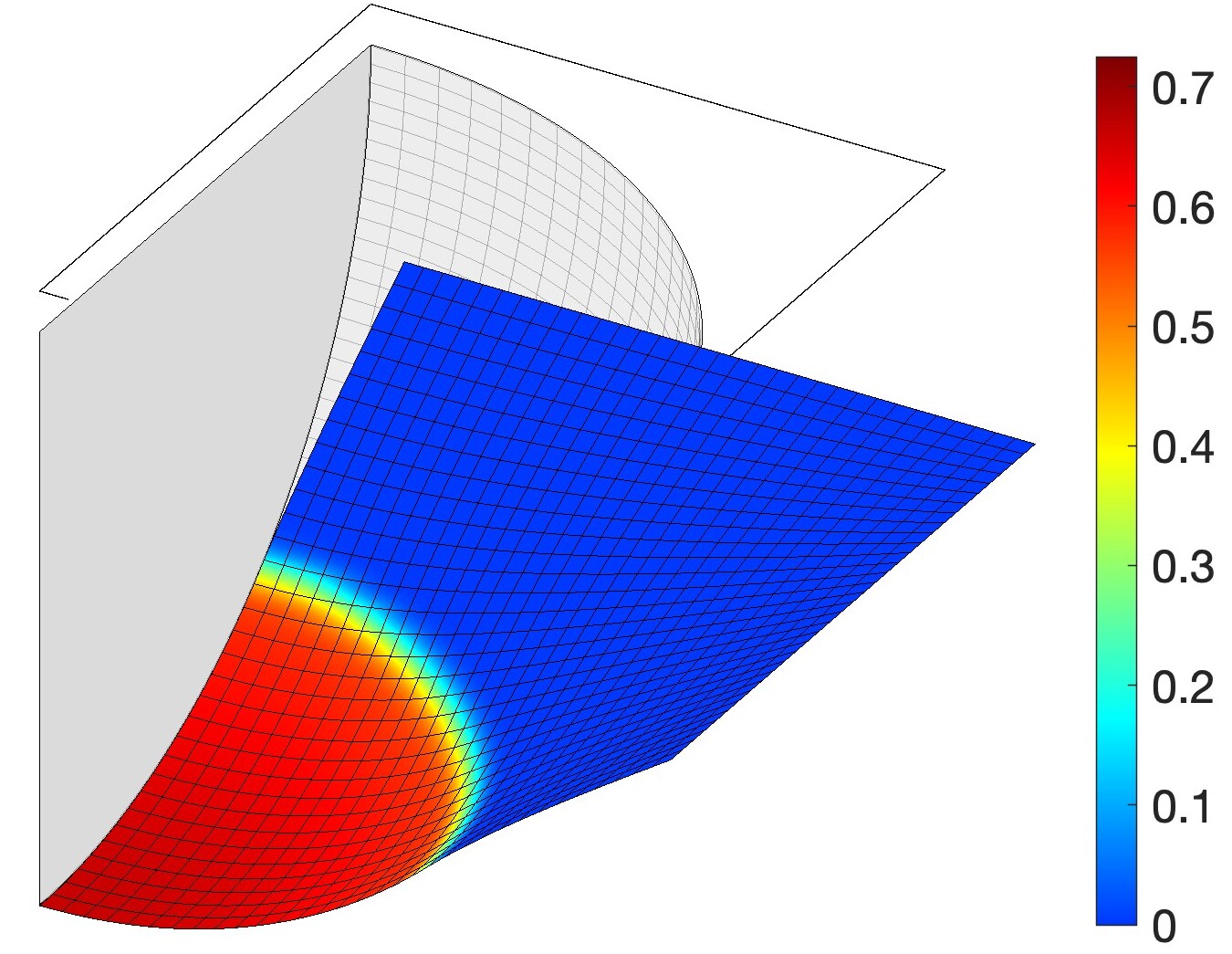}}
\put(-7.9,6.49){\includegraphics[height=34mm]{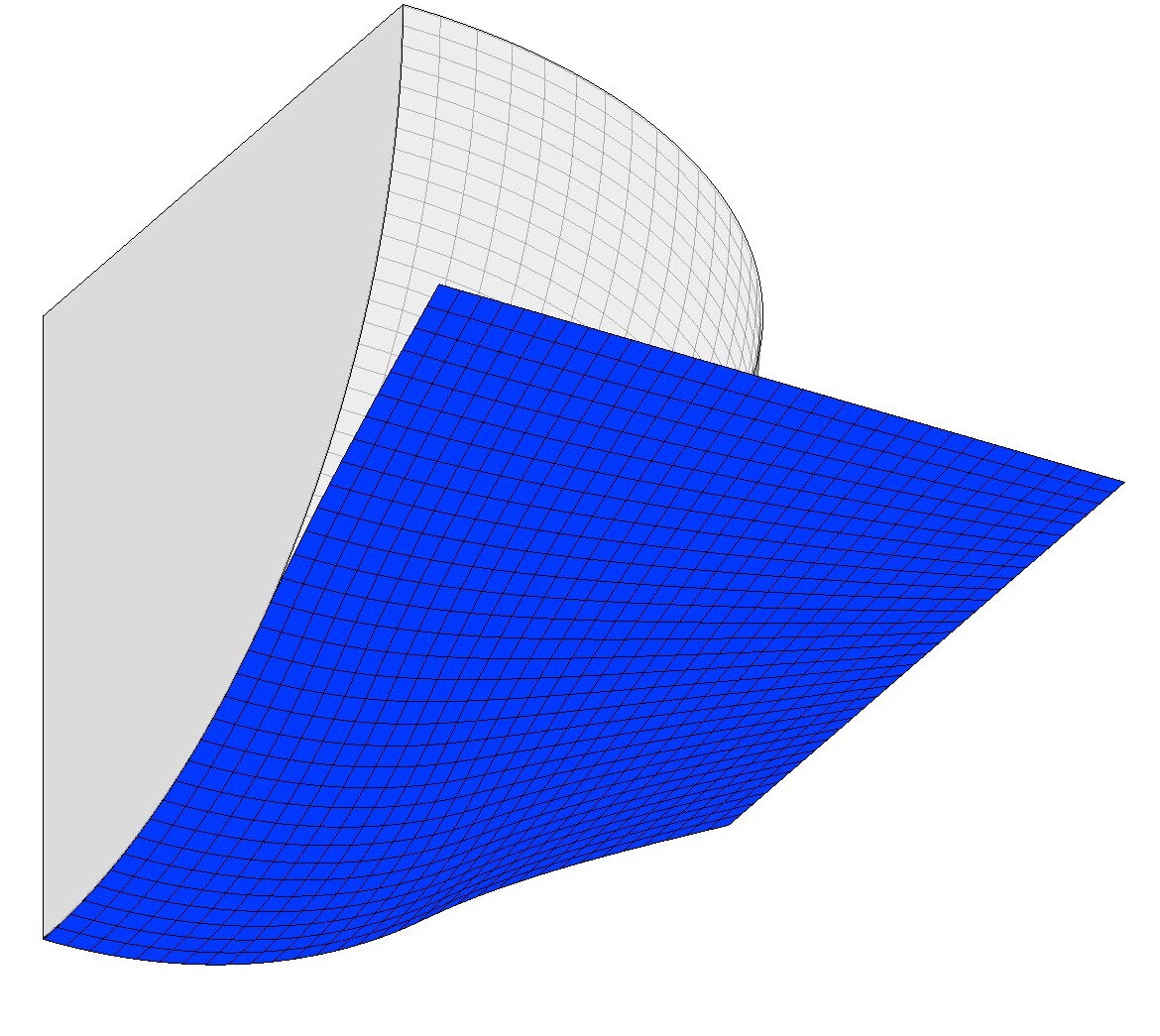}}
\put(-4.1,6.49){\includegraphics[height=34mm]{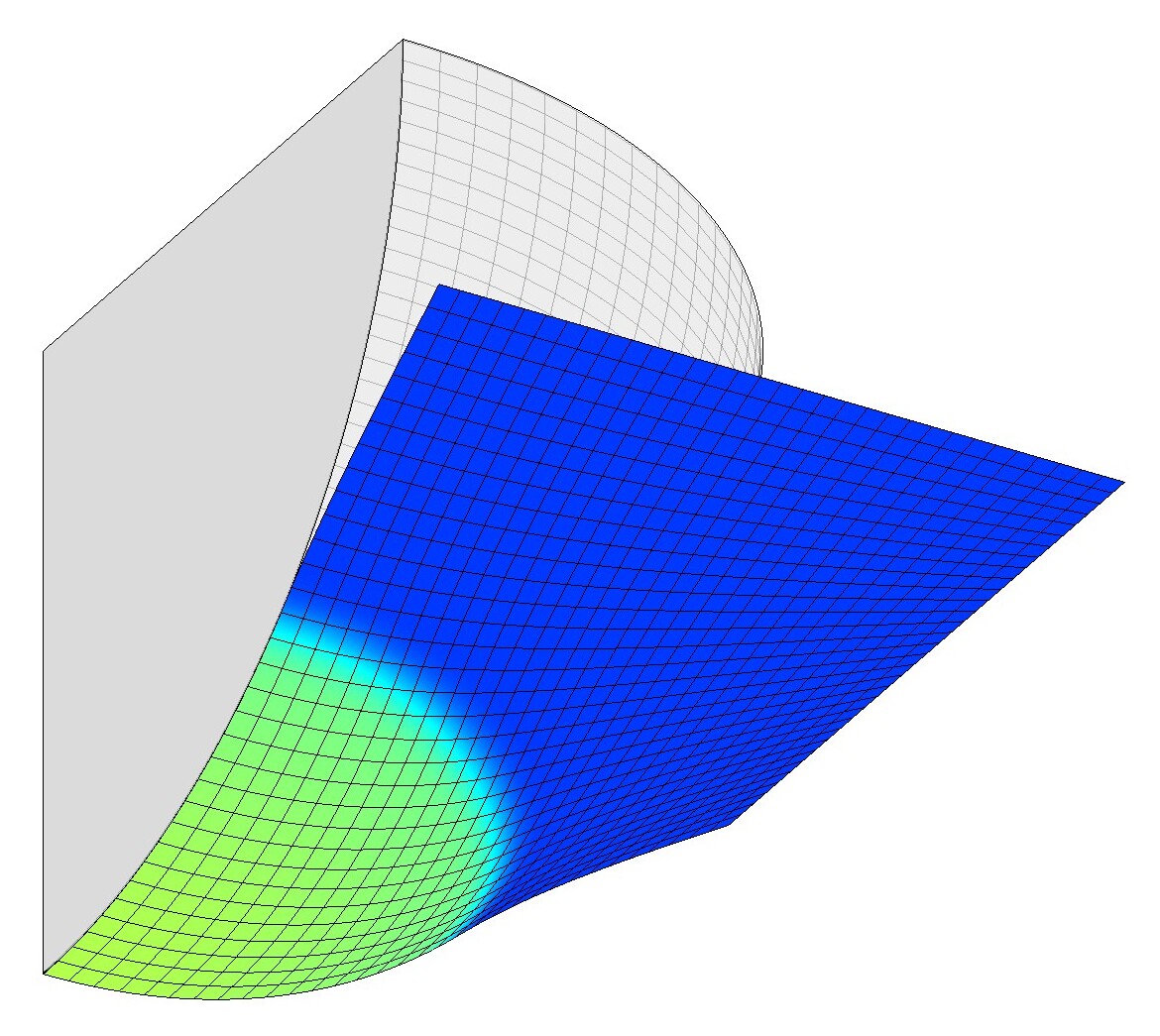}}
\put(-.3,6.49){\includegraphics[height=34mm]{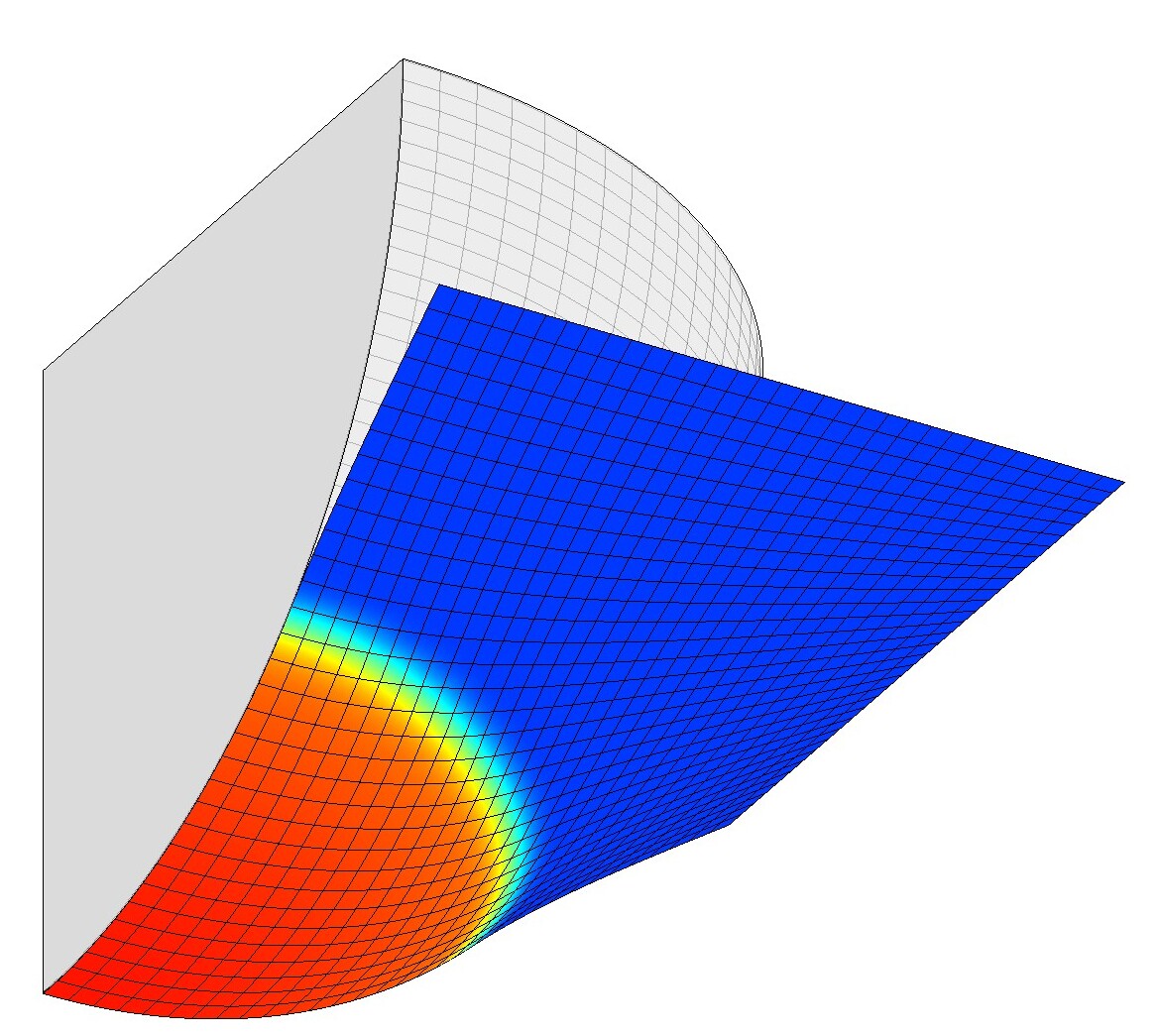}}
\put(3.5,6.49){\includegraphics[height=34mm]{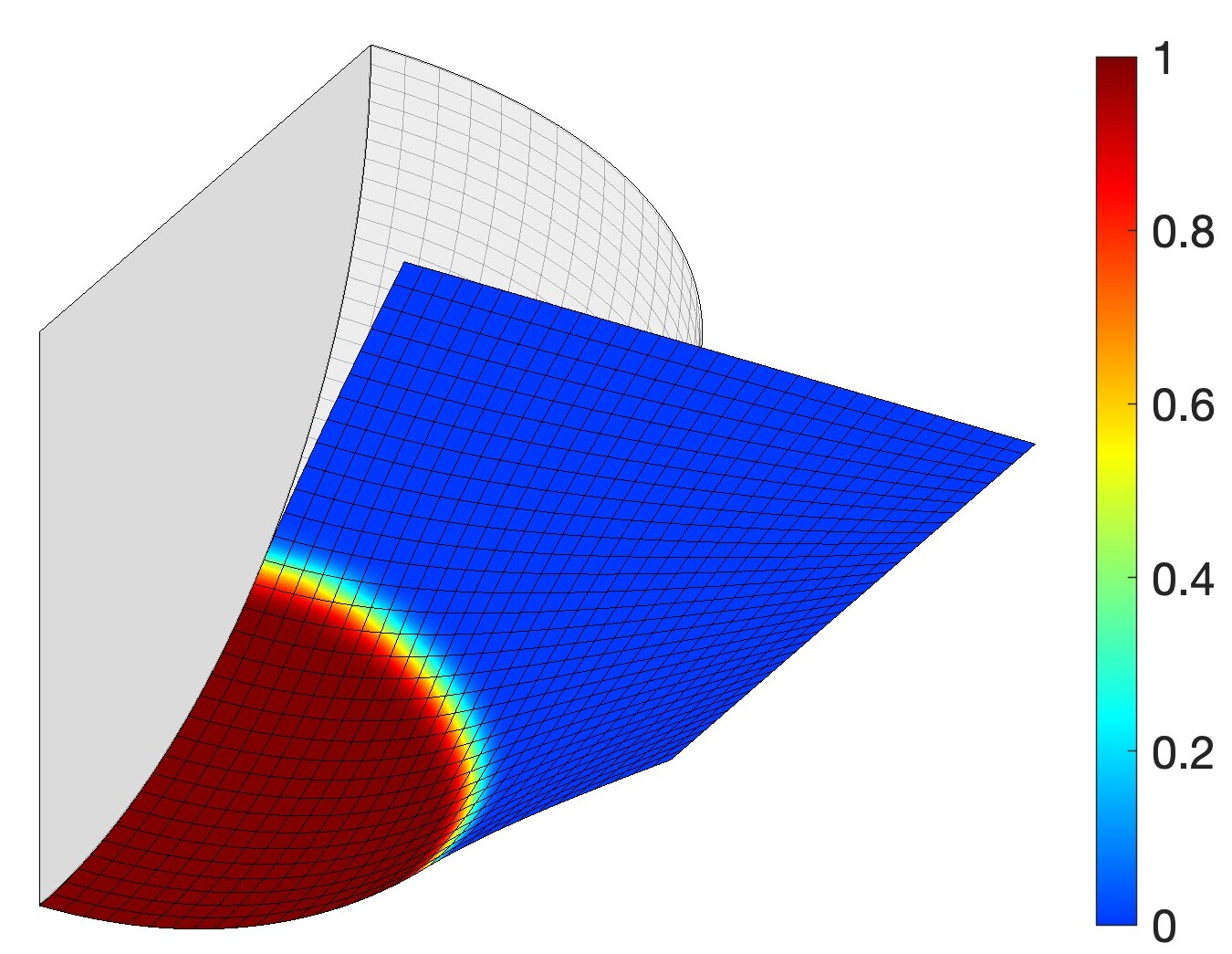}}
\put(-7.9,3.12){\includegraphics[height=34mm]{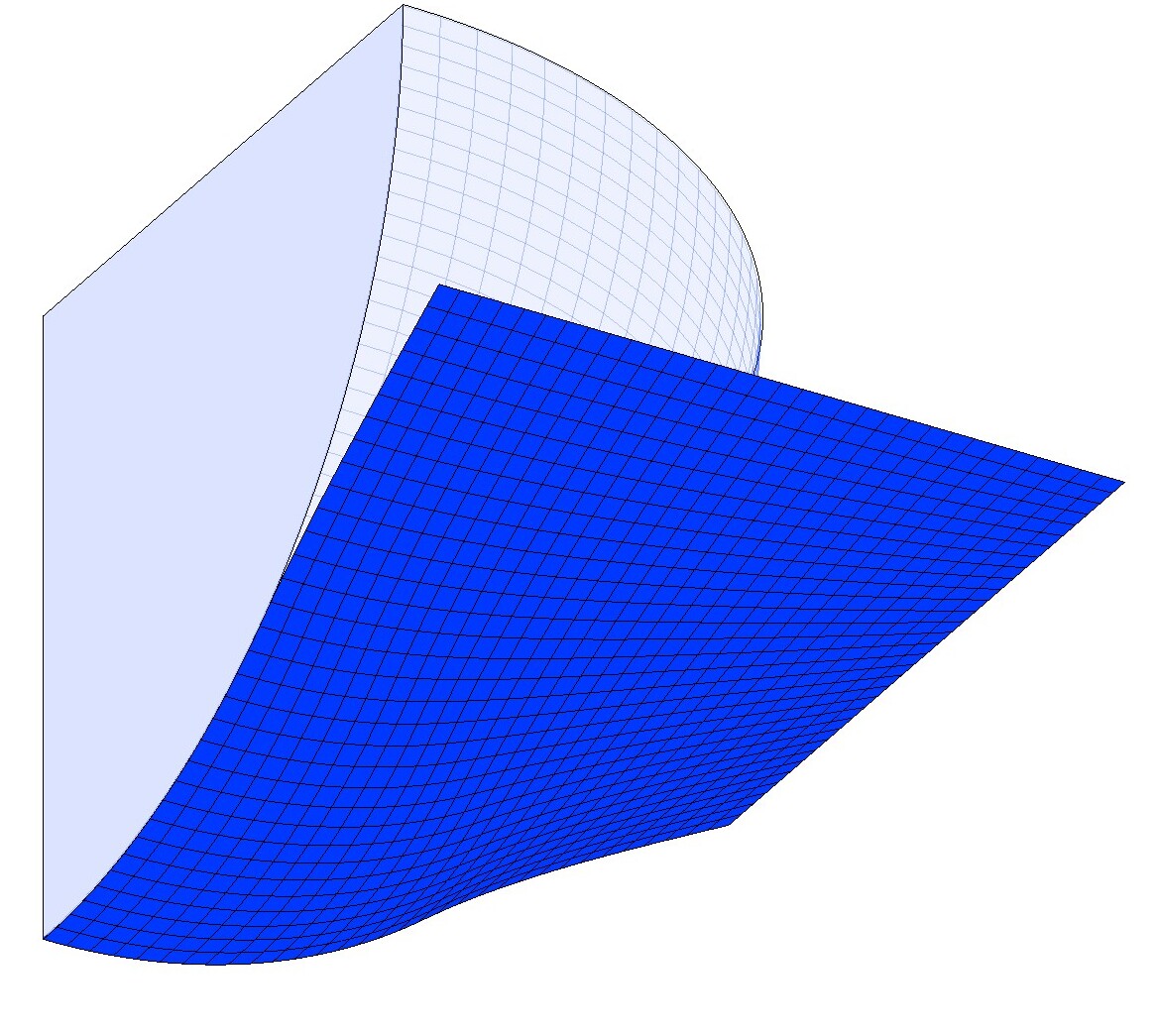}}
\put(-4.1,3.12){\includegraphics[height=34mm]{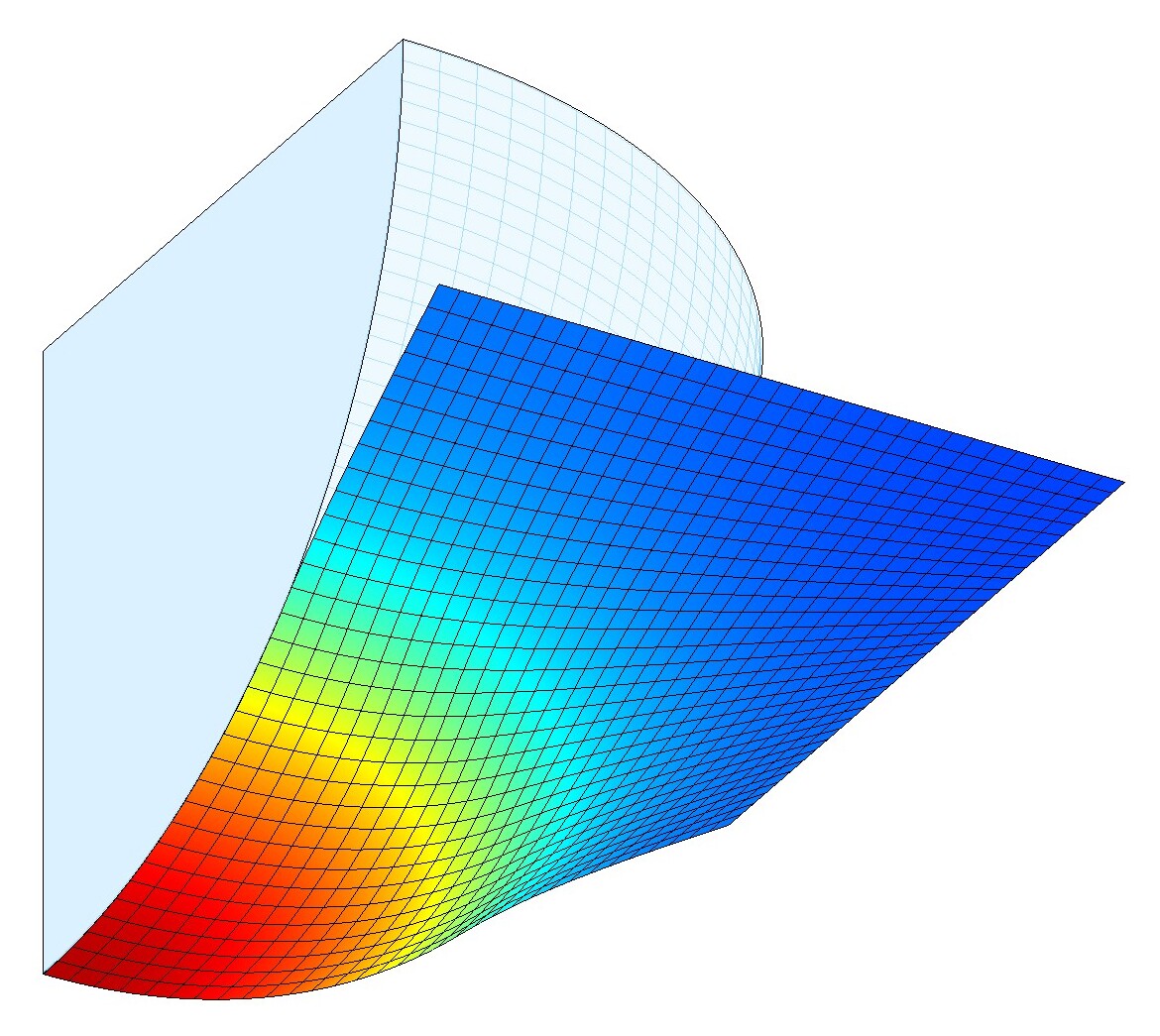}}
\put(-.3,3.12){\includegraphics[height=34mm]{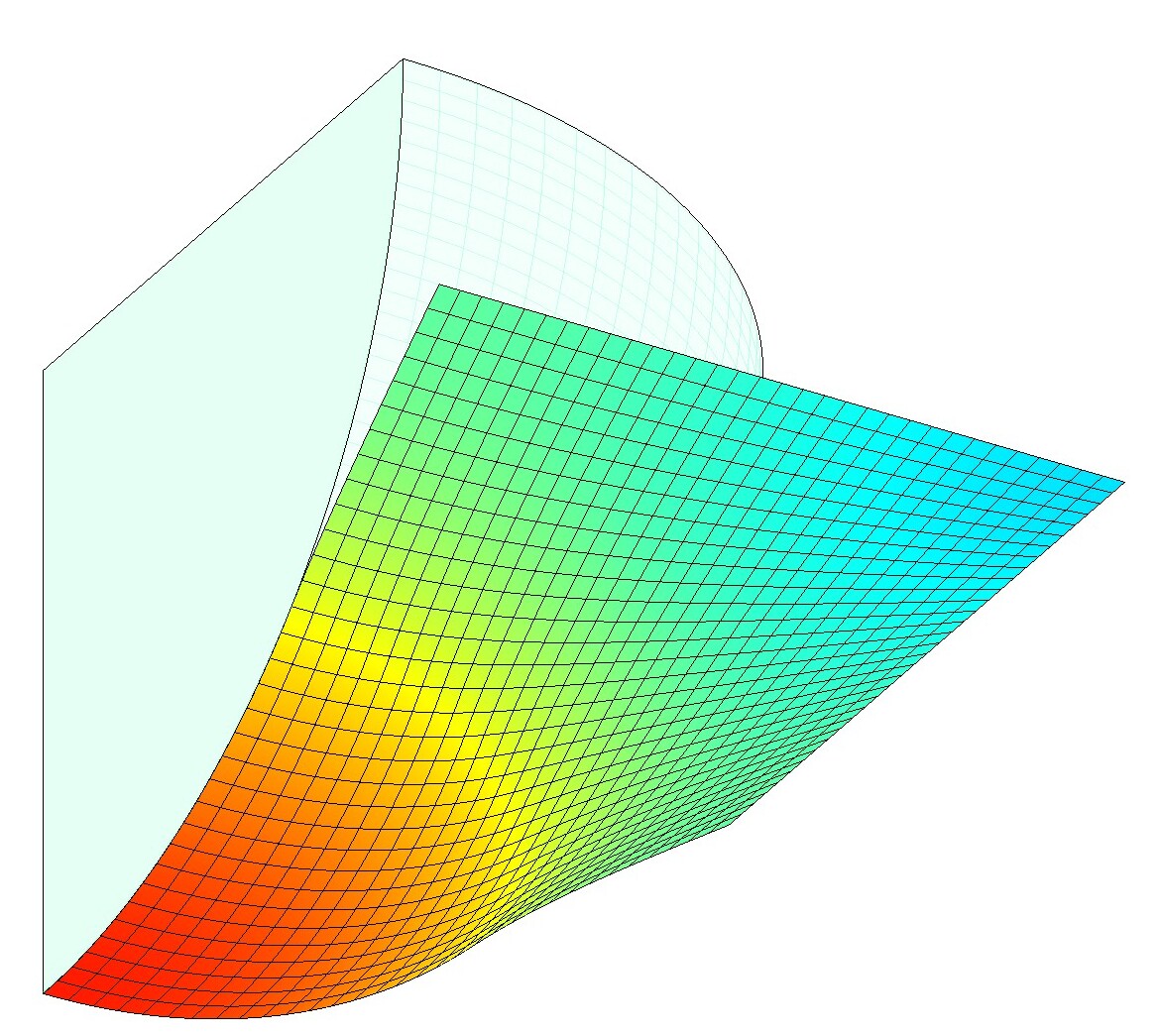}}
\put(3.5,3.12){\includegraphics[height=34mm]{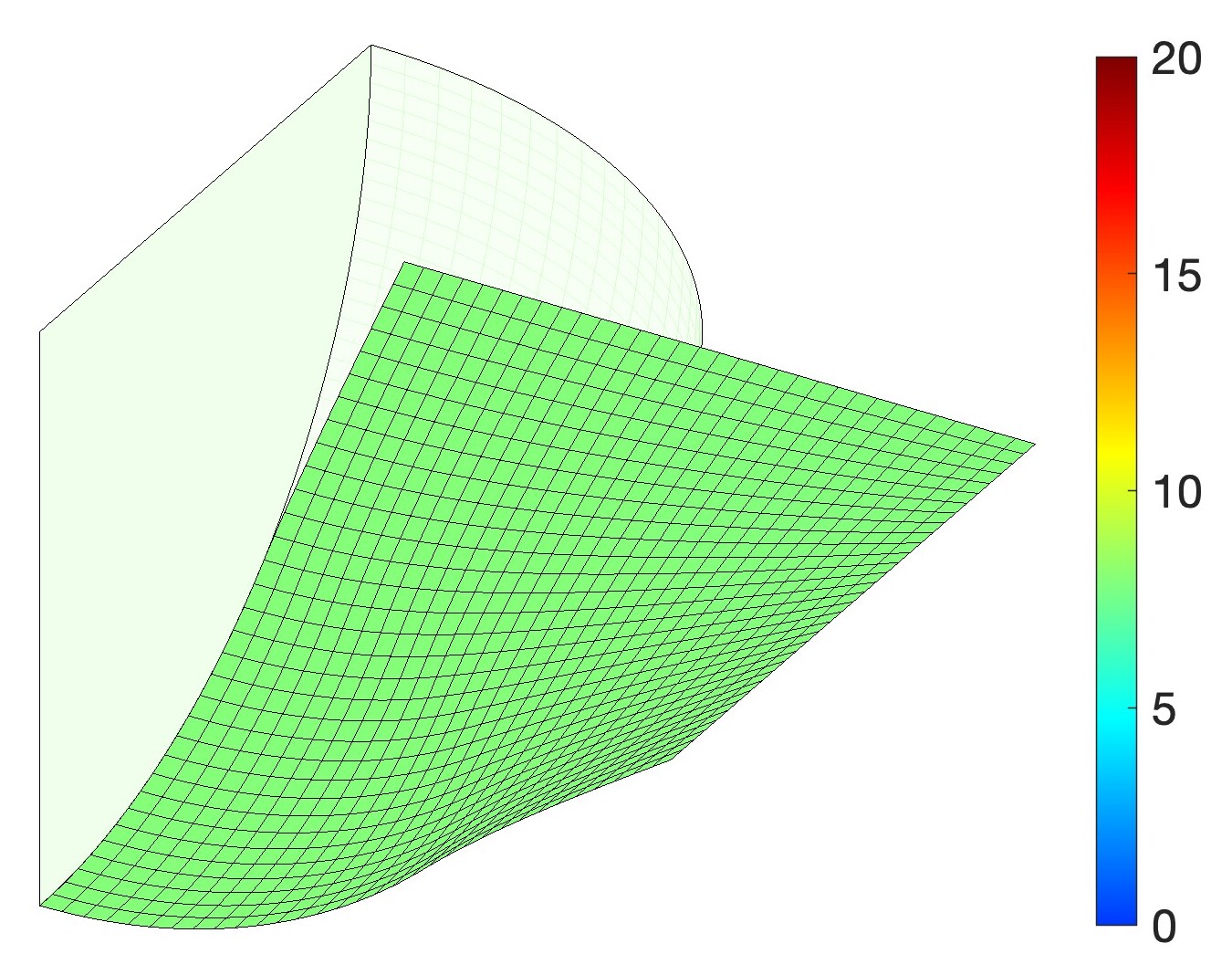}}
\put(-7.9,-.25){\includegraphics[height=34mm]{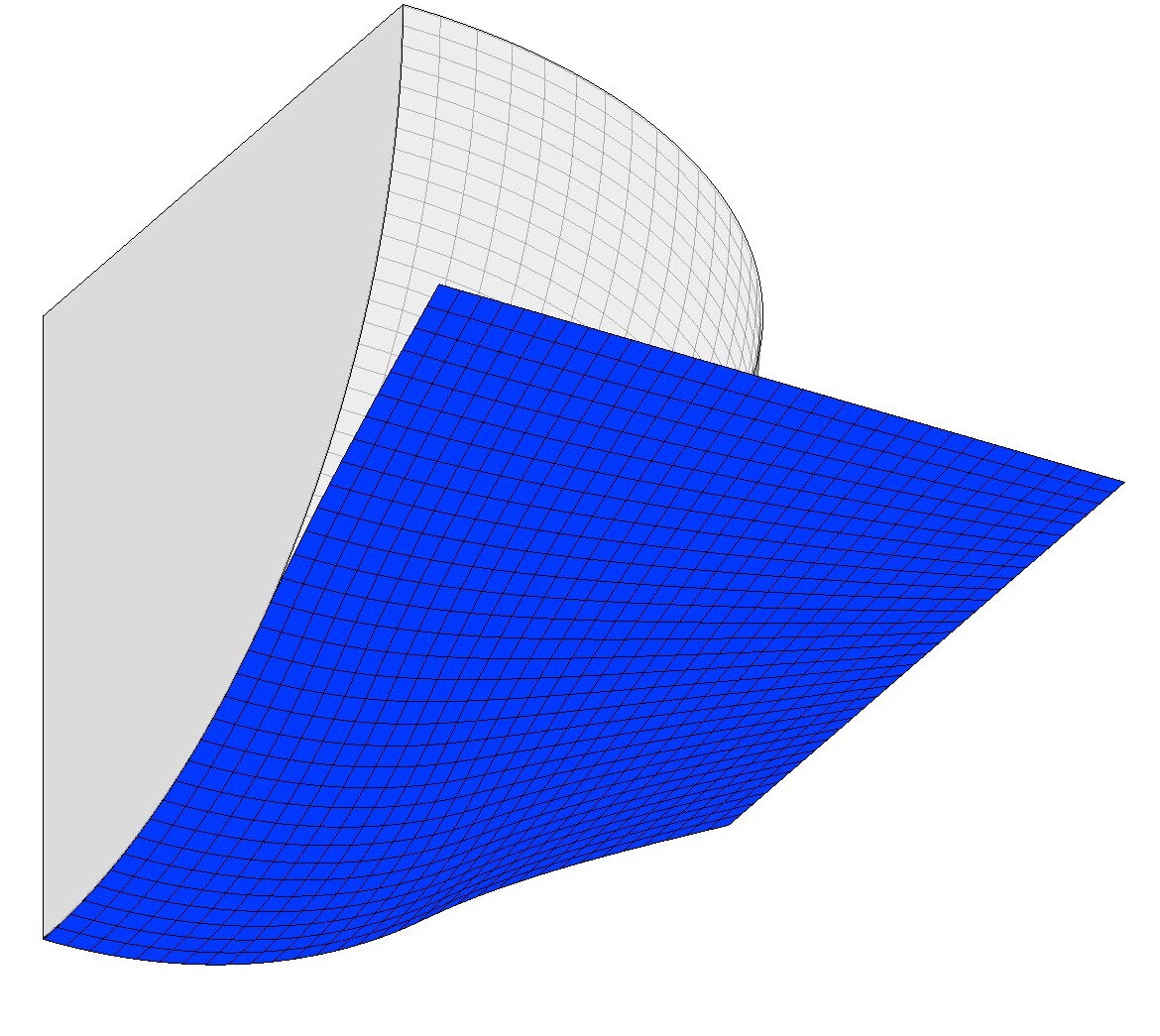}}
\put(-4.1,-.25){\includegraphics[height=34mm]{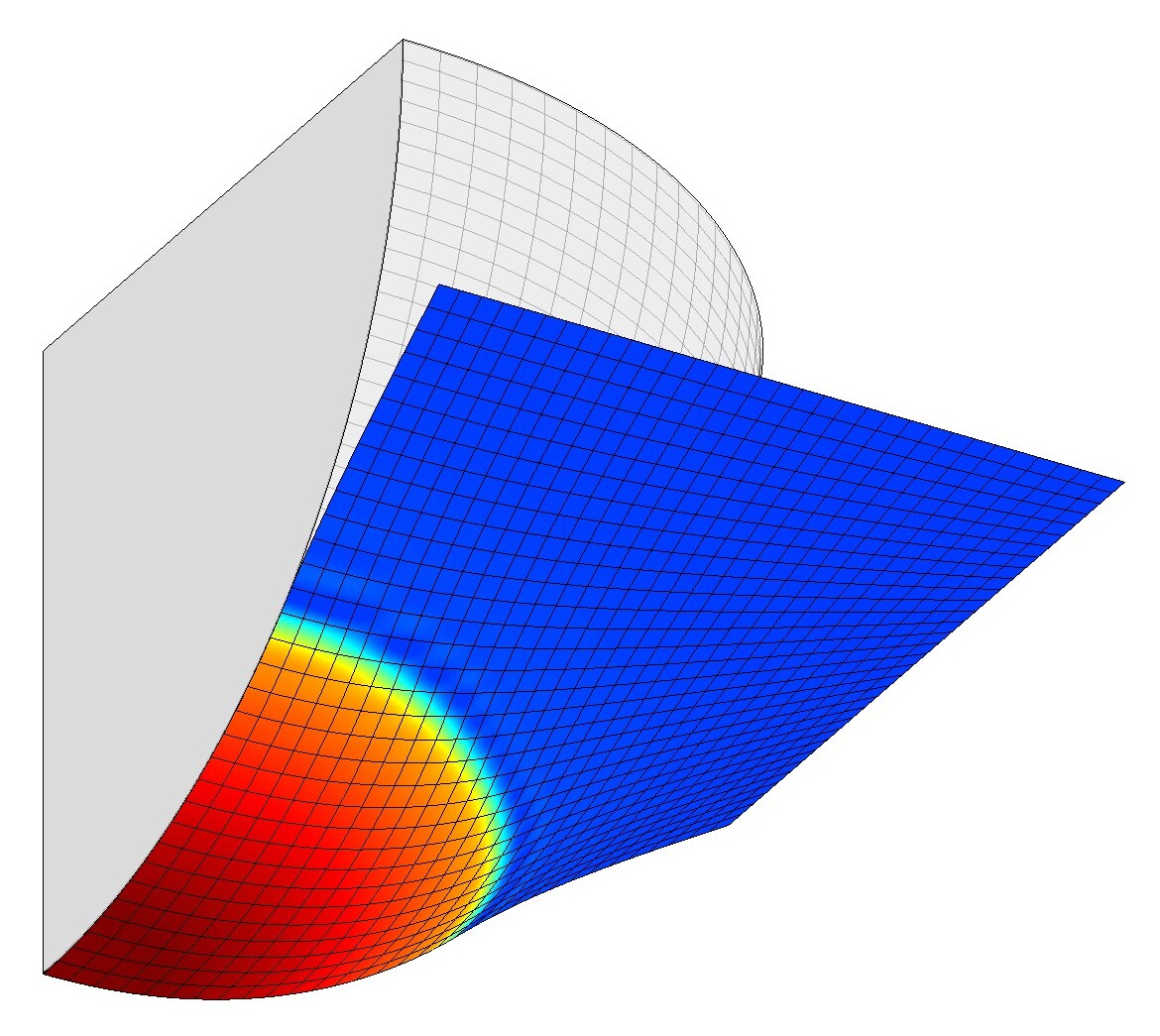}}
\put(-.3,-.25){\includegraphics[height=34mm]{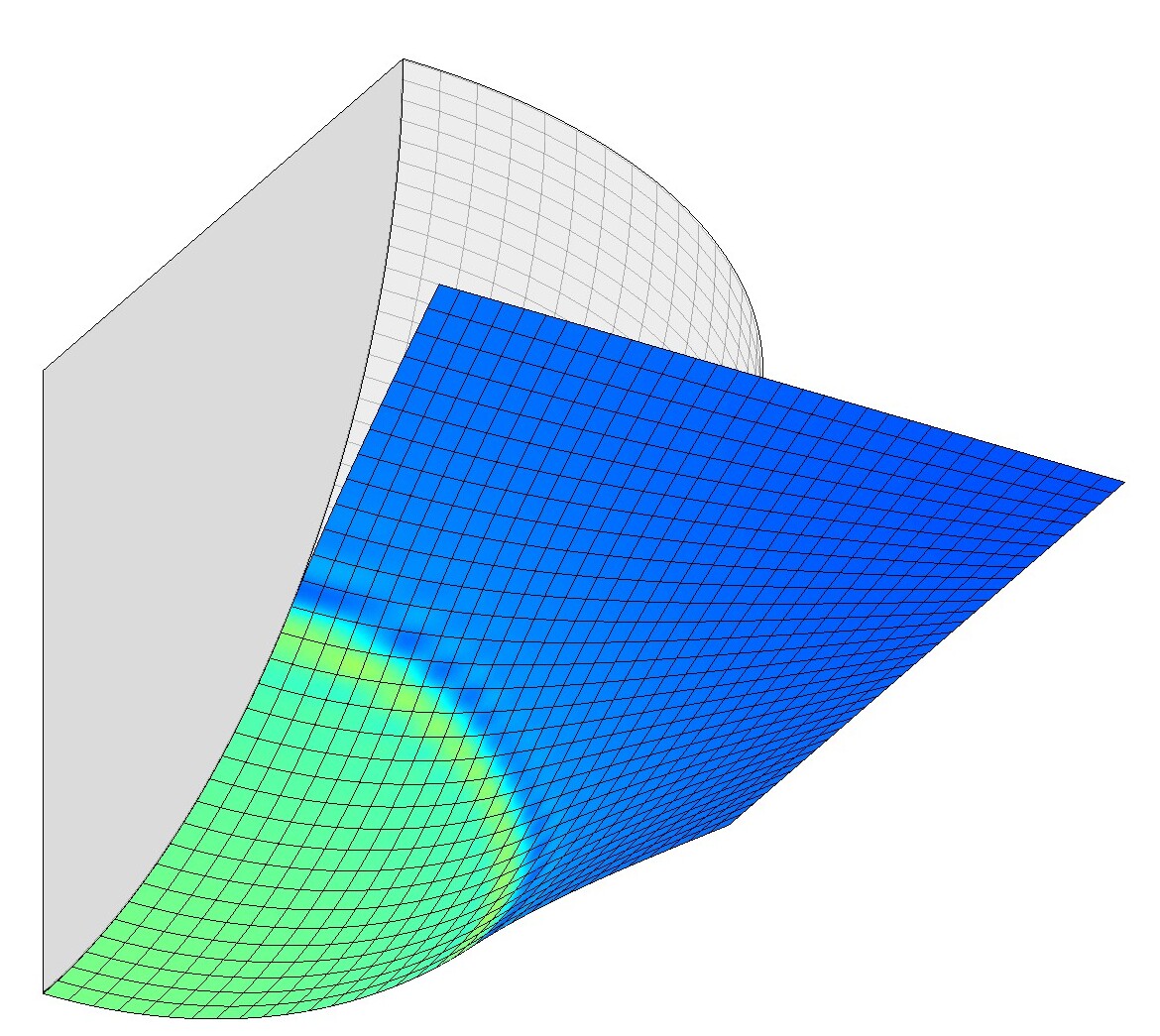}}
\put(3.5,-.25){\includegraphics[height=34mm]{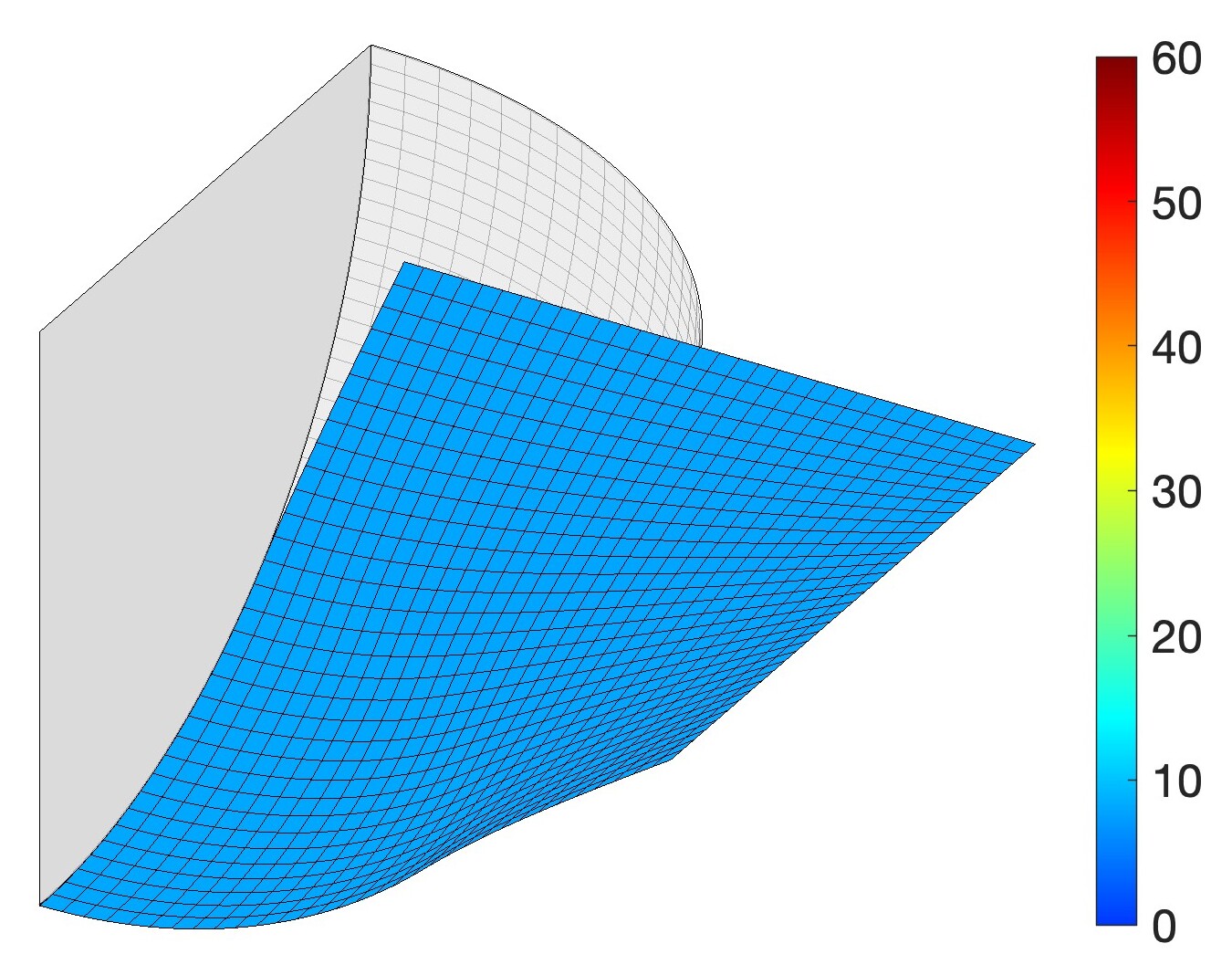}}
\put(7.03,12.7){\scriptsize $p_\mrc$}
\put(7.1,9.35){\scriptsize $\phi$}
\put(6.85,6.0){\scriptsize $\Delta T$}
\put(6.75,2.65){\scriptsize $\Delta T_\mrc$}
\put(-5.8,9.95){\scriptsize $t=0$}
\put(-2.0,9.95){\scriptsize $t=0.5$}
\put(1.8,9.95){\scriptsize $t=1.5$}
\put(5.6,9.95){\scriptsize $t=10$}
\put(-5.8,6.65){\scriptsize $t=0$}
\put(-2.0,6.65){\scriptsize $t=0.5$}
\put(1.8,6.65){\scriptsize $t=1.5$}
\put(5.6,6.65){\scriptsize $t=10$}
\put(-5.8,3.3){\scriptsize $t=0$}
\put(-2.0,3.3){\scriptsize $t=0.5$}
\put(1.8,3.3){\scriptsize $t=1.5$}
\put(5.6,3.3){\scriptsize $t=10$}
\put(-5.8,-.05){\scriptsize $t=0$}
\put(-2.0,-.05){\scriptsize $t=0.5$}
\put(1.8,-.05){\scriptsize $t=1.5$}
\put(5.6,-.05){\scriptsize $t=10$}
\end{picture}
\cprotect\caption{Six-field bonding between membrane and heatable sphere: Evolution of the contact pressure~$p_\mrc$ [$F_0/L_0^2$] (top row), contact bonding state $\phi$ (second row), membrane \& sphere temperature increase $\Delta T$ [K] (third row) and interface temperature increase $\Delta T_\mrc$ [K] (bottom row) for $\unde{C_c} = 10^{-7}F_0/(L_0$K).
The downward motion of the sphere is also shown, thus illustrating the evolution of all six fields.
See also supplementary movie \verb|cobo_sphere_six_field.mp4| at \href{https://doi.org/10.5446/74036}{doi.org/10.5446/74036}.}  
\label{f:CoboT9}
\end{center}
\end{figure}

The final example considers the full six-field problem, where also the interface temperature $T_\mrc$ is an unknown, see Fig.~\ref{f:CoboEx}f.
Fig.~\ref{f:CoboT8} shows the evolution of the bonded area and temperatures for the two interfacial heat capacities $\unde{C_c} = 10^{-8}F_0/(L_0$K) and $\unde{C_c} = 10^{-7}F_0/(L_0$K).
Both are for\footnote{Note that $\unde{C_2}$ and $\unde{C_c}$ are expressed per area, whereas $\unde{C_1}$ is not.} $\unde{C_1} = 10^{-6}F_0L_0/$K and all other parameters as in Table \ref{t:CoboSheet}.
The first $\unde{C_c}$ is so small that the solution is very close to the five-field solution of Fig.~\ref{f:CoboT7}a \& c -- as it should be -- and hence serves as a verification of the FE implementation and of Eq.~\eqref{e:5fTc}.
For the second $\unde{C_c}$, a significant difference appears: 
The larger heat capacity stores a significant amount of heat in the interface, thus reducing and delaying the heating of both membrane and sphere.  
Another consequence is that the interfacial temperature $T_\mrc$ now peaks at a later time instead of being highest initially. 
(Now it is lowest initially.)
The evolution of $T_\mrc$ is also seen in Fig.~\ref{f:CoboT9} together with the evolution of $p_\mrc$, $\phi$ and $T_2$.
The sharp jump in $T_\mrc$ at the contact boundary is due to the vanishing interfacial heat conductivity ($\unde{k_c} = 0$) assumed in the present formulation.
Here it is assumed that the interfacial medium (e.g.~a glue) extends across the entire membrane sheet and is in thermal contact with it.
The interfacial medium therefore also heats up outside the bonding area, since heat generated in the bonding interface can flow into the sheet, then be conducted across the sheet, and finally flow back into the interfacial medium.
Fig.~\ref{f:CoboT9} also shows the downward motion and temperature increase of the sphere, thus illustrating the evolution of all six fields.

\section{Conclusion}\label{s:concl}

This work presented a large deformation contact model for coupled chemical, mechanical and thermal contact interactions,
that is based on six interacting fields: the two deformations and temperatures of the two contacting bodies and an interfacial bonding field and an interfacial temperature.
The model's behavior is described by a quadratic contact potential that depends on the reversible deformation gap $\bg_\mre$, the interfacial temperature $T_\mrc$, the bonding state $\phi$ and debonding state $1-\phi$.
The corresponding material parameters $\unde{E_c} = \unde{E_c}(T_\mrc,\phi)$, $\unde{C_c} = \unde{C_c}(\bg_\mre,\phi)$, $\overrightarrow{\unde{K_c}} = \overrightarrow{\unde{K_c}}(\bg_\mre,T_\mrc)$ and $\overleftarrow{\unde{K_c}} = \overleftarrow{\unde{K_c}}(\bg_\mre,T_\mrc)$ are generally functions of the remaining fields.
Three functional examples are considered here -- including separation-dependent debonding, temperature-dependent bonding, and pressure-dependent bonding -- but the versatility of the formulation allows for many more cases.
The coupled contact model is discretized with nonlinear finite elements using classical and isogeometric shape functions, implicit time integration, and monolithic coupling.
As the bonding variable is described by an ODE, it can be condensed-out at the finite element level, hence eliminating one field in the numerical implementation.
The model applies to a wide range of natural and man-made bonding and debonding problems.
This is illustrated by six examples with increasing degree of coupling.
They exhibit at least cubic convergence rates for quadratic finite elements.

The current finite element examples consider all six interacting fields, but for uniform master body displacements and temperatures.
An extension to varying displacements and temperatures can be included in future work following established thermo-mechanical contact formulations.
Further, the current FE implementation has direct $\bu\leftrightarrow\phi$, $\phi\rightarrow T$ and $T\rightarrow\bu$ coupling, but no direct $T\rightarrow\phi$ and $\bu\rightarrow T$ coupling, cf.~Fig.~\ref{f:CoboEx}f.
The first can be incorporated in future work using the model of Eq.~\eqref{e:ft1}.
The latter is present in the thermoelastic heating term that can also be included in future work.
Also planned for future studies with the proposed model are tangential debonding, frictional heating and binder mobility.
The latter occurs for instance in cellular membranes and requires a diffusion term in the interfacial debonding equation, which then becomes a PDE instead of an ODE.

\bigskip

{\Large{\bf Acknowledgements}}

The author acknowledges initial support of this work by the German Research Foundation through project SA1822/8-1, and thanks Katharina Immel for her help with literature and drawings and Hoang Lam Nguyen for proofreading the manuscript.

\appendix

\section{Linearization of the gap vector}\label{s:lin-ii}

The linearization of the gap vector in \eqref{e:ccasesi} leads to the increment
\eqb{l}
\Delta\bg_\mre = \left\{\begin{array}{llll}
\!\!\Delta\bg_\mre^\mathrm{(i)} \is \Delta\bx_2 - \Delta\bx_1  & $for sticking contact (i)$\,, \\[1mm]
\!\!\Delta\bg_\mre^\mathrm{(ii)} \is \Delta\bx_2 - \Delta\bx_1 - \ba_\alpha^\mrp\,\Delta\xi^\alpha_\mrp~ & $for frictionless contact (ii)$\,,
\end{array}\right.
\label{e:Dge0}\eqe
cf.~Eq.~(17) in \citet{cobo}.
Here, 
\eqb{l}
\ba_\alpha^\mrp := \ds\pa{\bx_1}{\xi^\alpha}\Big|_{\xi^\alpha = \xi^\alpha_\mrp}\,,\quad \alpha = 1,2\,,
\eqe
denotes the two tangent vectors on the master surface at the closest projection point $\bx_\mrp = \bx_1(\xi_\mrp^\alpha)$, see Fig.~\ref{f:cpp}. 
The $\Delta\xi^\alpha_\mrp$-term, only present in Case (ii), comes from changes in the closest projection point coordinate $\xi^\alpha_\mrp$ due to changes in the deformation of master and slave surfaces. 
Using Eq.~\eqref{e:Dge0}, the gap increment of Case (ii) can be expressed through the gap increment of Case (i), i.e.
\eqb{l}
\Delta\bg_\mre^\mathrm{(ii)} = \Delta\bg_\mre^\mathrm{(i)} - \ba_\alpha^\mrp\,\Delta\xi^\alpha_\mrp\,,
\label{e:Dge0a}\eqe
Increment $\Delta\xi^\alpha_\mrp$ is given in Eq.~\eqref{e:Dxi0} of Appendix~\ref{s:Dxi} for the discrete case.
The continuous counterpart following from \eqref{e:Dxi0} for FE approximations
\eqref{e:Dbah} is
\eqb{l}
\Delta\xi^\alpha_\mrp 
	= c^{\alpha\beta}_\mrp\,\ba^\mrp_\beta\cdot\Delta\bg_\mre^\mathrm{(i)} + g_\mrn\,c^{\alpha\beta}_\mrp\,\bn_\mrp\cdot\Delta\ba^\mrp_\beta\,.	
\label{e:Dxic}\eqe
Here $c^{\alpha\beta}_\mrp$ is the inverse of $c_{\alpha\beta}:=a_{\alpha\beta}-g_\mrn b_{\alpha\beta}$ at the projection point, which is found to be
\eqb{l}
c^{\alpha\beta}_\mrp = \ds\frac{(1-2Hg_\mrn)a^{\alpha\beta} + g_\mrn\,b^{\alpha\beta}}{1-2Hg_\mrn + \kappa\,g_\mrn^2}\,,
\eqe
with $H$, $\kappa$, $a^{\alpha\beta}$ and $b^{\alpha\beta}$ being the mean curvature, Gaussian curvature, surface metric and  curvature tensor components, respectively, evaluated at the projection point.
Inserting \eqref{e:Dxic} into \eqref{e:Dge0a} gives\footnote{Note, that at $g_\mrn =0$, one has $c^{\alpha\beta}_\mrp = a^{\alpha\beta}$, and consequently \eqref{e:Dge0b} simplifies to $\Delta\bg_\mre^\mathrm{(ii)} = \big(\bn_\mrp\otimes\bn_\mrp\big)\Delta\bg_\mre^\mathrm{(i)}$. But in a penalty formulation, where generally $g_\mrn\neq0$, the full expression in \eqref{e:Dge0b} should be used for consistency.}
\eqb{l}
\Delta\bg_\mre^\mathrm{(ii)} = \big(\bone - c^{\alpha\beta}_\mrp\,\ba^\mrp_\alpha\otimes\ba^\mrp_\beta \big)\,\Delta\bg_\mre^\mathrm{(i)} 
	- g_\mrn\,c^{\alpha\beta}_\mrp\,\big(\ba_\alpha^\mrp\otimes\bn_\mrp\big)\,\Delta\ba_\beta^\mrp	\,.
\label{e:Dge0b}\eqe
Given $\Delta\bg_\mre$ for both cases, the increments in \eqref{e:DtMSc} are reexamined next.
For sticking (Case (i)), the contact traction takes the form
\eqb{l}
\undetc^\mathrm{(i)} = \undeEc^\unde{eff}\bg_\mre^\mathrm{(i)},
\eqe
and has the derivative
\eqb{l}
\ds\pa{\undetc^\mathrm{(i)}}{\bg_\mre^\mathrm{(i)}} = \undeEc^\unde{eff}. 
\label{e:tigi}\eqe
An example is $\undeEc^\unde{eff} = \undeEc + \overleftarrow{\unde{K_0}}\bone\phi^2/g_0^2$ according to the debonding model of Sec.~\ref{s:con1}, see Eq.~(\ref{e:ptM1}.1).  
For frictionless contact (Case (ii)) the traction takes the form
\eqb{l}
\undetc^\mathrm{(ii)} = \unde{E_n^{eff}}\bg_\mre^\mathrm{(ii)},
\label{e:tcii}\eqe
where an example for $\unde{E_n^{eff}}$ is given in Eq.~\eqref{e:Eneff} for the bonding model of Sec.~\ref{s:con3}.
The increment of this is
\eqb{l}
\Delta\undetc^\mathrm{(ii)} = \unde{E_n^{eff}}\Delta\bg_\mre^\mathrm{(ii)} 
	+ \ds\pa{\undetc^\mathrm{(ii)}}{\phi}\,\Delta\phi + \ds\pa{\undetc^\mathrm{(ii)}}{T_\mrc}\,\Delta T_\mrc\,.
\eqe
Inserting \eqref{e:Dge0b} then gives
\eqb{l}
\Delta\undetc^\mathrm{(ii)} = \ds\pa{\undetc^\mathrm{(ii)}}{\bg_\mre^\mathrm{(i)}}\,\Delta\bg_\mre^\mathrm{(i)} 
	+ \pa{\undetc^\mathrm{(ii)}}{\ba^\mrp_\beta} \,\Delta\ba_\beta^\mrp 
	+ \ds\pa{\undetc^\mathrm{(ii)}}{\phi}\,\Delta\phi + \ds\pa{\undetc^\mathrm{(ii)}}{T_\mrc}\,\Delta T_\mrc\,,
\label{e:Dtii}\eqe
for the definitions
\eqb{l}
\ds\pa{\undetc^\mathrm{(ii)}}{\bg_\mre^\mathrm{(i)}} := \unde{E_n^{eff}}\big(\bone - c^{\alpha\beta}_\mrp\,\ba^\mrp_\alpha\otimes\ba^\mrp_\beta \big)\,,\qquad 
\pa{\undetc^\mathrm{(ii)}}{\ba^\mrp_\beta} := - \unde{E_n^{eff}} g_\mrn\,c^{\alpha\beta}_\mrp\,\ba_\alpha^\mrp\otimes\bn_\mrp\,.
\label{e:tiigi}\eqe
Both cases thus combine to
\eqb{l}
\Delta\undetc = \ds\pa{\undetc}{\bg_\mre^\mathrm{(i)}}\,\Delta\bg_\mre^\mathrm{(i)} + \pa{\undetc}{\ba^\mrp_\beta} \,\Delta\ba_\beta^\mrp
+ \ds\pa{\undetc}{\phi}\,\Delta\phi + \ds\pa{\undetc}{T_\mrc}\,\Delta T_\mrc\,,
\label{e:Dtc}\eqe
where $\partial\undetc/\partial\bg_\mre^\mathrm{(i)}$ is given by either \eqref{e:tigi} or (\ref{e:tiigi}.1), and $\partial\undetc/\partial\ba^\mrp_\beta$ is either zero or given by (\ref{e:tiigi}.2) for the two respective cases.

Unlike $\Delta\undetc$, increments $\Delta\unde{M_c}$ and $\Delta\unde{S_c}$ do not pick up terms in Case (ii) for the example considered here, which is shown next.
The Case (ii) example is the one from Sec.~\ref{s:con3},
and according to that, both derivatives $\partial\unde{M_c}/\partial\bg_\mre$ and $\partial\unde{S_c}/\partial\bg_\mre$ are proportional to surface normal $\bn_\mrp$, cf.~\eqref{e:ptMS} and \eqref{e:dfm3}. 
Thus, in their scalar product with $\Delta\bg_\mre^\mathrm{(ii)}$ from \eqref{e:Dge0b}, only $\bone\,\Delta\bg_\mre^\mathrm{(i)}$ survives.
Hence, for both Case (i) and (ii),
\eqb{rlrlrlr}
\Delta\unde{M_c} \is \ds\pa{\unde{M_c}}{\bg_\mre}\,\Delta\bg_\mre^\mathrm{(i)} 
	\plus \ds\pa{\unde{M_c}}{\phi}\,\Delta\phi \plus \ds\pa{\unde{M_c}}{T_\mrc}\,\Delta T_\mrc\,, \\[4mm]
\Delta\unde{S_c} \is \ds\pa{\unde{S_c}}{\bg_\mre}\,\Delta\bg_\mre^\mathrm{(i)} 
	\plus \ds\pa{\unde{S_c}}{\phi}\,\Delta\phi \plus \ds\pa{\unde{S_c}}{T_\mrc}\,\Delta T_\mrc\,.	
\label{e:DMSii}\eqe
\vspace{-1mm}

\section{Analytical solution of the bonding ODE}\label{s:anaODE}

\begin{figure}[h]
\begin{center} \unitlength1cm
\begin{picture}(0,5.6)
\put(-4,-.2){\includegraphics[height=58mm]{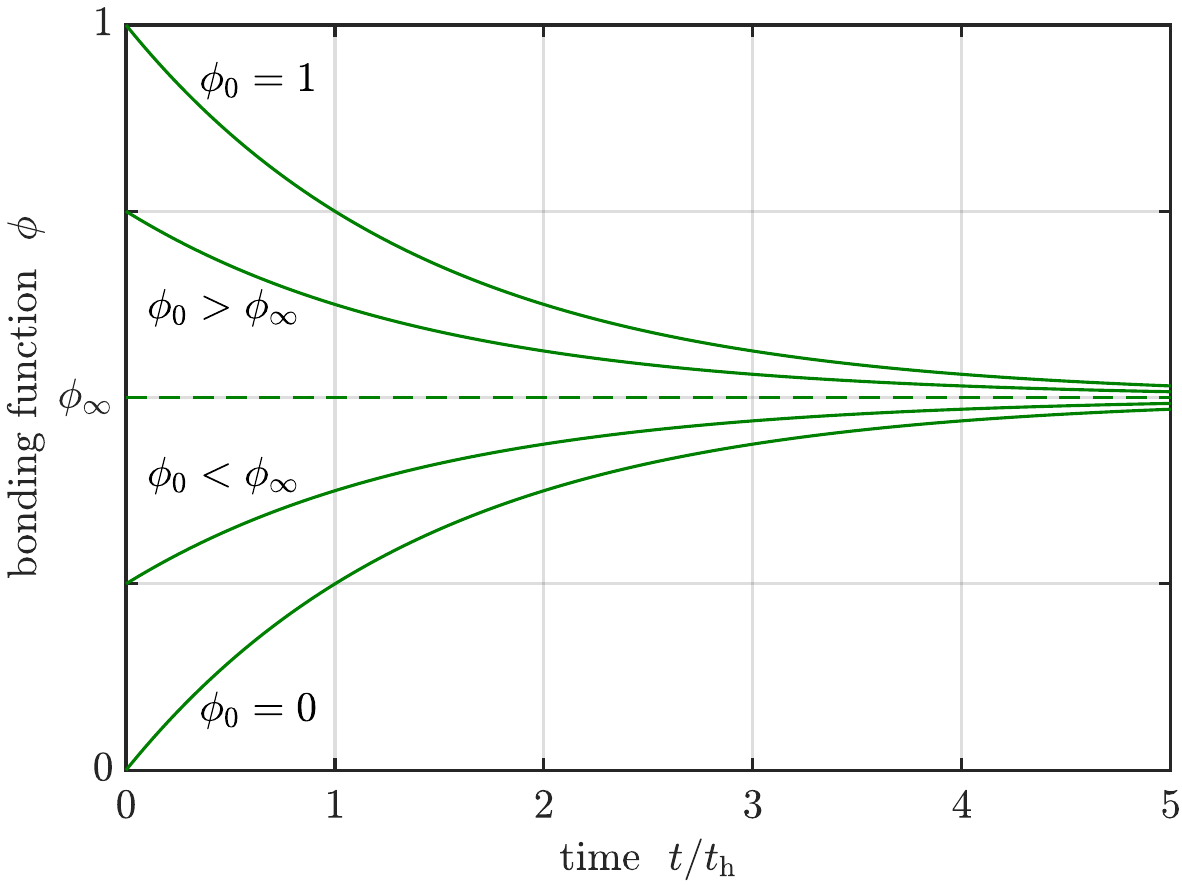}}
\end{picture}
\caption{Analytical solution of the bonding ODE for various initial conditions $\phi_0$.}
\label{f:phi}
\end{center}
\end{figure}

This section provides the analytical solution of the bonding ODE for time-independent bonding parameters.
From \eqref{e:Mc} and \eqref{e:Kc} follows
\eqb{l}
\unde{M_c} = \overrightarrow{\unde{K_c}}\,(\phi-1) + \overleftarrow{\unde{K_c}}\,\phi
= \unde{K_c}\,(\phi-\phi_\infty)\,,
\label{e:McB}\eqe
where $\unde{K_c} := \overrightarrow{\unde{K_c}} + \overleftarrow{\unde{K_c}}$ and $\phi_\infty := \overrightarrow{\unde{K_c}}/\unde{K_c}$ have been introduced.
Bonding ODE \eqref{e:bondODE} can then be written as
\eqb{l}
\tau\,\dot\phi + \phi - \phi_\infty = 0\,,
\label{e:bondODEa}\eqe
where
\eqb{l}
\tau := \ds\frac{\unde{n_c}}{c_\mrr\,\unde{K_c}}
\eqe
is its time scale.
If $c_\mrr$, $\overrightarrow{\unde{K_c}}$ and $\overleftarrow{\unde{K_c}}$ are constant in time, the bonding ODE has the analytical solution
\eqb{l}
\phi(t) = (\phi_0-\phi_\infty)\,e^{-t/\tau} + \phi_\infty\,,
\label{e:anaphi}\eqe
where $\phi_0 := \phi(0)$ is the initial bonding state.
For $t\rightarrow\infty$ the bonding state approaches $\phi_\infty$, which in turn contains the limit cases
\eqb{l}
\phi_\infty = \left\{\begin{array}{lll} 
0 & $for$~~\overrightarrow{\unde{K_c}}\Big/\overleftarrow{\unde{K_c}} \rightarrow 0 & $(no bonding)$, \\[2mm]
1 & $for$~~\overleftarrow{\unde{K_c}}\Big/\overrightarrow{\unde{K_c}} \rightarrow 0 & $(no debonding)$.
\end{array}\right.
\eqe
For $0\leq\phi_0<\phi_\infty$ the bonding state is increasing, while it decreases for $1\geq\phi_0>\phi_\infty$.
The changes halve for the half life $t_\mrh := \tau\ln2$, as seen in Fig.~\ref{f:phi}. 

\section{Mechano-sensitive osseointegration: Alternative formulation}\label{s:con4a}

The osseointegration model of Sec.~\ref{s:con4} can also be formulated in an alternative way.
Instead of considering $\overrightarrow{\unde{K_c}}=$ const.~and $c_\mrr=c_\mrr(\bg_\mre)$ according to \eqref{e:crosseo},
one can also set $c_\mrr=$ const.~and use the pressure- and separation-dependent bonding function
\eqb{l}
\overrightarrow{\unde{f_m}}(\bg_\mre)  = \ds\frac{1}{2} \left\{\begin{array}{ll}
\cos\bigg(\ds\frac{g_\mrn-g_\mathrm{opt}}{g_\mathrm{lim}-g_\mathrm{opt}}\pi\bigg) + 1 ~ & 
	$for$~~g_\mathrm{opt} \leq g_\mrn \leq g_\mathrm{lim}\,, \\[4mm]	
2 & $for$~~g_\mrn \leq g_\mathrm{opt}~~\&~~p_\mrc \leq p_\mathrm{opt}\,, \\[1mm]
\cos\bigg(\ds\frac{p_\mrc-p_\mathrm{opt}}{p_\mathrm{lim}-p_\mathrm{opt}}\pi\bigg) + 1 ~ & 
	$for$~~p_\mathrm{opt} \leq p_\mrc \leq p_\mathrm{lim}\,, \\[3mm]	
0 & $else$\,,
\end{array}\right.
\label{e:fosseo}\eqe
together with \eqref{e:Kc} and $\overrightarrow{\unde{f_T}}\equiv1$.
This alternative leads to exactly the same bonding ODE in \eqref{e:bondODE}.
But the influence on mechanical contact changes. 
While \eqref{e:crosseo} has no influence on it, Eq.~\eqref{e:fosseo} causes contact forces according to \eqref{e:tc} and hence affects mechanical contact.
These contact forces can become very large, as is shown in the following.

Introducing
\eqb{llllll}
s_g \dis \sin\bigg(\ds\frac{g_\mrn-g_\mathrm{opt}}{g_\mathrm{lim}-g_\mathrm{opt}}\pi\bigg), &
s_p \dis \sin\bigg(\ds\frac{p_\mrc-p_\mathrm{opt}}{p_\mathrm{lim}-p_\mathrm{opt}}\pi\bigg), \\[4mm]
c_g \dis \cos\bigg(\ds\frac{g_\mrn-g_\mathrm{opt}}{g_\mathrm{lim}-g_\mathrm{opt}}\pi\bigg), &
c_p \dis \cos\bigg(\ds\frac{p_\mrc-p_\mathrm{opt}}{p_\mathrm{lim}-p_\mathrm{opt}}\pi\bigg),
\eqe
leads to 
\eqb{l}
\ds\pa{\overrightarrow{\unde{f_m}}}{\bg_\mre}  = \frac{1}{2} \left\{\begin{array}{ll}
\ds\frac{-s_g\,\pi}{g_\mathrm{lim}-g_\mathrm{opt}} \,\bn_\mrp~  & $for$~~g_\mathrm{opt} \leq g_\mrn \leq g_\mathrm{lim}\,, \\[4mm]
\ds\frac{s_p\,\pi\,E_\mrn}{p_\mathrm{lim}-p_\mathrm{opt}}\,\bn_\mrp ~ & $for$~~p_\mathrm{opt} \leq p_\mrc \leq p_\mathrm{lim}\,, \\[4mm]
\mathbf{0} & $else$\,,
\end{array}\right.
\eqe
and
\eqb{l}
\ds\paqq{\overrightarrow{\unde{f_m}}}{\bg_\mre}{\bg_\mre} = -\frac{1}{2} \left\{\begin{array}{ll}
\ds\frac{c_g\,\pi^2}{(g_\mathrm{lim}-g_\mathrm{opt})^2}\,\bn_\mrp\otimes \bn_\mrp 
	+ \frac{s_g\,\pi}{g_\mathrm{lim}-g_\mathrm{opt}}\pa{\bn_\mrp}{\bg_\mre}~  & $for$~~g_\mathrm{opt} \leq g_\mrn \leq g_\mathrm{lim}\,, \\[4mm]
\ds\frac{c_p\,\pi^2\,E_\mrn^2}{(p_\mathrm{lim}-p_\mathrm{opt})^2}\,\bn_\mrp\otimes\bn_\mrp
	- \frac{s_p\,\pi\,E_\mrn}{p_\mathrm{lim}-p_\mathrm{opt}}\pa{\bn_\mrp}{\bg_\mre}~ & 
	$for$~~p_\mathrm{opt} \leq p_\mrc \leq p_\mathrm{lim}\,, \\[4mm]
\mathbf{0} & $else$\,,
\end{array}\right.
\label{e:ddfosseo}\eqe
since $g_\mrn = \bg_\mre\cdot\bn_\mrp$ and $p_\mrc = -E_\mrn\,g_\mrn$.
The first part of expression \eqref{e:ddfosseo} describes the chemical contact stiffness in the normal direction, and its magnitude is bounded by
\eqb{l}
\unde{f''_n} := 
\max\bigg(\ds\frac{\pi^2}{2\,(g_\mathrm{lim}-g_\mathrm{opt})^2}\,,\,\frac{\pi^2\,E_\mrn^2}{2\,(p_\mathrm{lim}-p_\mathrm{opt})^2} \bigg).
\eqe
From \eqref{e:ptMS} thus follows that the normal chemical contact forces are only negligible if 
\eqb{l}
\ds\frac{\overrightarrow{\unde{K_0}}\,\unde{f''_n}}{2\,\unde{E_n}} \ll 1\,.
\label{e:cond_n}\eqe
The second part of expression \eqref{e:ddfosseo} describes the chemical contact stiffness in the tangential direction.
The tensor $\partial\bn_\mrp/\partial\bg_\mre = (\bone-\bn_\mrp\otimes\bn_\mrp)/g_\mrn$ has the magnitude $1/g_\mrn$, which is bounded by $1/g_\mathrm{lim}$ and $E_\mrn/p_\mathrm{lim}$ for the two non-zero cases in Eq.~\eqref{e:ddfosseo}.
The magnitude of the second part of expression \eqref{e:ddfosseo} is therefore bounded by
\eqb{l}
\unde{f''_t} := 
\max\bigg(\ds\frac{\pi}{2g_\mathrm{lim}\,(g_\mathrm{lim}-g_\mathrm{opt})}\,,\,\frac{\pi\,E_\mrn^2}{2p_\mathrm{lim}\,(p_\mathrm{lim}-p_\mathrm{opt})} \bigg).
\eqe
From \eqref{e:ptMS} thus follows that the tangential chemical contact forces are only negligible if 
\eqb{l}
\ds\frac{\overrightarrow{\unde{K_0}}\,\unde{f''_t}}{2\,\unde{E_t}} \ll 1\,.
\label{e:cond_t}\eqe
For the parameters in Sec.~\ref{s:ex2} (see Table~\ref{t:RCSI-para}) we have $E_\mrn \approx \unde{E_n} = \unde{E_t} = \unde{E_c} = 200 E_\mrb/R$ such that
\eqb{l}
\ds\frac{\overrightarrow{\unde{K_0}}\,\unde{f''_n}}{2\,\unde{E_n}} 
= \frac{\overrightarrow{\unde{K_0}}\,\pi^2\,\unde{E_n}}{4\,p^2_\mathrm{lim}} 
= (15\pi)^2
\label{e:cond_na}\eqe
and
\eqb{l}
\ds\frac{\overrightarrow{\unde{K_0}}\,\unde{f''_t}}{2\,\unde{E_t}} 
= \frac{\overrightarrow{\unde{K_0}}\,\pi\,\unde{E_n}}{4\,p^2_\mathrm{lim}} 
= 225\,\pi\,.
\eqe
Hence, conditions \eqref{e:cond_n} and \eqref{e:cond_t} are NOT satisfied for the numerical example of Sec.~\ref{s:ex2}, and so violating the assumption of a one-way coupled problem.
The problem is essentially the penalty parameter in \eqref{e:cond_na}.
The larger it is picked, the worse the problem.
The only way to properly ensure one-way coupling during osseointegration, therefore, is to use the original model of Sec.~\ref{s:con4}, as it causes no chemical contact forces by construction.

\section{Projection point increment $\Delta\xi^\alpha_\mrp$}\label{s:Dxi}

The derivation of increment $\Delta\xi^\alpha_\mrp$ appearing in Eq.~\eqref{e:DNm} can for instance be found in \citet{spbc,spbf}.
It results in
\eqb{l}
\Delta\xi^\alpha_\mrp = \ds\pa{\xi^\alpha_\mrp}{\mx_\mrs^e}\Delta\mx_\mrs^e + \ds\pa{\xi^\alpha_\mrp}{\mx_\mrm^{\bar e}}\Delta\mx_\mrm^{\bar e}\,,
\label{e:Dxi0}\eqe
with
\eqb{lll}
\ds\pa{\xi^\alpha_\mrp}{\mx_\mrs^e} \is c^{\alpha\beta}_\mrp \ba^\mrp_\beta\,\mN_e \,,  \\[4mm]
\ds\pa{\xi^\alpha_\mrp}{\mx_\mrm^{\bar e}} \is -c^{\alpha\beta}_\mrp \big(\ba^\mrp_\beta\,\mN_{\bar e} - g_\mrn\,\bn_\mrp\,\mN_{\bar e,\beta}  \big) \,,  
\label{e:Dxi}\eqe
cf.~Eq.~(57)-(58) in \citet{spbf}.
It is noted that the dyadic product of the contact traction in \eqref{e:tcii} with \eqref{e:Dxi} results in
\eqb{lll}
\ds\undetc^\mathrm{(ii)}\otimes\pa{\xi^\alpha_\mrp}{\mx_\mrs^e} \is -\ds\bigg[\pa{\undetc^\mathrm{(ii)}}{\ba^\mrp_\beta}\bigg]^\mrT\mN_e\,, \\[4mm]
\ds\undetc^\mathrm{(ii)}\otimes\pa{\xi^\alpha_\mrp}{\mx_\mrm^{\bar e}} \is \ds\bigg[\pa{\undetc^\mathrm{(ii)}}{\ba^\mrp_\beta}\bigg]^\mrT\mN_{\bar e} 
	\,-\, E_n^\mathrm{eff}c^{\alpha\beta}_\mrp g_\mrn^2\bn_\mrp\otimes\bn_\mrp\,\mN_{\bar e,\beta} \,,
\label{e:tcxi}\eqe
due to (\ref{e:tiigi}.2).
This allows to show that the FE tangent matrix in Sec.~\ref{s:FEx} becomes symmetric.

\section{Energy released by bonding}\label{s:DPib}

The energy released by bonding can be obtained by integrating the specific release rate over space and time.
As noted in Sec.~\ref{s:FET} this rate is $-\mu_\mrc\,\unde{R_c}$ on the contact interface.
Its units are power per reference area.
The total energy released by bonding between time $t_1$ and $t_2$ thus is
\eqb{l}
\Delta\Pi_\mrb = -\ds\int_{\sS_{\mrc0}}\int_{t_1}^{t_2} \mu_\mrc\,\unde{R_c}\,\dif t\,\dif A\,.
\eqe
Inserting \eqref{e:bondODE0}, $\unde{M_c}=\unde{n_c}\mu_c$ and \eqref{e:McB} in case of no debonding ($\overleftarrow{\unde{K_c}}=0$) then gives
\eqb{l}
\Delta\Pi_\mrb = -\ds\int_{\sS_{\mrc0}}\int_{t_1}^{t_2} \overrightarrow{\unde{K_c}}(\phi-1)\dot\phi\,\dif t\,\dif A\,.
\label{e:DPit}\eqe
If $\overrightarrow{\unde{K_c}}$ is assumed constant, time integration can be performed analytically to give
\eqb{l}
\Delta\Pi_\mrb = -\ds\int_{\sS_{\mrc0}}\frac{\overrightarrow{\unde{K_c}}}{2}\Big[(\phi-1)^2\Big]_{t_1}^{t_2}\,\dif A\,,\qquad\big($for $\overrightarrow{\unde{K_c}}=$ const.$\big), 
\eqe
which results in expression \eqref{e:dPi3} for $t_1=0$ and $t_2=t$.
On the other hand, if $\overrightarrow{\unde{K_c}}$ changes in time, such as for the pressure-dependent bonding model of \eqref{e:Kc} and \eqref{e:fm3}, the precise energy release follows from \eqref{e:DPit}, possibly with the help of numerical integration.

\section{Newton-Raphson iteration performance}\label{s:NRconv}

Table \ref{t:NRiterations} shows the Newton-Raphson (NR) iteration error in the energy norm (NR residual times NR increment) for the multi-field examples of Sec.~\ref{s:ex}. 
The convergence is quadratic unless active contact set updates occur.
\begin{table}[h]
\centering
\small{
\begin{tabular}{|r|r|r|r|r|r|r|}
  \hline
   & 2-field prob.: & 3-field prob.: & 4-field prob.: & 4-field prob.: & 5-field prob.: & 6-field prob.: \\ 
   NR & balloon deb. & fixed sphere & falling sphere & rising sphere & heat.~sphere & interface $\unde{C_c}$ \\ 
   iteration & from Fig.~\ref{f:CoboBalloon5} & from Fig.~\ref{f:CoboT1} & from Fig.~\ref{f:CoboT3} & from Fig.~\ref{f:CoboT5} 
   	& in Fig.~\ref{f:CoboT7}a,c & in Fig.~\ref{f:CoboT8}b,d \\ \hline 
   & & & & & & \\[-4mm]   
   1 & 1.6274e--10 &  \!\!\underline{~~1.2359e--09$_{~}$}\!\! & \!\!\underline{~~4.7054e--09$_{~}$}\!\! & 1.1246e--07$_{~}$\!\! 
   	& \!\!\underline{~~5.3101e--09$_{~}$}\!\! & 1.1064e--08$_{~}$\!\! \\ 
   2 & 2.1994e--14 &  2.6779e--13$_{~}$\!\! & \!\!\underline{~~1.7775e--10$_{~}$}\!\! & \!\!\underline{~~3.2055e--10$_{~}$}\!\! 
   	& \!\!\underline{~~6.0256e--11$_{~}$}\!\! & \!\!\underline{~~1.5831e--10$_{~}$}\!\! \\  
   3 & 1.5090e--21 &  1.2241e--21$_{~}$\!\! & 1.5922e--13$_{~}$\!\! & \!\!\underline{~~1.2993e--10$_{~}$}\!\! & 8.8186e--14$_{~}$\!\! 
   	& \!\!\underline{~~1.6501e--11$_{~}$}\!\! \\
   4 & 1.9297e--32 &  9.9541e--33$_{~}$\!\! & 2.6620e--20$_{~}$\!\! & 4.1077e--13$_{~}$\!\! & 3.0353e--21 & 5.8371e--14$_{~}$\!\! \\ 
   5 &  &  & 1.3574e--32$_{~}$\!\! & 5.7108e--19$_{~}$\!\! & 1.0332e--32$_{~}$\!\! & 5.9416e--20$_{~}$\!\! \\
   6 &  &  &  & 4.3053e--32$_{~}$\!\! & & 1.2685e--32$_{~}$\!\! \\    
   \hline
\end{tabular}}
\vspace{-1mm}
\caption{Newton-Raphson iteration error in the energy norm for various examples.
The convergence rate is quadratic unless the active contact set updates, which is indicated by the horizontal lines.
Such updates are only allowed when the energy norm drops below 1e--08.
Reported are time step $t= 0.75T$ (balloon example with $m=16$) and $t=0.3125t_0$ (sphere examples with $m=32$), which are typical time steps.
The time step size in the sphere examples here is four times larger than the one used in Secs.~\ref{s:ex3a}-\ref{s:ex3d}.}
\label{t:NRiterations}
\end{table}

\bibliographystyle{apalike}
\bibliography{bibliography}

\end{document}